\documentclass[prb,twocolumn,showpacs,amsmath,amssymb,superscriptaddress]{revtex4-2}

\numberwithin{equation}{section}
\renewcommand\theequation{\arabic{section}.\arabic{equation}}

\usepackage{graphicx}
\usepackage{color}
\usepackage{braket}
\usepackage{amsmath}
\usepackage{amsfonts}

\usepackage{amsmath,amssymb,mathrsfs}
\usepackage{bbm}
\usepackage[dvipsnames]{xcolor}

\DeclareMathOperator{\Sign}{sgn}

\definecolor{amber}{rgb}{1.0, 0.75, 0.0}

\date{\today}                  % Activate to display a given date or no date

\begin{document}

\title{Universality of Abelian and non-Abelian Wannier functions in one dimension}

\author{Kiryl Piasotski}
\affiliation{Institut f\"ur Theorie der Statistischen Physik, RWTH Aachen, 
52056 Aachen, Germany and JARA - Fundamentals of Future Information Technology}

\author{Mikhail Pletyukhov}
\affiliation{Institut f\"ur Theorie der Statistischen Physik, RWTH Aachen, 
52056 Aachen, Germany and JARA - Fundamentals of Future Information Technology}

\author{Clara S. Weber}
\affiliation{Institut f\"ur Theorie der Statistischen Physik, RWTH Aachen, 
52056 Aachen, Germany and JARA - Fundamentals of Future Information Technology}

\author{Jelena Klinovaja}
\affiliation{Department of Physics, University of Basel, Klingelbergstrasse 82, 
CH-4056 Basel, Switzerland}

\author{Dante M. Kennes}
\email[Email: ]{dante.kennes@mpsd.mpg.de}
\affiliation{Institut f\"ur Theorie der Statistischen Physik, RWTH Aachen, 
52056 Aachen, Germany and JARA - Fundamentals of Future Information Technology}
\affiliation{Max Planck Institute for the Structure and Dynamics of Matter, Center for Free Electron Laser Science (CFEL), Luruper Chaussee 149, 22761 Hamburg, Germany}

\author{Herbert Schoeller}
%\email[Email: ]{schoeller@physik.rwth-aachen.de}
\affiliation{Institut f\"ur Theorie der Statistischen Physik, RWTH Aachen, 
52056 Aachen, Germany and JARA - Fundamentals of Future Information Technology}

\begin{abstract}
Within a Dirac model in $1+1$ dimensions, a prototypical model to describe low-energy physics for a wide class of lattice models, we propose a field-theoretical version for the representation of Wannier functions, the Zak-Berry connection, and the geometric tensor. In two natural Abelian gauges we present universal scaling of the Dirac Wannier functions in terms of four fundamental scaling functions that depend only on the phase $\gamma$ of the gap parameter and the charge correlation length $\xi$ in an insulator. The two gauges allow for a universal low-energy formulation of the surface charge and surface fluctuation theorem, relating the boundary charge and its fluctuations to bulk properties. Our analysis describes the universal aspects of Wannier functions for the wide class of one-dimensional generalized Aubry-Andr\'e-Harper lattice models. In the low-energy regime of small gaps we demonstrate universal scaling of all lattice Wannier functions and their moments in the corresponding Abelian gauges. In particular, for the quadratic spread of the lattice Wannier function,  we find the universal result $Za\xi/8$, where $Za$ is the length of the unit cell. This result solves a long-standing problem providing further evidence that an insulator is only characterized by the two fundamental length scales $Za$ and $\xi$. Finally, we discuss also non-Abelian lattice gauges and find  that lattice Wannier functions of maximal localization show universal scaling and are uniquely related to the Dirac Wannier function of the lower band. In addition, via the winding number of the determinant of the non-Abelian transformation, we establish a bulk-boundary correspondence for the number of edge states up to the bottom of a certain band, which requires no symmetry constraints. Our results present evidence that universal aspects of Wannier functions and of the boundary charge are uniquely related and can be elegantly described within universal low-energy theories.    
\end{abstract}

% \pacs{05.10.Cc, 05.30.-d, 05.30.Jp, 73.23.-b}
% \keywords{}

\maketitle

\section{Introduction}

Wannier functions \cite{wannier_pr_37} have matured to an invaluable tool in various fields of solid state physics. They provide a useful basis set of exponentially localized single-particle states \cite{kohn_pr_59} integral to  density functional theory \cite{marzari_etal_rmp_12}, are used in the definition of tight-binding models with short-ranged hoppings, can characterize the topological properties of crystals \cite{bradlyn_etal_nature_17,po_etal_natcom_17,po_etal_prl_18}, and are the basis for the modern theory of polarization and localization \cite{resta_rmp_94,sgiarovello_etal_prb_01,vanderbilt_book_18}. However, Wannier functions are gauge dependent, begging the question: which gauges are particularly useful? One of them has been identified as the gauge of maximally localized Wannier functions \cite{marzari_vanderbilt_prb_97}, which we call the ML gauge in the following. In this gauge the quadratic spread $\langle \Delta x^2\rangle^{1/2}$ of the Wannier function is related to the fluctuations of the bulk polarization and to the quasimomentum integral of the geometric tensor \cite{marzari_vanderbilt_prb_97,souza_etal_prb_00} quantifying important topological hallmarks of the system. Furthermore, the first moment of the Wannier function (its average) can be related to the Zak-Berry phase or the bulk polarization \cite{vanderbilt_book_18}, a quantity widely used to determine Chern numbers in two-dimensional topological systems.

Complementary to the bulk polarization the macroscopic boundary (or surface) charge $Q_B$ for a half-infinite insulating system has been studied extensively in a series of recent works  \cite{park_etal_prb_16,thakurathi_etal_prb_18,pletyukhov_etal_prbr_20,pletyukhov_etal_prb_20,lin_etal_prb_20,pletyukhov_etal_prr_20,weber_etal_prl_20}. Interestingly, the boundary charge \cite{vanderbilt_kingsmith_prb_93} and its fluctuations \cite{weber_etal_prl_20} can be related to the bulk polarization and its fluctuations by the so-called {\it surface charge} and {\it surface fluctuation} theorem. The surface charge theorem has first been discussed in Ref.~[\onlinecite{vanderbilt_kingsmith_prb_93}] relating $Q_B$ to the Zak-Berry phase of all occupied bands, up to an unknown integer, for  noninteracting, nondisordered systems. Recently, this theorem was further analysed in Ref.~[\onlinecite{pletyukhov_etal_prb_20}] for the wide class of generalized Aubry-Andr\'e-Harper models \cite{harper_pps_55,aubry_andre_aips_80,ganeshan_etal_prl_13,lahini_etal_prl_09,kraus_etal_prl_12,degottardi_etal_prl_13,schreiber_etal_science_15}. Here even the unknown integer of the surface charge theorem was determined fixing the precise gauge in the Wannier functions needed to relate its first moment to the Zak-Berry phase. In the small gap limit a low-energy Dirac model in $1+1$ dimensions  \cite{gangadharaiah_etal_prl_12,thakurathi_etal_prb_18,pletyukhov_etal_prr_20}  can be used to reveal a universal form of the boundary charge in terms of the phase of the gap parameter \cite{pletyukhov_etal_prr_20}. Complementing the surface charge theorem, recently the surface fluctuation theorem \cite{weber_etal_prl_20} was established as a relationship between the boundary charge fluctuations and the second moment of the density-density correlation function or, equivalently, the fluctuations of the bulk polarization.

These recent developments raise the important question whether Wannier functions themselves exhibit universal properties beyond their first two moments, in particularly fundamental gauges. To address this question we cover the wide class of generalized Aubry-Andr\'e-Harper models. We do so by considering a $2$-band Dirac model, which requires a field-theoretical generalization of Wannier functions, the Zak-Berry connection and the geometric tensor, which we provide. Any of such field theories must always be complemented by appropriate asymptotic conditions at high-energies to provide a well-defined relation to the lattice models falling into the universality class of the respective field theory. In addition to the above mentioned ML gauge, we define the asymptotically free (AF) gauge, which is fixed by requiring that the eigenfunctions turn into conventional plane waves at large momenta. Here, we show that, for a microscopic lattice model, this gauge corresponds precisely to the one used in Ref.~[\onlinecite{pletyukhov_etal_prb_20}] to fix the unknown integer in the surface charge theorem. As a result, the AF gauge is useful to formulate  a universal field-theoretical version of the surface charge theorem, relating the boundary charge to the first moment of the Dirac Wannier function and to the phase $\gamma$ of the gap parameter, which in turn is related to the field-theoretical version of the Zak-Berry phase. In contrast, we show that the  fundamental ML gauge is useful for the universal formulation of the field theoretical version of the surface fluctuation theorem, relating the boundary charge fluctuations to the second moment of the Wannier function or the momentum integral over the field-theoretical version of the geometric tensor. 
\begin{figure*}[t!]
    \centering
	 \includegraphics[width =0.95\columnwidth]{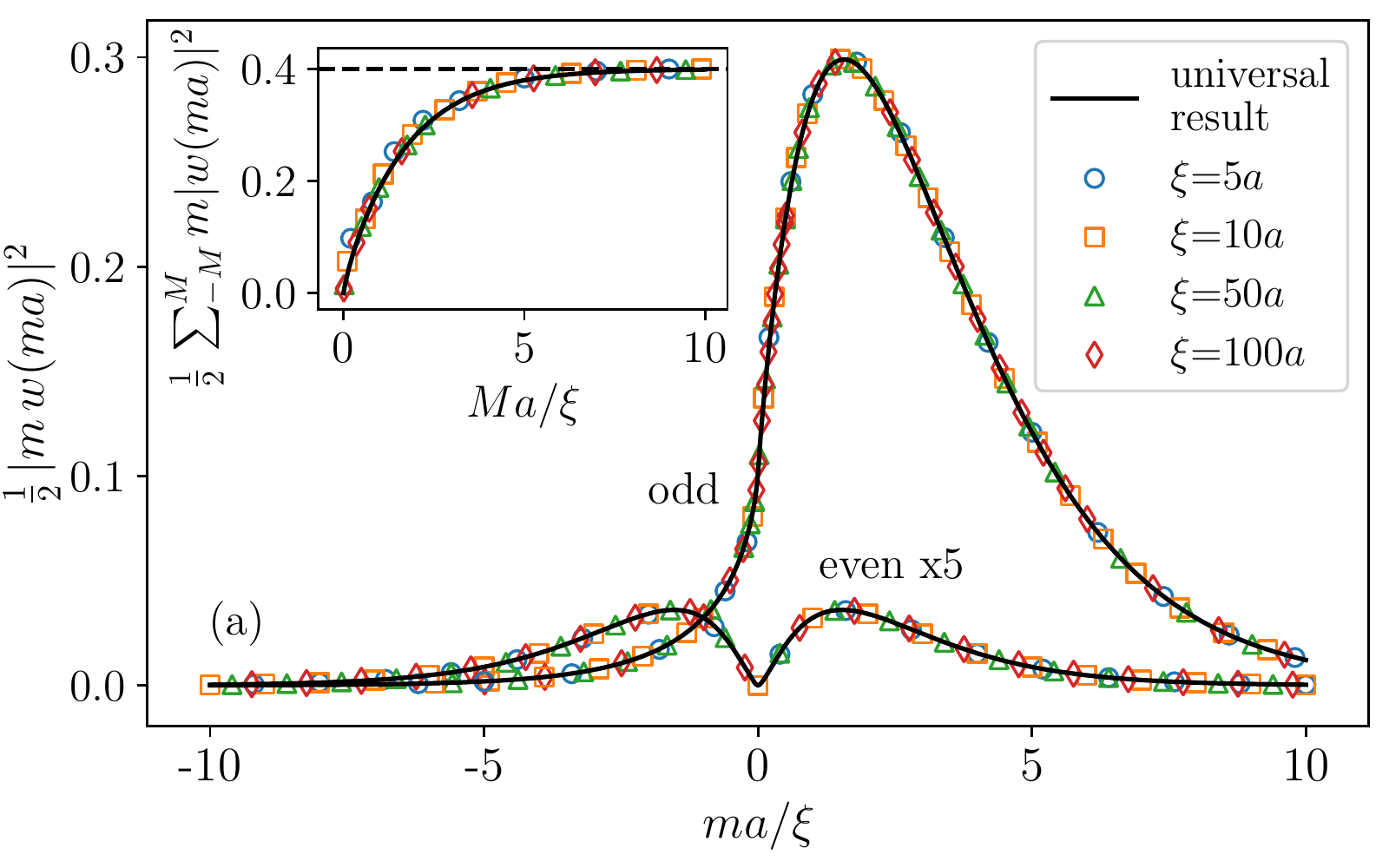}
\hfill	 \includegraphics[width =0.95\columnwidth]{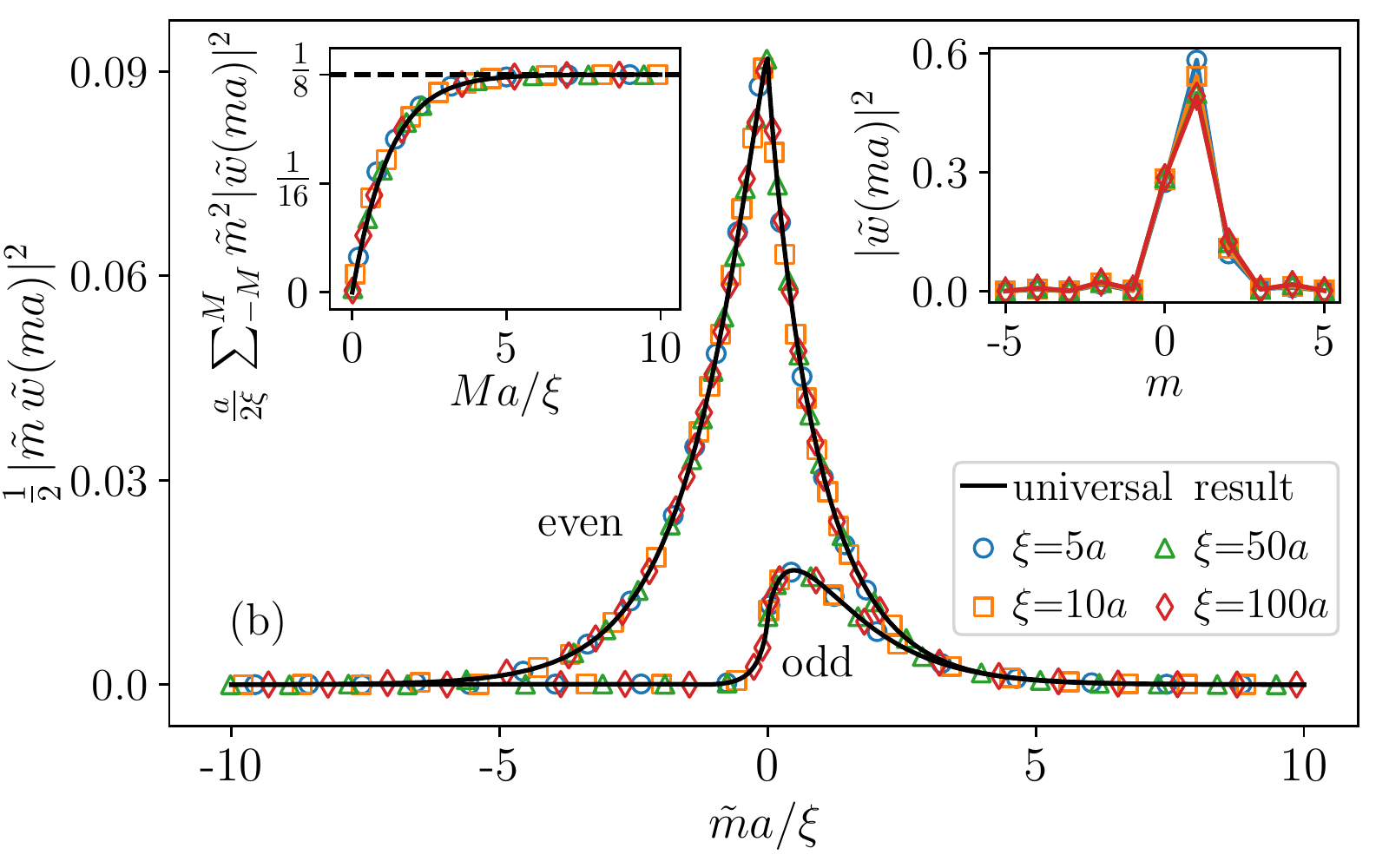}
	  \caption{Universal scaling of the lattice Wannier functions $w(ma)$ and $\tilde{w}(ma)$ for the lower band $\alpha=1$ of the Rice-Mele model in (a) AF gauge and (b) ML gauge, respectively, for different correlation lengths $\xi=ta/\Delta$ determined by the gap size $2\Delta$ and the average hopping $t$. The gap parameter is given by  $\Delta e^{i\gamma}=-\delta v - 2i\delta t$, where $v_{1,2}=\pm\delta v$ and $t_{1,2}= t\pm\delta t$ are the alternating on-site energies and hopping amplitudes, respectively. The phase of the gap parameter is chosen as $\gamma=0.2\,\pi$. Main panels: $\frac{1}{2}|m\,w(ma)|^2$ (AF gauge) and $\frac{1}{2}|\tilde{m}\,\tilde{w}(ma)|^2$ (ML gauge; with $\tilde{m}=m-\gamma_{1}/\pi$ and $\gamma_{1}=-\gamma + \pi\, \Sign\gamma=0.8\pi$ denoting the Zak-Berry phase) as function of $ma/\xi$ and $\tilde{m}a/\xi$ for different $\xi=5a,10a,50a,100a$. All discrete lattice points fall on top of the fundamental universal scaling functions of the Dirac model, $|F_{A/B}(ma/\xi)|^2$ (AF gauge) and $|\tilde{F}_{A/B}(\tilde{m}a/\xi)|^2$ (ML gauge), for $m$ odd ($F_A$ and $\tilde{F}_A$) and $m$ even ($F_B$ and $\tilde{F}_B$) (black solid lines; the even points in the AF gauge are multiplied by a factor of $5$ for visibility). Left inset of (a): The scaling of the first moment $C_1(Ma)/(2a)=\frac{1}{2}\sum_{m=-M}^M m |w(ma)|^2$ as function of $Ma/\xi$ for different $\xi=5a,10a,50a,100a$. All lattice points fall on top of a universal scaling function $\frac{1}{2}\int_{-Ma/\xi}^{Ma/\xi}dy\big[F_A(y)^2 + F_B(y)^2\big]/y$ (black solid line). They converge smoothly to the value $\gamma_{1}/(2\pi)=C_1/(2a)=0.4$. Left inset of (b): Scaling of the quadratic spread $\langle\Delta x^2\rangle(Ma)/(2a\xi)=\frac{a}{2\xi}\sum_{m=-M}^M \tilde{m}^2 |\tilde{w}(ma)|^2$ as function of $Ma/\xi$ for different $\xi=5a,10a,50a,100a$. All lattice points fall on top of the universal result $\int_0^{Ma/\xi}dy\big[ \tilde{F}_A(y)^2 + \tilde{F}_B(y)^2\big]$ (black solid line) and converge smoothly to the universal value $1/8$. Right inset of (b):  $|\tilde{w}(ma)|^2$ for $\xi=5a,10a,50a,100a$ shown as function of $m$, giving rise to the misleading visual impression of a localized state with $\xi$-independent spread of order $a$. A similar form appears for $|w(ma)|^2$ (not shown). 
	  } 
    \label{fig:scaling_Z_2}
\end{figure*}

Returning to the question of universality of Wannier functions beyond the first and second moment we find that strikingly the entire line shape of Wannier functions exhibits universal behavior in the small gap limit. Surprisingly, we show that for the whole class of generalized Aubry-Andr\'e-Harper models all lattice Wannier functions display universal scaling on the length scale $\xi$, which is the fundamental length scale of insulators on which charge correlations decay exponentially. $\xi$ is proportional to the inverse gap and is defined here  by the exponential decay length of the square of the Wannier functions. We show that all scaling functions can be related to four fundamental functions of the Dirac model which are fully characterized by the phase $\gamma$ of the gap parameter. 

In Figs.~\ref{fig:scaling_Z_2}(a,b) we illustrate these four fundamental scaling functions, denoted by ${F}_{A,B}$ and $\tilde{F}_{A,B}$ in the following, for the lower band Wannier functions in the AF and ML gauge of the Rice-Mele model, a fundamental one-dimensional lattice model with a $2$-site unit cell put forward in the context of topological insulators \cite{rice_mele_prl_82}. In Fig.~\ref{fig:scaling_Z_2}(a) we show that for the AF gauge, all Wannier functions multiplied with the lattice site index $m$ collapse upon scaling $m\to ma/\xi$ for different correlation lengths $\xi$ (here, $a$ denotes the lattice constant). For the ML gauge, as shown in Fig.~\ref{fig:scaling_Z_2}(b), the same applies but it turns out to be important to include a shift of the lattice site index $m$ by the first moment of the Wannier function. Surprisingly, the universal scaling holds even outside the strict range of low-energy theories, i.e., it persists for rather large gaps (where $\xi\sim O(a)$) and also for rather small length scales $|ma|\sim O(a)$. The inset of Fig.~\ref{fig:scaling_Z_2}(a) and the left inset of Fig.~\ref{fig:scaling_Z_2}(b) show the universality of the surface charge and surface fluctuation theorem discussed above, respectively, which follow from the universality of the Wannier functions.  We find that the first moment converges to the universal result $C_1/(Za)=\gamma_{1}/(2\pi)$, where $Za$ is the length of the unit cell and $\gamma_{1}$ is the Zak-Berry phase of the lower band, which is related to the phase of the gap parameter and to the boundary charge. For the quadratic spread in the ML gauge we find the universal result $\langle \Delta x^2\rangle/(Za)=\xi/8$, which is in full agreement with the universal result for the boundary charge fluctuations \cite{weber_etal_prl_20}. Interestingly, we obtain the scaling $\left \langle \Delta x^2\right\rangle/(Za)\sim O(\xi)$ in any gauge and for any band for arbitrary unit cell length $Za$. This shows that the spread is not an independent length scale but is universally linked to the exponential decay length $\xi$. In the small gap limit this solves a long-standing problem since in the previous literature only the inequality $\langle \Delta x^2\rangle \lesssim Za\,\xi$ has been stated \cite{souza_etal_prb_00,sgiarovello_etal_prb_01}, leaving it open whether the so-called localization length $\lambda=\langle \Delta x^2\rangle^{1/2}$ defines an independent length scale \cite{resta_sorella_prl_99}. 

Our analysis shows that universal aspects of Wannier functions and the boundary charge are ultimately related to generic effects arising at band anticrossing points with a small gap defining a corresponding universal length scale clearly separated from the lattice spacing. The reason why universal scaling was easily overlooked before is at least two-fold: On the one hand density functional theory calculations for specific materials are often applied to systems where the lattice spacing $a$ and the exponential decay length $\xi$ are not clearly separated. On the other hand, even in cases of clearly separated $\xi\gg a$, the visual impression of the square $|w(ma)|^2$ of the Wannier function does not reveal any universal scaling. As shown in the right inset of Fig.~\ref{fig:scaling_Z_2}(b), the visual impression is that of a rather boring Lorentzian with broadening $\sim a$. This form is only important for the correct normalization of the Wannier function but it does not reveal any universal scaling on length scale $\xi$. Universal scaling is only visible when multiplying the Wannier function by the spacial coordinate and plotting the square $|m\,w(ma)|^2$, such that the asymptotic $(1/m)^2$-behavior of the Lorentzian is cancelled. As a consequence, it is roughly the product of a Lorentzian on scale $a$ and a universal scaling function on scale $\xi$ which characterizes the subtle and universal line shape of Wannier functions revealed in the present work. 

Furthermore, we show that the fully universal scaling exemplified for the Rice-Mele model persists for the Wannier functions of all bands for any size $Z$ of the unit cell. It turns out that the lattice Wannier functions of a given band with two gaps at the bottom and top of the band show universal scaling in terms of two different correlation length scales referring to these two energetically adjacent gaps. Additionally, we show that each Wannier function can be naturally split in two additive contributions, one corresponding to the upper and one to the lower half of the band, each of them scaling only with a single length scale. Moreover, we also discuss a non-Abelian gauge of maximal localization (NA-ML) mixing the Wannier functions of all bands including and below a given one, which has been proposed in Ref.~\onlinecite{marzari_vanderbilt_prb_97} to maximally localize the total sum of the quadratic spreads of a certain set of Wannier functions. Interestingly, we find that all Wannier functions in the NA-ML gauge reveal universal scaling on the same length scale $\xi$ referring to the band gap between that given band and the next. Additionally, these Wannier functions can be uniquely related to the lower band Wannier function of the Dirac model. Furthermore, we will present an explicit construction of the NA-ML gauge in terms of the Wilson propagator and show that the winding number associated with the unitary non-Abelian gauge transformation allows for a formulation of a bulk-boundary correspondence (BBC) determining the number of edge states up to the  band chosen to define the NA-ML gauge via bulk properties. This form of the BBC is of particular interest since it is not restricted to zero-energy edge states or to particular symmetry properties of the system.   

Our work is organized as follows. In Section~\ref{sec:model} we start with the description of the lattice models under consideration and the relation of their eigenstates to the eigenstates of the Dirac model. The low-energy theory of the Zak-Berry connection, the Zak-Berry phase, and the geometric tensor, together with their relation to the corresponding lattice quantities is the subject of Section~\ref{sec:zak_boundary}. In this section we will also define the AF and ML gauge and formulate the low-energy version of the surface charge and surface fluctuation theorem in terms of the low-energy version of the Zak-Berry phase and the momentum integral of the geometric tensor, respectively. The definition of Wannier functions in Dirac theory and their precise relation to the lattice Wannier functions will be analysed in Section~\ref{sec:wannier}. Furthermore, we will describe in this section the relation of all moments of the Wannier functions in Dirac and lattice theory and explain how the surface charge and surface fluctuation theorem can be formulated in terms of the moments of the Dirac Wannier functions. The universal scaling of the Dirac and lattice Wannier functions is the central subject of Section~\ref{sec:wannier_universality}. We will calculate the Dirac Wannier functions together with their moments analytically in the AF and ML gauge, and state the definitions and properties of the fundamental scaling functions. The universal scaling of all lattice Wannier functions and their moments is then stated via their relation to the Dirac Wannier functions, and is demonstrated for two explicit examples $Z=2$ and $Z=3$. Finally, in Section~\ref{sec:nonabelian} we construct explicitly the NA-ML gauge and demonstrate that all non-Abelian Wannier functions show universal scaling corresponding to the Dirac theory of the highest gap defining the set of mixed bands. In addition, we set up an interesting bulk-boundary correspondence relating the winding number of the determinant of the non-Abelian gauge transformation to the number of edge states present in all gaps up to a certain one. We close with a summary and outlook in Section~\ref{sec:summary}. Some more involved technical details are presented in a series of Appendices.  

We use units $e=\hbar=1$.

\section{The model and eigenstates}
\label{sec:model}

In this section we will state the class of lattice models under consideration together with their relation to the Dirac field theory. In particular we will show the representation of the Bloch eigenstates in lattice theory and the eigenfunctions of the Dirac Hamiltonian, together with their precise relationship. In summary it will turn out that the Dirac field theory is an elegant way for a formulation of degenerate perturbation theory in terms of slowly varying right and left movers, applicable to lattice models with arbitrary size of the unit cell.

\subsection{Lattice model}
\label{sec:lattice_model}

We consider the following class of so-called generalized Aubry-Andr\'e-Harper lattice models in one dimension, characterized by nearest-neighbor hopping and a single orbital per site, together with any periodic on-site and hopping modulation
\begin{align}
    \label{eq:H}
    H &= H_0 + H'\,,\\
    \label{eq:H_0}
    H_0 &= -t \sum_m (|m+1\rangle\langle m| + {\rm h.c.}) \,,\\
    \label{eq:H'}
    H' &= \sum_m \delta v_m |m\rangle\langle m| - \sum_m \delta t_m(|m+1\rangle\langle m| + {\rm h.c.})\,.
\end{align}
Here, $m$ is the lattice site index, $t>0$, and all $\delta v_m=\delta v_{m+Z}$, $\delta t_m=\delta t_{m+Z}$ are periodic and real (possible phases of the hoppings can always be gauged away in one dimension). The number of sites in a unit cell is denoted by $Z$. We relate $m=Z(n-1)+j$ to the index $n$ labelling the unit cells and the index $j=1,\dots,Z$ labelling the sites within a unit cell.

\begin{figure}[t!]
    \centering
	 	 \includegraphics[width =0.9\columnwidth]{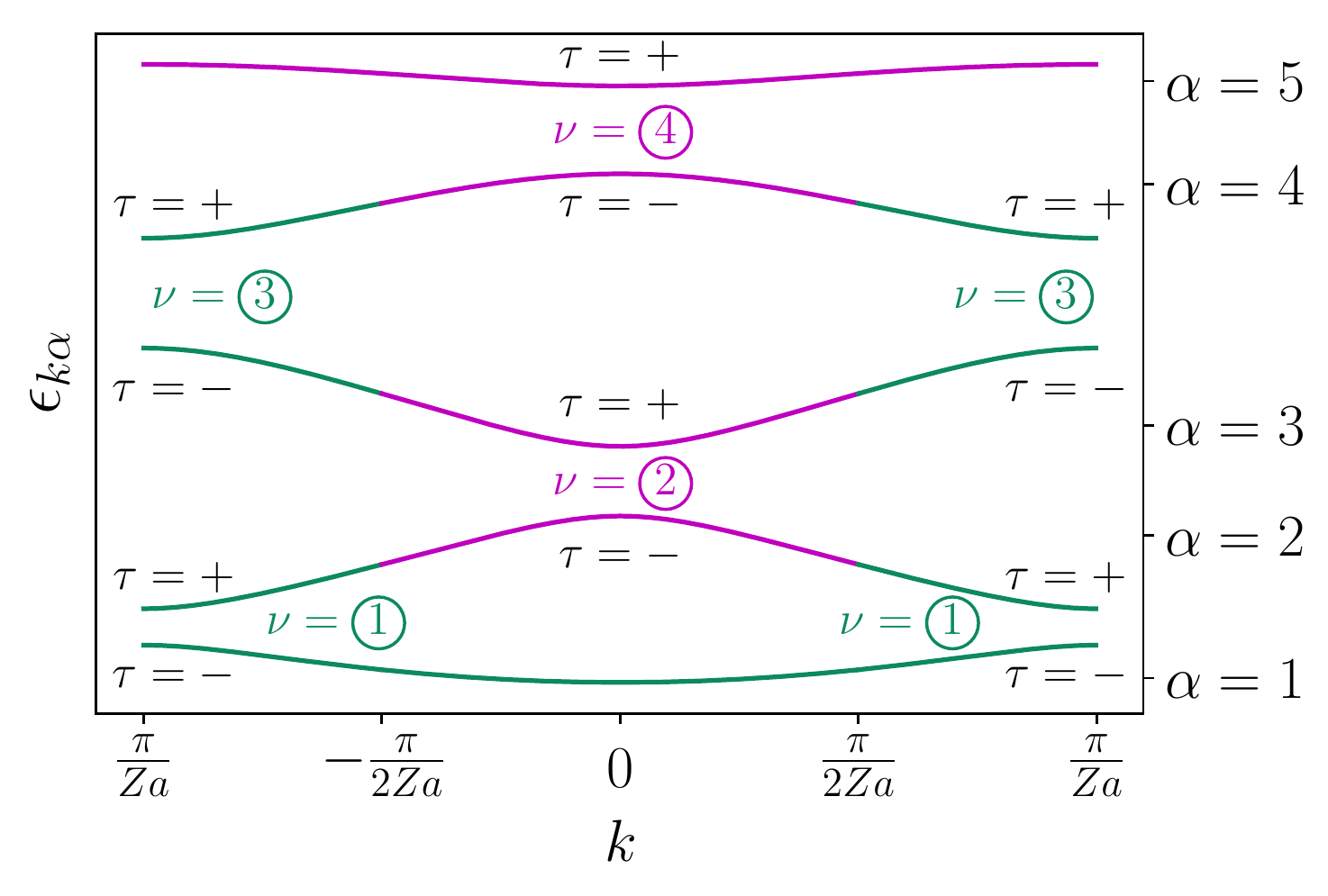}
	  \caption{Sketch of the band structure for $Z=5$. We find $Z-1=4$ gaps labelled by $\nu=1,\dots,4$ opening up at $k=0$ for $\nu=2,4$ even, and at $k=\pm\frac{\pi}{Za}$ for $\nu=1,3$ odd. All bands labelled by $\alpha=1,\dots,5$ are then naturally split in two halves, the upper and lower halves labelled by $(\nu,\tau)=(\alpha,-)$ and $(\nu,\tau)=(\alpha-1,+)$, respectively. An exception are the lowest $\alpha=1$ and highest $\alpha=5$ band. They are not split in two halves and labelled by $(\nu,\tau)=(1,-)$ and $(\nu,\tau)=(4,+)$. The Dirac theory for gap $\nu$ is a good description for all eigenstates referring to a pair of band parts labelled by $(\nu,\pm)$ (indicated by violet and green color for $\nu$ even or odd, respectively). The momentum $q$ in Dirac theory is the difference of $k$ to the point where the gap opens, i.e., $k=q$ for $\nu$ even, and $k=\pm\frac{\pi}{Za}+q$ for $\nu$ odd and $\Sign{q}=\mp$. The cutoff for $|q|$ is given by $\Lambda_\alpha=\frac{\pi}{2Za}$ for $\alpha=2,3,4$ and by $\Lambda_\alpha=\frac{\pi}{Za}$ for $\alpha=1,5$.
	  } 
    \label{fig:band_structure_1}
\end{figure}

We assume the condition
\begin{align}
    \label{eq:condition}
    |\delta v_m|, \ |\delta t_m| \ll t \,,
\end{align}
such that $H'$ can be considered as a small perturbation. Due to the presence of $H'$ there will be $Z$ bands labelled by the band index $\alpha = 1,\dots,Z$ (from bottom to top), with $Z-1$ gaps in between labelled by the gap index $\nu=1,\dots,Z-1$, see the sketch of the band structure shown in Fig.~\ref{fig:band_structure_1}. 

In standard solid state notation of the reduced zone scheme, the Bloch eigenfunctions on the lattice are written as
\begin{align}
    \label{eq:bloch}
    \psi_{k\alpha}(ma) = \sqrt{\frac{Za}{2\pi}} \,u_{k\alpha}(ma) e^{ikma}\,,
\end{align}
where $\alpha=1,\dots,Z$ is the band index (labelled from bottom to top), $ma$ denotes the lattice space position and $a$ is the lattice spacing. The length $Za$ defines the size of the unit cell. The quasimomentum $k$ is defined within the first Brillouin zone $-\frac{\pi}{Za}<k<\frac{\pi}{Za}$ and 
\begin{align}
    \label{eq:u_x_periodicity}
    u_{k\alpha}(ma) = u_{k\alpha}(ma+Za) 
\end{align}
is periodic with the unit cell size. We assume that the gauge is always chosen such that the total Bloch function is periodic in $k$ 
\begin{align}
    \label{eq:k_periodicity}
    \psi_{k\alpha}(ma) = \psi_{k+\frac{2\pi}{Za},\alpha}(ma)\,.
\end{align}
Note that this means that $u_{k\alpha}(ma)$ is not periodic in $k$ but fulfils the condition 
\begin{align}
    \label{eq:u_k_periodicity}
    u_{k+\frac{2\pi}{Za},\alpha}(ma) = u_{k\alpha}(ma) e^{-i\frac{2\pi}{Z}m}\,.
\end{align}
At fixed $k$ in the first Brillouin zone, the states $u_{k\alpha}(ja)=\langle j|u_{k\alpha}\rangle$ in unit cell space $j=1,\dots,Z$ are eigenfunctions of a Hamiltonian $h_k$ defined on the periodic continuation of the unit cell 
\begin{align}
    \label{eq:h_k_eigenstates}
    h_k |u_{k\alpha}\rangle &= \epsilon_{k\alpha} |u_{k\alpha}\rangle \,,\\
    \nonumber
    h_k &= \sum_{j=1}^Z \delta v_j |j\rangle\langle j| \\
    \label{eq:h_k_def}
    & - \sum_{j=1}^Z t_j\left(|j+1\rangle\langle j| e^{-ika} + |j\rangle\langle j+1| e^{ika} \right) \,,
\end{align}
where $t_j=t+\delta t_j$, and we identify $|Z+1\rangle\equiv |1\rangle$. In contrast to $h_k$ and $|u_{k\alpha}\rangle$, the dispersion relation is periodic in $k$: $\epsilon_{k+\frac{2\pi}{Za},\alpha}=\epsilon_{k\alpha}$.

We note the following normalization and completeness relations
\begin{align}
    \label{eq:psi_normalization}
    &\sum_m \psi_{k\alpha}(ma)^*\psi_{k'\alpha'}(ma) = \delta_{\alpha\alpha'}\delta(k-k')\,,\\
    \label{eq:psi_completeness}
    &\sum_\alpha\int_{-\pi/Za}^{\pi/Za}dk\,\psi_{k\alpha}(ma) \psi_{k\alpha}(m'a)^* = \delta_{mm'}\,.
\end{align}
This corresponds to the following relations for the periodic Bloch part $u_{k\alpha}(ja)$ at fixed quasimomentum $k$ and $j=1,\dots,Z$
\begin{align}
    \label{eq:u_normalization}
    &\sum_{j=1}^Z u_{k\alpha}(ja)^* u_{k\alpha'}(ja) = \delta_{\alpha\alpha'}\,,\\
    \label{eq:u_completeness}
    &\sum_\alpha u_{k\alpha}(ja) u_{k\alpha}(j'a)^* = \delta_{jj'}\,.
\end{align}
Introducing the following scalar product in unit cell space
\begin{align}
    \label{eq:scalar_lattice}
    \langle u_{k\alpha}|u_{k\alpha'}\rangle = \sum_{j=1}^Z u_{k\alpha}(ja)^* u_{k\alpha'}(ja) \,,
\end{align}
we can write (\ref{eq:u_normalization}) and (\ref{eq:u_completeness}) in compact form as
\begin{align}
    \label{eq:u_normalization_sp}
    &\langle u_{k\alpha}|u_{k\alpha'}\rangle = \delta_{\alpha\alpha'} \,,\\
    \label{eq:u_completeness_sp}
    &\sum_\alpha |u_{k\alpha}\rangle\langle u_{k\alpha}| = \mathbbm{1}_Z\,.
\end{align}

\subsection{Dirac model}
\label{sec:dirac_model}

For a small perturbation $H'$ compared to the typical band width set by the average hopping $t$, the states in the middle of the bands are nearly untouched and only close to the gaps the eigenstates of $H_0$ are coupled significantly by $H'$. We denote the single-particle eigenstates of $H_0$ in the extended zone scheme as 
\begin{align}
    \label{eq:eigenstates_H0}
    \langle m |k\rangle = \sqrt{\frac{a}{2\pi}}\,e^{ikma}\,,
\end{align}
with energy $\epsilon_k=-2t\cos(ka)$, and $-\pi/a < k < \pi/a$. Since $H'$ is periodic with the size $Za$ of the unit cell, the gap openings will happen in general via higher-order processes close to the two Fermi points $\pm k_F^{(\nu)}$ where
\begin{align}
    \label{eq:kF_def}
    k_F^{(\nu)} = \frac{\pi\nu}{Za}\quad,\quad \nu=1,\dots,Z-1 \,.
\end{align}
As outlined in detail in Refs.~\onlinecite{gangadharaiah_etal_prl_12,thakurathi_etal_prb_18,pletyukhov_etal_prr_20} the coupling of the low-energy states close to gap $\nu$ can be described via Brillouin-Wigner perturbation theory by the effective Hamiltonian 
\begin{align}
    \label{eq:H'_eff}
    H^{(\nu)}_{\text{eff}} = P^{(\nu)}\Bigg{(}H+H'Q^{(\nu)}{\frac{1}{\epsilon^{(\nu)}_F-Q^{(\nu)} H Q^{(\nu)}}}Q^{(\nu)} H'\Bigg{)}P^{(\nu)}\,,
\end{align}
where $P^{(\nu)}$ projects on the low-energy space, $Q^{(\nu)}=1-P^{(\nu)}$, and $\epsilon_F^{(\nu)}=-2t\cos(k_F^{(\nu)}a)$ is the Fermi energy where the gap opens. This coupling leads to a complex gap parameter $\Delta^{(\nu)} e^{i\gamma^{(\nu)}}$ defined by
\begin{align}
    \label{eq:H'_matrix_element}
    \langle k_F^{(\nu)} + k| H^{(\nu)}_{\text{eff}} |-k_F^{(\nu)} + k'\rangle \approx \Delta^{(\nu)} e^{i\gamma^{(\nu)}} \delta(k-k')\,.
\end{align}

In standard convention of low-energy theories it is very useful to split off the strongly oscillating parts $e^{\pm i k_F^{(\nu)}ma}$ of the eigenfunctions (\ref{eq:bloch}) and parametrize the Bloch states $\psi_{k\alpha}(ma)$ in terms of slowly varying right and left movers $\psi_{q\tau p}^{(\nu)}(x)$ in continuum notation as
\begin{align}
    \label{eq:dirac_lattice}
    \psi_{k\alpha}(ma) &= \sqrt{a}\, \psi_{q\tau}^{(\nu)}(ma) \,,\\
    \label{eq:psi_dirac}
    \psi_{q\tau}^{(\nu)}(x) &= \sum_{p=\pm} \psi_{q\tau p}^{(\nu)}(x) e^{ip k_F^{(\nu)}x}\,,\\
    \label{eq:psi_p_dirac}
    \psi_{q\tau p}^{(\nu)}(x) &= \chi_{q\tau p}^{(\nu)}\frac{1}{\sqrt{2\pi}} e^{iqx}\,.
\end{align}
Here, $p=\pm$ is the index for right/left movers, and $\tau=\pm$ is the index describing the bands above/below the gap $\nu$, see Fig.~\ref{fig:band_structure_1}
\begin{align}
    \label{eq:band_relation}
    \tau = \begin{cases} + & {\rm for} \quad \alpha=\nu+1 \\ - & {\rm for} \quad  \alpha=\nu \end{cases}  \,.
\end{align}
Thereby, the band part labelled by $(\nu,\tau)$ corresponds to the lower (upper) half of band $\alpha=\nu+1$ ($\alpha=\nu$) for $\tau=\pm$, see Fig.~\ref{fig:band_structure_1}. An exception are the lowest $\alpha=1$ and highest $\alpha=Z$ band which are not split in two halves and labelled by $(\nu,\tau)=(1,-)$ or $(\nu,\tau)=(Z-1,+)$, respectively.   

The momentum $q$ in Eq.~(\ref{eq:dirac_lattice}) corresponds to the difference of the quasimomentum $k$ in lattice theory to the quasimomentum defining the band bottom or band top at which the gap appears. This means that for even $\nu$ one expands around $k=0$ such that $q\equiv k$, and for odd $\nu$ one expands around $k=\pm \frac{\pi}{Za}$ such that $q\equiv k\mp\frac{\pi}{Za}$, see Fig.~\ref{fig:band_structure_1}. In both cases the momentum $q$ has to fulfil the condition 
\begin{align}
    \label{eq:cutoff_dirac}
    |q| < \Lambda_\alpha = \begin{cases} \frac{\pi}{2Za} & \text{for} \quad \alpha=2,\dots,Z-1 \\
    \frac{\pi}{Za} & \text{for} \quad \alpha=1,Z \end{cases}\,.
\end{align}

By linearizing the dispersion relation $\epsilon_k$ of $H_0$ around $\pm k_F^{(\nu)}$, and using (\ref{eq:H'_matrix_element}) one can set up an effective Dirac Hamiltonian $H^{(\nu)}_D$ corresponding to gap $\nu=1,\dots,Z-1$ with gap parameter $\Delta^{(\nu)} e^{i\gamma^{(\nu)}}$, which has the vectors $\underline{\psi}_{q\tau}^{(\nu)}(x)=(\psi_{q\tau +}^{(\nu)}(x),\psi_{q\tau -}^{(\nu)}(x))^T$ as eigenstates
\begin{align}
    \nonumber
    H^{(\nu)}_D &= -iv_F^{(\nu)}\partial_x \sigma_z \\
    \label{eq:dirac_h}
    & + \Delta^{(\nu)} \cos(\gamma^{(\nu)}) \sigma_x - \Delta^{(\nu)} \sin(\gamma^{(\nu)}) \sigma_y\,, 
\end{align}
where $v_F^{(\nu)} = 2ta\sin(k_F^{(\nu)}a)$ is the Fermi velocity. The energy of the Dirac eigenstates are given by  
\begin{align}
    \label{eq:dirac_energy}
    \epsilon_{q\tau}^{(\nu)} = \tau \epsilon_q^{(\nu)} \quad,\quad \epsilon_q^{(\nu)}=\sqrt{(v_F^{(\nu)}q)^2 + (\Delta^{(\nu)})^2}\,,
\end{align}
which provides a very good approximation to the true dispersion when $|q|\ll\frac{\pi}{2Za}$. 

In contrast to the dispersion, we note that the Dirac model provides a very good approximation to the exact eigenstates for all $|q|<\Lambda_\alpha$. For large $|q|\sim 1/(Za) \gg \Delta^{(\nu)}/v_F^{(\nu)}$, the eigenstates of the Dirac model are (up to a gauge factor) given by 
\begin{align}
    \label{eq:chi_large_q}
    \chi_{q\tau p}^{(\nu)} \approx \delta_{p,\Sign(q\tau)}\,.
\end{align}
Inserting this form in (\ref{eq:dirac_lattice}-\ref{eq:psi_p_dirac}), one obtains the correct eigenfunctions of the lattice model in the absence of a gap, see Appendix~\ref{app:vons_condition}. This means that for small gaps $\Delta^{(\nu)}\ll t$, the Dirac theory is a useful approximation for the description of all eigenstates of the lattice model. Therefore, all Wannier functions can be fully described by Dirac theory on all length scales. In contrast, dynamical quantities like Green's and correlation functions can only be described for low energies since the energy dispersion enters into these quantities.  

When comparing the Bloch eigenstates (\ref{eq:bloch}) with the parametrization (\ref{eq:dirac_lattice}-\ref{eq:psi_p_dirac}) one finds straightforwardly the following relations between $u_{k\alpha}(ja)$ and $\chi_{q\tau p}^{(\nu)}$ (note that $\alpha$ and $(\nu,\tau)$ are related via (\ref{eq:band_relation}))
\begin{align}
    \nonumber
    \underline{\nu \,\, {\rm even}}: & \quad k=q \\
    \label{eq:nu_even}
    & u_{k\alpha}(ja) = \frac{1}{\sqrt{Z}}\sum_{p=\pm} \chi_{q\tau p}^{(\nu)}\, e^{i\pi p \frac{\nu}{Z}j}\,,\\
    \nonumber
    \underline{\nu \,\, {\rm odd}}: & \quad k = \pm \frac{\pi}{Za} + q \quad,\quad \Sign(q) = \mp  \\
    \label{eq:nu_odd}
    & u_{k\alpha}(ja) = \frac{1}{\sqrt{Z}}\sum_{p=\pm} \chi_{q\tau p}^{(\nu)}\, e^{i\pi p \frac{\nu}{Z}j} e^{\mp i\frac{\pi}{Z}j}\,.
\end{align}
We note that both conditions (\ref{eq:u_x_periodicity}) and (\ref{eq:u_k_periodicity}) are respected by (\ref{eq:nu_even}) and (\ref{eq:nu_odd}). For the special case $j=Z$ we get for arbitrary $\nu$ 
\begin{align}
    \label{eq:u_k_Z}
    u_{k\alpha}(Za) = \frac{1}{\sqrt{Z}} \sum_{p=\pm} \chi_{q\tau p}^{(\nu)}\,.
\end{align}
Using this relation one can easily check whether the two local gauges of lattice and Dirac theory have been chosen identical.

Since the Dirac theory explicitly exhibits the slowly varying parts $\psi_{q\tau p}^{(\nu)}(x)$, one can interpret it as a continuum theory for all $x$ on the real axis and not only for $x=ma$, where the lattice theory is reproduced. Therefore, although the Dirac theory has only a physical meaning for $|q|<\Lambda_\alpha$, one takes all values for the momentum $q$ into account and writes the normalization and completeness relations as
\begin{align}
    \label{eq:psi_dirac_normalization}
    &\sum_p\int dx\, \psi_{q\tau p}^{(\nu)}(x)^* \psi_{q'\tau' p}^{(\nu)}(x) = \delta_{\tau\tau'}\delta(q-q')\,,\\
    \label{eq:psi_dirac_completeness}
    &\sum_\tau\int dq\,\psi_{q\tau p}^{(\nu)}(x) \psi_{q\tau p'}^{(\nu)}(x')^* = \delta_{pp'}\delta(x-x')\,.
\end{align}
This is equivalent to taking the following relations for the parts $\chi_{q\tau p}^{(\nu)}$ at fixed $q$
\begin{align}
    \label{eq:chi_p_dirac_normalization}
    \sum_p (\chi_{q\tau p}^{(\nu)})^* \chi_{q\tau' p}^{(\nu)} &= \delta_{\tau\tau'}\,,\\
    \label{eq:chi_p_dirac_completeness}
    \sum_\tau \chi_{q\tau p}^{(\nu)}(\chi_{q\tau p'}^{(\nu)})^* &= \delta_{pp'} \,.
\end{align}
Introducing the following scalar product in right/left space 
\begin{align}
    \label{eq:scalar_product_dirac}
    \langle \chi_{q\tau}^{(\nu)}|\chi_{q\tau'}^{(\nu)}\rangle = 
    \sum_{p=\pm}(\chi_{q\tau p}^{(\nu)})^* \chi_{q\tau' p}^{(\nu)} \,,
\end{align}
we can write (\ref{eq:chi_p_dirac_normalization}) and (\ref{eq:chi_p_dirac_completeness}) in compact form as
\begin{align}
    \label{eq:chi_p_dirac_normalization_sp}
    \langle \chi_{q\tau}^{(\nu)}|\chi_{q\tau'}^{(\nu)}\rangle &= \delta_{\tau\tau'}\,,\\
    \label{eq:chi_p_dirac_completeness_sp}
    \sum_{\tau=\pm} |\chi_{q\tau}^{(\nu)}\rangle\langle \chi_{q\tau}^{(\nu)}| &= \mathbbm{1}_2 \,.
\end{align}

It can be checked that the relations (\ref{eq:chi_p_dirac_normalization}) and (\ref{eq:chi_p_dirac_completeness}) for $\chi_{q\tau p}^{(\nu)}$, together with (\ref{eq:nu_even}) and (\ref{eq:nu_odd}), are consistent with the normalization condition (\ref{eq:u_normalization}) and the completeness relation (\ref{eq:u_completeness}) for $u_{k\alpha}(ja)$, see Appendix~\ref{app:vons_condition}. The normalization condition follows exactly from (\ref{eq:nu_even}) and (\ref{eq:nu_odd}), and is a special case of the useful relation (valid for any integers $r,s\ge 0$) 
\begin{align}
    \label{eq:dirac_lattice_identity}
    \langle \partial_k^r u_{k\alpha}|\partial_k^s u_{k\alpha'}\rangle 
    =\delta_{\nu\nu'}\langle\partial_q^r \chi_{q\tau}^{(\nu)}| \partial_q^s \chi_{q\tau'}^{(\nu')}\rangle\,.
\end{align}
The completeness relation is more subtle and uses in addition the fact that the Dirac theory can reproduce all eigenstates of the original lattice model under the perturbative condition (\ref{eq:condition}). 

The fact that the Dirac theory can capture all eigenstates of the lattice model leads to the following rigorous rule when considering any $k$-integral over a function $f_{k\alpha}$ depending on the eigenfunctions of band $\alpha$ of the lattice theory (like projectors on Bloch states for the completeness relation, Zak-Berry connection for the Zak-Berry phase, geometric tensor for boundary charge fluctuations, or Bloch states for Wannier functions) 
\begin{align}
    \nonumber
    \int_{|k|<\pi/Za} dk \,f_{k\alpha} &\approx \\
    \label{eq:k_integral_calculation_via_dirac}
    &\hspace{-2cm}
    \approx \int_{|q|<\Lambda_\alpha} dq\,\left(f_{q-}^{(\alpha)} + f_{q+}^{(\alpha-1)}\right)\,,
\end{align}
where $f_{q\tau}^{(\nu)}$ is the corresponding function in Dirac theory obtained by the replacements (\ref{eq:dirac_lattice}), (\ref{eq:nu_even}) or (\ref{eq:nu_odd}). Thereby, for $\alpha=1$ or $\alpha=Z$, we omit implicitly the terms with $\nu=0$ or $\nu=Z$ on the right hand side, respectively. The approximate sign in this relation is meant in a perturbative sense that all higher order corrections are of relative order $O(\Delta^{(\nu)}/t)$ and negligible for $\Delta^{(\nu)}\ll t$. However we note that (\ref{eq:k_integral_calculation_via_dirac}) is only valid when the gauges in lattice and Dirac theory have been chosen identical such that the relation (\ref{eq:dirac_lattice}) between the eigenfunctions holds globally for all quasimomenta in a certain band $\alpha$. 

Since a field theory does not know anything about scales of the order of the lattice spacing, the integrals over $q$ on the right hand side of (\ref{eq:k_integral_calculation_via_dirac}) are conveniently extended to infinity in the field theoretical version and properly regularized in case of divergences. Since the regime of large momenta refers to the physics in the absence of a gap their contribution can be analytically analysed quite easily. For the various physical quantities discussed in this work we will show that the field-theoretical contributions beyond the cutoff either vanish for each individual term on the right hand side of (\ref{eq:k_integral_calculation_via_dirac}) (as for Zak-Berry connection and geometric tensor) or, after a proper regularisation, the contributions of high momenta cancel between the two terms (as for Wannier functions in certain gauges). In this way a full equivalence between lattice and Dirac theory can be set up, provided that the gaps are small compared to the typical band width. Furthermore, for the bands $\alpha=2,\dots,Z$ the two terms on the right hand side will be shown to refer to the description of the universal behavior of the two halves of the bands, the upper one corresponding to the Dirac theory for $\nu=\alpha$ and the lower one to $\nu=\alpha-1$. However, for this interpretation to make sense we will see that non-universal terms arising from cutting the spectrum in the middle of the band have to be removed by extending the lattice theory for the two halves to infinity via an asymptotically free theory far away from the gap.  

When eigenfunctions are compared between the lattice and Dirac theory and identified via (\ref{eq:dirac_lattice}) one always has to guarantee that the gauges are chosen in precisely the same way (at least locally for a certain region in quasimomentum space). Therefore, it will turn out to be important to state specific gauges of particular interest. One of them is the so-called asymptotically free (AF) gauge, which is characterized by a real and positive value of $u_{k\alpha}(Za)$ 
\begin{align}
    \label{eq:AF_gauge}
    \text{\underline{AF gauge}:}\quad u_{k\alpha}(Za) > 0 \quad\Leftrightarrow\quad \sum_p \chi_{q\tau p}^{(\nu)} > 0 \,.
\end{align}
Here we used (\ref{eq:u_k_Z}) to formulate the equivalent condition in Dirac theory. This gauge has the particular advantage that it leads precisely to the asymptotic condition (\ref{eq:chi_large_q}) without any further gauge factors. As explained later in Section~\ref{sec:sct} the AF gauge leads to a unique relation between the Zak-Berry phase and the boundary charge both in lattice and Dirac theory. Another gauge will be introduced in Sections~\ref{sec:zak_lattice} and \ref{sec:zak_dirac} which is called the maximally localized (ML) gauge. In the ML gauge the Wannier functions in lattice and Dirac theory have minimal spread which can be related to the boundary charge fluctuations. It will turn out that the ML gauges in lattice and Dirac theory are {\it not} the same, such that the relation (\ref{eq:dirac_lattice}) involves additional phase factors depending on the quasimomentum. Finally, in Section~\ref{sec:nonabelian} we will also discuss non-Abelian lattice gauges where a mixing of the Bloch states from a set of bands $\alpha=1,\dots,\nu$ is involved, such that the non-Abelian Wannier functions turn out to have maximal localization. This gauge is called the non-Abelian gauge of maximal localization (NA-ML). It is a special lattice gauge which turns out to be constructed in such a way that the non-Abelian Wannier functions can be directly related to the Wannier function of the upper half of the highest band $\nu$ or, equivalently, to the Wannier function of the lower band of the Dirac model corresponding to the gap between the highest band $\nu$ and band $\nu+1$.

\section{Zak-Berry connection, geometric tensor and boundary charge}
\label{sec:zak_boundary}

As a prerequisite for the analysis of universal aspects of Wannier functions we develop in this section the low-energy theory for the Zak-Berry connection, the Zak-Berry phase, and the geometric tensor in Dirac theory, together with the relation to the corresponding objects in lattice theory. In this connection, we will also introduce the definition of the ML gauge in Dirac theory and relate it to the corresponding definition in lattice theory. Finally, we will show how an important physical observable, the boundary charge and its fluctuations, can be related via the surface charge and surface fluctuation theorem to the Zak-Berry phase and the momentum integral of the geometric tensor in low energy.

\subsection{Zak-Berry connection and geometric tensor for lattice model}
\label{sec:zak_lattice}

For the lattice the Zak-Berry connection and geometric tensor are defined by
\begin{align}
    \label{eq:zak_berry_lattice}
    (A_k)_{\alpha\beta} &= (A_k)_{\beta\alpha}^* = i\langle u_{k\alpha}|\partial_k u_{k\beta}\rangle  \,,\\
    \label{eq:geometric_tensor_lattice_1}
    \end{align}
    \begin{align}
    (\mathcal{Q}_k)_{\alpha\beta} &= (\mathcal{Q}_k)_{\beta\alpha}\\
    \label{eq:geometric_tensor_lattice_2}
     &= \langle \partial_k u_{k\alpha}|\partial_k u_{k\beta}\rangle \delta_{\alpha\beta}
    - |\langle u_{k\alpha}|\partial_k u_{k\beta}\rangle|^2\\
    \label{eq:geometric_tensor_lattice_3}
    &= \sum_{\alpha'\ne\alpha}|\langle u_{k\alpha'}|\partial_k u_{k\alpha}\rangle|^2
    (\delta_{\alpha\beta} - \delta_{\alpha'\beta})\,,
\end{align}
where we made use of the normalization and completeness relations (\ref{eq:u_normalization_sp}) and (\ref{eq:u_completeness_sp}) to derive the last equality. We note that the diagonal component $\mathcal{Q}_{k\alpha}\equiv(\mathcal{Q}_k)_{\alpha\alpha}$ of the geometric tensor can also be written as
\begin{align}
    \label{eq:geometric_tensor_lattice_diagonal_1}
    \mathcal{Q}_{k\alpha} &= \sum_{\alpha'\ne\alpha}|\langle u_{k\alpha'}|\partial_k u_{k\alpha}\rangle|^2\\
    \nonumber
    &= \partial_k\langle u_{k\alpha}|\partial_k u_{k\alpha}\rangle
    + \langle u_{k\alpha}|(i\partial_k)^2|u_{k\alpha}\rangle\\
    \label{eq:geometric_tensor_lattice_diagonal_2}
     &\hspace{2cm} - \langle u_{k\alpha}|i\partial_k |u_{k\alpha}\rangle^2 \,.
\end{align}

Taking a gauge transformation
\begin{align}
    \label{eq:gauge_trafo_lattice}
    \tilde{u}_{k\alpha} = e^{i\varphi_{k\alpha}} u_{k\alpha}\,,
\end{align}
we find from (\ref{eq:geometric_tensor_lattice_2}) that the geometric tensor is gauge invariant and from (\ref{eq:zak_berry_lattice}) that the Zak-Berry connection is gauge invariant for $\alpha\ne\beta$, whereas $A_{k\alpha}\equiv (A_k)_{\alpha\alpha}$ transforms as
\begin{align}
    \label{eq:A_trafo_lattice}
    \tilde{A}_{k\alpha} = A_{k\alpha} - \partial_k\varphi_{k\alpha}\,.
\end{align}
Furthermore,  
\begin{align}
    \label{eq:zak_berry_phase_lattice}
    \gamma_{\alpha} = \int_{-\pi/Za}^{\pi/Za}dk A_{k\alpha}
\end{align}
denotes the Zak-Berry phase (which should be distinguished from the notation $\gamma^{(\nu)}$ for the phase of the gap parameter in Dirac theory).  According to (\ref{eq:A_trafo_lattice}), the Zak-Berry phase transform under a gauge transformation by the winding number of the phase $\varphi_{k\alpha}$
\begin{align}
    \label{eq:gauge_trafo_zak_berry_phase}
    \tilde{\gamma}_{\alpha} = \gamma_{\alpha} - \varphi_{\pi/Za,\alpha}+\varphi_{-\pi/Za,\alpha}\,.
\end{align}

In the AF gauge defined by (\ref{eq:AF_gauge}) it is shown in Ref.~\onlinecite{pletyukhov_etal_prb_20} that
\begin{align}
    \label{eq:zb_minus_k_AF_lattice}
    u_{-k,\alpha} = (u_{k\alpha})^* \quad\Rightarrow\quad A_{-k,\alpha} = A_{k\alpha} \,.
\end{align}
The ML gauge is defined by a constant Zak-Berry connection
\begin{align}
    \label{eq:max_loc_lattice}
    \underline{\text{Lattice ML gauge}}: \quad \tilde{A}_{k\alpha} = \frac{Za}{2\pi}\tilde{\gamma}_{\alpha}\,.
\end{align}
According to (\ref{eq:A_trafo_lattice}) this can be achieved by the gauge factor $e^{i\varphi_{k\alpha}}$ with
\begin{align}
    \label{eq:gauge_max_loc_lattice}
    \varphi_{k\alpha} = \int_{-\pi/Za}^k dk' A_{k'\alpha} - k \frac{Za}{2\pi} \gamma_{\alpha} -\frac{1}{2}\gamma_{\alpha}\,,
\end{align}
which fulfils 
\begin{align}
    \varphi_{k\alpha} &=\varphi_{k+2\pi/Za,\alpha} \,,\\
    \label{eq:varphi_zero}
    \varphi_{\pm\pi/Za,\alpha} &= 0 \,,\\
    \label{eq:varphi_minus_k}
    \varphi_{-k,\alpha} &= -\varphi_{k\alpha}\,,
\end{align}
where we used (\ref{eq:zb_minus_k_AF_lattice}) for the last equality. From (\ref{eq:varphi_zero}) we find that the winding number is zero leaving the Zak-Berry phase invariant 
\begin{align}
    \label{eq:gamma_invariance}
    \tilde{\gamma}_{\alpha} =\gamma_{\alpha}\,.
\end{align}
Since the AF and ML gauge are the relevant gauges used in this work, we will implicitly indicate the AF gauge by symbols without a tilde and the ML gauge by a tilde symbol.

\subsection{Zak-Berry connection and geometric tensor for Dirac model}
\label{sec:zak_dirac}

In the Dirac theory we define the Zak-Berry connection and the geometric tensor by 
\begin{align}
    \label{eq:zak_berry_connection_dirac}
    (A_q^{(\nu)})_{\tau\tau'} &= i\langle\chi_{q\tau}^{(\nu)}|\partial_q\chi_{q \tau'}^{(\nu)}\rangle\,,\\
    \label{eq:geometric_tensor_dirac_1}
    (\mathcal{Q}_q^{(\nu)})_{\tau\tau'} &= 
    \langle \partial_q\chi_{q\tau}^{(\nu)}|\partial_q\chi_{q\tau}^{(\nu)}\rangle
    \delta_{\tau\tau'} - |\langle\chi_{q\tau}^{(\nu)}|\partial_q\chi_{q \tau'}^{(\nu)}\rangle|^2\\
    \label{eq:geometric_tensor_dirac_2}
    &= \Sign(\tau\tau')|\langle\chi_{q \bar{\tau}}^{(\nu)}|\partial_q\chi_{q\tau}^{(\nu)}\rangle|^2\,,
\end{align}
where we defined $\bar{\tau}=-\tau$ and used $\delta_{\tau\tau'}-\delta_{\bar{\tau}\tau'}=\Sign(\tau\tau')$ to get the last equation. This gives the important property
\begin{align}
    \label{eq:geometric_tensor_dirac_property}
    \sum_{\tau=\pm} (\mathcal{Q}_q^{(\nu)})_{\tau\tau'} = \sum_{\tau'=\pm} (\mathcal{Q}_q^{(\nu)})_{\tau\tau'} = 0\,.
\end{align}
Analog to (\ref{eq:geometric_tensor_lattice_diagonal_2}), the diagonal component $\mathcal{Q}_{q\tau}^{(\nu)}\equiv(\mathcal{Q}_q^{(\nu)})_{\tau\tau}$ can also be written as
\begin{align}
    \nonumber
    \mathcal{Q}_{q\tau}^{(\nu)} &= 
    \partial_q \langle\chi_{q\tau}^{(\nu)}|\partial_q\chi_{q\tau}^{(\nu)}\rangle \\
    \label{eq:geometric_tensor_dirac_diagonal}
    &+ \langle\chi_{q\tau}^{(\nu)}|(i\partial_q)^2|\chi_{q\tau}^{(\nu)}\rangle 
    - \langle\chi_{q\tau}^{(\nu)}|i\partial_q|\chi_{q\tau}^{(\nu)}\rangle^2
    \,.
\end{align}

Taking a gauge transformation
\begin{align}
    \label{eq:gauge_trafo_dirac}
    \tilde{\chi}_{q\tau p}^{(\nu)} = e^{i\phi_{q\tau}^{(\nu)}} \chi_{q\tau p}^{(\nu)}\,,
\end{align}
we find analog to the lattice that the geometric tensor and the nondiagonal components of the Zak-Berry connection are gauge invariant, whereas $A_{q\tau}^{(\nu)}=(A_q^{(\nu)})_{\tau\tau}$ transforms as
\begin{align}
    \label{eq:A_trafo_dirac}
    \tilde{A}_{q\tau}^{(\nu)} = A_{q\tau}^{(\nu)} - \partial_q\phi_{q\tau}^{(\nu)}\,.
\end{align}
In the following we allow only for gauges where the Zak-Berry connection vanishes for $|q|\rightarrow \infty$
\begin{align}
    \label{eq:zak_berry_asymptotics_dirac}
    \lim_{|q|\rightarrow\infty} A_{q\tau}^{(\nu)} = 0\,,
\end{align}
such that the Zak-Berry phase in Dirac theory, defined by 
\begin{align}
    \label{eq:zak_berry_phase_dirac}
    \gamma^{(\nu)}_\tau = \int dq A_{q\tau}^{(\nu)}  
\end{align}
is a well-defined quantity. We note that this is fulfilled for the AF gauge (\ref{eq:AF_gauge}) where we get from (\ref{eq:chi_large_q}) for momenta beyond the cutoff $\Lambda_c$ 
\begin{align}
    \label{eq:chi_q_large}
    \chi_{q\tau p}^{(\nu)} \approx \begin{cases} \delta_{\tau p} &\text{for}\quad q>\Lambda_c \\ \delta_{p,-\tau} &\text{for}\quad q<-\Lambda_c \end{cases}\,,
\end{align}
leading to a vanishing Zak-Berry connection for $|q|>\Lambda_c$.

To get the relation for the Zak-Berry connection and the geometric tensor between the lattice and Dirac definition, we use the identity (\ref{eq:dirac_lattice_identity}) and find
\begin{align}
    \label{eq:zak_berry_lattice_dirac}
    (A_k)_{\alpha\alpha'} &= \delta_{\nu\nu'} (A_q^{(\nu)})_{\tau\tau'}\,,\\
    \label{eq:geometric_tensor_lattice_dirac}
    (\mathcal{Q}_k)_{\alpha\alpha'} &= \delta_{\nu\nu'} (\mathcal{Q}_q^{(\nu)})_{\tau\tau'}\,.
\end{align}
Since the geometric tensor is gauge invariant in both the lattice and Dirac theory, Eq.(\ref{eq:geometric_tensor_lattice_dirac}) holds independent of the gauge choice in lattice and Dirac theory (they can be even different). The relation (\ref{eq:zak_berry_lattice_dirac}) for the Zak-Berry connection holds within any choice for the local gauge such that the relations (\ref{eq:nu_even}) and (\ref{eq:nu_odd}) hold between the eigenfunctions. 

Concerning the Zak-Berry phase of band $\alpha$ we find from (\ref{eq:zak_berry_phase_lattice}), (\ref{eq:zak_berry_lattice_dirac}), and (\ref{eq:k_integral_calculation_via_dirac}) 
\begin{align}
    \label{eq:zak_berry_alpha_dirac}
    \gamma_{\alpha} \approx \gamma^{(\alpha)}_- + \gamma^{(\alpha-1)}_+\,.
\end{align}
As discussed above this holds only when the Zak-Berry connection in Dirac theory vanishes beyond the cutoff $\Lambda_c$ which is fulfilled for the AF gauge. In the AF gauge we will furthermore show below via the explicit eigenstates of the Dirac model (see Section~\ref{sec:wannier_dirac_explicit}) that $\gamma^{(\nu)}_\tau$ is related to the phase $\gamma^{(\nu)}$ of the gap parameter $\Delta^{(\nu)} e^{i\gamma^{(\nu)}}$ by
\begin{align}
    \label{eq:zak_berry_phase_vs_gap_phase}
    \gamma^{(\nu)}_+ = \gamma^{(\nu)} \quad,\quad \gamma^{(\nu)}_- = - \gamma^{(\nu)} + \pi \,s_{\gamma^{(\nu)}} \,,
\end{align}
where we abbreviated the sign function by 
\begin{align}
    \label{eq:sign_function}
    s_{\gamma} = \Sign{\gamma}\,, 
\end{align}
and assumed $-\pi < \gamma^{(\nu)} < \pi$ with periodic continuation to the other regimes. 

In the following, we denote the eigenstates of lattice and Dirac theory in the AF gauge by $\psi_{k\alpha}(ma)$ and $\psi_{q\tau}^{(\nu)}(x)$, respectively, which are connected by the relation (\ref{eq:dirac_lattice}). Correspondingly we use the notation $u_{k\alpha}$ and $\chi_{q\tau p}^{(\nu)}$ in this gauge which are related by (\ref{eq:nu_even}) and (\ref{eq:nu_odd}). From (\ref{eq:zb_minus_k_AF_lattice}) we note that we get in the AF gauge
\begin{align}
    \label{eq:zb_minus_k_AF_dirac}
    A_{-q,\tau}^{(\nu)} = A_{q\tau}^{(\nu)}\quad.
\end{align}

The ML gauge in Dirac theory is defined by a vanishing Zak-Berry connection
\begin{align}
    \label{eq:max_loc_dirac}
    \underline{\text{Dirac ML gauge}}: \quad \tilde{A}_{q\tau}^{(\nu)} = 0\,,
\end{align}
such that the corresponding Zak-Berry phase is also zero
\begin{align}
    \label{eq:zak_berry_dirac_ML}
    \tilde{\gamma}^{(\nu)}_\tau = 0 \,.
\end{align}
According to (\ref{eq:A_trafo_dirac}) this corresponds to the choice of a gauge factor $e^{i\phi_{q\tau}^{(\nu)}}$ with
\begin{align}
    \label{eq:gauge_max_loc_dirac}
     \phi_{q\tau}^{(\nu)} = \int_0^q dq'\,A_{q'\tau}^{(\nu)} + \phi_{0,\tau}^{(\nu)} \,.
\end{align}
Note that this is not a contradiction to the corresponding condition (\ref{eq:max_loc_lattice}) for the lattice theory where it is not possible to remove the Zak-Berry phase via a global gauge transformation. The Dirac model for a given gap $\nu$ describes only those states in $k$-space which belong to the two halves of the bands separated by gap $\nu$. Therefore, a particular global gauge chosen in Dirac theory corresponds to a certain {\it local} gauge in lattice theory defined for all quasimomenta lying in one half of a given band. Such a gauge can not necessarily be extended to a global gauge for a certain band which is continuous and periodic in $k$. 

In order to show the explicit difference between the ML gauges in lattice and Dirac theory we start from the AF gauge and define a gauge transformation to the ML gauge by the transformed quantities
\begin{align}
    \label{eq:gauge_trafo_real_Z_max_loc}
    \tilde{u}_{k\alpha} = e^{i\varphi_{k\alpha}} u_{k\alpha} \quad,\quad
    \tilde{\chi}_{q \tau p}^{(\nu)} = e^{i\phi_{q\tau}^{(\nu)}} \chi_{q\tau p}^{(\nu)}\,.
\end{align}
From the conditions (\ref{eq:max_loc_lattice}), (\ref{eq:gauge_max_loc_lattice}), (\ref{eq:max_loc_dirac}), and (\ref{eq:gauge_max_loc_dirac}), defining the ML gauge in lattice and Dirac theory, we get
\begin{align}
    \label{eq:phase_lattice_1}
    \varphi_{k\alpha} &= \int_{-\pi/Za}^k dk' A_{k'\alpha} - (k+\frac{\pi}{Za})\frac{Za}{2\pi}\gamma_{\alpha}\\
    \label{eq:phase_lattice_2}
     &= \int_0^k dk' A_{k'\alpha} - k\frac{Za}{2\pi}\gamma_{\alpha}  \,,\\
    \label{eq:phase_dirac}
    A_{q\tau}^{(\nu)} &= \partial_q \phi_{q\tau}^{(\nu)}\,,
\end{align}
where we made use of (\ref{eq:zb_minus_k_AF_lattice}) in the AF gauge to get the form (\ref{eq:phase_lattice_2}). Using (\ref{eq:zak_berry_lattice_dirac}) we find from the two forms for $\varphi_{k\alpha}$ that, for both the lower and upper half of band $\alpha$, we get the following relation between the ML gauges of lattice and Dirac theory
\begin{align}
    \nonumber
    \varphi_{k\alpha} &= \int_0^q dq' A_{q'\tau}^{(\nu)} - q\frac{Za}{2\pi} \gamma_{\alpha} \\
    \label{eq:phase_lattice_dirac}
    &= \phi_{q\tau}^{(\nu)} - \phi_{0,\tau}^{(\nu)} - q\frac{Za}{2\pi}\gamma_{\alpha}\,,
\end{align}
leading to the following relation between the gauge factors
\begin{align}
    \label{eq:gauge_factor_lattice_dirac}
    e^{i\varphi_{k\alpha}} = e^{i\phi_{q\tau}^{(\nu)}} e^{-i\phi_{0,\tau}^{(\nu)}} e^{-iq\frac{Za}{2\pi}\gamma_{\alpha}} \,.
\end{align}
The last two factors on the right hand side define the difference between the ML gauges in lattice and Dirac theory, which have to be added to (\ref{eq:nu_even}) and (\ref{eq:nu_odd}) to get the precise relation between the eigenstates of lattice and Dirac theory in the ML gauge. Furthermore, they show that the ML gauge in Dirac theory is not smooth when crossing over the middle of a band $\alpha$ since the indices $(\nu\tau)$ and the relation between $q$ and $k$ change.

\subsection{The boundary charge}
\label{sec:boundary_charge}

\subsubsection{The surface charge theorem}
\label{sec:sct}

The charge accumulated at the boundary, or the boundary charge\cite{park_etal_prb_16,thakurathi_etal_prb_18,pletyukhov_etal_prbr_20,pletyukhov_etal_prb_20,lin_etal_prb_20,pletyukhov_etal_prr_20,weber_etal_prl_20}, is defined by restricting the tight-binding Hamiltonian (\ref{eq:H}) to the half-infinite space $m>0$ and averaging the excess charge with a macroscopic envelope function $f_m$ via 
\begin{align}
    \label{eq:QB_nu_def}
    Q_B^{(\nu)} &= \sum_{m=1}^\infty \left(\rho^{(\nu)}_m - \bar{\rho}^{(\nu)}\right) f_m \,.
\end{align}
Here, $\rho^{(\nu)}_{m}$ is the average charge at lattice site $m$ for a half-infinite system if the chemical potential is located in gap $\nu$, including all edge states up to this energy. The average charge per site in the bulk is denoted by $\bar{\rho}^{(\nu)}=\nu/Z$. The macroscopic and probe specific envelope function is denoted by $f_m$ which must have certain properties. In particular, it has the value 1 in the first range of $m$, which far exceeds the localization length $\xi$, and then it smoothly crosses over to 0 over the second range of $m$, which is also $\gg \xi$. The importance of including these features into the definition of $f_m$ for the universality of the boundary charge properties as well as their experimental relevance has been elucidated in Refs.~\onlinecite{park_etal_prb_16,thakurathi_etal_prb_18,pletyukhov_etal_prbr_20,pletyukhov_etal_prb_20}. If only a single band $\alpha$ is occupied we denote the corresponding boundary charge by $Q_{B,\alpha}$, such that the total boundary charge can be decomposed as
\begin{align}
    \label{eq:QB_nu_decomposition_1}
    Q_B^{(\nu)} &= \sum_{\alpha=1}^\nu Q_{B,\alpha} + Q_E^{(\nu)}\,, \\
    \label{eq:QB_nu_decomposition_2}
    &= \sum_{\alpha=1}^\nu \left(Q_{B,\alpha} + Q_{E,\alpha}\right)\,, 
\end{align}
where $Q_E^{(\nu)}$ denotes the integer contribution from all occupied edge states up to the chemical potential, and $Q_{E,\alpha}$ denotes the contribution of an edge state present in gap $\alpha$. The latter is either unity or zero and we assume for the validity of (\ref{eq:QB_nu_decomposition_2}) that, for the last gap $\alpha=\nu$, the chemical potential is located at the bottom of the conduction band $\alpha=\nu+1$.  

As shown in Ref.~\onlinecite{pletyukhov_etal_prb_20}, one finds in the AF gauge of the lattice theory that the Zak-Berry phase is related to the boundary charge of a single band in the following way
\begin{align}
    \label{eq:QB_zak_berry_phase}
    Q_{B,\alpha} = -\frac{\gamma_{\alpha}}{2\pi} - \frac{Z-1}{2Z} \,.
\end{align}
Inserting (\ref{eq:zak_berry_alpha_dirac}) and (\ref{eq:zak_berry_phase_vs_gap_phase}) in the expression (\ref{eq:QB_zak_berry_phase}) for the boundary charge we get 
\begin{align}
    \label{eq:QB_alpha_dirac}
    Q_{B,\alpha} = \frac{\gamma^{(\alpha)}-\gamma^{(\alpha-1)}}{2\pi} + \frac{1}{2Z} - \theta(\gamma^{(\alpha)})\,.
\end{align}
Noting that an edge state appears in gap $\nu$ for $0<\gamma^{(\nu)}<\pi$ (see Ref.~\onlinecite{pletyukhov_etal_prr_20}), we find that the sum of the boundary charge of band $\alpha$ and the charge $Q_{E,\alpha}$ of the edge state in gap $\alpha$ is given by 
\begin{align}
    \label{eq:QB_alpha_edge_dirac}
    Q_{B,\alpha} + Q_{E,\alpha} = \frac{\gamma^{(\alpha)}-\gamma^{(\alpha-1)}}{2\pi} + \frac{1}{2Z} \,.
\end{align}
Assuming that the chemical potential is located at the bottom of the conduction band, this gives the following result for the boundary charge $Q_B^{(\nu)}$ when the lowest $\nu$ bands are filled
\begin{align}
    \label{eq:QB_nu}
    Q_B^{(\nu)} = \frac{\gamma^{(\nu)}}{2\pi} + \frac{\nu}{2Z} \,.
\end{align}
Here we have used the fact that the lower half of the lowest band gives a negligible contribution to the Zak-Berry phase since we can approximate the eigenstates by the free ones in this regime leading to a vanishing Zak-Berry connection (see Appendix~\ref{app:vons_condition}). The result (\ref{eq:QB_nu}) states the low-energy version of the {\it surface charge theorem} relating the boundary charge to the phase $\gamma^{(\nu)}$ of the gap parameter which is related via (\ref{eq:zak_berry_phase_vs_gap_phase}) to the low-energy version of the Zak-Berry phase. We note that the result (\ref{eq:QB_nu}) is fully consistent with the direct calculation of the boundary charge for a half-infinite Dirac model as shown in Ref.~\onlinecite{pletyukhov_etal_prr_20}. 

An essential point for the formulation of the surface charge theorem in low energy is the fact that the sum of the two Zak-Berry phases $\gamma^{(\nu)}_\tau$ over $\tau=\pm$ gives in the AF gauge 
\begin{align}
    \label{eq:zak_berry_sum_dirac}
    \sum_{\tau=\pm} \gamma^{(\nu)}_\tau = \pi \,s_{\gamma^{(\nu)}} \,.
\end{align}
This gives rise to the effect that all other gaps $1,\dots,\nu-1$ contribute only constants to the boundary charge $Q_B^{(\nu)}$ and the final result (including the edge states) can be written in terms of the Zak-Berry phase $\gamma^{(\nu)}_-$ of the top of the valence band alone. We note that the result $\sum_\tau \gamma^{(\nu)}_\tau=\pi \text{mod}(2\pi)$ can be shown on quite general principles without using the explicit forms of the eigenfunctions of the Dirac model, see Appendix~\ref{app:zb_cancellation}.

\subsubsection{The surface fluctuation theorem}
\label{sec:sft}

In lattice theory the $k$-integral over the geometric tensor is of particular interest since it can be related to the boundary charge fluctuations \cite{weber_etal_prl_20}
\begin{align}
    \label{eq:QB_fluctuations_geometric_tensor} 
    l_p (\Delta Q_B^{(\nu)})^2 = \sum_{\alpha,\beta=1}^\nu \int_{-\pi/Za}^{\pi/Za}\frac{dk}{2\pi} (\mathcal{Q}_k)_{\alpha\beta}\,,
\end{align}
where $(\Delta Q_B^{(\nu)})^2$ are the fluctuations when the chemical potential lies in gap $\nu$ (the precise position is unimportant). Here, $l_p\gg\xi$ is a length scale on which the charge measurement probe looses smoothly the contact to the sample, see Ref.~\onlinecite{weber_etal_prl_20} for details. Since the geometric tensor is gauge-invariant, this relation holds in any gauge, in particular in the AF and ML gauge. Summing the geometric tensor $(\mathcal{Q}_k)_{\alpha\beta}$ over $\alpha$ or $\beta$ up to some $\nu$ in the low-energy regime, we find from (\ref{eq:geometric_tensor_lattice_dirac}) and (\ref{eq:geometric_tensor_dirac_property}) that all contributions vanish where two bands are separated by a gap. Therefore, we obtain
\begin{align}
    \label{eq:geometric_tensor_sum_alpha}
    \sum_{\alpha=1}^\nu (\mathcal{Q}_k)_{\alpha\beta} = \delta_{\beta\nu} \mathcal{Q}_{q-}^{(\nu) } 
    \begin{cases} \delta_{\nu,\text{even}} & \text{for}\quad k\approx 0 \\
    \delta_{\nu,\text{odd}} & \text{for}\quad k\approx \pm \frac{\pi}{Za} \end{cases}\,,
\end{align}
and
\begin{align}
    \label{eq:geometric_tensor_sum_double}
    \sum_{\alpha,\beta=1}^\nu (\mathcal{Q}_k)_{\alpha\beta} = \mathcal{Q}_{q-}^{(\nu) } 
    \begin{cases} \delta_{\nu,\text{even}} & \text{for}\quad k\approx 0 \\
    \delta_{\nu,\text{odd}} & \text{for}\quad k\approx \pm \frac{\pi}{Za} \end{cases}\,.
\end{align}
These relations are also independent of the gauge choice in Dirac theory since the geometric tensor is gauge invariant in Dirac theory as well. Using (\ref{eq:k_integral_calculation_via_dirac}) and the fact that the geometric tensor vanishes in Dirac theory in the asymptotically free region $|q|>\Lambda_c$, we obtain in the low-energy regime for the fluctuations 
\begin{align}
    \label{eq:QB_fluctuations_geometric_tensor_low_energy_1} 
    l_p (\Delta Q_B^{(\nu)})^2 &\approx \int \frac{dq}{2\pi} \mathcal{Q}_{q-}^{(\nu) }
    = \int \frac{dq}{2\pi} |(A_q^{(\nu)})_{+-}|^2\\
    \label{eq:QB_fluctuations_geometric_tensor_low_energy_2} 
    &= \int \frac{dq}{2\pi}\Big\{\langle\chi_{q-}^{(\nu)}|(i\partial_q)^2|\chi_{q-}^{(\nu)}\rangle 
    - (A_{q-}^{(\nu)})^2\Big\} \, ,
\end{align}
where we made use of (\ref{eq:geometric_tensor_dirac_2}), (\ref{eq:geometric_tensor_dirac_diagonal}), and (\ref{eq:zak_berry_asymptotics_dirac}) for the last two steps. 

The result (\ref{eq:QB_fluctuations_geometric_tensor_low_energy_1}) is the low-energy form of the {\it surface fluctuation theorem} relating the boundary charge fluctuations to the momentum integral of the geometric tensor. As shown only the geometric tensor of the valence band enters. This is physically intuitive since one expects no fluctuations of the charge from the low-lying bands. As will be shown below in Section~\ref{sec:wannier_universality} we obtain explicitly via the eigenfunctions of the Dirac model
\begin{align}
    \label{eq:fluctuations_dirac_explicit}
    \int \frac{dq}{2\pi} \mathcal{Q}_{q-}^{(\nu) } = \xi^{(\nu)}/8 \quad,\quad \xi^{(\nu)} = \frac{v_F^{(\nu)}}{2\Delta^{(\nu)}}\,, 
\end{align}
showing that the result (\ref{eq:QB_fluctuations_geometric_tensor_low_energy_1}) is fully consistent with the direct calculation of the boundary charge fluctuations within the Dirac model via the second momentum of the correlation function \cite{weber_etal_prl_20}. 

Most importantly, we note that the fluctuations of the total boundary charge can {\it not} be written as the sum over the fluctuations of the individual bands. If only band $\alpha$ is occupied the fluctuations are given by 
\begin{align}
    \label{eq:fluctuations_single_band}
    l_p \Delta (Q_{B,\alpha})^2 &= \int_{-\pi/Za}^{\pi/Za}\frac{dk}{2\pi} \mathcal{Q}_{k\alpha} \,.
\end{align}
When summing this expression over $\alpha$ one does not obtain the fluctuations $l_p (\Delta Q_B^{(\nu)})^2$ when all bands $\alpha=1,\dots,\nu$ are occupied, as given by Eq.~(\ref{eq:QB_fluctuations_geometric_tensor}) involving the summation over all matrix elements of the geometric tensor. In particular one looses the important cancellation property (\ref{eq:geometric_tensor_dirac_property}) which is the central ingredient that all contributions from gaps between occupied bands do not contribute to the fluctuations of the total boundary charge, rendering the fluctuations to depend only on the low-energy properties of the model.

\section{Wannier functions}
\label{sec:wannier}

In this section we present the definition of the central objects of our work, the Wannier functions in Dirac theory and their corresponding moments, together with their precise relation to the lattice Wannier functions. In Section~\ref{sec:wannier_lattice} we present a short summary of the definition of Wannier functions in lattice theory. The Dirac Wannier functions are defined in Section~\ref{sec:wannier_dirac} and it is shown how their first and second moments can be related to the Zak-Berry phase and the momentum integral of the geometric tensor as defined within the Dirac theory in Section~\ref{sec:zak_dirac}. At the end of this section we show how the surface charge and surface fluctuation theorem can be formulated very elegantly in terms of the moments of the Dirac Wannier functions. The precise relation of the Dirac Wannier functions and their moments to the corresponding lattice quantities is the subject of Sections~\ref{sec:wannier_lattice_dirac_AF} and \ref{sec:wannier_lattice_dirac_ML} in the AF and ML gauge, respectively.

\subsection{Wannier functions for lattice model}
\label{sec:wannier_lattice}

On the lattice the dimensionless Wannier functions for band $\alpha$ are defined by
\begin{align}
    \label{eq:wannier_lattice}
    w_{R,\alpha}(ma) = \sqrt{\frac{Za}{2\pi}}\,\int_{-\pi/Za}^{\pi/Za}dk \, \psi_{k\alpha}(ma) e^{-ikR}\,,
\end{align}
where $R=Za n$, with $n$ integer, denotes a lattice vector. Using (\ref{eq:psi_normalization}) and (\ref{eq:psi_completeness}) one finds that the Wannier functions labelled by $\alpha$ and $R$ form an orthonormal and complete set of states
\begin{align}
    \label{eq:wannier_normalization}
    &\sum_m w_{R,\alpha}(ma)^* w_{R',\alpha'}(ma) = \delta_{\alpha\alpha'}\delta_{RR'}\,,\\
    \label{eq:wannier_completeness}
    &\sum_\alpha\sum_R w_{R,\alpha}(ma) w_{R,\alpha}(m'a)^* = \delta_{mm'}\,.
\end{align}
Defining by $w_{\alpha}(ma)\equiv w_{0,\alpha}(ma)$ the Wannier function centered at zero 
\begin{align}
    \label{eq:w_lattice}
    w_{\alpha}(ma) &= \sqrt{\frac{Za}{2\pi}}\,\int_{-\pi/Za}^{\pi/Za}dk \, \psi_{k\alpha}(ma) \,,\\
    \label{eq:w_lattice_2}
     &= Za\,\int_{-\pi/Za}^{\pi/Za}\frac{dk}{2\pi}\, u_{k\alpha}(ma) e^{ikma} \,,
\end{align}
we find with the Bloch form (\ref{eq:bloch}) and the periodicity property (\ref{eq:u_x_periodicity}) that the Wannier function $w_{R,\alpha}(ma)$ follows from shifting $w_{\alpha}(ma)$ by the lattice vector $R$
\begin{align}
    \label{eq:wannier_shift}
    w_{R,\alpha}(ma) = w_{\alpha}(ma - R)\,.
\end{align}
Therefore, all properties of the Wannier functions follow from studying the properties of $w_{\alpha}(ma)$.

The Wannier functions depend on the gauge in a non-trivial way. If (\ref{eq:w_lattice_2}) is the Wannier function in the AF gauge, we obtain in the ML gauge
\begin{align}
    \label{eq:w_tilde_lattice}
    \tilde{w}_{\alpha}(ma) &= Za\,\int_{-\pi/Za}^{\pi/Za}\frac{dk}{2\pi}\, e^{i\varphi_{k\alpha}} u_{k\alpha}(ma) e^{ikma} \,,
\end{align}
where the phase $\varphi_{k\alpha}$ is given by (\ref{eq:gauge_max_loc_lattice}). We note that due to the properties (\ref{eq:zb_minus_k_AF_lattice}) and (\ref{eq:varphi_minus_k}), both the Wannier functions in AF and ML gauge are real. 

For the bands $\alpha=2,\dots,Z-1$ we show in Appendix~\ref{app:wannier_splitting} that the Wannier functions in the AF or ML gauge can be split into two contributions corresponding to the upper and lower half of the band
\begin{align}
    \label{eq:wannier_splitting}
    w_{\alpha}(ma) &= w_{\text{u},\alpha}(ma) + w_{\text{d},\alpha}(ma) \,,\\ 
    \label{eq:wannier_ud}
    w_{\text{u}/\text{d},\alpha}(ma) &= w^\prime_{\text{u}/\text{d},\alpha}(ma) \pm \delta w_{\alpha}(ma) \,, 
\end{align}
where $w^\prime_{\text{u}/\text{d},\alpha}(ma)$ denotes the part from the corresponding integration regions 
\begin{align}
    w^\prime_{\text{u}/\text{d},\alpha}(ma) &= \frac{Za}{2\pi}\cdot\\
    &\hspace{-1.5cm}
    \cdot\begin{cases}\int_{|k|<\frac{\pi}{2Za}} dk \,u_{k\alpha}(ma) e^{ikma} & \text{for $\alpha$ even/odd}\\ 
    \int_{\frac{\pi}{2Za}<|k|<\frac{\pi}{Za}} dk \,u_{k\alpha}(ma) e^{ikma} & \text{for $\alpha$ odd/even}\end{cases}\,,
\end{align}
and $\delta w_{\alpha}(ma)$ arises from the extension of the quasimomentum  integrations to $\pm\infty$ taking the free solutions of $u_{k\alpha}(ma)$ in the gapless case, see Appendix~\ref{app:wannier_splitting} for details. In the AF and ML gauge one obtains explicitly
\begin{align}
    \label{eq:delta_w_AF}
    \delta w_{\alpha}(ma) &= \frac{\sqrt{Z}}{\pi m} \sin\big\{\frac{\pi}{Z}(\alpha-\frac{1}{2})m\big\}\,,\\
    \nonumber
    \delta \tilde{w}_{\alpha}(ma) &= \\
    \label{eq:delta_w_ML}
    &\hspace{-1.5cm}
    =\frac{\sqrt{Z}}{\pi \tilde{m}_\alpha} 
    \sin\big\{\frac{\pi}{Z}(\alpha-\frac{1}{2})m +\frac{1}{4}(\gamma_+^{(\alpha-1))}-\gamma_-^{(\alpha)})\big\}\,,
\end{align}
where we defined the shifted variable
\begin{align}
    \label{eq:m_shift}
    \tilde{m}_\alpha = m - \frac{Z}{2\pi}\gamma_{\alpha}\,.
\end{align}
Whereas the corrections $\delta w_{\alpha}(ma)$ cancel out when considering the total Wannier function $w_{\alpha}(ma)$ of a band, we will see in Section~\ref{sec:wannier_universality} that the Wannier functions $w_{\text{u}/\text{d},\alpha}(ma)$ and $\tilde{w}_{\text{u}/\text{d},\alpha}(ma)$ show only universal scaling if the corrections $\pm\delta w_{\alpha}(ma)$ and $\pm\delta\tilde{w}_{\alpha}(ma)$ are taken into account. 

For the bands $\alpha=1,Z$ the splitting in upper and lower part makes no sense since the lower/upper half of the band $\alpha=1/Z$ is already described by a free theory for small gap. Therefore, we use the convention 
\begin{align}
    \label{eq:wannier_1Z_ud}
    w_{\text{d},1}(ma) = w_{\text{u},Z}(ma) = 0 \quad,\quad \delta w_{1/Z}(ma) = 0\,.
\end{align}

The moments of the Wannier functions are defined by 
\begin{align}
    \label{eq:moments_wannier_lattice}
    C_{r\alpha} = \langle x^r \rangle_{\alpha} = \sum_m (ma)^r |w_{\alpha}(ma)|^2 \,.
\end{align}
A corresponding definition is used for the moments $C_{\text{u}/\text{d},r\alpha}$ of $w_{\text{u}/\text{d},\alpha}$. Inserting (\ref{eq:w_lattice}) and (\ref{eq:bloch}) we find after some straightforward manipulations
\begin{align}
    \label{eq:moments_wannier_lattice_u}
    C_{r\alpha} = \frac{Za}{2\pi} \int_{-\pi/Za}^{\pi/Za} dk \,\langle u_{k\alpha}|(i\partial_k)^r|u_{k\alpha}\rangle\,.
\end{align}
For $r=1$ and $r=2$ we find with (\ref{eq:zak_berry_lattice}) and (\ref{eq:geometric_tensor_lattice_diagonal_2})
\begin{align}
    \label{eq:first_moment_wannier_lattice}
    C_{1\alpha} &= \frac{Za}{2\pi} \int_{-\pi/Za}^{\pi/Za} dk \,A_{k\alpha} = \frac{Za}{2\pi} \gamma_{\alpha}\,,\\
    \label{eq:second_moment_wannier_lattice}
    C_{2\alpha} &= \frac{Za}{2\pi} \int_{-\pi/Za}^{\pi/Za} dk \,
    \Big\{\mathcal{Q}_{k\alpha} + (A_{k\alpha})^2\Big\}\,,
\end{align}
and for the quadratic spread
\begin{align}
    \nonumber
    \langle\Delta x^2\rangle_{\alpha} &= C_{2\alpha} - (C_{1\alpha})^2\\
    \label{eq:quadratic_spread}
    &\hspace{-1cm}
    = Za \int_{-\pi/Za}^{\pi/Za} \frac{dk}{2\pi} \, \Big\{\mathcal{Q}_{k\alpha} 
    + (A_{k\alpha} - \frac{Za}{2\pi}\gamma_{\alpha})^2\Big\} \,.
\end{align}
Since the geometric tensor is gauge invariant, the condition for maximal localization is given by a constant Zak-Berry connection corresponding to the ML gauge (\ref{eq:max_loc_lattice}). Since, according to (\ref{eq:gamma_invariance}), the Zak-Berry phase does not change in this gauge, the first moment stays invariant 
\begin{align}
    \label{eq:C_1_invariance}
    \tilde{C}_{1\alpha} = C_{1\alpha} = \frac{Za}{2\pi}\gamma_{\alpha}\,.
\end{align}

In the ML gauge we get for the minimal quadratic spread $\langle\Delta x^2\rangle_{\alpha,\text{min}}$ from (\ref{eq:quadratic_spread}) and (\ref{eq:QB_fluctuations_geometric_tensor}) the result
\begin{align}
    \label{eq:quadratic_spread_min}
    \langle\Delta x^2\rangle_{\alpha,\text{min}} = Za \int_{-\pi/Za}^{\pi/Za} \frac{dk}{2\pi} \, \mathcal{Q}_{k\alpha} 
    = Za\, l_p \Delta (Q_{B,\alpha})^2\,,
\end{align}
where $l_p \Delta (Q_{B,\alpha})^2$ are the fluctuations when only the band $\alpha$ is occupied. 

So in summary we can formulate the surface charge and surface fluctuation theorem for a singly occupied band in terms of the first and second moment of the Wannier function as
\begin{align}
    \label{eq:sct_wannier}
    Q_{B,\alpha} &= - \frac{C_{1\alpha}}{Za} - \frac{Z-1}{2Z}\,,\\
    \label{eq:sft_wannier}
    l_p \Delta (Q_{B,\alpha})^2 &= \frac{1}{Za}\,\tilde{D}_{2\alpha} \,,
\end{align}
where $C_{r\alpha}$ refers to the AF gauge, and $\tilde{D}_{r\alpha}$ are the moments in the ML gauge defined relative to the first moment $C_{1\alpha}=\frac{Za}{2\pi}$ 
\begin{align}
    \label{eq:D_moment}
    \tilde{D}_{r\alpha} = \sum_m (\tilde{m}_\alpha a)^r |\tilde{w}_{\alpha}(ma)|^2 \,,
\end{align}
with $\tilde{m}_\alpha$ defined in (\ref{eq:m_shift}). These moments are related to the moments $\tilde{C}_r^{(\alpha)}$. E.g., for $r=1,2$, one obtains by using the normalization $\sum_m |\tilde{w}_{\alpha}(ma)|^2=1$ and (\ref{eq:C_1_invariance})
\begin{align}
    \label{eq:CD_1}
    \tilde{D}_{1\alpha} &= \tilde{C}_1^{(\alpha)} - \frac{Za}{2\pi}\gamma_{\alpha} = 0 \,,\\
    \label{eq:CD_2}
    \tilde{D}_{2\alpha} &= \tilde{C}_{2\alpha} - \left(\tilde{C}_{1\alpha}\right)^2  \,.
\end{align}

As discussed at the end of Section~\ref{sec:sft} the fluctuations $l_p \Delta (Q_{B,\alpha})^2$ of the individual bands are not sufficient to calculate the fluctuations $l_p (\Delta Q_B^{(\nu)})^2$ when all bands $\alpha=1,\dots,\nu$ are occupied. However, as we will see in the next two sections for the low-energy regime of small gaps, the surface charge and surface fluctuation theorem for $\nu$ occupied bands can be written entirely in terms of the moments of the upper component $w_{\text{u},\nu}$ and $\tilde{w}_{\text{u},\nu}$ of the Wannier functions for the highest valence band as
\begin{align}
    \label{eq:sct_wannier_u}
    Q_B^{(\nu)} &\approx -\frac{1}{Za}C_{u,1\nu} + \frac{1}{2} s_{\gamma^{(\nu)}} + \frac{\nu}{2Z}\,,\\
    \label{eq:sft_wannier_u}
    l_p (\Delta Q_B^{(\nu)})^2 &\approx \frac{1}{Za}\tilde{C}_{u,2\nu} \,.
\end{align}
This physically intuitive result reflects the fact that the boundary charge is a low-energy property only which is insensitive to the properties of low-lying and occupied bands. Alternatively, as we will discuss in Section~\ref{sec:nonabelian}, it is also possible to define maximally localized Wannier functions in a non-Abelian gauge where all occupied bands are mixed. As we will see these Wannier functions are closely related to the Wannier function $\tilde{w}_{\text{u},\nu}$ in the Abelian ML gauge and shows precisely the same universal scaling as the Dirac Wannier function of the valence band.

\subsection{Wannier functions for Dirac model}
\label{sec:wannier_dirac}

Within the Dirac theory we define the Wannier functions by
\begin{align}
    \label{eq:wannier_dirac}
    w_{y,\tau}^{(\nu)}(x) = \sum_{p=\pm} w^{(\nu)}_{y,\tau p}(x) e^{ip k_F^{(\nu)}(x-y)} = w_\tau^{(\nu)}(x-y)\,,
\end{align}
where 
\begin{align}
    \label{eq:wannier_p_dirac}
    w^{(\nu)}_{y,\tau p}(x) = \int \frac{dq}{\sqrt{2\pi}} \,\psi_{q\tau p}^{(\nu)}(x) e^{-iqy} = w^{(\nu)}_{\tau p}(x-y)\,,
\end{align}
and 
\begin{align}
    \label{eq:wannier_zero}
    w^{(\nu)}_\tau(x) &= w_{y=0,\tau}^{(\nu)}(x) \quad,\quad w^{(\nu)}_{\tau p}(x) = w^{(\nu)}_{y=0,\tau p}(x) \,,\\
    \label{eq:wannier_p_zero}
    w^{(\nu)}_{\tau p}(x) &= \int\frac{dq}{\sqrt{2\pi}} \,\psi_{q\tau p}^{(\nu)}(x)
    =  \int\frac{dq}{2\pi} \,\chi_{q\tau p}^{(\nu)} e^{iqx}\,.
\end{align}
Correspondingly, for the ML gauge, we define $\tilde{w}^{(\nu)}_{\tau p}(x)$ by 
\begin{align}
    \label{eq:wannier_p_ML}
    \tilde{w}^{(\nu)}_{\tau p}(x) &= \int\frac{dq}{2\pi} \,\tilde{\chi}_{q\tau p}^{(\nu)} e^{iqx}\,.
\end{align}
In contrast to the lattice Wannier functions, the Dirac Wannier functions are complex quantities both in the AF and ML gauge. 

Since the Wannier functions in the Dirac model contain a continuous shift $y$ their dimension is given by inverse length and the normalization and completeness relation follow from (\ref{eq:psi_dirac_normalization}) and (\ref{eq:psi_dirac_completeness}) as
\begin{align}
    \label{eq:wannier_dirac_normalization}
    &\sum_p\int dx\, w_{y,\tau p}^{(\nu)}(x)^* w_{y',\tau' p}^{(\nu)}(x) = \delta_{\tau\tau'}\delta(y-y')\,,\\
    \label{eq:wannier_dirac_completeness}
    &\sum_\tau\int dy\,w_{y,\tau p}^{(\nu)}(x) w_{y,\tau p'}^{(\nu)}(x')^* = \delta_{pp'}\delta(x-x')\,.
\end{align}

The moments $C^{(\nu)}_{r\tau}$ of the Dirac Wannier functions are defined by
\begin{align}
    \label{eq:moments_wannier}
    C^{(\nu)}_{r\tau} = \int dx\, x^r \sum_{p=\pm}|w^{(\nu)}_{\tau p}(x)|^2\,,
\end{align}
and, correspondingly, we denote the moments in the ML gauge by $\tilde{C}^{(\nu)}_{r\tau}$. Note that these moments have dimension $(\text{length})^{r-1}$ in contrast to the moments $C_{r\alpha}$ defined within the lattice theory. This is due to the different normalization in the continuum Dirac theory. Nevertheless, we will see below that the first and second moment are related to the boundary charge and the fluctuations, respectively, in a similar way as in (\ref{eq:sct_wannier}) and (\ref{eq:sft_wannier}) but without the denominator $Za$, see below. Note that the first moment does {\it not} play the role of the position of the Wannier function in Dirac theory since it is dimensionless. A finite value of the first moment $C^{(\nu)}_{1\tau}$ indicates an asymmetry of $\sum_{p=\pm}|w^{(\nu)}_{\tau p}(x)|^2$ for positive and negative $x$. 
%Furthermore note that the low-energy theory produces divergences of the Wannier function for $|x|\rightarrow 0$ which can be cut off below $a$ if one introduces a momentum cutoff $\sim 1/a$. 

Using the form (\ref{eq:wannier_p_zero}) we find after some straightforward manipulations 
\begin{align}
    \label{eq:moments_wannier_fourier}
    C^{(\nu)}_{r\tau} = \int \frac{dq}{2\pi} \langle\chi_{q\tau}^{(\nu)}|(i\partial_q)^r|\chi_{q\tau}^{(\nu)}\rangle \,.
\end{align}
Using (\ref{eq:zak_berry_phase_dirac}), (\ref{eq:geometric_tensor_dirac_2}) and (\ref{eq:geometric_tensor_dirac_diagonal}) we can thus write for the first and second moment
\begin{align}
    \label{eq:first_moment_zak_berry_phase_dirac}
    C^{(\nu)}_{1\tau} &= \frac{\gamma^{(\nu)}_\tau}{2\pi}\,,\\
    \label{eq:second_moment_geometric_tensor_dirac}
    C^{(\nu)}_{2\tau} &= \int\frac{dq}{2\pi}|(A_q^{(\nu)})_{\bar{\tau}\tau}|^2 +\int\frac{dq}{2\pi} (A_{q\tau}^{(\nu)})^2\,.
\end{align}
Since the first term on the right hand side of (\ref{eq:second_moment_geometric_tensor_dirac}) is gauge invariant and the second one has a positive integrand it follows that the gauge of maximally localized Wannier functions is given by the ML gauge (\ref{eq:max_loc_dirac}) where the Zak-Berry connection vanishes. We note that there is a fundamental difference to the lattice theory where the quadratic spread is defined by $C_{2\alpha}-(C_{1\alpha})^2$. The analog formula is not possible in low energy since the dimension of $C_{r\tau}^{(\nu)}$ is $(\text{length})^{r-1}$.

Taking (\ref{eq:first_moment_zak_berry_phase_dirac}) and (\ref{eq:second_moment_geometric_tensor_dirac}) together with (\ref{eq:QB_nu}), (\ref{eq:zak_berry_phase_vs_gap_phase}), and (\ref{eq:QB_fluctuations_geometric_tensor_low_energy_2}), we can formulate the surface charge and  surface fluctuation theorem in low energy via the first and second moment of the Dirac Wannier functions as 
\begin{align}
    \label{eq:sct_wannier_dirac}
    Q_B^{(\nu)} &\approx -C_{1-}^{(\nu)} + \frac{1}{2} s_{\gamma^{(\nu)}} + \frac{\nu}{2Z}\,,\\
    \label{eq:sft_wannier_dirac}
    l_p (\Delta Q_B^{(\nu)})^2 &\approx \tilde{C}_{2-}^{(\nu)} \,.
\end{align}
Here, the moments $C_{r-}^{(\nu)}$ refer to the AF gauge, whereas $\tilde{C}_{r-}^{(\nu)}$ are the moments in the ML gauge. Note that, in contrast to (\ref{eq:sct_wannier}) and (\ref{eq:sft_wannier}) (where a single band has been considered), we consider here the total boundary charge and its fluctuations when the lowest $\nu$ bands are filled and the chemical potential is located at the bottom of the conduction band (which is only important to calculate the edge state contribution for the boundary charge in the last gap). This result shows that the boundary charge and its fluctuations are quantities probing only low energy features and, therefore, can be related in a universal way to the first and second moment of the Dirac Wannier functions corresponding to the top of the highest valence band.

\subsection{The relation between Wannier functions in lattice and Dirac theory}
\label{sec:wannier_lattice_dirac}

We now consider the precise relation of the Dirac Wannier functions to the Wannier function defined within the lattice theory. We start with the AF gauge where the relation (\ref{eq:dirac_lattice}) between eigenfunctions of the lattice and the Dirac theory hold globally for all quasimomenta in a certain band $\alpha$ since the gauge is the same. For the ML gauge it is more subtle since there is difference between the gauges in lattice and Dirac theory, see Eq.~(\ref{eq:gauge_factor_lattice_dirac}). Therefore one has to add gauge factors depending on whether one considers the contribution of the upper or lower half of the band to the Wannier function.

\subsubsection{The AF gauge}
\label{sec:wannier_lattice_dirac_AF}

For the AF gauge, we can insert (\ref{eq:dirac_lattice}) in (\ref{eq:w_lattice}) and use (\ref{eq:k_integral_calculation_via_dirac}) to get
\begin{align}
    \nonumber
    w_{\alpha}(ma) &= a\sqrt{Z}\int_{|q|<\Lambda_\alpha} \frac{dq}{\sqrt{2\pi}}\\
    \label{eq:wannier_dirac_lattice_zw}
    &\cdot\Big\{\psi_{q-}^{(\alpha)}(ma) + \psi_{q+}^{(\alpha-1)}(ma)\Big\}\,,
\end{align}
where as usual we omit the terms with $\psi_{q+}^{(0)}$ and $\psi_{q-}^{(Z)}$ if $\alpha=1$ or $\alpha=Z$, respectively. As shown in Appendix~\ref{app:wannier_splitting} the region $|q|>\Lambda_\alpha$ contributes $\pm\delta w_{\text{u}/\text{d},\alpha}(ma)$ to the first/second term on the right hand side, i.e., precisely the same contribution (\ref{eq:delta_w_AF}) we used to extend half of a band for $\alpha=2,\dots,Z-1$ (for $\alpha=1,Z$ there is no contribution from $|q|>\frac{\pi}{Za}$). Therefore, after inserting (\ref{eq:psi_dirac}) and (\ref{eq:wannier_p_zero}), we obtain the following relation between the Wannier functions in lattice and Dirac theory
\begin{align}
    \label{eq:wannier_AF_dirac_lattice_u}
    w_{\text{u},\alpha}(ma) &= a\sqrt{Z} \sum_p \,w_{-,p}^{(\alpha)}(ma)e^{ipk_F^{(\alpha)}ma} \,,\\
    \label{eq:wannier_AF_dirac_lattice_d}
    w_{\text{d},\alpha}(ma) &= a\sqrt{Z} \sum_p \, w_{+,p}^{(\alpha-1)}(ma)e^{ipk_F^{(\alpha-1)}ma}\,.
\end{align}
Thereby, we will show in Appendix~\ref{app:wannier_splitting} that the momentum integral to define the Wannier functions 
$w_{\tau p}^{(\nu)}(x)$ via (\ref{eq:wannier_p_zero}) has to be regularized in such a way that unphysical contributions $\sim\delta(x)$ are absent, see the explicit calculation in Section~\ref{sec:wannier_universality}. 

To understand the relation of the various moments in the AF gauge, we need to evaluate $|w_{\alpha}(ma)|^2$ using the Wannier function from the sum of (\ref{eq:wannier_AF_dirac_lattice_u}) and (\ref{eq:wannier_AF_dirac_lattice_d}). The simplest cases are $\alpha=1$ or $\alpha=Z$, where only one term contributes: $w_{1}(ma)=w_u^{(1)}(ma)$ and $w^{(Z)}(ma)=w_d^{(1)}(ma)$, see (\ref{eq:wannier_1Z_ud}). For $\alpha=1$ we get with $k_F^{(1)}a=\pi/Z$
\begin{align}
    \nonumber
    |w_1(ma)|^2 &= a^2 Z \sum_p |w_{-,p}^{(1)}(ma)|^2 \\
    \label{eq:w_1_squared_exact}
    &\hspace{-1cm}
    + a^2 Z \sum_p w_{-,p}^{(1)}(ma) \, (w_{-,-p}^{(1)}(ma))^* e^{i2\pi pm/Z}\,, 
\end{align}
and a similar result for $\alpha=Z$ with $w_{-,p}^{(1)}\rightarrow w_{+,p}^{(Z-1)}$ and $e^{i2\pi pm/Z}\rightarrow e^{-i2\pi pm/Z}$. When inserting this formula into the definition (\ref{eq:moments_wannier_lattice}) of the moments and neglecting the variation of the slowly varying function $w_{-,p}^{(1)}(ma)$ over the unit cell, we find that the strongly oscillating terms of (\ref{eq:w_1_squared_exact}) can be neglected when averaging them over a unit cell. This leads to the result
\begin{align}
    \nonumber
    C_{r1}&\approx a^2 Z \sum_m (ma)^r \sum_p |w_{-,p}^{(1)}(ma)|^2 \\
    \label{eq:C_1_lattice_dirac}
    &\approx Za \int dx \,x^r \sum_p |w_{-,p}^{(1)}(x)|^2 =Za \,C_{r-}^{(1)}\,,\\
    \nonumber
    C_{rZ}&\approx a^2 Z \sum_m (ma)^r \sum_p |w_{+,p}^{(Z-1)}(ma)|^2 \\
    \label{eq:C_Z_lattice_dirac}
    &\approx Za \int dx \,x^r \sum_p |w_{+,p}^{(Z-1)}(x)|^2 = Za \,C_{r+}^{(Z-1)}\,.
\end{align}
For the bands $\alpha=2,\dots,Z-1$, both terms (\ref{eq:wannier_AF_dirac_lattice_u}) and (\ref{eq:wannier_AF_dirac_lattice_d}) contribute to the Wannier function. This leads to more oscillating terms occurring for the moments involving also $e^{i\pi p (2\alpha-1)m/Z}$ and $e^{i\pi p m/Z}$. Such terms give again a negligible contribution to the moments when averaging them over two unit cells, leading to the general result
\begin{align}
    \label{eq:C_r_1_lattice_dirac}
    C_{r\alpha}\approx  Za \left\{C_{r-}^{(\alpha)} + C_{r+}^{(\alpha-1)}\right\}\,,
\end{align}
together with
\begin{align}
    \label{eq:C_ud_r_1_lattice_dirac}
    C_{\text{u},r\alpha}\approx  Za \,C_{r-}^{(\alpha)} \quad,\quad 
    C_{\text{d},r\alpha}\approx  Za \,C_{r+}^{(\alpha-1)}\,.
\end{align}
Inserting (\ref{eq:C_ud_r_1_lattice_dirac}) in (\ref{eq:sct_wannier_dirac}) we find (\ref{eq:sct_wannier_u}), which states the surface charge theorem in terms of the upper component $w_{\text{u},\nu}$ of the lattice Wannier function for the highest valence band.

As discussed later in all detail in Section~\ref{sec:all_scales}, we note that the divergence of the Dirac Wannier functions $|w^{(\nu)}_{\tau p}(x)|^2\sim 1/x^2$ for $|x|\ll \xi^{(\nu)}$ is not important for the calculation of the moments for $r>0$ since the contributions from the region $|m|\sim O(1)$ can be neglected in the universal limit $\xi\gg a$. The only exception is $r=0$ where the normalization of the lattice Wannier function is fully determined by the region $|m|\sim O(1)$.

\subsubsection{The ML gauge}
\label{sec:wannier_lattice_dirac_ML}

In the ML gauge we have shown via the two last factors on the right hand side of (\ref{eq:gauge_factor_lattice_dirac}) that there is a difference between the gauges of maximally localized Wannier functions in lattice and Dirac theory. The last factor leads to a trivial shift of the position of the Wannier function $w^{(\nu)}_{\tau p}(x)$
\begin{align}
    \label{eq:w_shift}
    w^{(\nu)}_{\tau p}(x) \rightarrow w^{(\nu)}_{\tau p}(x - \frac{Za}{2\pi}\gamma_{\alpha})\,.
\end{align}
Including this shift and the phase factor $e^{-i\phi_{0,\tau}^{(\nu)}}$ we have to modify (\ref{eq:wannier_AF_dirac_lattice_u}) and (\ref{eq:wannier_AF_dirac_lattice_d}) in the following way (see Appendix~\ref{app:wannier_splitting} for details) to get the correct relationship between the maximally localized Wannier functions in lattice and Dirac theory 
\begin{align}
    \nonumber
    \tilde{w}_{\text{u},\alpha}(ma) &= \\ 
    \label{eq:wannier_ML_dirac_lattice_u}
    &\hspace{-1cm}
    = a\sqrt{Z} \sum_p e^{-i\phi_{0,-}^{(\alpha)}} \tilde{w}_{-,p}^{(\alpha)}(\tilde{m}_\alpha a)e^{ipk_F^{(\alpha)}ma} \,,\\
    \nonumber
    \tilde{w}_{\text{d},\alpha}(ma) &= \\
    \label{eq:wannier_ML_dirac_lattice_d}
    &\hspace{-1cm}
    = a\sqrt{Z} \sum_p e^{-i\phi_{0,+}^{(\alpha-1)}} \tilde{w}_{+,p}^{(\alpha-1)}(\tilde{m}_\alpha a)e^{ipk_F^{(\alpha-1)}ma}\,,
\end{align}
where $\tilde{m}_\alpha$ is defined in (\ref{eq:m_shift}). 

Using (\ref{eq:wannier_ML_dirac_lattice_u}) and (\ref{eq:wannier_ML_dirac_lattice_d}), we obtain analog to (\ref{eq:C_r_1_lattice_dirac}) from the definition (\ref{eq:D_moment}) for the moments in the ML gauge
\begin{align}
    \label{eq:tilde_D_r_1_lattice_dirac}
    \tilde{D}_{r\alpha}\approx  Za \left\{\tilde{C}^{(\alpha)}_{r-} + \tilde{C}^{(\alpha-1)}_{r+}\right\}\,,
\end{align}
together with
\begin{align}
    \label{eq:tilde_D_ud_r_1_lattice_dirac}
    \tilde{D}_{\text{u},r\alpha}\approx  Za \,\tilde{C}^{(\alpha)}_{r-} \quad,\quad
    \tilde{D}_{\text{d},r\alpha}\approx  Za \,\tilde{C}^{(\alpha-1)}_{r+}\,.
\end{align}
Inserting (\ref{eq:tilde_D_ud_r_1_lattice_dirac}) in (\ref{eq:sft_wannier_dirac}) we find (\ref{eq:sft_wannier_u}), which states the surface fluctuation theorem in terms of the upper component $\tilde{w}_{\text{u},\nu}$ of the lattice Wannier function for the highest valence band.

As discussed in more detail in Section~\ref{sec:all_scales}, the relations (\ref{eq:tilde_D_r_1_lattice_dirac}) and (\ref{eq:tilde_D_ud_r_1_lattice_dirac}) should be used only for even values of $r$. As shown in Section~\ref{sec:wannier_moments} the Dirac moments $\tilde{C}^{(\nu)}_{r\tau}$ are exactly zero for all odd values of $r$. However, this does not mean that the lattice moments $\tilde{D}_{r\alpha}$ and $\tilde{D}_{\text{u/d},r\alpha}$ are exactly zero for odd values of $r$ (except for $\tilde{D}_{1,\alpha}=0$ which is zero by definition). The smallness of the odd moments in lattice theory has to be understood by an order of magnitude analysis compared to the even moments. The odd moments are of subleading order $O(a^2\xi^{r-2})$ as compared to the even moments which are of $O(a\,\xi^{r-1})$. Therefore, in the limit of small gaps $\xi\gg a$, the odd moments are of no interest in the ML gauge and can be neglected. 

We note that corresponding relations of $\tilde{C}_{r\alpha}$ or $\tilde{C}_{\text{u/d},r\alpha}$ to the Dirac theory are not possible since the Dirac Wannier functions have a divergence $|\tilde{w}_{\tau p}^{(\nu)}(x)|^2\sim 1/x^2$ for $x\rightarrow 0$, see Section~\ref{sec:wannier_dirac_explicit}. Therefore, it is important to define the moments of the Dirac Wannier functions around the reference point $x=0$, otherwise they contain a divergence.

\section{Universality of Wannier functions}
\label{sec:wannier_universality}

This section is devoted to the most important result of this work that, in the case of small gaps, all lattice Wannier functions in AF or ML gauge show universal scaling in terms of a small set of universal scaling functions determining the shape of the Dirac Wannier functions in the corresponding gauges. To obtain this result we first calculate all eigenstates and the Wannier functions of the Dirac model analytically in Section~\ref{sec:wannier_dirac_explicit}. The fundamental universal scaling functions in the Dirac theory are then introduced in Section~\ref{sec:scaling_functions}, together with a summary of all their symmetry properties and asymptotic forms. The moments of the Dirac Wannier functions are discussed analytically in Section~\ref{sec:wannier_moments}, both in the AF and ML gauge. In Section~\ref{sec:wannier_scaling} we derive the universal scaling form of all lattice Wannier functions and their moments in terms of the fundamental scaling functions of the Dirac theory. The universal scaling is demonstrated explicitly for two examples with unit cells of size $2a$ and $3a$ in Sections~\ref{sec:universal_scaling_Z_2} and \ref{sec:universal_scaling_Z_3}, respectively. Finally, Section~\ref{sec:all_scales} contains a qualitative discussion of the properties of lattice Wannier functions and the scaling of their moments on all length scales, in particular including the one at small scales of the order of the lattice spacing. This analysis shows clearly that the visual impression of the square of lattice Wannier functions reveals only the trivial gapless limit, leading to the misleading visual impression of a localized wave function with spread determined by the lattice spacing. In contrast, the whole universal scaling properties on length scale $\xi$ are only visible when multiplying the lattice Wannier function with the spatial coordinate and then squaring it.

\subsection{Explicit evaluation of Dirac Wannier functions}
\label{sec:wannier_dirac_explicit}

The eigenfunctions (\ref{eq:psi_dirac}) of the Dirac model (\ref{eq:dirac_h}) in the AF and ML gauge are given by 
\begin{align}
    \label{eq:chi_tilde}
    |\tilde{\chi}_{q\tau}^{(\nu)}\rangle = e^{i\phi_{q\tau}^{(\nu)}}|\chi_{q\tau}^{(\nu)}\rangle = 
    \frac{1}{\sqrt{2\epsilon_q^{(\nu)}}} \left(\begin{array}{c} g_{q\tau}^{(\nu)} e^{i\gamma^{(\nu)}} \\ \tau g_{q\bar{\tau}}^{(\nu)} 
    \end{array}\right)\,,
\end{align}
where $\bar{\tau}=-\tau$, 
\begin{align}
    \label{eq:eps_g}
    \epsilon_q^{(\nu)} = \sqrt{(v_F^{(\nu)}q)^2 + (\Delta^{(\nu)})^2}\quad,\quad
    g_{q\tau}^{(\nu)} = \sqrt{\epsilon_q^{(\nu)} + \tau v_F^{(\nu)}q} \,,
\end{align}
and the gauge factor is given by
\begin{align}
    \label{eq:phi}
    e^{i\phi_{q\tau}^{(\nu)}} = \frac{g_{q\tau}^{(\nu)}e^{i\gamma^{(\nu)}}+\tau g_{q\bar{\tau}}^{(\nu)}}
    {|g_{q\tau}^{(\nu)}e^{i\gamma^{(\nu)}}+\tau g_{q\bar{\tau}}^{(\nu)}|}\,,
\end{align}
where
\begin{align}
    \label{eq:denominator}
    |g_{q\tau}^{(\nu)}e^{i\gamma^{(\nu)}}+\tau g_{q\bar{\tau}}^{(\nu)}| = \sqrt{2(\epsilon_q^{(\nu)} + \tau\Delta^{(\nu)}\cos{\gamma^{(\nu)}})}\,.
\end{align}
We note the helpful properties
\begin{align}
    \label{eq:g_properties}
    g_{q\tau}^{(\nu)}g_{q\bar{\tau}}^{(\nu)} = \Delta^{(\nu)} \quad,\quad g_{-q,\tau}^{(\nu)}=g_{q\bar{\tau}}^{(\nu)}\,.
\end{align}
As required for the AF gauge by (\ref{eq:AF_gauge}), we find with (\ref{eq:u_k_Z}) a positive $Z$-component for $u_{k\alpha}(Za)$ 
\begin{align}
    \label{eq:u_k_Z_explicit}
        u_{k\alpha}(Za) = \frac{1}{\sqrt{Z}} \sum_{p=\pm} \chi_{q\tau p}^{(\nu)}
        =\frac{|g_{q\tau}^{(\nu)}e^{i\gamma^{(\nu)}}+\tau g_{q\bar{\tau}}^{(\nu)}|}{\sqrt{2Z\epsilon_q^{(\nu)}}} > 0\,.
\end{align}

We note that the gauge factor (\ref{eq:phi}) allows for a unique and analytic definition of the phase $\phi_{q\tau}^{(\nu)}$ since
\begin{align}
    \label{eq:phi_def}
    \Sign\{\text{Im}(g_{q\tau}^{(\nu)}e^{i\gamma^{(\nu)}}+\tau g_{q\bar{\tau}}^{(\nu)})\} = s_{\gamma^{(\nu)}} \,.
\end{align}
Therefore, the phase $\phi_{q\tau}^{(\nu)}$ can be chosen to have the same sign as $\gamma^{(\nu)}$ for all $q$ (note that we take $-\pi < \gamma^{(\nu)} < \pi$). Using
\begin{align}
    \label{eq:phi_+_asymptotic}
    e^{i\phi_{q+}^{(\nu)}} &\rightarrow 
    \begin{cases} e^{i\gamma^{(\nu)}} &\text{for} \quad q\rightarrow\infty \\ 1 &\text{for} \quad q\rightarrow -\infty \end{cases}\,,\\
    \label{eq:phi_-_asymptotic}
    e^{i\phi_{q-}^{(\nu)}} &\rightarrow 
    \begin{cases} -1 &\text{for} \quad q\rightarrow\infty \\ e^{i\gamma^{(\nu)}} &\text{for} \quad q\rightarrow -\infty \end{cases}\,,
\end{align}
we get from 
\begin{align}
    \label{eq:dirac_berry_connection}
    A_{q\tau}^{(\nu)}=\partial_q \phi_{q\tau}^{(\nu)} = \frac{v_F^{(\nu)} \Delta^{(\nu)} \sin \gamma^{(\nu)}}{2 \epsilon_q^{(\nu)} (\epsilon_q^{(\nu)} + \tau \Delta^{(\nu)} \cos \gamma^{(\nu)})}
\end{align}
the result (\ref{eq:zak_berry_phase_vs_gap_phase}) for the Zak-Berry phases.

We note that $\sin{\gamma^{(\nu)}}=0$ is a special point where the Dirac Zak-Berry connection is zero. For a half-infinite system with $x>0$ it can be shown \cite{pletyukhov_etal_prr_20} that an edge state is present in the gap for $\gamma^{(\nu)}>0$ with  energy $\epsilon^{(\nu)}_{\text{E}}=-\Delta^{(\nu)}\cos{\gamma^{(\nu)}}$. Therefore, at this special point, the edge state energy touches either the higher (for $\gamma^{(\nu)}=\pm\pi$) or the lower band (for $\gamma^{(\nu)}=0$), and the Wannier functions of the corresponding bands have very special properties, they do not decay exponentially (for $w^{(\nu)}_{\tau p}(x)$ in AF gauge) or change discontinuously (for $\tilde{w}^{(\nu)}_{\tau p}(x)$ in ML gauge). Therefore, we exclude the cases $\gamma^{(\nu)}=0, \tau=-$ and $\gamma^{(\nu)}=\pm\pi, \tau=+$ in the following and discuss them separately in Appendix~\ref{app:sin_gamma_zero}. 

For the phase factors $e^{-i\phi_{0,\tau}^{(\nu)}}$ occurring in (\ref{eq:wannier_ML_dirac_lattice_u}) and (\ref{eq:wannier_ML_dirac_lattice_d}) we get from (\ref{eq:phi}) the result
\begin{align}
    \nonumber
    e^{-i\phi_{0,\tau}^{(\nu)}} &= \frac{e^{-i\gamma^{(\nu)}}+\tau}{|e^{-i\gamma^{(\nu)}}+\tau|}\\
    \label{eq:phi_0_explicit}
    &= e^{-i\gamma^{(\nu)}/2} \begin{cases} 1 & \text{for} \quad \tau=+ \\ -i\, s_{\gamma^{(\nu)}} & \text{for} \quad \tau=- \end{cases} \,,
\end{align}
where we used the convention $-\pi < \gamma^{(\nu)} < \pi$.

Using (\ref{eq:chi_tilde}), (\ref{eq:phi}) and (\ref{eq:denominator}), we find for $\chi_{q\tau p}^{(\nu)}$ the form
\begin{align}
    \label{eq:chi}
    \chi_{q\tau p}^{(\nu)} = \frac{\epsilon_q^{(\nu)} + p\tau v_F^{(\nu)} q + \tau \Delta^{(\nu)} e^{ip\gamma^{(\nu)}}} {2\sqrt{\epsilon_q^{(\nu)}(\epsilon_q^{(\nu)}+\tau\Delta^{(\nu)}\cos\gamma^{(\nu)})}}\,.
\end{align}
Inserting the forms (\ref{eq:chi_tilde}) and (\ref{eq:chi}) for $\tilde{\chi}_{q\tau}^{(\nu)}$ and $\chi_{q\tau}^{(\nu)}$
in (\ref{eq:wannier_p_zero}), and introducing the dimensionless variables
\begin{align}
    \label{eq:dimensionless}
    \bar{q}=2\xi^{(\nu)} q \,\,,\,\, \bar{x}=\frac{x}{2\xi^{(\nu)}}\,\,,\,\,\bar{\epsilon}_{\bar{q}}=\frac{\epsilon_q^{(\nu)}}{\Delta^{(\nu)}}=\sqrt{1+\bar{q}^2}\,,
\end{align}
with $\xi^{(\nu)}=v_F^{(\nu)}/(2\Delta^{(\nu)})$, we find for the Wannier functions the integral representation
\begin{align}
    \label{eq:w_dirac}
    w^{(\nu)}_{\tau p}(x)&=\frac{1}{4\pi\xi^{(\nu)}}\int d\bar{q} \,e^{-\bar{\eta}|\bar{q}|} e^{i\bar{q}\bar{x}} 
    \chi_{\tau p}^{(\nu)}(\bar{q})\,, \\
    \label{eq:tilde_w_dirac}
    \tilde{w}^{(\nu)}_{\tau p}(x)&=\frac{1}{4\pi\xi^{(\nu)}}\int d\bar{q}\,e^{-\bar{\eta}|\bar{q}|} e^{i\bar{q}\bar{x}}
    \tilde{\chi}_{\tau p}^{(\nu)}(\bar{q})\,,
\end{align}
where
\begin{align}
    \label{eq:chi_bar_q}    
    \chi_{\tau p}^{(\nu)}(\bar{q}) &= \frac{\bar{\epsilon}_{\bar{q}} + p\tau\bar{q} + \tau e^{ip\gamma^{(\nu)}}} {2\sqrt{\bar{\epsilon}_{\bar{q}}}\sqrt{\bar{\epsilon}_{\bar{q}}+\tau\cos\gamma^{(\nu)}}}\,,\\
    \label{eq:tilde_chi_bar_q}    
    \tilde{\chi}_{\tau p}^{(\nu)}(\bar{q}) &= 
    \frac{\sqrt{\bar{\epsilon}_{\bar{q}} + p\tau\bar{q}}}{\sqrt{2 \bar{\epsilon}_{\bar{q}}}}
    \begin{cases} e^{i\gamma^{(\nu)}} & \text{for}\quad p=+ \\ \tau & \text{for}\quad p=- \end{cases}\,.
\end{align}
We have included a convergence factor $e^{-\bar{\eta}|\bar{q}|}$, with $\bar{\eta}\sim a/\xi^{(\nu)}$, since the integrals diverge for large $|\bar{q}|$. This divergence occurs since a low-energy theory can only describe the universal regime $|x|\gg a$. As shown in Appendix~\ref{app:wannier_integral}, one can also find a more convenient integral representation by closing the integration contour for $s_x=\Sign(x)=\pm$ in the upper/lower half, respectively, leading to convergent integrals such that the limit $\bar{\eta}\rightarrow 0$ can be performed under the integral. This regularisation is possible for all $x\ne 0$ and removes an unphysical contribution $\sim\delta(x)$ from the Dirac Wannier function, see Appendices~\ref{app:wannier_splitting} and \ref{app:wannier_scaling_real_q}. In this way we obtain the universal representation
\begin{align}
    \label{eq:w_conv}
    w^{(\nu)}_{\tau p}(x)&=\frac{is_x}{4\pi\xi^{(\nu)}}\int_0^\infty d\kappa \,e^{-|\bar{x}|\kappa} 
    \delta \chi_{\tau p}^{(\nu)}(is_x\kappa)\,,\\ 
    \label{eq:tilde_w_conv}
    \tilde{w}^{(\nu)}_{\tau p}(x)&=\frac{is_x}{4\pi\xi^{(\nu)}}\int_0^\infty d\kappa \,e^{-|\bar{x}|\kappa} 
    \delta\tilde{\chi}_{\tau p}^{(\nu)}(is_x\kappa)\,,
\end{align}
where we defined $s_x=\Sign(x)$, and 
\begin{align}
    \label{eq:dI_def}
    \delta \chi_{\tau p}^{(\nu)}(is_x\kappa) &= \chi_{\tau p}^{(\nu)}(is_x\kappa + 0^+) - \chi_{\tau p}^{(\nu)}(is_x\kappa - 0^+)\,, \\
    \label{eq:tilde_dI_def}
    \delta\tilde{\chi}_{\tau p}^{(\nu)}(is_x\kappa) &= \tilde{\chi}_{\tau p}^{(\nu)}(is_x\kappa + 0^+) - 
    \tilde{\chi}_{\tau p}^{(\nu)}(is_x\kappa - 0^+)\,, 
\end{align}
describe the jump of the integrands across the branch cut along the imaginary axis which emerge from the various square roots. Alternatively, we show in Appendix~\ref{app:wannier_scaling_real_q} how the Wannier functions can be represented via convergent momentum integrals on the real axis. 

The results (\ref{eq:w_conv}) and (\ref{eq:tilde_w_conv}) provide the form of the Wannier functions in the universal regime $|x|,\xi^{(\nu)}\gg a$. However, as long as $\xi^{(\nu)}\gg a$, the lattice Wannier functions at $x=ma$ can be reproduced for {\it all} $m$ (even for $m=0$) from the Dirac Wannier functions via the sum of (\ref{eq:wannier_AF_dirac_lattice_u}) and (\ref{eq:wannier_AF_dirac_lattice_d}) (in the AF gauge) or the sum of (\ref{eq:wannier_ML_dirac_lattice_u}) and (\ref{eq:wannier_ML_dirac_lattice_d}) (in the ML gauge). As shown in Appendix~\ref{app:wannier_splitting}, this arises from the fact that the high-momentum region $|q|>\Lambda_\alpha$ does not contribute to the total Wannier function of a certain band $\alpha$. 

\subsection{The universal scaling functions}
\label{sec:scaling_functions}

It is convenient to express the complex Dirac Wannier functions $w^{(\nu)}_{\tau p}(x)$ and $\tilde{w}^{(\nu)}_{\tau p}(x)$ in terms of real functions $w_{A/B,\tau}^{(\nu)}(x)$ and $\tilde{w}_{A/B,\tau}^{(\nu)}(x)$ by writing
\begin{align}
    \label{eq:w_p_decomposition}
    w^{(\nu)}_{\tau p}(x) &= \frac{1}{2}\Big[w_{B,\tau}^{(\nu)}(x) + i p \, w_{A,\tau}^{(\nu)}(x)\Big]\,,\\
    \label{eq:tilde_w_p_decomposition}
    \tilde{w}^{(\nu)}_{\tau p}(x) &= e^{i\phi_{0,\tau}^{(\nu)}} \frac{1}{2}\Big[\tilde{w}_{B,\tau}^{(\nu)}(x) 
    + i p \tilde{w}_{A,\tau}^{(\nu)}(x)\Big]\,,
\end{align}
which corresponds to the definitions
\begin{align}    
    \label{eq:w_A_def}
    w_{A,\tau}^{(\nu)}(x) &= -i\sum_p p\,w^{(\nu)}_{\tau p}(x) = w_{A,\tau}^{(\nu)}(x)^*\,,\\
    \label{eq:w_B_def}
    w_{B,\tau}^{(\nu)}(x) &= \sum_p w^{(\nu)}_{\tau p}(x) = w_{B,\tau}^{(\nu)}(x)^*\,,\\
    \label{eq:tilde_w_A_def}
    \tilde{w}_{A,\tau}^{(\nu)}(x) &= -i\,e^{-i\phi_{0,\tau}^{(\nu)}}\sum_p p\,\tilde{w}^{(\nu)}_{\tau p}(x) = \tilde{w}_{A,\tau}^{(\nu)}(x)^* \,,\\
    \label{eq:tilde_w_B_def}
    \tilde{w}_{B,\tau}^{(\nu)}(x) &= e^{-i\phi_{0,\tau}^{(\nu)}} \sum_p \tilde{w}^{(\nu)}_{\tau p}(x) = \tilde{w}_{B,\tau}^{(\nu)}(x)^*\,.
\end{align}
The fundamental universal and dimensionless scaling functions are then defined by
\begin{align}
    \label{eq:F_AB_scaling}
    F_{A/B,\tau}(x/\xi^{(\nu)};\gamma^{(\nu)}) &= x \,w_{A/B,\tau}^{(\nu)}(x) \,,\\ 
    \label{eq:tilde_F_AB_scaling}
    \tilde{F}_{A/B,\tau}(x/\xi^{(\nu)};\gamma^{(\nu)}) &= x \,\tilde{w}_{A/B,\tau}^{(\nu)}(x) \,,
\end{align}
in terms of which the moments (\ref{eq:moments_wannier}) can be expressed as
\begin{align}
    \label{eq:moments_AF_FAB}
    C_{r\tau}^{(\nu)} &= \frac{1}{2} (\xi^{(\nu)})^{r-1}\int dy y^{r-2}\left\{F_{A,\tau}(y)^2 + F_{B,\tau}(y)^2\right\} \,,\\
    \label{eq:moments_ML_FAB}
    \tilde{C}_{r\tau}^{(\nu)} &= \frac{1}{2} (\xi^{(\nu)})^{r-1}\int dy y^{r-2}
    \left\{\tilde{F}_{A,\tau}(y)^2 + \tilde{F}_{B,\tau}(y)^2\right\} \,.
\end{align}
The explicit form of the scaling functions in terms of integrations along the real momentum axis are derived in Appendix~\ref{app:wannier_scaling_real_q} and are explicitly presented in (\ref{eq:F_A_comapct}-\ref{eq:tilde_F_B_+_comapct}). These useful forms allow for a direct numerical evaluation. 

In Appendix~\ref{app:dirac_wannier_properties} we have listed a number of properties of the Dirac Wannier functions $w^{(\nu)}_{\tau p}(x)$ and $\tilde{w}^{(\nu)}_{\tau p}(x)$ under inversion of $p$, $\tau$, or $x$ and certain transformations of $\gamma^{(\nu)}$, together with corresponding properties of $w_{A/B,\tau}^{(\nu)}(x)$ and $\tilde{w}_{A/B,\tau}^{(\nu)}(x)$. As a consequence we note the following useful properties of the universal scaling functions $F_{A/B,\tau}(y;\gamma)$ and $\tilde{F}_{A/B,\tau}(y;\gamma)$ under inversion of $y$, $\tau$ or $\gamma$ and under the transformation $\gamma\rightarrow -\gamma + \pi s_\gamma$ (alternatively, these relations follow directly from the explicit forms (\ref{eq:F_A_comapct}-\ref{eq:tilde_F_B_+_comapct}) of the scaling functions)
\begin{align}
    \label{eq:F_A_-y_trafo}
    F_{A,\tau}(-y;\gamma) &= F_{A,\tau}(y;-\gamma) \,,\\
    \label{eq:F_B_-y_trafo}
    F_{B,\tau}(-y;\gamma) &= - F_{B,\tau}(y;\gamma) = - F_{B,\tau}(y;-\gamma) \,,\\
    \nonumber
    \tilde{F}_{A,\tau}(-y;\gamma) &= \\
    \label{eq:tilde_F_A_gamma_trafo}
    &\hspace{-1cm}
    =\tilde{F}_{A,\tau}(y;-\gamma) = -\tau s_\gamma \tilde{F}_{B,\tau}(y;-\gamma+\pi s_\gamma) \\
    \label{eq:tilde_F_A_-y_trafo}
    &\hspace{-1cm}
    =\tau\left\{\cos\gamma \, \tilde{F}_{A,\tau}(y;\gamma)
    - \sin\gamma \, \tilde{F}_{B,\tau}(y;\gamma)\right\}\,,\\
    \nonumber
    \tilde{F}_{B,\tau}(-y;\gamma) &= \\
    \label{eq:tilde_F_B_gamma_trafo}
    &\hspace{-1cm}
    = - \tilde{F}_{B,\tau}(y;-\gamma) = -\tau s_\gamma \tilde{F}_{A,\tau}(y;-\gamma+\pi s_\gamma) \\
    \label{eq:tilde_F_B_-y_trafo}
    &\hspace{-1cm}
    =-\tau\left\{\sin\gamma \, \tilde{F}_{A,\tau}(y;\gamma)
    + \cos\gamma \, \tilde{F}_{B,\tau}(y;\gamma)\right\}\,,\\
    \label{eq:F_A_-tau_trafo}
    F_{A,-\tau}(y;\gamma) &= - F_{A,\tau}(y;-\gamma+\pi s_{\gamma}) \,,\\
    \label{eq:F_B_-tau_trafo}
    F_{B,-\tau}(y;\gamma) &= F_{B,\tau}(y;-\gamma+\pi s_{\gamma}) \,,\\
    \label{eq:tilde_F_A_-tau_trafo}
    \tilde{F}_{A,-\tau}(y;\gamma) &=  \tau \, s_{\gamma} \, \tilde{F}_{B,\tau}(-y;\gamma) \,,\\
    \label{eq:tilde_F_B_-tau_trafo}
    \tilde{F}_{B,-\tau}(y;\gamma) &=  -\tau \, s_{\gamma} \, \tilde{F}_{A,\tau}(-y;\gamma) \,.
\end{align}
For the special case $\sin{\gamma}=0$ we list all properties in Appendix~\ref{app:sin_gamma_zero}.

\begin{figure}[t!]
    \centering
	 \includegraphics[width =1.0\columnwidth]{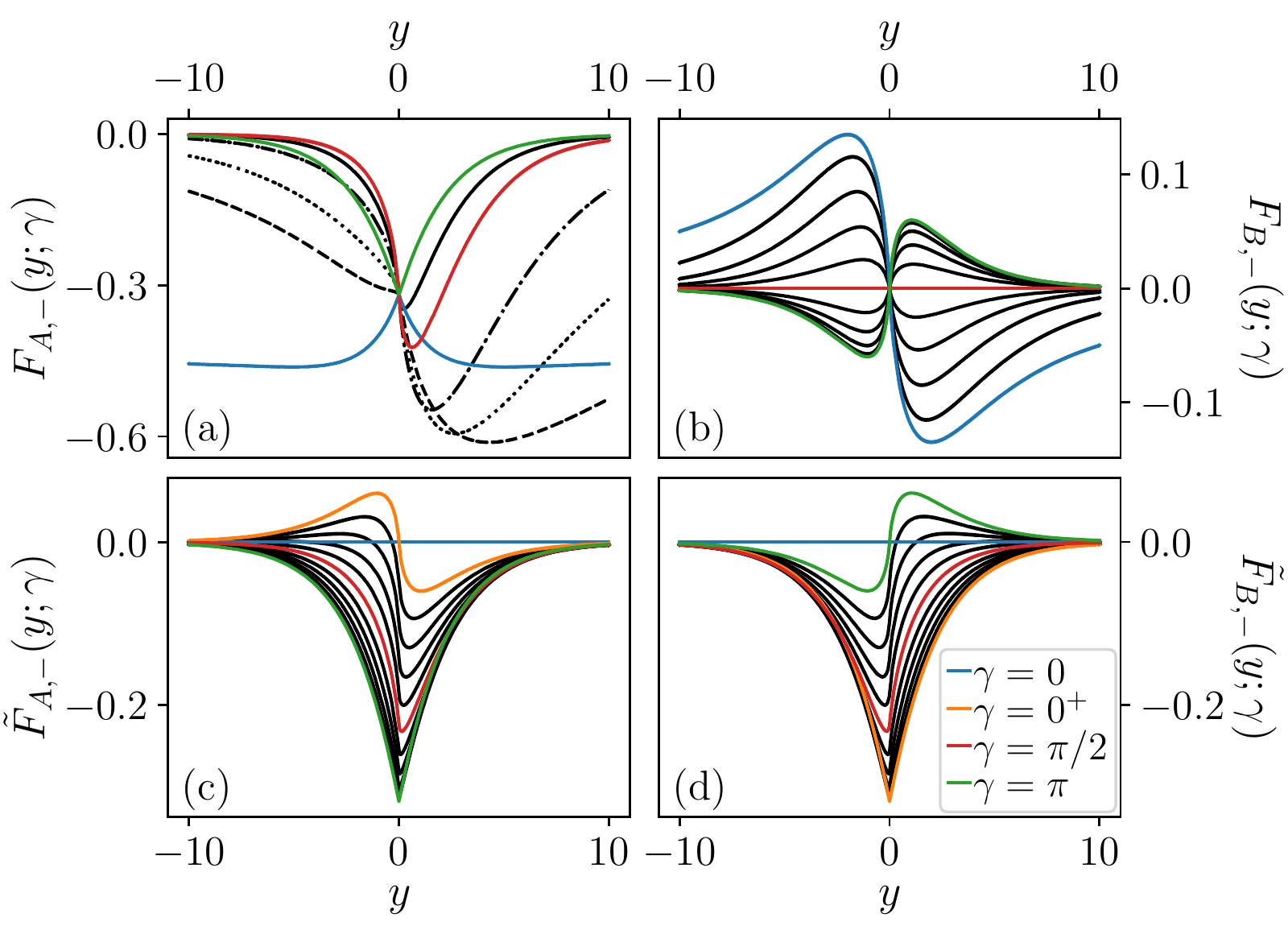}
	 %{scaling_functions.jpg}
	  \caption{The scaling functions $F_{A/B,-}(y;\gamma)$ and $\tilde{F}_{A/B,-}(y;\gamma)$ for various phases $\gamma$. We show only $\tau=-$ and consider $0\le\gamma\le \pi$ since all other cases follow from the transformation laws (\ref{eq:F_A_-y_trafo}-\ref{eq:tilde_F_B_-tau_trafo}) as $F_{A/B,+}(y;\gamma)=\mp F_{A/B,-}(y;\pi-\gamma)$, $\tilde{F}_{A/B,+}(y;\gamma)= \mp\tilde{F}_{A/B,-}(-y;\gamma)$, 
	  $F_{A/B,\tau}(y;-\gamma)=\pm F_{A/B,\tau}(-y;\gamma)$, and $\tilde{F}_{A/B,\tau}(y;-\gamma)= \pm\tilde{F}_{A/B,\tau}(-y;\gamma)$. According to (\ref{eq:tilde_F_A_gamma_trafo}), we note the relation $\tilde{F}_{B,-}(y;\gamma)=\tilde{F}_{A,-}(-y;\pi-\gamma)$. All scaling functions are exponentially decaying for $|y|\gg 1$, except for $F_{A,-}(y;0)\rightarrow - \sqrt{2}/\pi$ and $F_{B,-}(y;0)\rightarrow -\sqrt{2}/(\pi y)$, see (\ref{eq:F_A_asymptotic_sin_gam_zero}) and (\ref{eq:F_B_asymptotic_sin_gam_zero}). The scaling functions $\tilde{F}_{A/B,-}(y;\gamma)$ change discontinuously at $\gamma=0$ by a sign change, see (\ref{eq:tilde_F_A_gam_zero}) and (\ref{eq:tilde_F_B_gam_zero}). The cases $\gamma=0, 0^+, \pi/2, \pi$ are indicated by blue, orange, red and green, respectively.
	  The grid for the values of $\gamma$ is defined by $0.1\pi\cdot n$ with $n=0,1,...10$ in (b)-(d). 
	  In (a) the black curves show $\gamma=0.05 \pi$ (dashed line), $0.1 \pi$ (dotted), $0.2 \pi$ (dashdotted) and $0.75 \pi$ (solid).
	  } 
    \label{fig:scaling_functions}
\end{figure}
The typical form of the four fundamental scaling functions $F_{A/B,-}(y;\gamma)$ and $\tilde{F}_{A/B,-}(y;\gamma)$ are illustrated in Fig.~\ref{fig:scaling_functions}. We show only the lower band $\tau=-$ and positive values of the phase $\gamma$ since a sign change of $\tau$ or $\gamma$ is covered by the above properties. A special case is $\cos{\gamma}=0$, where 
\begin{align}
    \label{eq:F_B_cos_zero}
    F_{B,\tau}(y;\gamma) = 0 \quad \text{for}\,\cos\gamma=0\,.
\end{align}
The asymptotic forms for small $|y|\ll 1$ and large $|y|\gg 1$ can be analysed analytically, see Appendix~\ref{app:wannier_integral}, where we derive the corresponding asymptotic behavior of the Dirac Wannier functions in the AF and ML gauge for small $|x|\ll\xi^{(\nu)}$ and large $|x|\gg\xi^{(\nu)}$. For $y=0$ we get from (\ref{eq:w_small_x}) and (\ref{eq:tilde_w_small_x})
\begin{align}
    \label{eq:F_A_small_y}
    F_{A,\tau}(0;\gamma) &= \frac{\tau}{\pi}\,,\\
    \label{eq:F_B_small_y}
    F_{B,\tau}(0;\gamma) &= 0\,,\\
    \label{eq:tilde_F_A_small_y}
    \tilde{F}_{A,\tau}(0;\gamma) &= \frac{\tau}{2\pi}|e^{i\gamma}+\tau|\,,\\
    \label{eq:tilde_F_B_small_y}
    \tilde{F}_{B,\tau}(0;\gamma) &= -\frac{s_{\gamma}}{2\pi}|e^{i\gamma}-\tau|\,.
\end{align}
The asymptotic behavior for large $|y|\gg 1$ follows from (\ref{eq:w_large_x_1}), (\ref{eq:w_large_x_2}) and (\ref{eq:w_large_x_3})  as
\begin{align}
    \nonumber
    &\underline{\tau=\Sign(\cos\gamma)}:\\
    \label{eq:F_A_large_y_1}
    & F_{A,\tau}(y;\gamma)\rightarrow 
    \frac{\tau (1 + s_y\sin\gamma)\Gamma\left(\frac{3}{4}\right)}{2\pi\sqrt{|\cos\gamma|}}
    |y|^{\frac{1}{4}} e^{-\frac{|y|}{2}}\,,\\
    \label{eq:F_B_large_y_1}
    & F_{B,\tau}(y;\gamma)\rightarrow 
    \frac{\Gamma\left(\frac{3}{4}\right)\sqrt{|\cos\gamma|}}{2\pi}
    s_y|y|^{\frac{1}{4}} e^{-\frac{|y|}{2}}\,,\\
    \nonumber
    &\underline{\tau=-\Sign(\cos\gamma)}:\\
    \nonumber
    & F_{A,\tau}(y;\gamma)\rightarrow 
    \frac{\tau}{\sqrt{2\pi|\sin\gamma|}}|y|^{-\frac{1}{2}} e^{-|\sin\gamma|\frac{|y|}{2}}\\
    \label{eq:F_A_large_y_2}
    &\hspace{1cm}
    +\tau(1+s_ys_{\gamma})\sqrt{\frac{|\sin\gamma|}{2\pi}}\,
    |y|^{\frac{1}{2}} e^{-|\sin\gamma|\frac{|y|}{2}}\,,\\
    \label{eq:F_B_large_y_2}
    & F_{B,\tau}(y;\gamma)\rightarrow 
    -\frac{\sqrt{|\sin\gamma|}}{\sqrt{2\pi}\,|\cos\gamma|}
    s_y|y|^{-\frac{1}{2}} e^{-|\sin\gamma|\frac{|y|}{2}}\,,\\
    \nonumber
    &\underline{\cos\gamma=0}:\\
    \label{eq:F_A_large_y_3}
    & F_{A,\tau}(y;\gamma)\rightarrow \frac{\tau}{2\sqrt{\pi}}\left(|y|^{-\frac{1}{2}} + (1 + s_y s_{\gamma})|y|^{\frac{1}{2}}\right) 
    e^{-\frac{|y|}{2}} \,,\\
    \label{eq:F_B_large_y_3}
    & F_{B,\tau}(y;\gamma) = 0\,.
\end{align}
We note that the last result (\ref{eq:F_B_large_y_3}) holds for all $y$ in the special case $\cos\gamma=0$, see (\ref{eq:F_B_cos_zero}). From (\ref{eq:tilde_w_large_x}) we get for all cases
\begin{align}
    \nonumber
    & \tilde{F}_{A,\tau}(y;\gamma)\rightarrow 
    \frac{\Gamma\left(\frac{3}{4}\right)}{4\pi}|y|^{\frac{1}{4}} e^{-\frac{|y|}{2}}\\
    \label{eq:tilde_F_A_large_y}
    &\hspace{2cm}\cdot\tau\left(s_y  s_{\gamma}\, |e^{i\gamma}-\tau| + |e^{i\gamma}+\tau|\right) \,,\\
    \nonumber
    & \tilde{F}_{B,\tau}(y;\gamma)\rightarrow 
    \frac{\Gamma\left(\frac{3}{4}\right)}{4\pi}|y|^{\frac{1}{4}} e^{-\frac{|y|}{2}}\\
    \label{eq:tilde_F_B_large_y}
    &\hspace{2cm}\cdot\left(s_y |e^{i\gamma}+\tau| - s_{\gamma} \,|e^{i\gamma}-\tau|\right) \,.
\end{align}

As discussed after (\ref{eq:dirac_berry_connection}), we remind that the cases $\gamma=0, \tau=-$ and $\gamma=\pm\pi, \tau=+$ are excluded. At these points the scaling functions $F_{A/B,\tau}(y;\gamma)$ are not exponentially decaying
\begin{align}
    \label{eq:F_A_asymptotic_sin_gam_zero}
    F_{A,-}(y;0) &= - F_{A,+}(y;\pi) \rightarrow - \frac{\sqrt{2}}{\pi} \,,\\
    \label{eq:F_B_asymptotic_sin_gam_zero}
    F_{B,-}(y;0) &=  F_{B,+}(y;\pi) \rightarrow - \frac{\sqrt{2}}{\pi}\,\frac{1}{y} \,,
\end{align}
and the scaling functions $\tilde{F}_{A/B,\tau}(y;\gamma)$ change discontinuously by a sign change (with vanishing value exactly at $\gamma=0,\pi$)
\begin{align}
    \label{eq:tilde_F_A_gam_zero}
    \tilde{F}_{A,-}(y;0^\pm) &= \mp \tilde{F}_{B,+}(y;0) \,,\\
    \label{eq:tilde_F_B_gam_zero}
    \tilde{F}_{B,-}(y;0^\pm) &= \mp \tilde{F}_{A,+}(y;0) \,,\\
    \label{eq:tilde_F_A_gam_pi}
    \tilde{F}_{A,+}(y;\pi \pm 0^+) &= \mp \tilde{F}_{B,-}(y;\pi) \,,\\
    \label{eq:tilde_F_B_gam_pi}
    \tilde{F}_{B,+}(y;\pi \pm 0^+) &= \mp \tilde{F}_{A,-}(y;\pi) \,,
\end{align}
see Appendix~\ref{app:sin_gamma_zero} for details. Although it might have been expected that the Wannier functions in the ML gauge change discontinuously as function of $\gamma$ when an edge state touches the band edge, it is quite remarkable that the Wannier functions in the AF gauge stay continuous but show a non-exponential decay exactly at the touching point although the gap is still present. Up to our best knowledge this has not been reported before.   

In summary we find an exponential decay with a pre-exponential power law for $|y|\gg 1$ (similar power-laws for the pre-exponential functions have been obtained in Ref.~\onlinecite{he_vanderbilt_prl_01} for a variety of other localized single-particle wave functions) whereas, for $|y|\lesssim 1$, we obtain the scaling
\begin{align}
|y|\lesssim 1 \quad\Rightarrow\quad F_{A/B,\tau}(y;\gamma), \tilde{F}_{A/B,\tau}(y;\gamma)\sim O(1)\,. 
\end{align}
This scaling together with the exponential behavior at large $y$ is the essential reason why the moments scale as
\begin{align}
    C_{r\tau}^{(\nu)}, \tilde{C}_{r\tau}^{(\nu)} \sim (\xi^{(\nu)})^{r-1}\,,
\end{align}
leading via (\ref{eq:C_r_1_lattice_dirac}) and (\ref{eq:tilde_D_r_1_lattice_dirac}) to the universal scaling $a\xi$ (with $\xi\sim\xi^{(\alpha)},\xi^{(\alpha-1)}$) for the quadratic spread of the lattice Wannier functions. This is in contrast to the visual impression of the lattice Wannier functions leading to the incorrect scaling $a^2$, as we will discuss in all detail in Section~\ref{sec:all_scales}.

\subsection{Moments of Dirac Wannier functions}
\label{sec:wannier_moments}

The moments in the ML gauge are given by
\begin{align}
    \nonumber
    \tilde{C}_{r\tau}^{(\nu)} &= \sum_{p=\pm} \int \frac{dq}{2\pi} (\tilde{\chi}_{q\tau p}^{(\nu)})^*
    (i\partial_q)^r \tilde{\chi}_{q \tau p}^{(\nu)}  \\
    &\hspace{-1cm}
    = \frac{i^r (2 \xi^{(\nu)})^{r-1}}{4 \pi} \sum_{p=\pm} \int d \bar{q} \sqrt{1 + \frac{p \bar{q}}{\bar{\epsilon}_{\bar{q}}}}\partial_{\bar{q}}^r \sqrt{1 + \frac{p \bar{q}}{\bar{\epsilon}_{\bar{q}}}}.
\end{align}
Noticing that only even moments ($r=2 l$) are nonzero, we rewrite this expression in the symmetrized form
\begin{align}
    \label{eq:tilde_w_mom_scaling}  
    \tilde{C}_{2l,\tau}^{(\nu)} &= (\xi^{(\nu)})^{2l-1} \tilde{c}_{2l}\,,  \\
    \label{eq:tilde_w_mom_coef}
    \tilde{c}_{2l} &= \frac{2^{2l-2}}{2 \pi} \sum_{p=\pm} \int d \bar{q} \left[ \partial_{\bar{q}}^l \sqrt{1 + \frac{p \bar{q}}{\bar{\epsilon}_{\bar{q}}}} \right]^2 > 0\,.
\end{align}
In particular we find that the moments are independent of $\tau$ and depend on $\nu$ only via the decay length $\xi^{(\nu)}$.

\begin{figure}[b!]
    \centering
	 \includegraphics[width =0.9\columnwidth]{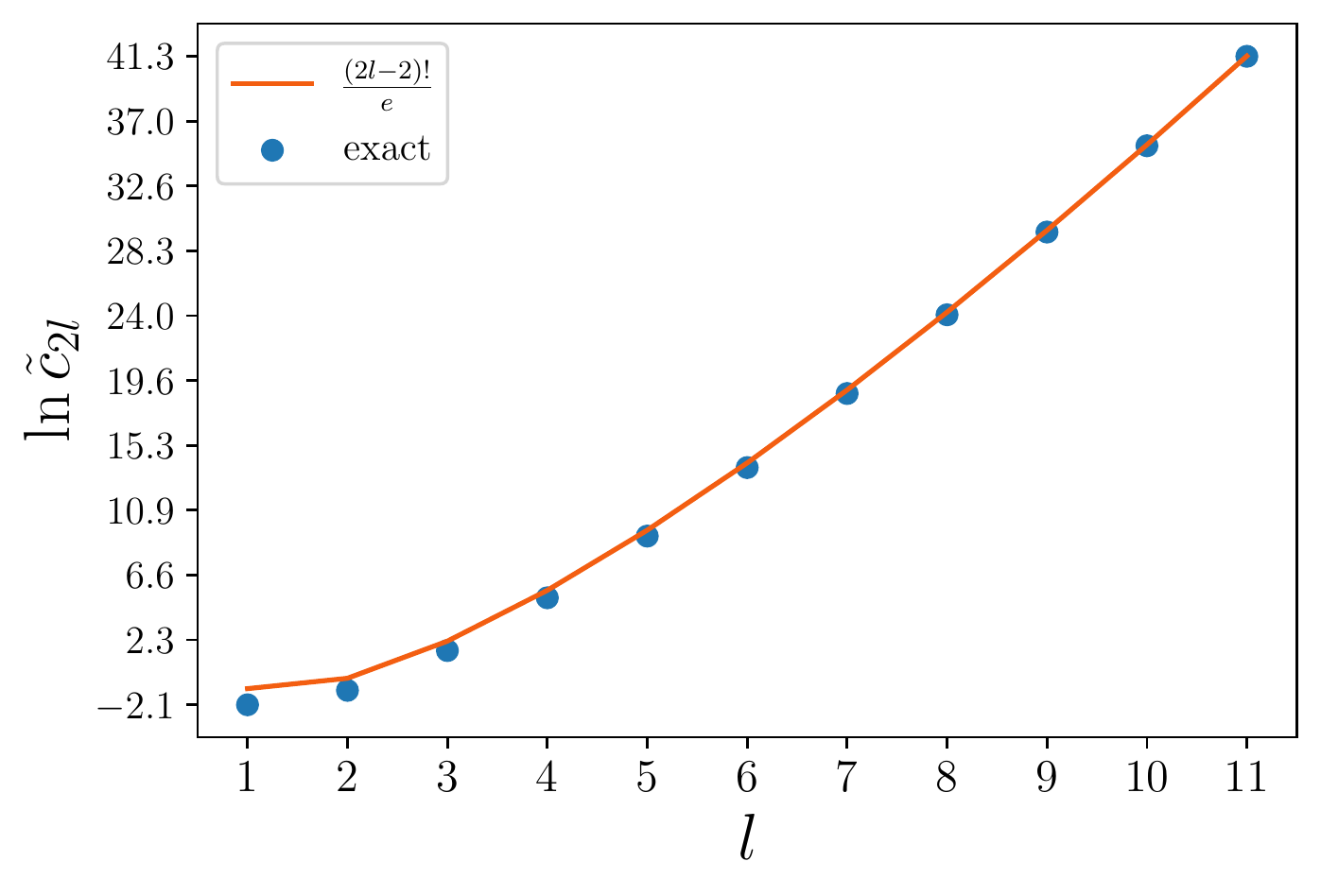}
	  \caption{Logarithmic plot of first $10$ coefficients $\bar{c}_{2l}$ and of their asymptotic value $\frac{(2l-2)!}{e}$. As one can notice, the sequence is very well approximated by its asymptotic even for moderate values of $l$. 
	  } 
    \label{fig:moments}
\end{figure}

The scaling behavior of $\tilde{C}_{2l,\tau}^{(\nu)}$ is given by \eqref{eq:tilde_w_mom_scaling}, while Eq.~\eqref{eq:tilde_w_mom_coef} provides positive dimensionless coefficients. Making the integration variable change $\bar{q} = \tan \theta$, we conveniently represent them in the form
\begin{align}
    \label{eq:til_c2l}
    \tilde{c}_{2l}&=\frac{1}{4 \pi}\int_{-\pi}^{\pi} d \theta  \left[  \cos \theta \left(  \frac{\partial}{\partial \theta} 2 \cos^2 \theta \right)^{l-1} \sin \frac{\theta}{2}  \right]^2.
\end{align}
In particular, the second moment is recovered from
\begin{align}
    \label{eq:tilde_C_2_Dirac}
    \tilde{C}^{(\nu)}_{2\tau}&=\frac{\xi^{(\nu)}}{4 \pi}\int_{-\pi}^{\pi} d \theta  \cos^2 \theta  \sin^2 \frac{\theta}{2} = \frac{\xi^{(\nu)}}{8}. 
\end{align}
At large $l$ the sequence \eqref{eq:til_c2l} is well approximated by $\frac{(2l-2)!}{e}$, see Fig.~\ref{fig:moments}. 
%Replacing $\frac{1}{e} \to (1+\frac{1}{2l-1})^{-(2l+1)}$, which leads to the approximation $\tilde{c}_{2l} \approx (2l)! \frac{(2l-1)^{2l}}{(2l)^{2l+2}}$, we can even get the right value of the second moment with the same large-$l$ asymptotic behavior. However, the fourth and higher  moments are not exactly reproduced. \mikpl{(The approximate moments $\frac{(2l-1)!}{e}$ correspond to the density distribution $f (x) = \frac{e^{-|x|}}{2 e x^2}$, see the file "Wan\_moments.tex". Up to the multiplicative constant this result is consistent with \eqref{eq:tilde_w_small_x}, which suggests to use $\frac{1}{\pi^2}$ instead of $\frac{1}{e}$ above. Note that these values give the second moment in the corresponding distributions of this functional form. The correct value $\frac18$ lies in between, $\frac{1}{\pi^2}<\frac18 < \frac{1}{e}$.)}

In the AF gauge, the moments \eqref{eq:moments_wannier_fourier} are more complicated and can be expressed as
\begin{align}
    \nonumber
    C_{r\tau}^{(\nu)} &=  \sum_{p=\pm} \int \frac{dq}{2\pi} (\chi_{q\tau p}^{(\nu)})^*(i\partial_q)^r \chi_{q\tau p}^{(\nu)} \\
    &= i^r \sum_{p=\pm} \int \frac{dq}{2\pi} (\tilde{\chi}_{q\tau p}^{(\nu)})^* e^{i \phi_{q\tau}^{(\nu)}} \partial_q^r 
    \left[ e^{-i \phi_{q\tau}^{(\nu)}} \tilde{\chi}_{q \tau p}^{(\nu)} \right] \,,
\end{align}
such that 
\begin{align}
    \label{eq:w_mom_scaling}
    C_{r\tau}^{(\nu)} &= (\xi^{(\nu)})^{r-1} c_{r\tau}^{(\nu)}\,,\\
    \nonumber
    c_{r\tau}^{(\nu)} &= i^r 2^{r-1} \sum_{p=\pm} \int \frac{d \bar{q}}{4\pi} \sqrt{1 + \frac{p \bar{q}}{\bar{\epsilon}_{\bar{q}}}} 
    e^{i \phi_{q\tau}^{(\nu)}} \\
    \label{eq:w_mom_coef}
    & \times \sum_{s=0}^r \frac{r!}{s! (r-s)!} \left[ \partial_{\bar{q}}^{r-s} e^{-i \phi_{q\tau}^{(\nu)}} \partial_{\bar{q}}^s \sqrt{1 + \frac{p \bar{q}}{\bar{\epsilon}_{\bar{q}}}}  \right]\,.
\end{align}
In contrast to the ML gauge, the dimensionless coefficients $c_{r\tau}^{(\nu)}$ depend on $\tau$ and $\nu$ in the AF gauge via the phase factors involving $\phi_{q\tau}^{(\nu)}$.

In particular, we find
\begin{align}
    \nonumber
    C_{1\tau}^{(\nu)}  &=  \int \frac{dq}{2\pi}  \partial_q \phi_{q\tau}^{(\nu)} \\
    \label{eq:first_mom_w_dirac}
    &= \begin{cases} \frac{\gamma^{(\nu)}}{2 \pi}, & \tau=+ ,\\
    - \frac{\gamma^{(\nu)}}{2 \pi} + \frac12 s_{\gamma^{(\nu)}}, & \tau=- , \end{cases}
\end{align}
which agrees with \eqref{eq:first_moment_zak_berry_phase_dirac} and (\ref{eq:zak_berry_phase_vs_gap_phase}), and
\begin{align}
     C_{2\tau}^{(\nu)}  &= \tilde{C}_{2\tau}^{(\nu)} + \int \frac{dq}{2\pi} \left[\partial_q \phi_{q\tau}^{(\nu)} \right]^2  \nonumber \\
     &= \frac{\xi^{(\nu)}}{8} +  \frac{\xi^{(\nu)} }{4} \left[\tan^2 \gamma^{(\nu)} + \frac{ 4 \cot 2 \gamma^{(\nu)} 
     \Theta (- \tau)}{ \cos \gamma^{(\nu)}} s_{\gamma^{(\nu)}} \right. \nonumber  \\
     & \left. \qquad +   \frac{2 \tau (2 \gamma^{(\nu)}  \cot 2 \gamma^{(\nu)} -1)}{\pi \cos \gamma^{(\nu)}} \right].
\end{align}
One finds that $C^{(\nu)}_{2\tau}>\tilde{C}_{2\tau}^{(\nu)}=\xi^{(\nu)}/8$ as expected and a divergence when the first moment (\ref{eq:first_mom_w_dirac}) jumps (for $\tau=+$ at $\gamma^{(\nu)}=\pm\pi$, and for $\tau=-$ at $\gamma^{(\nu)}=0$), compare with the discussion after (\ref{eq:dirac_berry_connection}).

\subsection{Scaling properties of lattice Wannier functions}
\label{sec:wannier_scaling}

In this Section we reveal the scaling properties of the lattice Wannier functions as given by (\ref{eq:wannier_AF_dirac_lattice_u}), (\ref{eq:wannier_AF_dirac_lattice_d}), (\ref{eq:wannier_ML_dirac_lattice_u}), and (\ref{eq:wannier_ML_dirac_lattice_d}) via the Dirac Wannier functions. Inserting the decompositions (\ref{eq:w_p_decomposition}) and (\ref{eq:tilde_w_p_decomposition}) of the Dirac Wannier functions in real and imaginary parts, and using 
\begin{widetext}
\begin{align}
    \label{eq:cos_sin_term}
    \cos\left(\frac{\nu\pi}{Z}m\right) = (-1)^{\nu(n-1)} \cos\left(\pi\frac{\nu j}{Z}\right) \quad,\quad
    \sin\left(\frac{\nu\pi}{Z}m\right) = (-1)^{\nu(n-1)} \sin\left(\pi\frac{\nu j}{Z}\right)\,,
\end{align}
where $m=Z(n-1)+j$, we obtain
\begin{align}
    \label{eq:w_u_scaling}
    \frac{1}{\sqrt{Z}}(-1)^{\alpha(n-1)} m \,w_{\text{u},\alpha}(n,j) &= 
    F_{-}(\frac{ma}{\xi^{(\alpha)}};\frac{\alpha j}{Z},\gamma^{(\alpha)}) \,,\\
    \label{eq:w_d_scaling}
    \frac{1}{\sqrt{Z}}(-1)^{(\alpha-1)(n-1)} m \,w_{\text{d},\alpha}(n,j) &= 
    F_{+}(\frac{ma}{\xi^{(\alpha-1)}};\frac{(\alpha-1) j}{Z},\gamma^{(\alpha-1)}) \,,\\
    \label{eq:tilde_w_u_scaling}
    \frac{1}{\sqrt{Z}}(-1)^{\alpha(n-1)} \tilde{m}_\alpha \,\tilde{w}_{\text{u},\alpha}(n,j) &= 
    \tilde{F}_{-}(\frac{\tilde{m}_\alpha a}{\xi^{(\alpha)}};\frac{\alpha j}{Z},\gamma^{(\alpha)}) \,,\\
    \label{eq:tilde_w_d_scaling}
    \frac{1}{\sqrt{Z}}(-1)^{(\alpha-1)(n-1)} \tilde{m}_\alpha \,\tilde{w}_{\text{d},\alpha}(n,j) &= 
    \tilde{F}_{+}(\frac{\tilde{m}_\alpha a}{\xi^{(\alpha-1)}};\frac{(\alpha-1) j}{Z},\gamma^{(\alpha-1)}) \,,
\end{align}
where $\alpha=1,\dots,Z-1$ for $w_{\text{u},\alpha}$ and $\tilde{w}_{\text{u},\alpha}$, and 
$\alpha=2,\dots,Z$ for $w_{\text{d},\alpha}$ and $\tilde{w}_{\text{d},\alpha}$. The scaling functions $F_\tau$ and $\tilde{F}_\tau$ are defined in the following way in terms of the fundamental scaling functions $F_{A/B,\tau}$ and $\tilde{F}_{A/B,\tau}$ introduced in Section~\ref{sec:scaling_functions}
\begin{align}
    \label{eq:F_scaling}
    F_{\tau}(y;s,\gamma) = F_{B,\tau}(y;\gamma)\cos(\pi s) - F_{A,\tau}(y;\gamma) \sin(\pi s) \quad,\quad
    \tilde{F}_{\tau}(y;s,\gamma) = \tilde{F}_{B,\tau}(y;\gamma)\cos(\pi s) - \tilde{F}_{A,\tau}(y;\gamma) \sin(\pi s)\,.
\end{align}
As a result, for given value $j=1,\dots,Z$ of the site index within a unit cell, the up/down parts of the lattice Wannier functions reveal universal scaling with a single length scale. By adding the up and down parts we get for the total Wannier function of band $\alpha$ 
\begin{align}
    \label{eq:w_scaling}
    \frac{1}{\sqrt{Z}}(-1)^{\alpha(n-1)} m \,w_{\alpha}(n,j) &= 
    {F}_{-}(\frac{m a}{\xi^{(\alpha)}};\frac{\alpha j}{Z},\gamma^{(\alpha)}) 
    + (-1)^{n-1} {F}_{+}(\frac{m a}{\xi^{(\alpha-1)}};\frac{(\alpha-1) j}{Z},\gamma^{(\alpha-1)}),\\
    \label{eq:tilde_w_scaling}
    \frac{1}{\sqrt{Z}}(-1)^{\alpha(n-1)} \tilde{m}_\alpha \,\tilde{w}_{\alpha}(n,j) &= 
    \tilde{F}_{-}(\frac{\tilde{m}_\alpha a}{\xi^{(\alpha)}};\frac{\alpha j}{Z},\gamma^{(\alpha)}) 
    + (-1)^{n-1} \tilde{F}_{+}(\frac{\tilde{m}_\alpha a}{\xi^{(\alpha-1)}};\frac{(\alpha-1) j}{Z},\gamma^{(\alpha-1)}) \,,
\end{align}
where we leave out the second (first) term on the right hand side for $\alpha=1$ ($\alpha=Z$). For $\alpha=2,\dots,Z-1$ universal scaling appears with two different length scales corresponding to the gaps at the bottom and the top of the band. Therefore, to reveal universal scaling, one has to keep the ratio of the two length scales fixed. 

For the universal scaling of the moments
\begin{align}
    \label{eq:CD_r_ud_scaling}
    C_{\text{\text{u}/\text{d}},r\alpha}(Ma) = \sum_{m=-M}^M (ma)^r |w_{\text{\text{u}/\text{d}},\alpha}(ma)|^2 \quad,\quad
    \tilde{D}_{\text{\text{u}/\text{d}},r\alpha}(Ma) = \sum_{m=-M}^M (\tilde{m}_\alpha a)^r |\tilde{w}_{\text{\text{u}/\text{d}},\alpha}(ma)|^2\,,
\end{align}
we get from the above equations after neglecting the strongly oscillating terms
\begin{align}
    \label{eq:C_r_ud_scaling}
    \frac{C_{\text{u},r\alpha}(Ma)}{Za (\xi^{(\alpha)})^{r-1}} \approx \int_{\frac{-Ma}{\xi^{(\alpha)}}}^{\frac{Ma}{\xi^{(\alpha)}}}
    dy\,y^{r-2} G_{-}(y;\gamma^{(\alpha)}) \quad &,\quad
    \frac{C_{\text{d},r\alpha}(Ma)}{Za(\xi^{(\alpha-1)})^{r-1}} \approx \int_{\frac{-Ma}{\xi^{(\alpha-1)}}}^{\frac{Ma}{\xi^{(\alpha-1)}}}
    dy\,y^{r-2} G_{+}(y;\gamma^{(\alpha-1)})\,,\\
    \label{eq:tilde_D_r_ud_scaling}
    \frac{\tilde{D}_{\text{u},r\alpha}(Ma)}{Za(\xi^{(\alpha)})^{r-1}} \approx \int_{\frac{-Ma}{\xi^{(\alpha)}}}^{\frac{Ma}{\xi^{(\alpha)}}}
    dy\,y^{r-2} \tilde{G}_{-}(y;\gamma^{(\alpha)}) \quad &,\quad
    \frac{\tilde{D}_{\text{d},r\alpha}(Ma)}{Za(\xi^{(\alpha-1)})^{r-1}} \approx  \int_{\frac{-Ma}{\xi^{(\alpha-1)}}}^{\frac{Ma}{\xi^{(\alpha-1)}}}
    dy\,y^{r-2} \tilde{G}_{+}(y;\gamma^{(\alpha-1)})\,,
\end{align}
where
\begin{align}
    \label{eq:G_scaling}
    G_{\tau}(y;\gamma) = \frac{1}{2}\left\{|F_{A,\tau}(y;\gamma)|^2 + |F_{B,\tau}(y;\gamma)|^2\right\} \quad,\quad
    \tilde{G}_{\tau}(y;\gamma) &= \frac{1}{2}\left\{|\tilde{F}_{A,\tau}(y;\gamma)|^2 + |\tilde{F}_{B,\tau}(y;\gamma)|^2\right\}\,, 
\end{align}
\end{widetext}
or explicitly in terms of the right/left moving Dirac Wannier functions
\begin{align}
    \label{eq:G_w_scaling}
    G_{\tau}(x/\xi^{(\nu)};\gamma^{(\nu)}) &= x^2\sum_p |w^{(\nu)}_{\tau p}(x)|^2 \,,\\
    \label{eq:tilde_G_w_scaling}
    \tilde{G}_{\tau}(x/\xi^{(\nu)};\gamma^{(\nu)}) &= x^2\sum_p |\tilde{w}^{(\nu)}_{\tau p}(x)|^2 \,.
\end{align}
By using (\ref{eq:tilde_F_A_-y_trafo}) and (\ref{eq:tilde_F_B_-y_trafo}) we note that $\tilde{G}_\tau$ is a symmetric function
\begin{align}
    \label{eq:tilde_G_symmetry}
    \tilde{G}_\tau(-y;\gamma) = \tilde{G}_\tau(y;\gamma)\,.
\end{align}
As a result the right hand sides of (\ref{eq:tilde_D_r_ud_scaling}) are exactly zero for odd values of $r$. As mentioned already at the end of Section~\ref{sec:wannier_lattice_dirac_ML}, this does not mean that the lattice moments $\tilde{D}_{\text{u/d},r\alpha}(Ma)$ are zero for odd values of $r$. It only means that by dividing them by the leading order $a\xi^{r-1}$ (with $\xi\equiv\xi^{(\alpha)}$ for $\tilde{D}_{\text{u},r\alpha}(Ma)$ and $\xi\equiv\xi^{(\alpha-1)}$ for $\tilde{D}_{\text{d},r\alpha}(Ma)$), one gets zero in the limit $\xi\rightarrow\infty$. This is in contrast to the even moments in ML gauge which stay finite in this limit after divided by this order. Therefore, the odd moments in ML gauge are negligible and are not considered in the following. In the AF gauge the function $G_\tau(y;\gamma)$ is asymmetric due to the asymmetry of the scaling function $F_{A,\tau}(y;\gamma)$, see (\ref{eq:F_A_-y_trafo}) and Fig.~\ref{fig:scaling_functions}, with a significant larger part for either positive or negative values of $y$. Therefore, in the AF gauge, all moments are of order $a\,\xi^{r-1}$ and stay finite in the limit $\xi\rightarrow \infty$ when divided by this order. 

The total moment for $\alpha=2,\dots,Z-1$ is obtained from the sum (neglecting strongly oscillating terms involving $(-1)^n$)
\begin{align}
    \label{eq:C_r_tot_scaling}
    C_{r\alpha}(Ma) &\approx C_{\text{u},r\alpha}(Ma) + C_{\text{d},r\alpha}(Ma)\,,\\
    \label{eq:tilde_D_r_tot_scaling} 
    \tilde{D}_{r\alpha}(Ma) &\approx \tilde{D}_{\text{u},r\alpha}(Ma) + \tilde{D}_{\text{d},r\alpha}(Ma) \,,
\end{align}
where we leave out the second (first) term on the right hand side for $\alpha=1$ ($\alpha=Z$).

According to (\ref{eq:C_r_1_lattice_dirac}) and (\ref{eq:tilde_D_r_1_lattice_dirac}), we get for the asymptotic value $M\rightarrow\infty$
\begin{align}
    \label{eq:C_r_u_asymp}
    C_{\text{u},r\alpha}(Ma) &\rightarrow Za\,C_{r-}^{(\alpha)} = Za (\xi^{(\alpha)})^{r-1}\,c_{r-}^{(\alpha)}\,,\\
    \label{eq:C_r_d_asymp}
    C_{\text{d},r\alpha}(Ma) &\rightarrow Za\,C_{r+}^{(\alpha-1)} = Za (\xi^{(\alpha-1)})^{r-1}\,c_{r+}^{(\alpha-1)}\,,\\
    \label{eq:tilde_D_r_u_asymp}
    \tilde{D}_{\text{u},2l,\alpha}(Ma) &\rightarrow Za\,\tilde{C}_{2l,-}^{(\alpha)} 
    = Za (\xi^{(\alpha)})^{2l-1}\,\tilde{c}_{2l}\,,\\
    \label{eq:tilde_D_r_d_asymp}
    \tilde{D}_{\text{d},2l,\alpha}(Ma) &\rightarrow Za\,\tilde{C}_{2l,+}^{(\alpha-1)}
    = Za (\xi^{(\alpha-1)})^{2l-1}\,\tilde{c}_{2l}\,,
\end{align}
where we used (\ref{eq:w_mom_scaling}) and (\ref{eq:tilde_w_mom_scaling}). According to (\ref{eq:first_mom_w_dirac}) and (\ref{eq:tilde_C_2_Dirac}) we get for $r=1,2$
\begin{align}
    \label{eq:C_1_u_asymp}
    \frac{C_{\text{u},1\alpha}(Ma)}{Za} &\rightarrow -\frac{\gamma^{(\alpha)}}{2\pi} + \frac{1}{2}s_{\gamma^{(\alpha)}}\,,\\
    \label{eq:C_1_d_asymp}
    \frac{C_{\text{d},1\alpha}(Ma)}{Za} &\rightarrow \frac{\gamma^{(\alpha-1)}}{2\pi}\,,\\
    \label{eq:tilde_D_2_u_asymp}
    \frac{\tilde{D}_{\text{u},2\alpha}(Ma)}{Za\xi^{(\alpha)}} &\rightarrow \frac{1}{8}\,,\\
    \label{eq:tilde_D_2_d_asymp}
    \frac{\tilde{D}_{\text{d},2\alpha}(Ma)}{Za\xi^{(\alpha-1)}} &\rightarrow \frac{1}{8}\,.
\end{align}
Using (\ref{eq:C_r_tot_scaling}) and (\ref{eq:tilde_D_r_tot_scaling}) together with (\ref{eq:zak_berry_alpha_dirac}) and (\ref{eq:zak_berry_phase_vs_gap_phase}), we find for the asymptotic values of the total moments
\begin{align}
    \label{eq:C_1_tot_asymp}
    \frac{C_{1\alpha}(Ma)}{Za} &\rightarrow \frac{\gamma_\alpha}{2\pi} \,,\\
    \label{eq:tilde_D_2_tot_asymp}
    \frac{\tilde{D}_{2\alpha}(Ma)}{Za\xi^{(\alpha)}} &\rightarrow \frac{1}{8}\left(1+\frac{\xi^{(\alpha-1)}}{\xi^{(\alpha)}}\right)\,.
\end{align}

In the following two subsections we will demonstrate the universal scaling of the lattice Wannier functions and their moments for two examples with $Z=2$ and $Z=3$. For simplicity we assume that the dominant contribution to the gap parameter (\ref{eq:H'_matrix_element}) is the term $H'$ of the effective Hamiltonian (\ref{eq:H'_eff}). This means that all Fourier components
\begin{align}
    \label{eq:v_fourier}
    \delta\tilde{v}_\nu &= \frac{1}{Z}\sum_{j=1}^Z e^{-i2\pi j\nu/Z} \delta v_j \,,\\
    \label{eq:t_fourier}
    \delta\tilde{t}_\nu &= \frac{1}{Z}\sum_{j=1}^Z e^{-i2\pi j\nu/Z} \delta t_j \,,
\end{align}
have approximately the same order of magnitude. By convention we set
\begin{align}
    \label{eq:zero_fourier}
    \delta\tilde{v}_0 = \delta\tilde{t}_0 = 0 \,.
\end{align}
By inserting the form (\ref{eq:H'}) of $H'$ and using (\ref{eq:eigenstates_H0}) for the eigenstates of $H_0$, it is then straightforward to show that
\begin{align}
    \label{eq:gap_equation}
    \Delta^{(\nu)} e^{i\gamma^{(\nu)}} = \delta\tilde{v}_\nu - 2 e^{-i\pi\nu/Z}\,\delta\tilde{t}_\nu \,.
\end{align}
The $2(Z-1)$ lattice parameters $\delta v_j$ and $\delta t_j$, with $j=1,\dots,Z$ and $\sum_j \delta v_j = \sum_j \delta t_j = 0$, are then fixed via the $Z-1$ complex gap parameters $\Delta^{(\nu)}e^{i\gamma^{(\nu)}}$, with $\nu=1,\dots,Z-1$, by using (\ref{eq:gap_equation}) and the inverse Fourier transform of (\ref{eq:v_fourier}) and (\ref{eq:t_fourier})
\begin{align}
    \label{eq:v_inverse_fourier}
    \delta v_j &= \sum_{\nu=1}^{Z-1} e^{i2\pi j\nu/Z} \delta \tilde{v}_\nu \,,\\
    \label{eq:t_inverse_fourier}
    \delta t_j &= \sum_{\nu=1}^{Z-1} e^{i2\pi j\nu/Z} \delta \tilde{t}_\nu \,.
\end{align}

\subsubsection{Universal scaling for Z=2}
\label{sec:universal_scaling_Z_2}

\begin{figure}[t!]
    \centering
    \includegraphics[width =\columnwidth]{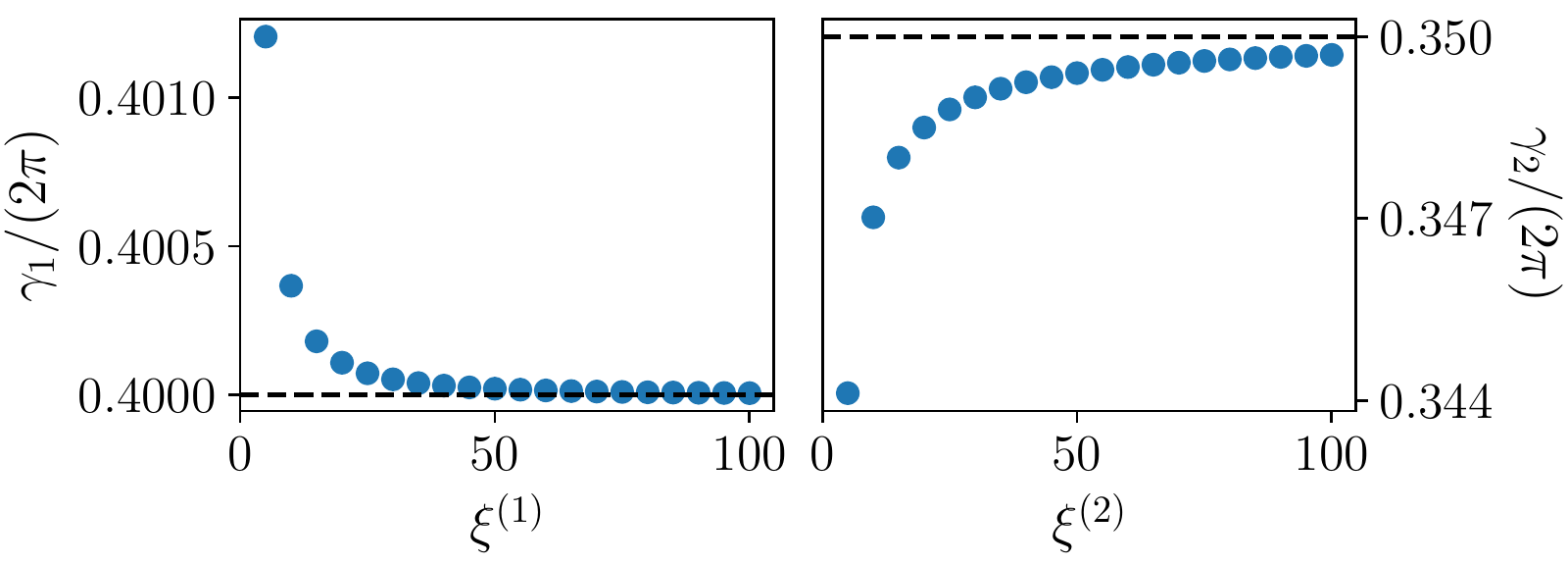}
	  \caption{The Zak-Berry phase $\gamma_\alpha/(2\pi)$ calculated from (\ref{eq:zak_berry_phase_lattice}) as function of $\xi^{(\alpha)}$ for $\alpha=1$, $Z=2$ (left panel) and for $\alpha=2$, $Z=3$ (right panel). The parameters are the same as in Fig.~\ref{fig:scaling_Z_2} and Fig.~\ref{fig:w_AF_Z_3}, respectively. For all $\xi^{(\alpha)}$ used in these figures we find that the deviation from the approximation $\gamma_1/(2\pi)\approx 0.4$ (left panel, see (\ref{eq:zak_berry_phase_Z_2})) and $\gamma_2\approx \frac{7}{20}=0.35$ (right panel, see (\ref{eq:zak_berry_phase_Z_3_alpha_2})) is less than $\approx 0.02\%$. 
	  } 
    \label{fig:zak_berry_phase}
\end{figure}

For $Z=2$ (the so-called Rice-Mele model \cite{rice_mele_prl_82,heeger_rmp_01} which, for $\delta v_{1/2}=0$, turns into the Su-Schrieffer-Heeger model \cite{su_etal_prl_79}) we show in Fig.~\ref{fig:scaling_Z_2}(a) and Fig.~\ref{fig:scaling_Z_2}(b) the universal scaling of the Wannier functions $w(ma)\equiv w_1(ma)=w_{\text{u},1}(ma)$ and $\tilde{w}(ma)\equiv\tilde{w}_1(ma)=\tilde{w}_{\text{u},1}(ma)$ of the lowest band $\alpha=1$, together with the scaling of their first and second moment, respectively. The lattice parameters are fixed by the complex gap parameter $\Delta e^{i\gamma}\equiv\Delta^{(1)}e^{i\gamma^{(1)}}$. The lattice Wannier functions are calculated numerically from their definitions (\ref{eq:w_lattice_2}) and (\ref{eq:w_tilde_lattice}) via the Bloch states $u_{k1}(ma)$ and the gauge phase $\varphi_{k1}$. The latter follows from (\ref{eq:gauge_max_loc_lattice})  via the Zak-Berry connection $A_{k1}$. The Bloch states and the Zak-Berry connection are calculated analytically in Appendix~\ref{app:bloch_Z_23}. The Zak-Berry phase $\gamma_1$ of the lowest band can either be calculated from its definition (\ref{eq:zak_berry_phase_lattice}) via numerical integration over the Zak-Berry connection or from the approximate formulas (\ref{eq:zak_berry_alpha_dirac}) and (\ref{eq:zak_berry_phase_vs_gap_phase}) as
\begin{align}
    \label{eq:zak_berry_phase_Z_2}
    \gamma_1 \approx \gamma^{(1)}_- = - \gamma^{(1)} + \pi s_{\gamma^{(1)}} \,.
\end{align}
As shown in Fig.~\ref{fig:zak_berry_phase} the two ways to calculate the Zak-Berry phase coincide quite well in the low-energy regime of small gaps. Therefore, it makes no visible difference in Fig.~\ref{fig:scaling_Z_2}(b) which choice is taken to calculate the shift variable $\tilde{m}_1=m-\gamma_1/\pi$.

With $m=2(n-1)+j$ and $j=1,2$, we find from (\ref{eq:w_u_scaling}) and (\ref{eq:F_scaling}) for odd/even values of $m$ (i.e., for $j=1,2$)
\begin{align}
    \label{eq:w_scaling_Z_2}
    \frac{1}{2} |m \,w(ma)|^2 &\approx \begin{cases} |F_{A,-}(y;\gamma)|^2 & \text{for}\quad $m$\,\text{odd} \\ 
    |F_{B,-}(y;\gamma)|^2 & \text{for}\quad $m$\,\text{even} \end{cases} \,,\\    
    \label{eq:tilde_w_scaling_Z_2}
    \frac{1}{2} |\tilde{m}\,\tilde{w}(ma)|^2 &\approx \begin{cases} |\tilde{F}_{A,-}(\tilde{y};\gamma)|^2 & \text{for}\quad $m$\,\text{odd} \\ 
    |\tilde{F}_{B,-}(\tilde{y};\gamma)|^2 & \text{for}\quad $m$\,\text{even} \end{cases} \, ,
\end{align}
where we used the abbreviations 
\begin{align}
 y\equiv ma/\xi \quad,\quad \tilde{y}\equiv \tilde{m}a/\xi \,, 
\end{align}
with $\tilde{m}\equiv\tilde{m}_1=m - \gamma_1/\pi$. As a consequence, the four fundamental scaling functions $F_{A/B,-}(y;\gamma)$ and $\tilde{F}_{A/B,-}(y;\gamma)$ show up naturally in the scaling behavior of the lower band (in AF or ML gauge, respectively) of the Rice-Mele model for odd (A) and even (B) sites. Similarly, for the upper band the corresponding scaling functions with $\tau=+$ appear, which are related via (\ref{eq:F_A_-tau_trafo}-\ref{eq:tilde_F_B_-tau_trafo}) to the scaling functions with $\tau=-$. For larger values of $Z>2$, the same scaling functions appear but in subtle combinations, as we will demonstrate in the next subsection for $Z=3$.

In the main panels of Fig.~\ref{fig:scaling_Z_2}(a) and Fig.~\ref{fig:scaling_Z_2}(b) we reveal the universal scaling properties by plotting the left hand side of Eqs.~(\ref{eq:w_scaling_Z_2}) and (\ref{eq:tilde_w_scaling_Z_2}) for different values of $\xi$ as function of $y=ma/\xi$ or $\tilde{y}=\tilde{m}a/\xi$, and find that all curves fall on top of the universal functions $|F_{A/B,-}(y;\gamma)|^2$ and $|\tilde{F}_{A/B,-}(y;\gamma)|^2$ (for $m$ odd/even), respectively (note that we omitted the subindex $\tau=-$ and the superindex $(1)$ for all quantities used in the caption of these figures). Although (\ref{eq:w_scaling_Z_2}) and (\ref{eq:tilde_w_scaling_Z_2}) are expected to hold only for large $|m|\gg 1$ and for small gaps (where $\xi\gg a$), it is remarkable that the coincidence is even quite well for $|m|\sim O(1)$ and for large gaps (where $\xi\sim O(a)$). We will come back to this point in Section~\ref{sec:all_scales}, where we discuss the behavior for small scales $m\sim O(1)$. All the universal scalings are in accordance with the symmetries and asymptotic behaviors discussed for the scaling functions in Section~\ref{sec:scaling_functions}, compare with Fig.~\ref{fig:scaling_functions}. For even $m$, the function $|m\,w(ma)|^2$ scales symmetrically with zero value at $m=0$, all other cases are asymmetric and have a finite value at $m=0$ (or $\tilde{m}=0$), in accordance with (\ref{eq:F_A_small_y}-\ref{eq:tilde_F_B_small_y}). For large $|m|\gg 1$ all functions are exponentially decaying with a pre-exponential power-law $\sim |ma/\xi|^r \exp(-|ma|/\xi)$. Since $\gamma=0.2\pi$ and $\tau=-$, we get $\tau=-\Sign(\cos\gamma)\ne 0$, leading via (\ref{eq:F_A_large_y_2}) and (\ref{eq:F_B_large_y_2}) to $r=\pm 1$ for $|m\,w(ma)|^2$ and odd $m\gtrless 0$, and to $r=-1$ for $|m\,w(ma)|^2$ and even $m$. For $|\tilde{m}\,\tilde{w}(ma)|^2$ we get $r=0.5$ for both $m$ even or odd, see (\ref{eq:tilde_F_A_large_y}) and (\ref{eq:tilde_F_B_large_y}). 

In the insets of Fig.~\ref{fig:scaling_Z_2}(a) and Fig.~\ref{fig:scaling_Z_2}(b) we show the universal scaling of the first moment $C_1(Ma)/(2a)$ of $w(ma)$ and of the quadratic spread $\tilde{D}_2(Ma)/(2a\xi)$ of $\tilde{w}(ma)$ (with $C_1\equiv C_{11}= C_{\text{u},11}$ and $\tilde{D}_2\equiv \tilde{D}_{21}=\tilde{D}_{\text{u},21}$) when plotted against $Ma/\xi$ for different $\xi$. As demonstrated, all discrete points fall on top of the universal curves (\ref{eq:C_r_ud_scaling}) (with $r=1$) and (\ref{eq:tilde_D_r_ud_scaling}) (with $r=2$), i.e., 
\begin{align}
    \label{eq:C_1_scaling_Z_2}
    \frac{C_1(Ma)}{Za} &\approx \int_{\frac{-Ma}{\xi}}^{\frac{Ma}{\xi}}
    dy\,\frac{1}{y} G_{-}(y;\gamma) \,,\\
    \label{eq:tilde_D_2_scaling_Z_2}
    \frac{\tilde{D}_2(Ma)}{Za\xi} &\approx \int_{\frac{-Ma}{\xi}}^{\frac{Ma}{\xi}}
    dy\,\tilde{G}_{-}(y;\gamma) \,,
\end{align}
with $\tilde{G}_-(y;\gamma)=\tilde{G}_-(-y;\gamma)$ due to (\ref{eq:tilde_G_symmetry}). In accordance with (\ref{eq:C_1_u_asymp}) and (\ref{eq:tilde_D_2_u_asymp}) they converge smoothly to the values
\begin{align}
    \label{eq:C_1_asymptotic_Z_2}
    \frac{C_1}{2a} &= \lim_{M\rightarrow\infty} \frac{C_1(Ma)}{2a} = \frac{\gamma_{1}}{2\pi}\,,\\
    \label{eq:tilde_D_2_asymptotic_Z_2}
    \frac{\tilde{D}_2}{2a\xi} &= \lim_{M\rightarrow\infty} \frac{\tilde{D}_2(Ma)}{2a\xi} = \frac{1}{8}\,.
\end{align}
The last result gives the universal value for the quadratic spread $\langle \Delta x^2 \rangle = \tilde{D}_2 = 2a\xi/8$ of the Wannier function $\tilde{w}(ma)$ in the ML gauge (for arbitrary size $Za$ of the unit cell it turns into $Za\xi/8$ for the lowest band).

\subsubsection{Universal scaling for Z=3}
\label{sec:universal_scaling_Z_3}
 
For $Z=3$ the Bloch states and the Zak-Berry connection are calculated analytically in Appendix~\ref{app:bloch_Z_23}. In contrast to the case $Z=2$, two gap parameters $\Delta^{(\nu)} e^{i\gamma^{(\nu)}}$, with $\nu=1,2$, are needed to fix the lattice parameters. They correspond to the two gaps and contain different phases $\gamma^{(\nu)}$ and different length scales $\xi^{(\nu)}=v_F^{(\nu)}/(2\Delta^{(\nu)})$. For the Zak-Berry phases of the three bands we obtain approximately from (\ref{eq:zak_berry_alpha_dirac}) and (\ref{eq:zak_berry_phase_vs_gap_phase})
\begin{align}
    \label{eq:zak_berry_phase_Z_3_alpha_1}
    \gamma_1 &\approx \gamma^{(1)}_- = - \gamma^{(1)} + \pi s_{\gamma^{(1)}}\,,\\
    \label{eq:zak_berry_phase_Z_3_alpha_2}
    \gamma_2 &\approx \gamma^{(1)}_+ + \gamma^{(2)}_- 
    = \gamma^{(1)} - \gamma^{(2)} + \pi s_{\gamma^{(2)}}\,,\\
    \label{eq:zak_berry_phase_Z_3_alpha_3}
    \gamma_3 &\approx \gamma^{(2)}_+ = \gamma^{(2)}\,,
\end{align}
where, as shown in Fig.~\ref{fig:zak_berry_phase}, it makes a negligible difference for small gaps whether one takes these approximate formulas to calculate the shift variables $\tilde{m}_\alpha = m - \frac{3}{2\pi}\gamma_\alpha$ or the precise definition (\ref{eq:zak_berry_phase_lattice}) by integrating over the lattice Zak-Berry connection.  

According to (\ref{eq:w_u_scaling}) and (\ref{eq:tilde_w_u_scaling}), the scaling of the Wannier functions of the first band involve the scaling functions $F_{-}(ma/\xi^{(1)};j/3;\gamma^{(1)})$ and $\tilde{F}_{-}(\tilde{m}_1 a/\xi^{(1)};j/3;\gamma^{(1)})$, with $j=1,2,3$. Similar to the first band for $Z=2$, only a single length scale $\xi^{(1)}$ appears, but the scaling is different since, for $j\ne 3$, linear combinations of the scaling functions $F_{A,-}(y;\gamma^{(1)})$ and $F_{B,-}(y;\gamma^{(1)})$ (or ($\tilde{F}_{A,-}(y;\gamma^{(1)})$) and $\tilde{F}_{B,-}(y;\gamma^{(1)})$) will occur, see Eq.~(\ref{eq:F_scaling}) and Eqs.~(\ref{eq:F_scaling_1}-\ref{eq:F_scaling_3}) below. However, besides this no other change of the scaling appears.

Of particular interest is the scaling of the second band $\alpha=2$ where two different length scales occur for the up and down components. 
From (\ref{eq:w_u_scaling}-\ref{eq:tilde_w_d_scaling}) we find
\begin{align}
    \label{eq:w_u_scaling_Z_3}
    \frac{1}{3}|m \,w_{\text{u},2}(n,j)|^2 &\approx 
    |F_{-}(\frac{ma}{\xi^{(2)}};\frac{2j}{3},\gamma^{(2)})|^2 \,,\\
    \label{eq:w_d_scaling_Z_3}
    \frac{1}{3} |m \,w_{\text{d},2}(n,j)|^2 &\approx
    |F_{+}(\frac{ma}{\xi^{(1)}};\frac{j}{3},\gamma^{(1)})|^2 \,,\\
    \label{eq:tilde_w_u_scaling_Z_3}
    \frac{1}{3}|\tilde{m}_2 \,\tilde{w}_{\text{u},2}(n,j)|^2 &\approx 
    |\tilde{F}_{-}(\frac{\tilde{m}_2 a}{\xi^{(2)}};\frac{2j}{3},\gamma^{(2)})|^2 \,,\\
    \label{eq:tilde_w_d_scaling_Z_3}
    \frac{1}{3}|\tilde{m}_2 \,\tilde{w}_{\text{d},2}(n,j)|^2 &\approx
    |\tilde{F}_{+}(\frac{\tilde{m}_2 a}{\xi^{(1)}};\frac{j}{3},\gamma^{(1)})|^2 \,,
\end{align}
\begin{figure}[b!]
    \centering
	 \includegraphics[width =0.8\columnwidth]{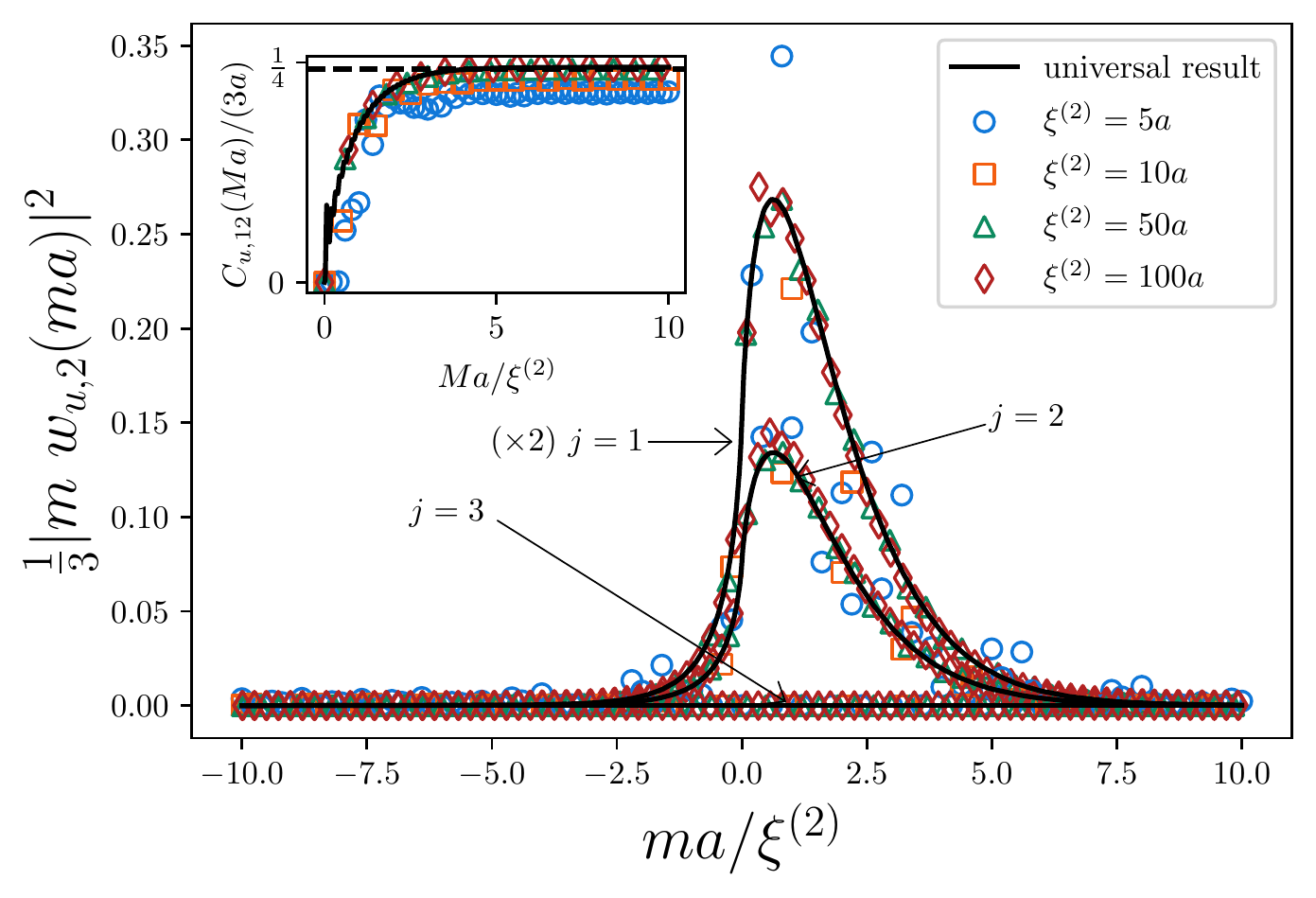}
	  \caption{Scaling of $\frac{1}{3}|m\,{w}_{\text{u},\alpha}(n,j)|^{2}$ for $\alpha=2$, $Z=3$, $\gamma^{(1)}=0.2\pi$, $\gamma^{(2)}=0.5\pi$, and $\xi^{(1)}=30 a$, as function of ${m}a/\xi^{(2)}$ for different values of $\xi^{(2)}$ and $j$. According to (\ref{eq:w_u_scaling_Z_3}) we obtain the universal curve $|F_{-}(ma/\xi^{(2)};2j/3,\gamma^{(2)})|^{2}$. The inset shows the scaling of the first moment $C_{\text{u},1\alpha}(Ma)/(Za)$ for $\alpha=2, Z=3$ as function of $s_2=Ma/\xi^{(2)}$. According to (\ref{eq:C_1_u_scaling_Z_3}) we obtain the universal curve $\int_{-s_2}^{s_2} dy \,y^{-1} G_{-}(y;\gamma^{(2)})$ with saturation at the correct value $-\frac{\gamma^{(2)}}{2\pi}+\frac{1}{2}s_{\gamma^{(2)}}=\frac{1}{4}$.
	  } 
    \label{fig:AF_up} 
\end{figure}
\begin{figure}[b!]
    \centering
	 \includegraphics[width =0.8\columnwidth]{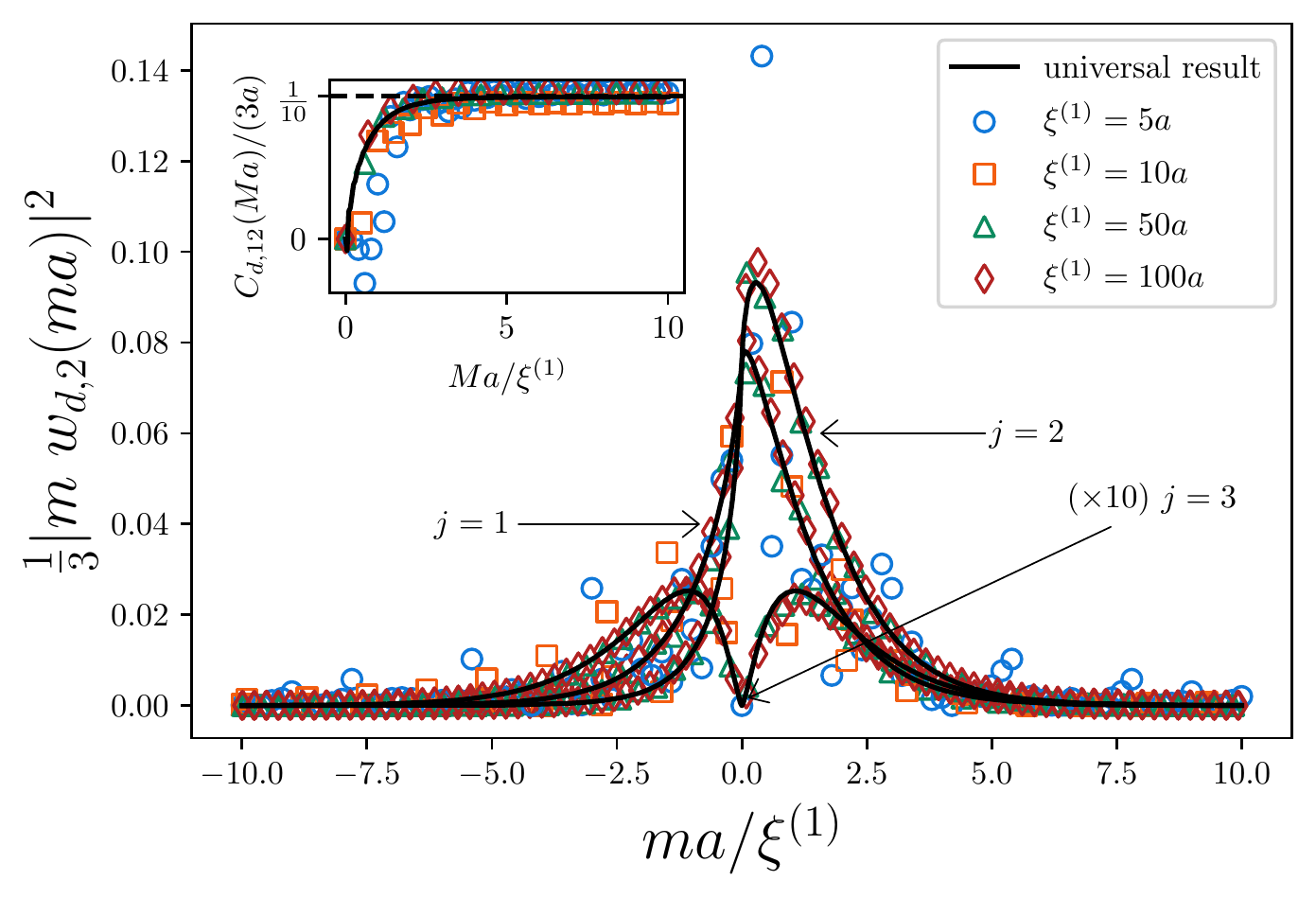}
	  \caption{Scaling of $\frac{1}{3}|m\,{w}_{\text{d},\alpha}(n,j)|^{2}$ for $\alpha=2$, $Z=3$, $\gamma^{(1)}=0.2\pi$, $\gamma^{(2)}=0.5\pi$, and $\xi^{(2)}=30 a$, as function of ${m}a/\xi^{(1)}$ for different values of $\xi^{(1)}$ and $j$. According to (\ref{eq:w_d_scaling_Z_3}) we obtain the universal curve $|F_{+}(ma/\xi^{(1)};j/3,\gamma^{(1)})|^{2}$. The inset shows the scaling of the first moment $C_{\text{d},1\alpha}(Ma)/(Za)$ for $\alpha=2, Z=3$ as function of $s_1=Ma/\xi^{(1)}$. According to (\ref{eq:C_1_d_scaling_Z_3}) we obtain the universal curve $\int_{-s_1}^{s_1} dy \,y^{-1} G_{+}(y;\gamma^{(1)})$ with saturation at the correct value $\frac{\gamma^{(1)}}{2\pi}=\frac{1}{10}$.
	  } 
    \label{fig:AF_down} 
\end{figure}
\begin{figure}[b!]
    \centering
	 \includegraphics[width =0.8\columnwidth]{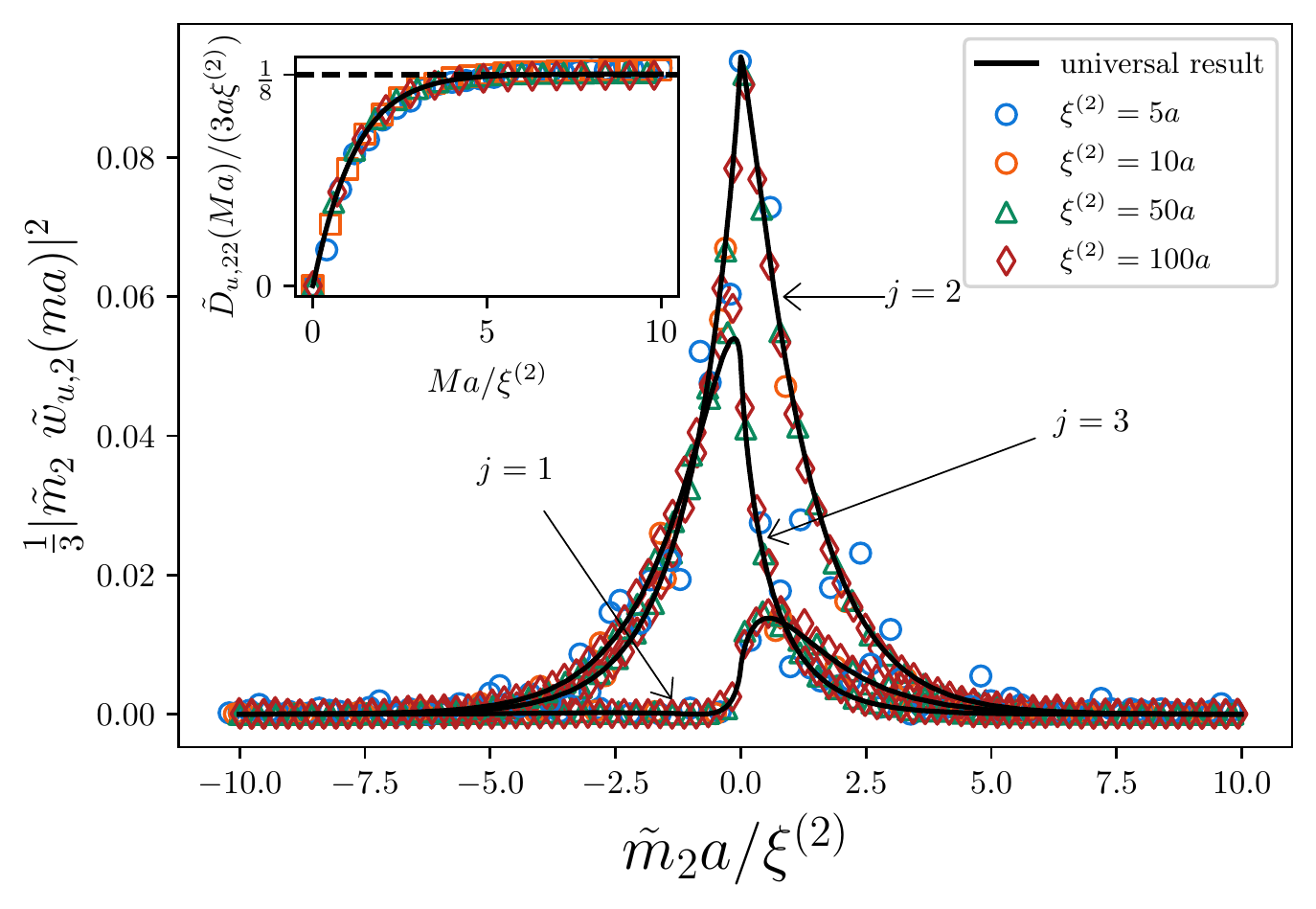}
	  \caption{Scaling of $\frac{1}{3}|\tilde{m}_\alpha\,\tilde{w}_{\text{u},\alpha}(n,j)|^{2}$ for $\alpha=2$, $Z=3$, $\gamma^{(1)}=0.2\pi$, $\gamma^{(2)}=0.5\pi$, and $\xi^{(1)}=30 a$, as function of $\tilde{m}_{2}a/\xi^{(2)}$ for different values of $\xi^{(2)}$ and $j$. According to (\ref{eq:tilde_w_u_scaling_Z_3}) we obtain the universal curve 
	  $|\tilde{F}_{-}(\tilde{m}_2 a/\xi^{(2)};2j/3,\gamma^{(2)})|^{2}$. The inset shows the scaling of the second moment $\tilde{D}_{\text{u},2\alpha}(Ma)/(Za\,\xi^{(\alpha)})$ for $\alpha=2, Z=3$ as function of $s_2=Ma/\xi^{(2)}$. According to (\ref{eq:tilde_D_2_u_scaling_Z_3}) we obtain the universal curve $\int_{-s_2}^{s_2} dy \,\tilde{G}_{-}(y;\gamma^{(2)})$ with saturation at the correct value $\frac{1}{8}$.
	  } 
    \label{fig:ML_up} 
\end{figure}
\begin{figure}[t!]
    \centering
	 \includegraphics[width =0.8\columnwidth]{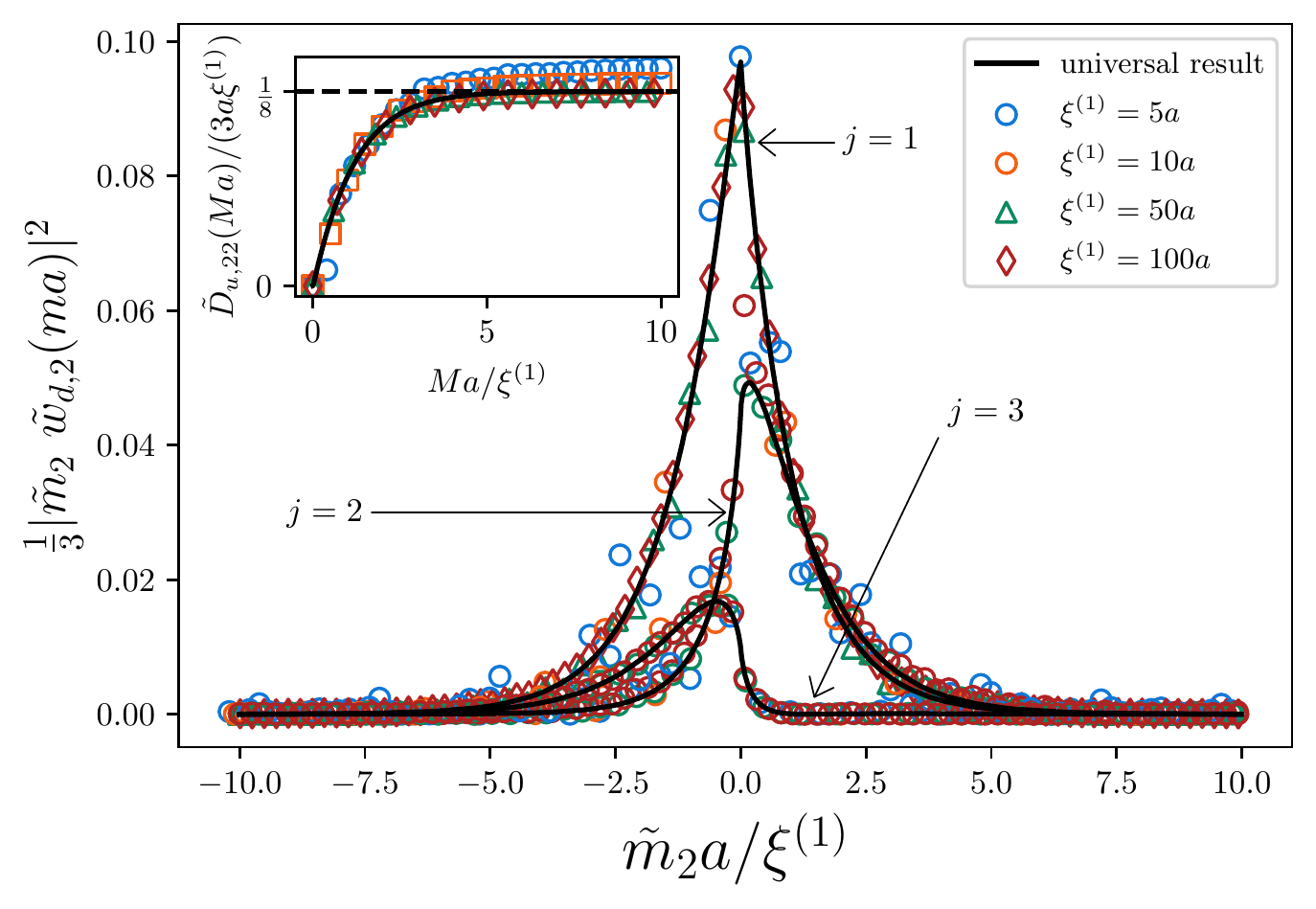}
	  \caption{Scaling of $\frac{1}{3}|\tilde{m}_\alpha\,\tilde{w}_{\text{d},\alpha}(n,j)|^{2}$ for $\alpha=2$, $Z=3$, $\gamma^{(1)}=0.2\pi$, $\gamma^{(2)}=0.5\pi$, and $\xi^{(2)}=30 a$, as function of $\tilde{m}_{2}a/\xi^{(1)}$ for different values of $\xi^{(1)}$ and $j$. According to (\ref{eq:tilde_w_d_scaling_Z_3}) we obtain the universal curve 
	  $|\tilde{F}_{+}(\tilde{m}_2 a/\xi^{(1)};j/3,\gamma^{(1)})|^{2}$. The inset shows the scaling of the second moment $\tilde{D}_{\text{d},2\alpha}(Ma)/(Za\,\xi^{(\alpha-1)})$ for $\alpha=2, Z=3$ as function of $s_1=Ma/\xi^{(1)}$. According to (\ref{eq:tilde_D_2_d_scaling_Z_3}) we obtain the universal curve $\int_{-s_1}^{s_1} dy \,\tilde{G}_{+}(y;\gamma^{(1)})$ with saturation at the correct value $\frac{1}{8}$.
	  } 
    \label{fig:ML_down} 
\end{figure}
i.e., the up and down components scale with $\xi^{(2)}$ and $\xi^{(1)}$, respectively, corresponding to the gap at the top and bottom of the second band. From (\ref{eq:F_scaling}) we find that $F_\tau$ involves the following combinations of the scaling functions $F_{A,\tau}$ and $F_{B,\tau}$
\begin{align}
    \nonumber
    F_{\tau}(y;\frac{1}{3},\gamma) &= - F_{\tau}(y;\frac{4}{3},\gamma) \\
    \label{eq:F_scaling_1}
    &=\frac{1}{2}\left\{F_{B,\tau}(y;\gamma) - \sqrt{3} \,F_{A,\tau}(y;\gamma)\right\}\,,\\
    \label{eq:F_scaling_2}
    F_{\tau}(y;\frac{2}{3},\gamma) &= -\frac{1}{2}\left\{F_{B,\tau}(y;\gamma) + \sqrt{3} \,F_{A,\tau}(y;\gamma)\right\}\,,\\
    \label{eq:F_scaling_3}
    F_{\tau}(y;1,\gamma) &= - F_{\tau}(y;2,\gamma) = - F_{B,\tau}(y;\gamma)\,,
\end{align}
and similiar equations for $F_\tau\rightarrow\tilde{F}_\tau$ and $F_{A/B,\tau}\rightarrow \tilde{F}_{A/B,\tau}$. The scaling of the four functions (\ref{eq:w_u_scaling_Z_3}-\ref{eq:tilde_w_d_scaling_Z_3}) is demonstrated in Fig.~\ref{fig:AF_up} (up component in AF gauge), Fig.~\ref{fig:AF_down} (down component in AF gauge), Fig.~\ref{fig:ML_up} (up component in ML gauge), and Fig.~\ref{fig:ML_down} (down component in ML gauge), for all components $j=1,2,3$, using the choice $\gamma^{(1)}=0.2\pi$ and $\gamma^{(2)}=\pi/2$. Due to the special choice  $\gamma^{(2)}=\pi/2$ it turns out that \begin{align}
    \frac{1}{3}|m \,w_{\text{u},2}(n,3)|^2 \approx |F_{B,-}(y;\pi/2)|^2 =0 
\end{align}
due to (\ref{eq:w_u_scaling_Z_3}), (\ref{eq:F_scaling_3}) and (\ref{eq:F_B_large_y_3}), leading to the vanishing $j=3$ component in Fig.~\ref{fig:AF_up}. 

\begin{figure}[b!]
    \centering
	 \includegraphics[width =0.8\columnwidth]{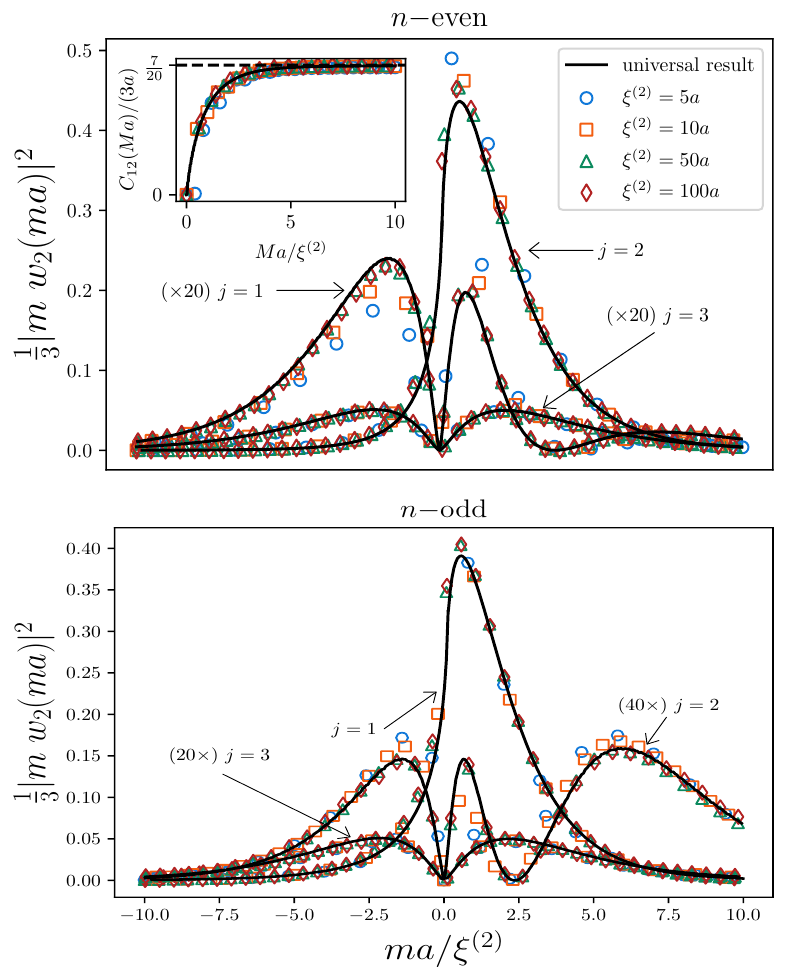}
	  \caption{Scaling of $\frac{1}{3}|m\,{w}_{\alpha}(n,j)|^{2}$ for $\alpha=2$, $Z=3$, $\gamma^{(1)}=0.2\pi$, $\gamma^{(2)}=0.5\pi$, and $\xi^{(1)}=2\xi^{(2)}$, as function of $ma/\xi^{(2)}$ for different values of $\xi^{(2)}$ and $j$, with $n$ even/odd for upper/lower panel. According to (\ref{eq:w_scaling_Z_3}) we obtain the universal curve $|{F}_{-}(\frac{ma}{\xi^{(2)}};\frac{2 j}{3},\gamma^{(2)}) \pm {F}_{+}(\frac{ma}{\xi^{(1)}};\frac{ j}{3},\gamma^{(1)})|^{2}$ for $n$ even/odd. The inset shows the scaling of the first moment $C_{1\alpha}(Ma)/(Za)$ for $\alpha=2, Z=3$ as function of $s=Ma/\xi^{(2)}$. According to (\ref{eq:C_1_scaling_Z_3}), we obtain the universal curve $\int_{-s}^s dy \,y^{-1}\tilde{G}_{-}(y;\gamma^{(2)}) + \int_{-s/2}^{s/2} dy \,y^{-1} \tilde{G}_{-}(y;\gamma^{(1)})$ with saturation at the correct value $\frac{\gamma_2}{2\pi}=\frac{7}{20}$, where $\gamma_2$ has been evaluated from (\ref{eq:zak_berry_phase_Z_3_alpha_2}).
	  }
    \label{fig:w_AF_Z_3}
\end{figure}

\begin{figure}[b!]
    \centering
	 \includegraphics[width =0.8\columnwidth]{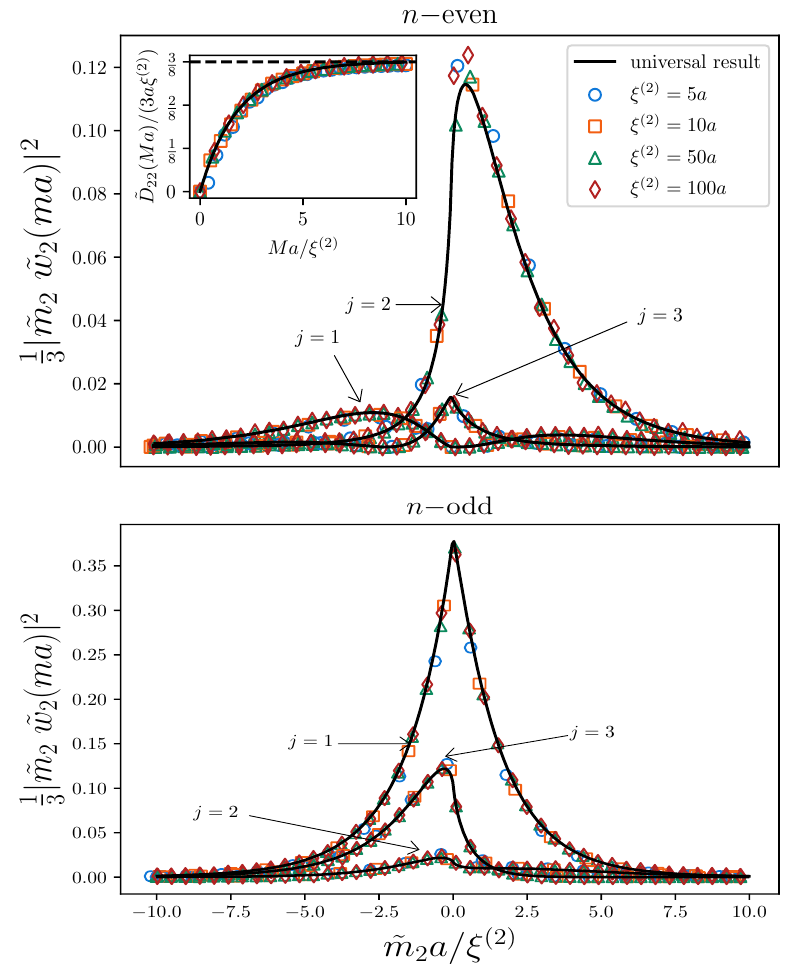}
	  \caption{Scaling of $\frac{1}{3}|\tilde{m}_{\alpha}\,\tilde{w}_{\alpha}(n,j)|^{2}$ for $\alpha=2$, $Z=3$, $\gamma^{(1)}=0.2\pi$, $\gamma^{(2)}=0.5\pi$, and $\xi^{(1)}=2\xi^{(2)}$, as function of $\tilde{m}_{2}a/\xi^{(2)}$ for different values of $\xi^{(2)}$ and $j$, with $n$ even/odd for upper/lower panel. According to (\ref{eq:tilde_w_scaling_Z_3}) we obtain the universal curve $|\tilde{F}_{-}(\frac{\tilde{m}_2 a}{\xi^{(2)}};\frac{2 j}{3},\gamma^{(2)}) \pm \tilde{F}_{+}(\frac{\tilde{m}_2 a}{\xi^{(1)}};\frac{ j}{3},\gamma^{(1)})|^{2}$ for $n$ even/odd. The inset shows the scaling of the second moment $\tilde{D}_{2\alpha}(Ma)/(Za\,\xi^{(\alpha)})$ for $\alpha=2, Z=3$ as function of $s=Ma/\xi^{(2)}$. According to (\ref{eq:tilde_D_2_scaling_Z_3}), we obtain the universal curve $\int_{-s}^s dy \,\tilde{G}_{-}(y;\gamma^{(2)}) + 2 \int_{-s/2}^{s/2} dy \,\tilde{G}_{-}(y;\gamma^{(1)})$ with saturation at the correct value $\frac{1}{8}(1+\xi^{(1)}/\xi^{(2)})=\frac{3}{8}$. 
	  }
    \label{fig:tilde_w_ML_Z_3}
\end{figure}

In Fig.~\ref{fig:w_AF_Z_3} and Fig.~\ref{fig:tilde_w_ML_Z_3} we show the scaling of the total Wannier functions $|m\,w_2(ma)|^2$ and $|\tilde{m}_2\,\tilde{w}_2(ma)|^2$ of the second band, respectively. Since two different length scales appear, the ratio $\xi^{(1)}/\xi^{(2)}$ has to be kept fixed and universality is demonstrated for different choices of $\xi^{(2)}$. Furthermore, since the scaling of the up and down components contain a different sign factor $(-1)^{n-1}$ (see Eqs.~(\ref{eq:w_u_scaling}-\ref{eq:tilde_w_d_scaling})), the total Wannier function is the sum or difference of the scaling functions of the up and down components depending on the parity of $n$, see Eqs.~(\ref{eq:w_scaling}-\ref{eq:tilde_w_scaling}). For $Z=3$ and $\alpha=2$ one obtains
\begin{widetext}
\begin{align}
    \label{eq:w_scaling_Z_3}
    \frac{1}{3} |m \,w_{2}(n,j)|^2 &= 
    |{F}_{-}(\frac{m a}{\xi^{(2)}};\frac{2j}{3},\gamma^{(2)}) 
    + (-1)^{n-1} {F}_{+}(\frac{m a}{\xi^{(1)}};\frac{j}{3},\gamma^{(1)})|^2\,,\\
    \label{eq:tilde_w_scaling_Z_3}
    \frac{1}{3} |\tilde{m}_2 \,\tilde{w}_{2}(n,j)|^2 &= 
    |\tilde{F}_{-}(\frac{\tilde{m}_2 a}{\xi^{(2)}};\frac{2j}{3},\gamma^{(2)}) 
    + (-1)^{n-1} \tilde{F}_{+}(\frac{\tilde{m}_2 a}{\xi^{(1)}};\frac{j}{3},\gamma^{(1)})|^2 \,,
\end{align}
\end{widetext}
leading to different scaling behavior for $n$ even or odd.

In the insets of Figs.~\ref{fig:AF_up}-\ref{fig:ML_down} we show the scaling of the first moment (for AF gauge) and the second moment (for ML gauge) of the up and down components of the second band. According to (\ref{eq:C_r_ud_scaling}) and (\ref{eq:tilde_D_r_ud_scaling}), they follow the following scaling laws with asymptotic behavior according to (\ref{eq:C_1_u_asymp}-\ref{eq:tilde_D_2_d_asymp})
\begin{align}
    \nonumber
    \frac{C_{\text{u},12}(Ma)}{Za} &\approx \int_{\frac{-Ma}{\xi^{(2)}}}^{\frac{Ma}{\xi^{(2)}}}
    dy\,y^{-1} G_{-}(y;\gamma^{(2)}) \,,\\
    \label{eq:C_1_u_scaling_Z_3}
    &\rightarrow - \frac{\gamma^{(2)}}{2\pi} + \frac{1}{2}s_{\gamma^{(2)}}\,,\\
    \label{eq:C_1_d_scaling_Z_3}
    \frac{C_{\text{d},12}(Ma)}{Za} &\approx \int_{\frac{-Ma}{\xi^{(1)}}}^{\frac{Ma}{\xi^{(1)}}}
    dy\,y^{-1} G_{+}(y;\gamma^{(1)})\rightarrow \frac{\gamma^{(1)}}{2\pi}\,,\\
    \label{eq:tilde_D_2_u_scaling_Z_3}
    \frac{\tilde{D}_{\text{u},22}(Ma)}{Za\xi^{(2)}} &\approx \int_{\frac{-Ma}{\xi^{(2)}}}^{\frac{Ma}{\xi^{(2)}}}
    dy\,\tilde{G}_{-}(y;\gamma^{(2)}) \rightarrow \frac{1}{8} \,,\\ 
    \label{eq:tilde_D_2_d_scaling_Z_3}
    \frac{\tilde{D}_{\text{d},22}(Ma)}{Za\xi^{(1)}} &\approx  \int_{\frac{-Ma}{\xi^{(1)}}}^{\frac{Ma}{\xi^{(1)}}}
    dy\,\tilde{G}_{+}(y;\gamma^{(1)}) \rightarrow \frac{1}{8} \,,
\end{align}
The scaling of the total moments of the second band in AF and ML gauge are shown in the insets of Figs.~\ref{fig:w_AF_Z_3} and Figs.~\ref{fig:tilde_w_ML_Z_3}, respectively. According to (\ref{eq:C_r_tot_scaling}) and (\ref{eq:tilde_D_r_tot_scaling}) they follow from the sum of the up and down components as
\begin{widetext}
\begin{align}
    \label{eq:C_1_scaling_Z_3}
    \frac{C_{12}(Ma)}{Za} &\approx \int_{-s_2}^{s_2} dy\,y^{-1} G_{-}(y;\gamma^{(2)}) + \int_{-s_2/r_{12}}^{s_2/r_{12}} dy\,y^{-1} G_{+}(y;\gamma^{(1)}) 
    \rightarrow \frac{\gamma_2}{2\pi}\,,\\
    \label{eq:tilde_D_2_scaling_Z_3}
    \frac{\tilde{D}_{22}(Ma)}{Za\xi^{(2)}} &\approx \int_{-s_2}^{s_2} dy\,\tilde{G}_{-}(y;\gamma^{(2)}) + r_{12}\int_{-s_2/r_{12}}^{s_2/r_{12}} dy\,\tilde{G}_{+}(y;\gamma^{(1)}) \rightarrow \frac{1}{8}(1+r_{12}) \,,
\end{align}
\end{widetext}
where we defined $s_2=Ma/\xi^{(2)}$ and $r_{12}=\xi^{(1)}/\xi^{(2)}$. As a result the ratio $r_{12}$ of the two length scales has to be kept fixed to see universal scaling by varying $\xi^{(2)}$.

\subsection{Properties of lattice Wannier functions on different scales}
\label{sec:all_scales}

In this section we exhibit and compare the important properties of lattice Wannier functions on different scales, always taking the case of small gaps where a clear separation of length scales is present $\xi\gg a$ (here, $\xi$ denotes some typical order of all $\xi^{(\nu)}$, $\nu=1,\dots,Z-1$). We distinguish three different regimes, called (F) (for free or gapless case), (S) (for scaling region where the length scale $\xi$ appears), and (E) (for exponentially decaying region):
\begin{align}
    \label{eq:regime_F}
    &\underline{\text{regime (F)}}:\quad |ma|\ll\xi \,,\\
    \label{eq:regime_S}
    &\underline{\text{regime (S)}}:\quad |ma|\lesssim\xi \,,\\
    \label{eq:regime_E}
    &\underline{\text{regime (E)}}:\quad |ma|\gg\xi \,.
\end{align}
Regime (E) is the regime where all Wannier functions decay exponentially with some universal pre-exponential power law, see Eqs.~(\ref{eq:F_A_large_y_1}-\ref{eq:tilde_F_B_large_y}). Therefore, in this regime the Wannier functions have a negligible contribution to the moments and the normalization. Regime (S) is the most important issue of our work where the universal scaling on length scale $\xi$ is visible. As shown in the previous Section~\ref{sec:wannier_scaling} (see also the insets of Figs.~\ref{fig:AF_up}-\ref{fig:tilde_w_ML_Z_3}) all moments $C_{r\alpha}(Ma)$ (with $r\ge 1$) and $\tilde{D}_{r\alpha}(Ma)$ (with $r\ge 2$ even) approach their asymptotic values on the length scale $\xi$. Finally, the region (F) is the regime of small length scales where the presence of the gap does not play any role. As a consequence, the Wannier functions are identical to the zero gap limit in this regime (although the phases $\gamma^{(\nu)}$ from the gap parameter occur in the ML gauge due to the special choice of this gauge). As we will see in the following the regime (F) is only important for the correct normalization of the Wannier function but is of no significance for the scaling and provides a misleading visual impression of the Wannier function.  

\begin{figure}[b!]
    \centering
	 \includegraphics[width =\columnwidth]{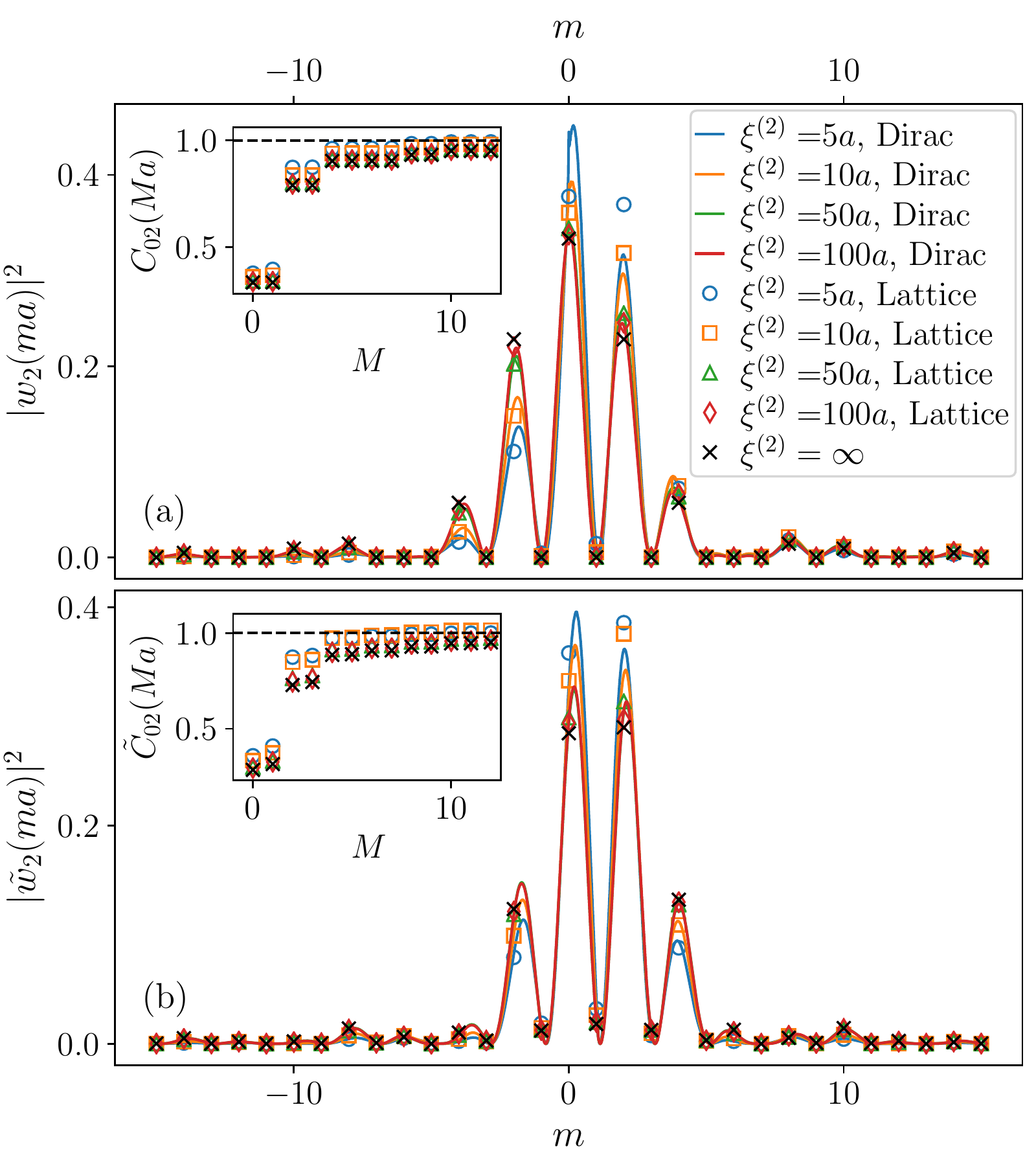}
	  \caption{Comparison between $|w_\alpha(ma)|^2$ (upper panel) and $|\tilde{w}_\alpha(ma)|^2$ (lower panel) as function of $m$ for $Z=3$ and $\alpha=2$ with $\xi^{(1)}=2\xi^{(2)}$, $\gamma^{(1)}=0.2\pi$, and $\gamma^{(2)}=0.5\pi$ with their Dirac counterpart for a variety of $\xi^{(2)}$. The results for large $\xi^{(2)}\gg a$ are in perfect agreement with the analytical results (\ref{eq:w_free}) and (\ref{eq:tilde_w_free}) at zero gap. In the two insets we show the scaling of the normalization $C_{0\alpha}(Ma)=\sum_{m=-M}^M |w_\alpha(ma)|^2$ and $\tilde{C}_{0\alpha}(Ma)=\sum_{m=-M}^M |\tilde{w}_\alpha(ma)|^2$ for $\alpha=2$. As can be seen the normalization reaches unity already on very small scales $M\sim Z$ for large $\xi^{(2)}\gg a$.
	  }
    \label{fig:w_small_scales}
\end{figure}

In regime (F) of small scales $|y|\sim|ma|/\xi\ll 1$, we can use the $y=0$ limit of the universal scaling functions, as given by Eqs.~(\ref{eq:F_A_small_y}-\ref{eq:tilde_F_B_small_y}). Using $|e^{i\gamma}-1|=2 s_{\gamma} \,\sin(\gamma/2)$ and $|e^{i\gamma}+1|=2\cos(\gamma/2)$ we find for the scaling functions (\ref{eq:F_scaling}) the free result at zero gap
\begin{align}
    \label{eq:F_free}
    F_\tau(y;s,\gamma) &\approx -\frac{\tau}{\pi}\sin(\pi s)\,,\\
    \label{eq:tilde_F_+_free}
    \tilde{F}_+(y;s,\gamma) &\approx -\frac{1}{\pi}\sin(\pi s + \gamma/2)\,,\\
    \label{eq:tilde_F_-_free}
    \tilde{F}_-(y;s,\gamma) &\approx -\frac{1}{\pi} s_{\gamma}\,\cos(\pi s + \gamma/2)\,.
\end{align}
Inserting these scaling functions in (\ref{eq:w_scaling}) and (\ref{eq:tilde_w_scaling}) we get for the lattice Wannier functions on small scales $|ma|\ll\xi$
\begin{widetext}
\begin{align}
    \label{eq:w_free}
    \frac{1}{\sqrt{Z}} \, m \, w_\alpha(ma) &\approx \frac{1}{\pi}\left\{\sin\left(\frac{\pi\alpha}{Z}m\right) - \sin\left(\frac{\pi(\alpha-1)}{Z}m\right)\right\}\,,\\
    \label{eq:tilde_w_free}
    \frac{1}{\sqrt{Z}} \, \tilde{m}_\alpha \, \tilde{w}_\alpha(ma) &\approx -\frac{1}{\pi}\left\{s_{\gamma^{(\alpha)}}\,\cos\left(\frac{\pi\alpha}{Z}m + \frac{1}{2}\gamma^{(\alpha)}\right) + \sin\left(\frac{\pi(\alpha-1)}{Z}m + \frac{1}{2}\gamma^{(\alpha-1)}\right)\right\}\,.
\end{align}
\end{widetext}
It is a straightforward exercise to see that this are indeed the exact results for the Wannier functions in the absence of a gap. Only due to the special choice of the ML gauge, the phases of the gap parameters enter in (\ref{eq:tilde_w_free}). In Fig.~\ref{fig:w_small_scales}  we show the square $|w_\alpha(ma)|^2$ and $|\tilde{w}_\alpha(ma)|^2$ of the Wannier functions in the AF and ML gauge as function of $m$ for the special case $Z=3$ and $\alpha=2$. As can be seen the numerical lattice result agrees perfectly with the analytical low-energy result (\ref{eq:w_scaling}) and (\ref{eq:tilde_w_scaling}) for large $\xi\gg a$, as well as with the gapless result (\ref{eq:w_free}) and (\ref{eq:tilde_w_free}) in the small scale regime $|ma|\ll\xi$. The scaling region (S) is not visible at all in this figure since the Wannier functions are of order $1/m^2\sim (a/\xi)^2\ll 1$ in this regime. Therefore, the visible impression of the square of the Wannier functions is approximately the result in the absence of the gap. As shown in the left insets of Fig.~\ref{fig:w_small_scales} the region (F) covers almost completely the correct normalization of the Wannier functions, such that the scaling of the normalizations
\begin{align}
    \label{eq:normalization_scaling}
    C_{0\alpha}(Ma) &= \sum_{m=-M}^M |w_\alpha(ma)|^2 \,,\\
    \label{eq:tilde_normalization_scaling}
    \tilde{C}_{0\alpha}(Ma) &= \sum_{m=-M}^M |\tilde{w}_\alpha(ma)|^2 
\end{align}
approaches unity already on scales $M\sim O(Z)$. In contrast, the region (S) contributes only the negligible order $\sim (\xi/a)(a/\xi)^2\sim a/\xi\ll 1$ to the normalization.  

We note that even the value at $m=0$ is perfectly reproduced by the free results (\ref{eq:w_free}) and (\ref{eq:tilde_w_free}) and gives an important contribution to the correct normalization. We obtain (note that the right hand side of (\ref{eq:w_free}) has to be expanded up to linear order in $m$ to get the correct result for $m=0$)
\begin{align}
    \label{eq:w_m=0}
    w_\alpha(0) &= \frac{1}{\sqrt{Z}} \,,\\
    \nonumber
    \tilde{w}_\alpha(0) &= \frac{2}{\gamma_\alpha\sqrt{Z}} \left\{s_{\gamma^{(\alpha)}}\,\cos\left(\frac{\gamma^{(\alpha)}}{2}\right) + \right. \\
    \label{eq:tilde_w_m=0}
    &\hspace{1cm} \left. + \sin\left(\frac{\gamma^{(\alpha-1)}}{2}\right)\right\}\,.
\end{align}
With $\gamma_\alpha=\gamma^{(\alpha-1)}-\gamma^{(\alpha)}+\pi s_{\gamma^{(\alpha)}}$ one finds a perfect agreement with the results in Fig.~\ref{fig:w_small_scales} for $\xi^{(\alpha-1)},\xi^{(\alpha)}\gg a$.

In contrast to the normalization, all the interesting scaling behavior on the length scale $\xi$ show up in the regime (S) and are only visible when multiplying the Wannier function with $m$ in the AF gauge or $\tilde{m}_\alpha$ in the ML gauge. In a nutshell one can express the subtle dependence of the Wannier functions on the two length scales $a$ and $\xi$ roughly as follows (in AF or ML gauge for any band). Replacing $ma$ by the continuous variable $x$ and rescaling the Wannier function via
\begin{align}
    x \equiv ma \quad,\quad w(x) \equiv \frac{1}{\sqrt{a}} w(ma)\,,
\end{align}
the qualitative form can be stated as follows (omitting strongly oscillating terms on scale $Za$ which contribute a negligible amount to the moments)
\begin{align}
    |w(x)|^2 \sim \delta_a(x) \,f(x/\xi) \,e^{-|x|/\xi} \,,
\end{align}
where $\delta_a(x)$ is a Lorentzian delta function on scale $a$, and $f(y)$ is a dimensionless function of order $O(1)$ 
\begin{align}
    \, \delta_a(x) = \frac{1}{\pi}\frac{a}{x^2+a^2}\quad,\quad f(y)\sim O(1)\,.
\end{align}
The most important fact is that the delta function covers the complete normalization on scales $x\sim a$
\begin{align}
    \sum_m |w(ma)|^2 \sim \int dx \,|w(x)|^2 \sim f(0) \sim O(1)\,,
\end{align}
whereas, due to the Lorentzian form of the delta function, the scaling of all moments is determined from the region $|x|\lesssim\xi$ as
\begin{align}
    \nonumber
    C_r &\sim \sum_m (ma)^r |w(ma)|^2 \sim \int dx \,x^r \,|w(x)|^r \\
    \nonumber
    &\sim a \int dx \, x^{r-2} \, f(x/\xi) \, e^{-|x|/\xi} \\
    \label{eq:C_r_estimate}
    &\sim a \,\xi^{r-1} \,\int_{|y|\lesssim 1} dy \,y^{r-2} \, f(y) \sim a \,\xi^{r-1}\,,
\end{align}
for all $r\ge 1$. Note that the region $|x|\sim a$ contributes only the negligible amount $a^r\ll a\,\xi^{r-1}$ to the moment for $r\ge 2$. Strictly speaking, this argument is only rigorous for even values of $r$ since in this case all terms of the integrand are positive. For odd values of $r$ it can happen that the prefactor in front of the leading term is zero. This happens for the scaling of the odd moments $\tilde{D}_{2l+1,\alpha}(Ma)$ and $\tilde{C}_{2l+1,\alpha}(Ma)$ in the ML gauge, which do not show universality but are of no importance since their asymptotic value is of negligible order $\sim a^2\,\xi^{2l-1}$ for $l>0$, see the detailed discussion after Eq.~(\ref{eq:tilde_G_symmetry}). In contrast, in the AF gauge the odd moments scale according to the estimate (\ref{eq:C_r_estimate}) for all $r\ge 1$ and their order of magnitude is fully determined by the scaling region (S) (even for $r=1$). They can be written as   
\begin{align}
    C_{2l+1} \sim a \,\int_{0<y\lesssim 1} dy \,y^{2l-1} \, \left\{f(y) - f(-y)\right\} \sim a \,\xi^{2l}\,,
\end{align}
such that convergence is guaranteed at $y=0$ even for $l=0$. Importantly, in the AF gauge, the function $f(y)\approx f(-y)$ is nearly symmetric for $|y|\ll 1$, see (\ref{eq:w_free}). Therefore, the region of small scales does not contribute and the odd moments show universal scaling from the regime (S) due to a significant asymmetry of the scaling function $f(y)$ for $|y|\sim O(1)$, see the discussion after Eq.~(\ref{eq:tilde_G_symmetry}). E.g., it is quite remarkable that the first moment in the AF gauge is fully determined by the nearly invisible asymmetry of the Wannier function on large scales $|ma|~\sim\xi\gg a$, whereas the visible impression of a symmetric shape on small scales $|ma|\sim O(a)$ would predict an incorrect vanishing first moment. 

The scaling behavior of the moments is demonstrated numerically in the insets of Figs.~\ref{fig:scaling_Z_2}(a), \ref{fig:scaling_Z_2}(b), \ref{fig:w_AF_Z_3}, and \ref{fig:tilde_w_ML_Z_3}, where it is shown that the first moments $C_{1\alpha}(Ma)$ in the AF gauge and the second moments $\tilde{D}_{2\alpha}(Ma)$ in the ML gauge scale indeed smoothly to their universal asymptotic values on the scale $Ma\sim\xi$ and obtain a negligible contribution from the small scale regime (F). Similar numerical evidence can be shown for all the other higher moments $C_{r\alpha}(Ma)$ with $r\ge 1$ and $\tilde{D}_{2l,\alpha}(Ma)$ with $l\ge 1$.

\section{Non-Abelian Wannier functions}
\label{sec:nonabelian}

In this final section we analyse another class of lattice Wannier functions which arise from a special non-Abelian gauge transformation which mixes the Bloch states from all bands $\alpha=1,\dots,\nu$ below a given one labelled by $\nu$. In particular the non-Abelian gauge of maximally localized Wannier functions has been proposed in the literature and the connection of their spread to the polarization fluctuations have been put forward. Here we analyse the universal scaling properties of the non-Abelian lattice Wannier functions and find the interesting result that they scale up to a surprisingly high precision according to the Dirac Wannier function of the lower band of the Dirac model corresponding to gap $\nu$ or, equivalently, similar to the upper part of the Wannier function of band $\nu$ as shown in Fig.~\ref{fig:ML_up}. In Section~\ref{sec:nonabelian_gauge_summary} we summarize the general definition of the non-Abelian gauge of maximal localization and show our central result of its universal scaling. The technical part of the construction of this non-Abelian gauge is outlined in Section~\ref{sec:NA-ML_gauge}, supplemented by three Appendices~\ref{app:nonabelian_splitting}, \ref{app:wilson_dirac_propagator} and \ref{app:spectrum_wilson_loop}. We show this construction in the main part of this work since it provides a very efficient way to analyse non-Abelian gauges analytically via the Wilson propagator which, up to our best knowledge, has not been reported before.

\subsection{Non-Abelian lattice gauge and summary of results}
\label{sec:nonabelian_gauge_summary}

So far we have discussed Abelian gauge transformations for a fixed band index $\alpha$ by allowing for a phase factor $e^{i\varphi_{k\alpha}}$ to transform the Bloch state as
\begin{align}
    |u_{k\alpha}\rangle\rightarrow |\tilde{u}_{k\alpha}\rangle = e^{i\varphi_{k\alpha}} |u_{k\alpha}\rangle \,,
\end{align}
where $\varphi_{k\alpha}=\varphi_{k+\frac{2\pi}{Za},\alpha}$ is a periodic phase variable. In multi-band systems more general non-Abelian gauge transformations have been discussed, see Ref.~\onlinecite{marzari_vanderbilt_prb_97} or reviews in Refs.~\onlinecite{marzari_etal_rmp_12,vanderbilt_book_18}. These gauge transformations mix the Bloch states of all band indices $\alpha=1,\dots,\nu$ up to a certain value $\nu$ (defining the valence band if a chemical potential is present in gap $\nu$) via a $k$-dependent unitary transformation as
\begin{align}
    \label{eq:u_nonabelian_trafo}
    |\hat{u}_{k\alpha}^{(\nu)}\rangle = \sum_{\alpha'=1}^\nu |u_{k\alpha'}\rangle (\hat{U}^{(\nu)}_k)_{\alpha'\alpha}\,.
\end{align}
Here, $\hat{U}^{(\nu)}_k$ is a unitary $\nu\times\nu$-matrix which is periodic under $k\rightarrow k+\frac{2\pi}{Za}$
\begin{align}
    \hat{U}^{(\nu)}_k\,(\hat{U}^{(\nu)}_k)^\dagger = 1 \quad,\quad
    \hat{U}^{(\nu)}_{k+\frac{2\pi}{Za}} = \hat{U}^{(\nu)}_k \,.
\end{align}
From the normalization (\ref{eq:u_normalization}) of the Bloch states and the unitarity of $\hat{U}^{(\nu)}_k$ we note the properties
\begin{align}
    \label{eq:hat_u_normalization}
    \langle \hat{u}_{k\alpha}^{(\nu)}|\hat{u}_{k\alpha'}^{(\nu)}\rangle &= \delta_{\alpha\alpha'}\,,\\
    \label{eq:hat_u_projector}
    \sum_{\alpha=1}^\nu |\hat{u}_{k\alpha}^{(\nu)}\rangle\langle \hat{u}_{k\alpha}^{(\nu)}| &=
    \sum_{\alpha=1}^\nu |u_{k\alpha}\rangle\langle u_{k\alpha}| \,.
\end{align}

Although the Bloch states $|\hat{u}_{k\alpha}^{(\nu)}\rangle$ are no longer eigenstates of the Bloch Hamiltonian $h_k$ as defined in (\ref{eq:h_k_def}), the corresponding non-Abelian Wannier functions
\begin{align}
    \label{eq:wannier_nonabelian}
    \hat{w}^{(\nu)}_\alpha(ma) = \frac{Za}{2\pi}\,\int_{-\frac{\pi}{Za}}^{\frac{\pi}{Za}} dk\,\hat{u}^{(\nu)}_{k\alpha}(ja)\,e^{ikma}
\end{align}
have interesting properties \cite{marzari_vanderbilt_prb_97} which we summarize in the following. 

Defining the Zak-Berry connection and the geometric tensor in the non-Abelian case analog to (\ref{eq:zak_berry_lattice}) and (\ref{eq:geometric_tensor_lattice_2}) as
\begin{align}
    \label{eq:zak_berry_connection_nonabelian}
    (\hat{A}^{(\nu)}_k)_{\alpha\beta} &= (\hat{A}^{(\nu)}_k)_{\beta\alpha}^* = i\langle \hat{u}^{(\nu)}_{k\alpha}|\partial_k \hat{u}^{(\nu)}_{k\beta}\rangle  \,,\\
    \label{eq:geometric_tensor_nonabelian_1}
    (\mathcal{\hat{Q}}^{(\nu)}_k)_{\alpha\beta} &= (\mathcal{\hat{Q}}_k^{(\nu)})_{\beta\alpha}\\
    \label{eq:geometric_tensor_nonabelian_2}
     &= \langle \partial_k \hat{u}^{(\nu)}_{k\alpha}|\partial_k \hat{u}^{(\nu)}_{k\beta}\rangle \delta_{\alpha\beta}
    - |\langle \hat{u}^{(\nu)}_{k\alpha}|\partial_k \hat{u}^{(\nu)}_{k\beta}\rangle|^2\,,
\end{align}
and introducing the following definition for the non-Abelian Zak-Berry phase matrix
\begin{align}
    \hat{\gamma}^{(\nu)}_{\alpha\beta} = \int_{-\frac{\pi}{Za}}^{\frac{\pi}{Za}} dk\,(\hat{A}^{(\nu)}_k)_{\alpha\beta}\,,
\end{align}
together with $\hat{\gamma}^{(\nu)}_\alpha = \hat{\gamma}^{(\nu)}_{\alpha\alpha}$ for the diagonal components, one finds the following useful transformation properties
\begin{align}
    \label{eq:zak_berry_nonabelian_trafo}
    &\hat{A}^{(\nu)}_k = (\hat{U}^{(\nu)}_k)^\dagger A^{(\nu)}_k \hat{U}^{(\nu)}_k
    + i (\hat{U}^{(\nu)}_k)^\dagger \partial_k \hat{U}^{(\nu)}_k \,,\\
    \label{eq:zak_berry_phase_nonabelian_trafo}
    &\sum_{\alpha=1}^\nu \hat{\gamma}^{(\nu)}_\alpha = \sum_{\alpha=1}^\nu \gamma_\alpha - 2\pi \,\text{wn} [\text{det}\hat{U}^{(\nu)}_k] \,,\\ 
    \label{eq:geometric_tensor_nonabelian_trafo}
    &\sum_{\alpha,\beta=1}^\nu (\hat{\mathcal{Q}}^{(\nu)}_k)_{\alpha\beta} =     
    \sum_{\alpha,\beta=1}^\nu (\mathcal{Q}_k)_{\alpha\beta}\,,
\end{align}
where $\text{wn}[\text{det}\hat{U}^{(\nu)}_k]$ denotes the winding number of the determinant of $\hat{U}_k^{(\nu)}$. Here, $A^{(\nu)}_k$ is a $\nu\times\nu$-matrix with matrix elements $(A^{(\nu)}_k)_{\alpha\beta}=(A_k)_{\alpha\beta}$ given by the Zak-Berry connection (\ref{eq:zak_berry_lattice}) from all band indices $\alpha,\beta=1,\dots,\nu$. 

Furthermore, defining the moments of the Wannier functions in the non-Abelian case analog to (\ref{eq:moments_wannier_lattice}) as
\begin{align}
    \label{eq:moments_nonabelian}
    \hat{C}^{(\nu)}_{r\alpha} &= \langle x^r \rangle^{(\nu)}_{\alpha} = \sum_m (ma)^r |\hat{w}^{(\nu)}_{\alpha}(ma)|^2 \\
    \label{eq:moments_nonabelian_u}
    &=\frac{Za}{2\pi} \int_{-\pi/Za}^{\pi/Za} dk \,\langle \hat{u}^{(\nu)}_{k\alpha}|(i\partial_k)^r|\hat{u}^{(\nu)}_{k\alpha}\rangle\,,
\end{align}
and introducing the following definitions for the sum over the positions and the quadratic spreads 
\begin{align}
    \label{eq:average_position_nonabelian}
    \langle x \rangle^{(\nu)} &= \sum_{\alpha=1}^\nu \hat{C}^{(\nu)}_{1\alpha}\,,\\
    \label{eq:spread_nonabelian}
    \langle \Delta x^2 \rangle^{(\nu)} &= \sum_{\alpha=1}^\nu \left(\hat{C}^{(\nu)}_{2\alpha} - (\hat{C}^{(\nu)}_{1\alpha})^2\right) \,,
\end{align}
we find from (\ref{eq:moments_nonabelian_u}) and the above definitions after some straightforward algebra
\begin{align}
    \label{eq:average_position_nonabelian_phase}
    \frac{\langle x \rangle^{(\nu)}}{Za} &= \sum_{\alpha=1}^\nu \frac{\hat{\gamma}^{(\nu)}_{\alpha}}{2\pi} \quad,\quad
    \hat{C}^{(\nu)}_{1\alpha} = \frac{Za}{2\pi}\,\hat{\gamma}^{(\nu)}_{\alpha}\,,\\
    \nonumber
    \langle \Delta x^2 \rangle^{(\nu)} &= \frac{Za}{2\pi}\,\sum_{\alpha,\beta=1}^\nu\,\int_{-\frac{\pi}{Za}}^{\frac{\pi}{Za}} dk\, 
    \Bigg\{(\hat{\mathcal{Q}}^{(\nu)}_k)_{\alpha\beta} \,+ \\
    \label{eq:spread_nonabelian_tensor}
    &\hspace{-2cm}
     +  \Big[(\hat{A}^{(\nu)}_k)_{\alpha\beta}-\frac{Za}{2\pi}\hat{\gamma}^{(\nu)}_{\alpha\beta}\Big]^2 
    + (1-\delta_{\alpha\beta})\Big(\frac{Za}{2\pi}\hat{\gamma}^{(\nu)}_{\alpha\beta}\Big)^2\Bigg\}\,.
\end{align}
Since the first term on the right hand side of (\ref{eq:spread_nonabelian_tensor}) is gauge invariant according to (\ref{eq:geometric_tensor_nonabelian_trafo}), and the last two terms are positive, the quadratic spread $\langle \Delta x^2 \rangle^{(\nu)}$ is minimal in the non-Abelian gauge of maximal localization (NA-ML), defined by a $k$-independent and band-diagonal Zak-Berry connection 
\begin{align}
    \label{eq:zak_berry_connection_NA-ML}
    (\hat{A}^{(\nu)}_k)_{\alpha\beta} = \frac{Za}{2\pi}\hat{\gamma}^{(\nu)}_{\alpha\beta} =
    \frac{Za}{2\pi}\hat{\gamma}^{(\nu)}_\alpha \delta_{\alpha\beta} \,.
\end{align}
As a consequence, using the surface fluctuation theorem (\ref{eq:QB_fluctuations_geometric_tensor}), we find that the minimal quadratic spread in the NA-ML gauge is related to the boundary charge fluctuations as
\begin{align}
    \label{eq:sft_nonabelian} 
    l_p (\Delta Q_B^{(\nu)})^2 &= \frac{1}{Za}\langle \Delta x^2 \rangle^{(\nu)}_\text{min} = 
    \sum_{\alpha=1}^\nu \hat{D}^{(\nu)}_{2\alpha} \,,
\end{align}
where we defined the moments  
\begin{align}    
\label{eq:hat_D}
    \hat{D}^{(\nu)}_{r\alpha} &= \sum_m (\hat{m}^{(\nu)}_\alpha a)^r |\hat{w}^{(\nu)}_{\alpha}(ma)|^2 
\end{align}
relative to the shift by the first moment, i.e., with
\begin{align}
    \hat{m}^{(\nu)}_\alpha = m - \hat{C}^{(\nu)}_{1\alpha} = m - \frac{Za}{2\pi}\hat{\gamma}^{(\nu)}_\alpha  \,.
\end{align}
Most importantly, the surface fluctuation theorem (\ref{eq:sft_nonabelian}) formulated in terms of the non-Abelian quadratic spread shows that the NA-ML gauge is unique in the sense that the boundary charge fluctuations can be written as the sum over the second moments of the Wannier functions of the individual bands. Within the Abelian ML gauge this is {\it not} possible, as already discussed after (\ref{eq:fluctuations_single_band}). For the boundary charge itself it seems to be similar at first sight to the Abelian AF and ML gauge, where it is also related to the sum over the Zak-Berry phases of the individual bands. However, as we have seen in Section~\ref{sec:sct}, a subtle cancellation procedure happens such that the sum over the Zak-Berry phases of the individual bands $\alpha=1,\dots,\nu$ is related to the phase $\gamma^{(\nu)}$ of the gap parameter of gap $\nu$, see the central equation (\ref{eq:zak_berry_sum_dirac}). As we will see in the following this is not needed in the non-Abelian gauge where all Zak-Berry phases $\hat{\gamma}_\alpha^{(\nu)}$ are equally spaced and related to $\gamma^{(\nu)}$ in the following universal way 
\begin{align}
    \nonumber
    \hat{\gamma}_\alpha^{(\nu)} &= - \frac{1}{\nu} \gamma^{(\nu)} + \\
    \label{eq:zak_berry_phases_nonabelian_result}
    & + \frac{\pi}{\nu}
    \begin{cases} \nu - 2\alpha +1 & \text{for}\quad \nu \,\text{even} \\ \nu - 2\alpha +1 + s_{\gamma^{(\nu)}} & \text{for}\quad \nu \,\text{odd} \end{cases}\,,
\end{align}
such that $-\pi < \hat{\gamma}^{(\nu)}_\alpha < \pi$. 

Consistent with the way the fluctuations are distributed among the bands in the non-Abelian case, we will find that all Wannier functions $\hat{w}_\alpha^{(\nu)}$ in the NA-ML gauge scale in the same way according to the upper component $\tilde{w}_{\text{u},\nu}$ of the Wannier function in the Abelian ML gauge for band $\nu$. The central result shown in the next Section~\ref{sec:NA-ML_gauge} is the precise relation
\begin{align}
    \nonumber
    \hat{w}^{(\nu)}_\alpha(ma) &\approx \frac{1}{\sqrt{\nu}}\,e^{i\theta^{(\nu)}_\alpha} \\
    \label{eq:wannier_abelian_vs_nonabelian}
    &\hspace{-1cm}
    \times\,a \sqrt{Z} \,\sum_p \tilde{w}^{(\nu)}_{-,p}(\hat{m}^{(\nu)}_\alpha a) \,e^{ipk_F^{(\nu)}ma} \,,
\end{align}
where $e^{i\theta^{(\nu)}_\alpha}$ is some unimportant phase factor. Up to this phase factor, we find in comparison to (\ref{eq:wannier_ML_dirac_lattice_u}) the central result that {\it all} Wannier functions $\hat{w}_\alpha^{(\nu)}(ma)$ in the NA-ML gauge scale precisely as $\frac{1}{\sqrt{\nu}}\,\tilde{w}_{\text{u},\nu}(ma)$ in the Abelian ML gauge, provided one replaces $\gamma_\nu \rightarrow \hat{\gamma}_\alpha^{(\nu)}$ in the shift variable such that $\tilde{m}_\nu\rightarrow\hat{m}^{(\nu)}_\alpha$. Using the scaling property (\ref{eq:tilde_w_u_scaling}) of the upper component of the Wannier function in the ML gauge, this gives the following scaling for all non-Abelian Wannier functions
\begin{align}
    \label{eq:hat_w_scaling}
    \frac{1}{Z}|\hat{m}^{(\nu)}_\alpha \,\hat{w}^{(\nu)}_{\alpha}(n,j)|^2 &\approx 
    \frac{1}{\nu}\,|\tilde{F}_{-}(\frac{\hat{m}^{(\nu)}_\alpha a}{\xi^{(\nu)}};\frac{\nu j}{Z},\gamma^{(\nu)})|^2 \,.
\end{align}
As a consequence, defining the following scaling functions for the non-Abelian moments
\begin{align}
    \label{eq:hat_D_r_scaling_def}
    \hat{D}^{(\nu)}_{r\alpha}(Ma) = \sum_{m=-M}^M (\hat{m}^{(\nu)}_\alpha a)^r \, |\hat{w}^{(\nu)}_\alpha(ma)|^2 \,,
\end{align}
we find from (\ref{eq:tilde_D_r_ud_scaling}) and (\ref{eq:tilde_D_2_u_asymp}) the universal scaling 
\begin{align}
    \label{eq:hat_D_r_scaling}
    \frac{\hat{D}^{(\nu)}_{r\alpha}(Ma)}{Za(\xi^{(\nu)})^{r-1}} \approx \frac{1}{\nu} \int_{-\frac{Ma}{\xi^{(\nu)}}}^{\frac{Ma}{\xi^{(\nu)}}}
    dy\,y^{r-2} \tilde{G}_{-}(y;\gamma^{(\nu)}) \,,
\end{align}
with the following asymptotic value for the second moment 
\begin{align}
    \label{eq:hat_D_2_asymptotic}
    \frac{\hat{D}^{(\nu)}_{2\alpha}(Ma)}{Za\xi^{(\nu)}} \rightarrow \frac{1}{8\nu} \,.
\end{align}

The universal relation (\ref{eq:wannier_abelian_vs_nonabelian}) of all non-Abelian Wannier functions $\alpha=1,\dots,\nu$ to the Dirac Wannier function of the lower band of the Dirac theory corresponding to gap $\nu$ is one of the most important results of this work. It shows that all non-Abelian Wannier functions scale independent of the band index in the same way and depend only on the low-energy properties of the model, provided that the condition of small gaps is fulfilled. Up to our best knowledge the non-Abelian Wannier functions have not been studied analytically in the literature and their universal scaling behavior for all generalized AAH models in terms of a single length scale $\xi^{(\nu)}$ has not been reported before.
\begin{figure}[b!]
    \centering
	 \includegraphics[width =0.9\columnwidth]{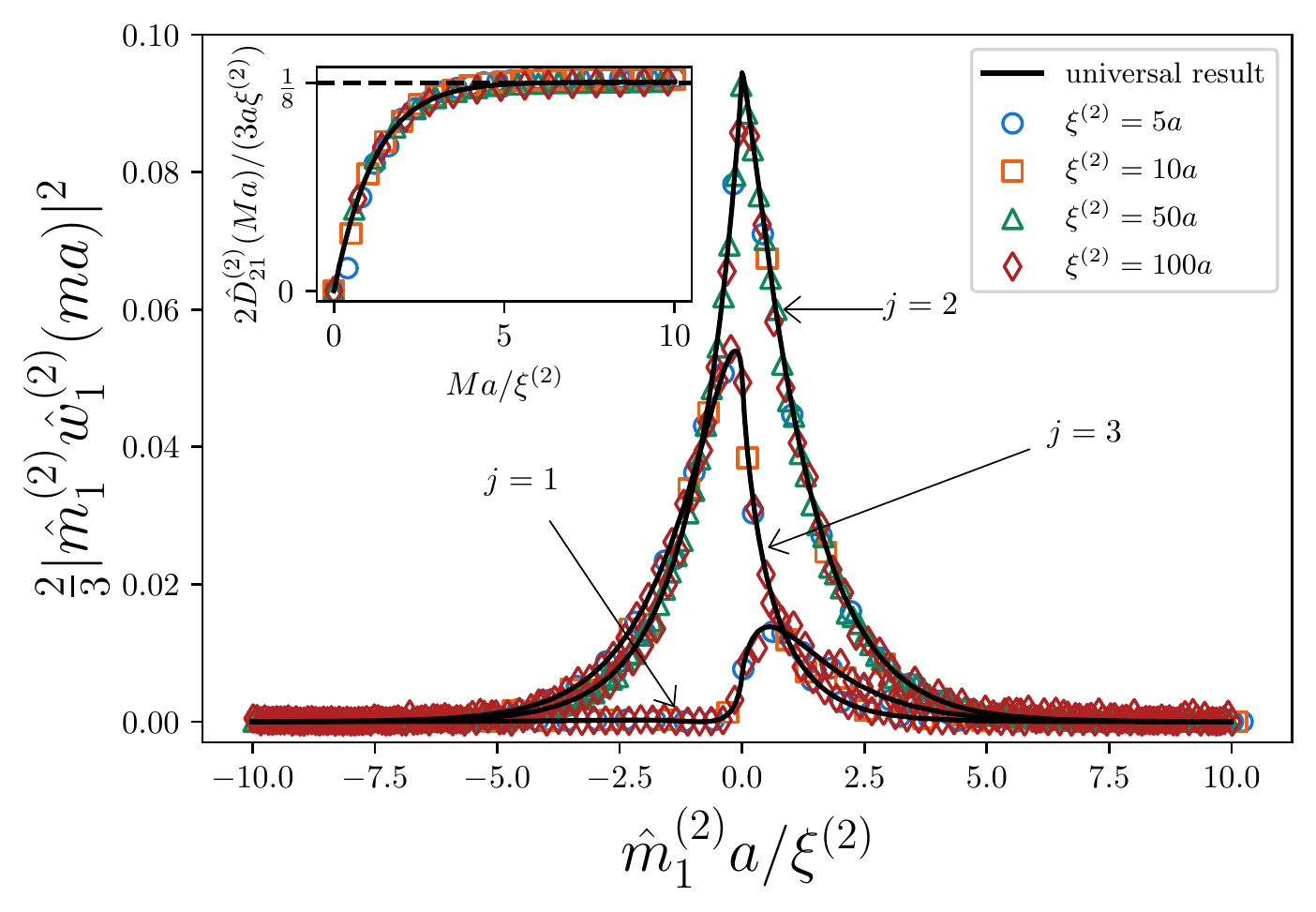}
	 	 \includegraphics[width =0.9\columnwidth]{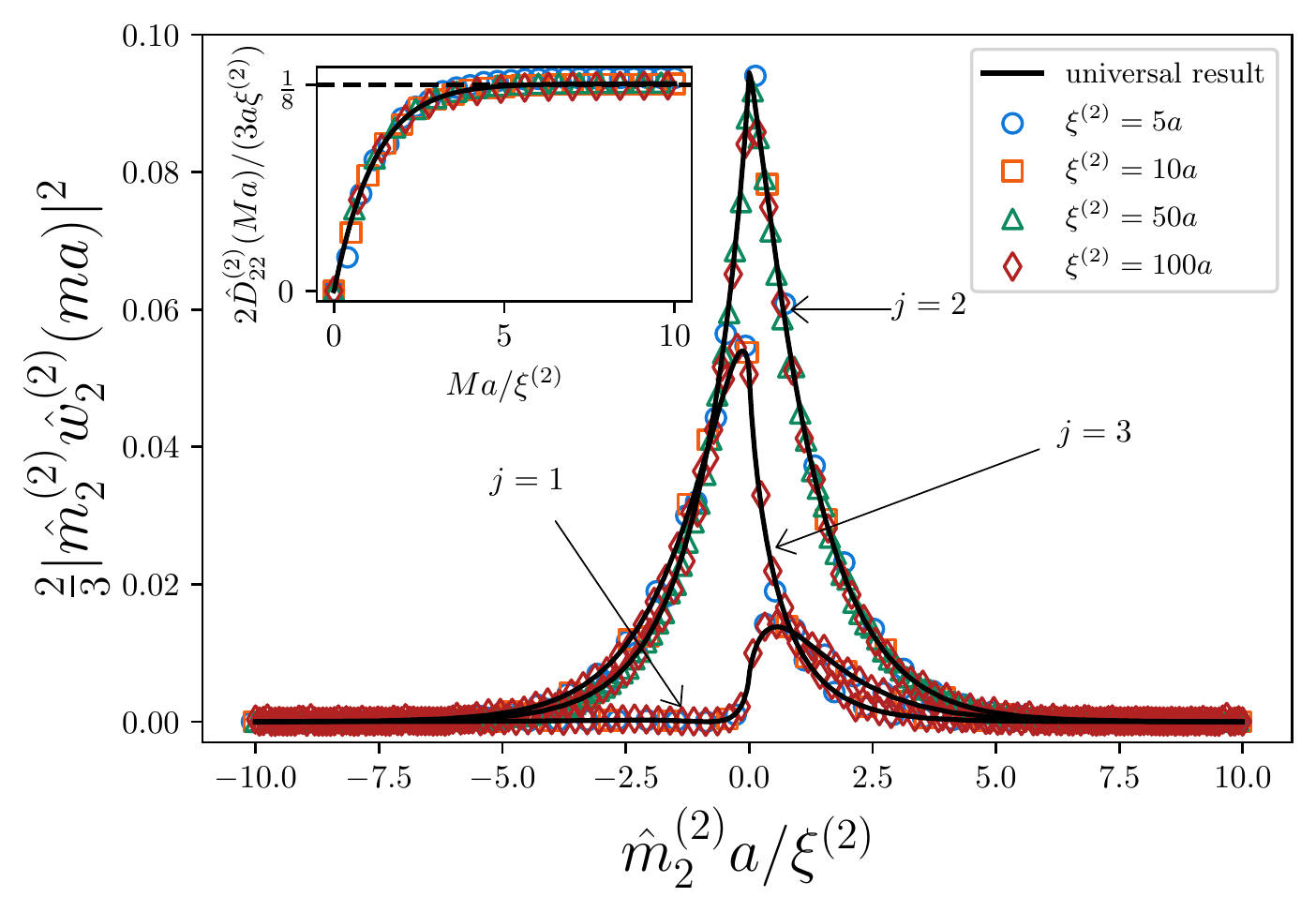}
	  \caption{Scaling of $\frac{\nu}{Z}|\hat{m}^{(\nu)}_\alpha \,\hat{w}^{(\nu)}_{\alpha}(n,j)|^2$ for $\alpha=1,2$ (upper/lower panel), $\nu=2$, $Z=3$, $\gamma^{(1)}=0.2\pi$, $\gamma^{(2)}=0.5\pi$, and $\xi^{(1)}=30 a$, as function of $\hat{m}^{(\nu)}_\alpha a/\xi^{(\nu)}$ for different values of $\xi^{(2)}$ and $j=1,2,3$. According to (\ref{eq:hat_w_scaling}) we obtain the universal curve 
	  $|\tilde{F}_{-}(\hat{m}_\alpha^{(2)} a/\xi^{(2)};2j/3,\gamma^{(2)})|^{2}$, i.e., the same scaling for both $\alpha=1,2$ as for $\frac{1}{3}|\tilde{m}_2\,\tilde{w}_{\text{u},2}(ma)|^2$ as function of $\tilde{m}_2 a/\xi^{(2)}$, see Fig.~\ref{fig:ML_up}. The two insets show the scaling of the corresponding second moments $\nu \hat{D}^{(\nu)}_{2\alpha}(Ma)/(Za\xi^{(\nu)})$ as function of $s_2=Ma/\xi^{(2)}$. According to (\ref{eq:hat_D_r_scaling}) and (\ref{eq:hat_D_2_asymptotic}) we obtain the universal curve $\int_{-s_2}^{s_2} dy \,\tilde{G}_{-}(y;\gamma^{(2)})$ with saturation at the correct value $\frac{1}{8}$. The scaling is independent of $\alpha=1,2$ and identical to the scaling of the second moment $\tilde{D}_{\text{u},22}(Ma)$ as shown in the inset of Fig.~\ref{fig:ML_up}.}
    \label{fig:NAML}
\end{figure}

For the special case $Z=3$ and $\nu=2$ we demonstrate in Fig.~\ref{fig:NAML} the universal scaling (\ref{eq:hat_w_scaling}) of the Wannier functions $\hat{w}_\alpha^{(\nu)}(ma)$ in NA-ML gauge and the corresponding scaling (\ref{eq:hat_D_r_scaling}) of the second moments $\hat{D}_{2\alpha}^{(\nu)}(Ma)$. Indeed, we find that the scaling is independent of $\alpha=1,2$ and follows the scaling of the upper part  $\tilde{w}_{\text{u},\nu}(ma)$ of the Wannier function and its second moment in ML gauge for band $\alpha=\nu$, compare with Fig.~\ref{fig:ML_up}. 

We note that also the winding number of the determinant $\text{det}\hat{U}^{(\nu)}_k$ contains interesting information about the total number of edge states $N^{(\nu)}_{\text{E}}$ present in all gaps $\nu'=1,\dots,\nu$. According to (\ref{eq:zak_berry_phase_nonabelian_trafo}), the winding number controls the change of the sum over all Zak-Berry phases from the bands $\alpha=1,\dots,\nu$ in the gauge transformation. On the other hand we can calculate the sum over the Zak-Berry phases in the AF and NA-ML gauge separately by using (\ref{eq:zak_berry_alpha_dirac}), (\ref{eq:zak_berry_phase_vs_gap_phase}) and (\ref{eq:zak_berry_phases_nonabelian_result})
\begin{align}
    \label{eq:sum_zbp_AF}
    \sum_{\alpha=1}^\nu \gamma_\alpha &= - \gamma^{(\nu)} + \pi \sum_{\nu'=1}^\nu s_{\gamma^{(\nu')}} \,,\\
    \label{eq:sum_zbp_NAMF}
    \sum_{\alpha=1}^\nu \hat{\gamma}^{(\nu)}_\alpha &= - \gamma^{(\nu)} + \pi \,\delta_{\nu,\text{odd}}s_{\gamma^{(\nu)}} \,.
\end{align}
Since an edge state is present in gap $\nu'$ if $s_{\gamma^{(\nu')}} > 0$, we get in addition for the total number of edge states the expression
\begin{align}
    \label{eq:number_edge_states}
    N^{(\nu)}_{\text{E}} = \frac{1}{2} \sum_{\nu'=1}^\nu (1+s_{\gamma^{(\nu')}}) 
    = \frac{1}{2}\left(\nu+\sum_{\nu'=1}^\nu s_{\gamma^{(\nu')}}\right)\,.
\end{align}
Comparing these equations we find the following relation between the number of edge states and the winding number of $\text{det}U^{(\nu)}_k$ 
\begin{align}
    N^{(\nu)}_{\text{E}} = \text{wn} [\text{det}\hat{U}^{(\nu)}_k] + \frac{1}{2}\left(\nu + \delta_{\nu,\text{odd}}s_{\gamma^{(\nu)}}\right) \,,
\end{align}
which is obviously an integer number for both $\nu$ even or odd. This relation can be viewed as a bulk-boundary correspondence relating the total number of edge states up to a certain gap $\nu$ to a winding number defined in terms of the bulk states. In contrast to other topological invariants defined in one-dimensional systems, this relation holds independent of any symmetry constraints and includes all edge states, independent of whether they are at zero energy or not.  

Using (\ref{eq:QB_zak_berry_phase}) and (\ref{eq:QB_nu}), we finally find for the total boundary charge without or with including the edge states
\begin{align}
    \label{eq:QB_bands}
    \sum_{\alpha=1}^\nu Q_{B,\alpha} &= -\frac{1}{2\pi}\sum_{\alpha=1}^\nu \gamma_\alpha - \frac{1}{2}\nu + \frac{\nu}{2Z} \,,\\
    \label{eq:sct_NAML}
    Q_B^{(\nu)} &= -\frac{1}{2\pi}\sum_{\alpha=1}^\nu \hat{\gamma}^{(\nu)}_\alpha + \frac{1}{2}\delta_{\nu,\text{odd}}s_{\gamma^{(\nu)}} 
    + \frac{\nu}{2Z} \,.
\end{align}
The last result states the surface charge theorem in terms of the non-Abelian Zak-Berry phases $\hat{\gamma}^{(\nu)}_\alpha$ or the first moments $\hat{C}_{1\alpha}^{(\nu)}=\frac{Za}{2\pi}\hat{\gamma}_\alpha^{(\nu)}$ of the non-Abelian Wannier functions.

\subsection{Explicit construction of the NA-ML gauge}
\label{sec:NA-ML_gauge}

We proceed in this section with the explicit construction of the NA-ML gauge and the analytical proof of the central identity (\ref{eq:wannier_abelian_vs_nonabelian}) relating all non-Abelian Wannier functions $\alpha=1,\dots,\nu$ to the upper part of the Wannier function in the Abelian ML gauge for the highest band $\nu$. The NA-ML gauge can be constructed explicitly via the unitary transformation
\begin{align}
    \label{eq:NAML_explicit}
    \hat{U}_k^{(\nu)} = U^{(\nu)}(k,k_0) V^{(\nu)} e^{-ik\frac{Za}{2\pi}\hat{\gamma}^{(\nu)}} \,, 
\end{align}
where 
\begin{align}
    U^{(\nu)}(k_1,k_2) = P e^{i\int_{k_2}^{k_1} dk A^{(\nu)}_k} 
\end{align}
denotes the Wilson propagator along the path from $k_2\rightarrow k_1$ (here, $P$ denotes the $k$-ordering operator analog to the time ordering operator), $k_0$ is an arbitrary reference point, and $V^{(\nu)}$ is the unitary transformation which diagonalizes the Wilson loop operator
\begin{align}
    \label{eq:V_def}
    e^{i \hat{\gamma}^{(\nu)}} &= (V^{(\nu)})^\dagger U^{(\nu)}_{L} \,V^{(\nu)}  \,,\\
    \label{eq:loop_def}
    U^{(\nu)}_{\text{L}}&\equiv U^{(\nu)}(k_0+\frac{2\pi}{Za},k_0)\,.
\end{align}
Indeed, using the differential equation
\begin{align}
    i\partial_k \hat{U}^{(\nu)}_k = - A_k^{(\nu)} \hat{U}_k^{(\nu)} + \hat{U}_k^{(\nu)} \frac{Za}{2\pi} \gamma^{(\nu)} \,,
\end{align}
and multiplying it from the left with $(U_k^{(\nu)})^\dagger$, one obtains directly the condition (\ref{eq:zak_berry_connection_NA-ML}) defining the NA-ML gauge. Furthermore, using the group property of the Wilson propagator together with its periodicity (following from the periodicity of the Zak-Berry connection $A_k^{(\nu)}=A_{k+\frac{2\pi}{Za}}^{(\nu)}$) 
\begin{align}
    \label{eq:prop_group}
    U^{(\nu)}(k_1,k_2) &= U^{(\nu)}(k_1,k_3) \, U^{(\nu)}(k_3,k_2) \,,  \\
    \label{eq:prop_periodicity}
    U^{(\nu)}(k_1,k_2) &= U^{(\nu)}(k_1+\frac{2\pi}{Za},k_2+\frac{2\pi}{Za})   \,,
\end{align}
we find periodicity of the non-Abelian gauge transformation
\begin{align}
    \nonumber
    \hat{U}^{(\nu)}_{k+\frac{2\pi}{Za}} &= U^{(\nu)}(k+\frac{2\pi}{Za},k_0+\frac{2\pi}{Za}) \hat{U}^{(\nu)}_{\text{L}}V^{(\nu)} e^{-i\hat{\gamma}^{(\nu)}} e^{-ik\frac{Za}{2\pi}\hat{\gamma}^{(\nu)}} \\
    &= U^{(\nu)}(k,k_0) V^{(\nu)}e^{-ik\frac{Za}{2\pi}\hat{\gamma}^{(\nu)}} = \hat{U}^{(\nu)}_{k}\,.
\end{align}
Here we used in the first step the group property (\ref{eq:prop_group}), and in the second step the periodicity (\ref{eq:prop_periodicity}) together with the definition (\ref{eq:V_def}) of the transformation $V^{(\nu)}$.

We note that the non-Abelian Zak-Berry phases $\hat{\gamma}^{(\nu)}_\alpha$ are independent of the choice $k_0$ of the reference point since the Wilson loop operator $(U^{(\nu)}_{\text{L}})^\prime$ for another reference point $k_0^\prime$ is related to $U^{(\nu)}_{\text{L}}$ via a unitary transformation
\begin{align}
    (U^{(\nu)}_{\text{L}})^\prime = 
    U^{(\nu)}(k_0,k_0^\prime) \, U^{(\nu)}_{\text{L}}\, U^{(\nu)}(k_0,k_0^\prime)^\dagger \,,
\end{align}
where we used the group property (\ref{eq:prop_group}) and the periodicity (\ref{eq:prop_periodicity}). This means that the unitary operator $(V^{(\nu)})^\prime$ with respect to the reference point $k_0^\prime$ is related to $V^{(\nu)}$ by 
\begin{align}
    (V^{(\nu)})^\prime = U^{(\nu)}(k_0^\prime,k_0) \, V^{(\nu)}\,.
\end{align}
As a consequence, up to trivial phase factors to define the unitary transformation $V^{(\nu)}$, we find that the unitary transformation $\hat{U}^{(\nu)}_k$ is unique and independent of $k_0$.

\begin{figure}[t!]
    \centering
	  \includegraphics[width =0.9\columnwidth]{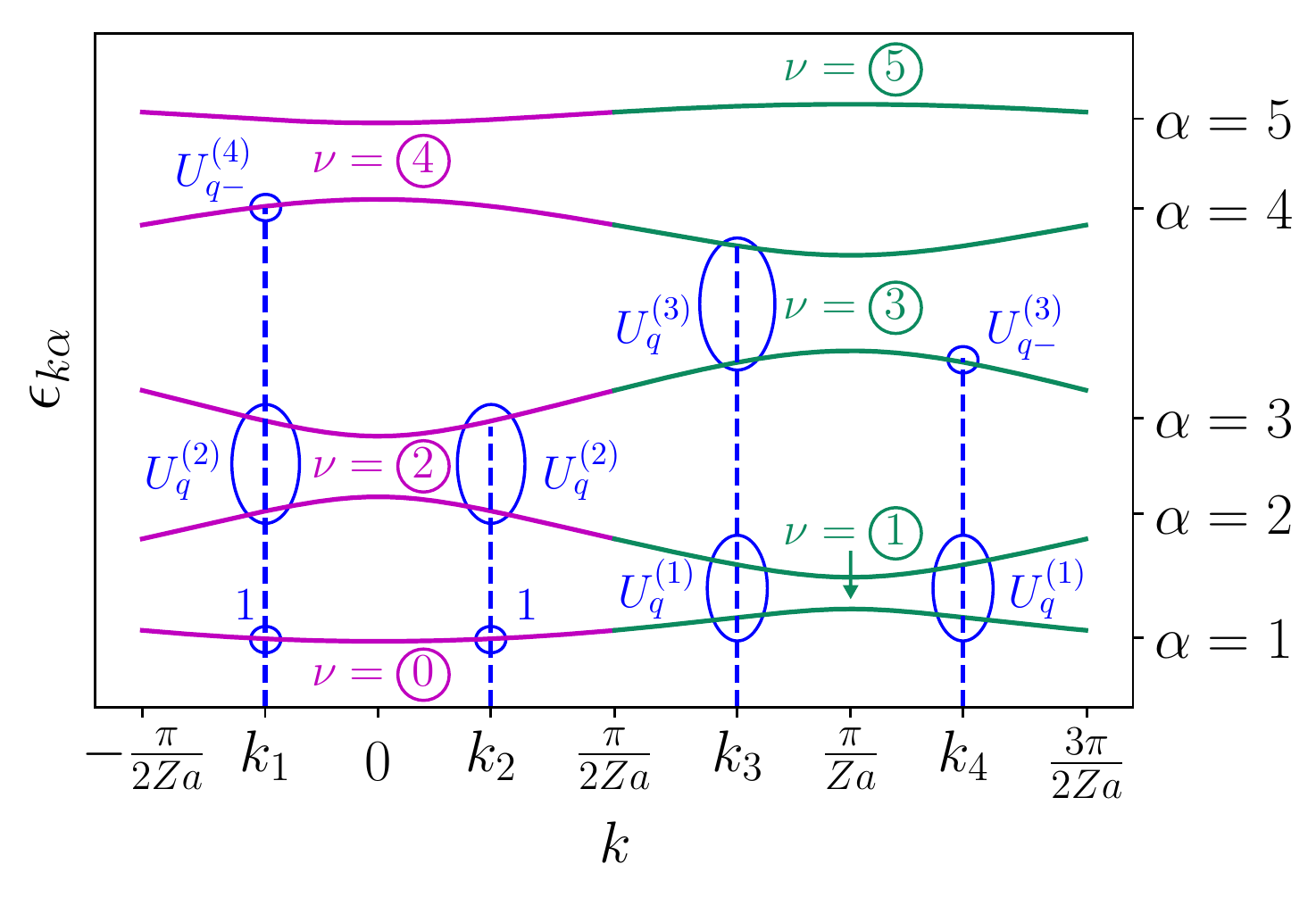}
	  \caption{Sketch of the typical band structure for the interval $-\frac{\pi}{Za}<k<\frac{3\pi}{2Za}$ for $Z=5$. The two subintervals (I) with $-\frac{\pi}{Za}<k<\frac{\pi}{2Za}$ and (II) with $\frac{\pi}{Za}<k<\frac{3\pi}{2Za}$ are indicated in violet and green colour, respectively. The lowest/highest band in subinterval (I)/(II) (corresponding to $\nu=0$ and $\nu=Z$, respectively) are characterized by $k$-independent Bloch states $u_{k\alpha}$. Four different $k$-values are shown, indicating the block structure (\ref{eq:block_structure_propagator_nu_even}) and (\ref{eq:block_structure_propagator_nu_odd}) of the lattice Wilson propagator for $\nu$ even or odd in the two subintervals. Here, $k_1$ corresponds to $U^{(\nu)}(k_1,-\frac{\pi}{2Za})$ for $\nu=4$ even in subinterval (I) (see the first equation of (\ref{eq:block_structure_propagator_nu_even})), $k_2$ to $U^{(\nu)}(k_2,-\frac{\pi}{2Za})$ for $\nu=3$ odd in subinterval (I) (see the first equation of (\ref{eq:block_structure_propagator_nu_odd})), $k_3$ to $U^{(\nu)}(k_3,\frac{\pi}{2Za})$ for $\nu=4$ even in subinterval (II) (see the second equation of (\ref{eq:block_structure_propagator_nu_even})), and $k_4$ to $U^{(\nu)}(k_4,\frac{\pi}{2Za})$ for $\nu=3$ odd in subinterval (II) (see the second equation of (\ref{eq:block_structure_propagator_nu_odd})). 
	  } 
    \label{fig:band_structure_2}
\end{figure}
For convenience, we will choose in the following $k_0=-\pi/(2Za)$ as reference point and discuss the two intervals (I) $k\in [-\pi/(2Za),\pi/(2Za)$ and (II) $k\in[\pi/(2Za),3\pi/(2Za)]$ separately since they are related to two different Dirac theories, see Fig.~\ref{fig:band_structure_2}. Using (\ref{eq:u_k_periodicity}), (\ref{eq:nu_even}) and (\ref{eq:nu_odd}), we get for the relation between the Bloch and Dirac states in this convention
\begin{align}
    \nonumber
    \underline{\text{(I)}}: &\quad \nu \,\,{\rm even}, \quad k=q \\
    \label{eq:Dirac_lattice_I}
    & u_{k\alpha}(ma) = \frac{1}{\sqrt{Z}}\sum_{p=\pm} \chi_{q\tau p}^{(\nu)}\, e^{i p k_F^{(\nu)}ma}\,,\\
    \nonumber
    \underline{\text{(II)}}: &\quad \nu \,\,{\rm odd}, \quad k=\frac{\pi}{Za}+q \\
    \label{eq:Dirac_lattice_II}
    & u_{k\alpha}(ma) = e^{-i\frac{\pi}{Z}m}\,\frac{1}{\sqrt{Z}}\,\sum_{p=\pm} \chi_{q\tau p}^{(\nu)}\, e^{i p k_F^{(\nu)}ma}\,,
\end{align}
where in both cases $|q|<\frac{\pi}{2Za}$. By convention we have added two trivial Dirac theories, one for $\alpha=1$ in subinterval (I) with $\nu=0$ and $\tau=+$, and another for $\alpha=Z$ in subinterval (II) with $\nu=Z$ and $\tau=-$. In these two cases the Bloch states are given by the gapless ones and are independent of $k$, such that the Zak-Berry connection is zero. 

The non-Abelian Wannier functions are then split as
\begin{align}
    \label{eq:hat_w_splitting}
    \hat{w}_\alpha^{(\nu)}(ma) &= \hat{w}_{I,\alpha}^{(\nu)}(ma) + \hat{w}_{II,\alpha}^{(\nu)}(ma) \,,\\  
    \label{eq:hat_w_I}
    \hat{w}_{I,\alpha}^{(\nu)}(ma) &= \frac{Za}{2\pi}\,\int_{-\frac{\pi}{2Za}}^{\frac{\pi}{2Za}} dk\,\hat{u}_{k\alpha}^{(\nu)}(ma)\,e^{ikma} \,,\\
    \label{eq:hat_w_II}
    \hat{w}_{II,\alpha}^{(\nu)}(ma) &= \frac{Za}{2\pi}\,\int_{\frac{\pi}{2Za}}^{\frac{3\pi}{2Za}} dk\,\hat{u}_{k\alpha}^{(\nu)}(ma)\,e^{ikma} \,,
\end{align}
where the transformed Bloch states follow from (\ref{eq:u_nonabelian_trafo}) and (\ref{eq:NAML_explicit}) as 
\begin{align}
    \nonumber
    \hat{u}_{k\alpha}^{(\nu)}(ma)\,e^{ikma} &= e^{ik\hat{m}^{(\nu)}_\alpha a} \\
    \label{eq:nonabelian_u_explicit}
    &\hspace{-1.5cm}
    \times\sum_{\alpha'=1}^\nu \,u_{k\alpha'}(ma)\,
    \left[U^{(\nu)}(k,-\frac{\pi}{2Za})V^{(\nu)}\right]_{\alpha'\alpha}\,.
\end{align}
It turns out to be useful to define the propagators when crossing the two subintervals as
\begin{align}
    \label{eq:prop_I}
    U^{(\nu)}_{\text{I}} &= U^{(\nu)}(\frac{\pi}{2Za},-\frac{\pi}{2Za})\,,\\
    \label{eq:prop_II}
    U^{(\nu)}_{\text{II}} &= U^{(\nu)}(\frac{3\pi}{2Za},\frac{\pi}{2Za})\,,
\end{align}
and use 
\begin{align}
    \label{eq:prop_splitting_II}
    U^{(\nu)}(k,-\frac{\pi}{2Za}) = U^{(\nu)}(k,\frac{\pi}{2Za}) \,U^{(\nu)}_{\text{I}}
\end{align}
for the representation of the propagator within subinterval (II). The Wilson loop operator defining the transformation $V^{(\nu)}$ via (\ref{eq:V_def}) can be expressed as  
\begin{align}
    \label{eq:wilson_loop_total}
    U_{\text{L}}^{(\nu)} = U^{(\nu)}_{\text{II}} \,U^{(\nu)}_{\text{I}} \,.
\end{align}

As outlined in detail in Appendix~\ref{app:nonabelian_splitting} one can extend the integration limits in (\ref{eq:hat_w_I}) and (\ref{eq:hat_w_II}) to $\pm\infty$ by using the $k$-independent values of the Bloch states and the propagator at the edges of the two subintervals. This means that the propagators for the two subintervals are extended as
\begin{align}
    \label{eq:prop_extension_I}
    \underline{(\text{I})}: & \, U^{(\nu)}(k,-\frac{\pi}{2Za}) \rightarrow 
    \begin{cases} U^{(\nu)}_{\text{I}} & \text{for}\,k>\frac{\pi}{2Za} \\
         1 &\text{for}\,k<-\frac{\pi}{2Za} \end{cases} \,,\\
    \label{eq:prop_extension_II}
    \underline{(\text{II})}: & \, U^{(\nu)}(k,\frac{\pi}{2Za}) \rightarrow 
    \begin{cases} U^{(\nu)}_{\text{II}} &\text{for}\,k>\frac{3\pi}{2Za} \\ 1 &\text{for}\,k<\frac{\pi}{2Za} \end{cases} \,.
\end{align}
The additional contributions are shown to cancel each other when taking the sum (\ref{eq:hat_w_splitting}) up to trivial delta functions which are subtracted as usual both in lattice and Dirac theory.

Since the Zak-Berry connection has the block structure (\ref{eq:zak_berry_lattice_dirac}) in terms of the Dirac Zak-Berry connection, we obtain a corresponding block structure for the propagator
\begin{align}
    \label{eq:prop_lattice_dirac_I}
    \underline{(\text{I})}: & \, U^{(\nu)}(k,-\frac{\pi}{2Za})_{\alpha_1\alpha_2} = \delta_{\nu_1\nu_2} (U^{(\nu_1)}_q)_{\tau_1\tau_2} \,,\\
    \label{eq:prop_lattice_dirac_II}
    \underline{(\text{II})}: & \, U^{(\nu)}(k,\frac{\pi}{2Za})_{\alpha_1\alpha_2} = \delta_{\nu_1\nu_2} (U^{(\nu_1)}_q)_{\tau_1\tau_2} \,\,,
\end{align}
where $U^{(\nu)}_q$ denotes the Wilson propagator in Dirac theory with respect to reference point $q=-\infty$
\begin{align}
    \label{eq:prop_dirac}
    U^{(\nu)}_q = P e^{i\int_{-\infty}^q dq' A^{(\nu)}_{q'}} \,.
\end{align}
Here, $A^{(\nu)}_q$ denotes the $2\times 2$-matrix of the Dirac Zak-Berry connection (\ref{eq:zak_berry_connection_dirac}). According to Fig.~\ref{fig:band_structure_2} we then get the following block structure of the two propagators for $\nu$ even and odd in the corresponding subintervals (where we use $k=q$ for subinterval (I) and $k=\frac{\pi}{Za}+q$ for subinterval (II))
\begin{widetext}
\begin{align}
    \nonumber
    &\underline{\nu\,\,{\rm even}}: \\
    \label{eq:block_structure_propagator_nu_even}
    & U^{(\nu)}(k,-\frac{\pi}{2Za}) =  
    \left(\begin{array}{cccccc} 
    U_{q-}^{(\nu)} & & & & &  \\
    & U_q^{(\nu-2)} & & & & \\ 
    & & U_q^{(\nu-4)} & & & \\ 
    & & & \ddots & & \\ 
    & & & & U_q^{(2)} & \\ 
    & & & & & 1 \\
    \end{array}\right)
    \,,\, U^{(\nu)}(k,\frac{\pi}{2Za}) =  
    \left(\begin{array}{cccc} 
    U_q^{(\nu-1)} & & &  \\
    & U_q^{(\nu-3)} & & \\ 
    & & \ddots & \\ 
    & & & U_q^{(1)} \\
    \end{array}\right) \,,\\
    \nonumber
    &\underline{\nu\,\,{\rm odd}}: \\
    \label{eq:block_structure_propagator_nu_odd}
    & U^{(\nu)}(k,-\frac{\pi}{2Za}) =  
    \left(\begin{array}{ccccc} 
    U_q^{(\nu-1)} & & & &  \\
    & U_q^{(\nu-3)} & & & \\ 
    & & \ddots & & \\ 
    & & & U_q^{(2)} & \\ 
    & & & & 1 \\
    \end{array}\right)
    \,,\, U^{(\nu)}(k,\frac{\pi}{2Za}) =  
    \left(\begin{array}{ccccc} 
    U_{q-}^{(\nu)} & & & &  \\
    & U_q^{(\nu-2)} & & & \\ 
    & & U_q^{(\nu-4)} & & \\ 
    & & & \ddots & \\ 
    & & & & U_q^{(1)} \\ 
    \end{array}\right)\,,
\end{align}
\end{widetext}
where $U^{(\nu)}_{q-}$ is the Abelian Dirac propagator for the upper part of the last band $\nu$, which can be expressed with the help of (\ref{eq:dirac_berry_connection}) and (\ref{eq:phi_-_asymptotic}) as
\begin{align}
    \label{eq:Dirac_prop_valence_band}
    U^{(\nu)}_{q-} = e^{i\int_{-\infty}^q dq' A^{(\nu)}_{q'-}} = e^{i\phi_{q-}^{(\nu)}} e^{-i\gamma^{(\nu)}}\,.
\end{align}
The $2\times 2$-matrix $U^{(\nu)}_q$ of the Wilson propagator in Dirac theory is calculated in Appendix~\ref{app:wilson_dirac_propagator}.
Most importantly, it is shown that this propagator transforms the Dirac states as
\begin{align}
    \label{eq:chi_trafo_dirac_nonabelian}
    \hat{\chi}^{(\nu)}_{q\tau p} \equiv \sum_{\tau'} \chi^{(\nu)}_{q\tau'p}\,(U_q^{(\nu)})_{\tau'\tau} = \delta_{p,-\tau}\,,
\end{align}
i.e., leads to a $q$-independent contribution. This has the effect that, after inserting (\ref{eq:Dirac_lattice_I}), (\ref{eq:Dirac_lattice_II}), (\ref{eq:nonabelian_u_explicit}), (\ref{eq:prop_splitting_II}), (\ref{eq:block_structure_propagator_nu_even}), and (\ref{eq:block_structure_propagator_nu_odd}) in (\ref{eq:hat_w_I}) and (\ref{eq:hat_w_II}) (with $k=q$ and $k=\frac{\pi}{Za}+q$, respectively, and the $q$-integrals extending to $\pm\infty$), all  propagators $U^{(\nu')}_q$ give rise to integrands where the $q$-dependence of the Dirac states disappears and only a purely oscillating exponential remains in the integrand. This leads to unphysical delta function contributions which can be disregarded, see Appendix~\ref{app:nonabelian_splitting}. The only remaining terms are those from the Abelian propagators $U^{(\nu)}_{q-}$, leading to the following result for the two subintervals
\begin{align}
    \nonumber
    \hat{w}^{(\nu)}_{\text{I},\alpha}(ma) &\approx \delta_{\nu,{\rm even}}\,\frac{Za}{2\pi} \int dq\,e^{iq\hat{m}_\alpha^{(\nu)}a} \\
    \label{eq:hat_w_I_zw}
    &\hspace{1cm}
    \times \,u_{q\nu}(ma)\,U^{(\nu)}_{q-}\,(V^{(\nu)})_{\nu\alpha} \,,\\
    \nonumber
    \hat{w}^{(\nu)}_{\text{II},\alpha}(ma) &\approx \delta_{\nu,{\rm odd}}\,\frac{Za}{2\pi}\,
    \int dq\,e^{i(\frac{\pi}{Za} + q)\hat{m}_\alpha^{(\nu)}a} \\
    \label{eq:hat_w_II_zw}
    &\hspace{0cm}
    \times \,u_{\frac{\pi}{Za}+q,\nu}(ma)\,U^{(\nu)}_{q-}\,(U_{\rm I}^{(\nu)}\,V^{(\nu)})_{\nu\alpha} \,.
\end{align}
Inserting (\ref{eq:Dirac_lattice_I}), (\ref{eq:Dirac_lattice_II}, (\ref{eq:Dirac_prop_valence_band}), and using the definition (\ref{eq:wannier_p_ML}) of the Dirac Wannier function in ML gauge, we find
\begin{align}
    \nonumber
    \hat{w}^{(\nu)}_{\text{I},\alpha}(ma) &\approx \delta_{\nu,{\rm even}}\, e^{-i\gamma^{(\nu)}}\,(V^{(\nu)})_{\nu\alpha}  \\
    \label{eq:hat_w_I_result}
    &\hspace{-1cm}
    \times \,a\sqrt{Z}\,\sum_p \,\tilde{w}_{-,p}^{(\nu)}(\hat{m}_\alpha^{(\nu)}) \,e^{ip k_F^{(\nu)}ma} \,,\\
    \nonumber
    \hat{w}^{(\nu)}_{\text{II},\alpha}(ma) &\approx \delta_{\nu,{\rm odd}}\, e^{-i\gamma^{(\nu)}}\,
    e^{-i\frac{1}{2}\hat{\gamma}^{(\nu)}_\alpha}\,(U_{\rm I}^{(\nu)}\,V^{(\nu)})_{\nu\alpha}  \\
    \label{eq:hat_w_II_result}
    &\hspace{-1cm}
    \times \,a\sqrt{Z}\,\sum_p \,\tilde{w}_{-,p}^{(\nu)}(\hat{m}_\alpha^{(\nu)}) \,e^{ip k_F^{(\nu)}ma} \,.
\end{align}
In Appendix~\ref{app:spectrum_wilson_loop} we analyse the unitary transformations $V^{(\nu)}$ and $U_{\rm I}^{(\nu)}\,V^{(\nu)}$, with the result that all matrix elements consist of phase factors multiplied with $1/\sqrt{\nu}$ (from the normalization). As a result, we find the central result (\ref{eq:wannier_abelian_vs_nonabelian}), relating all Wannier functions in NA-ML gauge to the Dirac Wannier function corresponding to the upper part of the highest band in ML gauge. In addition we will analyse in Appendix~\ref{app:spectrum_wilson_loop} the spectrum of the Wilson loop operator (\ref{eq:wilson_loop_total}) and find indeed that all non-Abelian Zak-Berry phases are equidistantly distributed according to (\ref{eq:zak_berry_phases_nonabelian_result}).

\section{Summary and outlook}
\label{sec:summary}

The universal scaling properties of Wannier functions found in the present work are fundamental in several respects. Of particular interest is the high precision value of the universal scaling even for rather large gaps, as it appears most prominently in the non-Abelian gauge of maximal localization (NA-ML), see Fig.~\ref{fig:NAML}. The NA-ML gauge has been put forward in the literature \cite{marzari_vanderbilt_prb_97,marzari_etal_rmp_12,vanderbilt_book_18} since in this gauge the total quadratic spread probes the fluctuations of the bulk polarization which, according to the surface fluctuation theorem \cite{weber_etal_prl_20}, is related to the fluctuations of a directly measurable observable, the boundary charge $Q_B^{(\nu)}$. Importantly, $Q_B^{(\nu)}$ is the {\it total} boundary charge corresponding to a chemical potential located in gap $\nu$, in contrast to the boundary charge $Q_{B,\alpha}$ of a single band $\alpha$ which is not measurable. The observable $Q_B^{(\nu)}$ probes essentially low-energy properties of the system since the low-lying states are occupied and contribute a fixed amount to the boundary charge which can not fluctuate. This characteristic feature of the boundary charge is directly linked to our result that the Wannier functions in the NA-ML gauge are probing low-energy properties and reveal universal scaling to high precision. This motivates further investigations of the NA-ML gauge in other multi-channel or higher-dimensional systems to find similar universal scaling assuming the small gap limit. 

Our proposal of how to define Wannier functions for field-theoretical Dirac models opens up pathways to describe universal scaling properties of Wannier functions for a whole class of multi-band lattice models very efficiently via a simplified continuum model with a low number of bands only. We have exemplified this here for a $2$-band Dirac model in $1+1$-dimension, covering the whole class of one dimensional tight-binding models with nearest-neighbor hopping, one orbital per site, and generic periodic on-site potential and hopping modulation. These are special models in the sense that the gap opens up at a single point in quasimomentum space (in our case either at $k=0$ or at $k=\pm\pi/(Za)$). This gives rise to a field-theoretical low-energy model consisting of one pair of a slowly varying right- and left-mover. In other more general multi-channel cases or models with longer-ranged hopping several anticrossings can appear in quasimomentum space each of them characterized by the presence of several channels. The same will apply to higher-dimensional systems. However, also for these more general cases, the boundary charge and the Wannier functions in the NA-ML gauge are expected to probe only the low-energy properties of the system provided that all gap openings close to the Fermi level remain small. Therefore, based on our proposal of how to define Wannier functions within continuum field theories, it will be of high interest to study the Wannier functions of more general field-theoretical models, consisting of multi-channel Dirac models or several Dirac models which are coupled to each other (each of them corresponding to one anticrossing point).

It is a quite remarkable result of our work that field-theoretical models also allow for the definition of the Zak-Berry connection and the geometric tensor, although these are quantities which are used in lattice models to characterize global topological properties of the whole band structure in a nonlocal way. In contrast, a field theory can only characterize the physics close to the Fermi level where the gap opens, i.e., it probes essentially local aspects of the band structure. However, this does not mean that a field theory is per se not capable of characterizing topological properties of lattice models. If the gaps at the bottom and the top of a given band are both small, it is possible to set up two independent field theories around the two gaps and supplement them by appropriate asymptotic conditions to connect them in a unique way. This allows to obtain full control over the whole Zak-Berry connection in the lattice theory via the Zak-Berry connection defined in the field theoretical models. For the lattice models studied in this work the band structure in the middle of the bands is characterized by free plane waves, leading naturally to the choice of asymptotically free plane waves in the corresponding field-theoretical models. For more general multi-channel models, as well as in the presence of spin-orbit interaction and Zeeman fields, generalizations have to be studied for small gaps by expanding around a different reference system (called $H_0$ in our work, see Eq.~(\ref{eq:H_0})). This reference system will again prescribe certain asymptotic conditions for the field-theory far away from the gap, which can be used to set up unique correlations between gaps opening up at different energies. Therefore, we expect that our way to describe universal properties of Wannier functions and the boundary charge via low-energy models will be also of interest for the description of global topological aspects in more general models.

\section{Acknowledgments}
\label{sec:acknowledgments}

We thank S. Bl\"ugel, E. Koch, and V. Meden for fruitful discussions. This work was supported by the Deutsche Forschungsgemeinschaft via RTG 1995, the Swiss National Science Foundation (SNSF) and NCCR QSIT and by the Deutsche Forschungsgemeinschaft (DFG, German Research Foundation) under Germany's Excellence Strategy - Cluster of Excellence Matter and Light for Quantum Computing (ML4Q) EXC 2004/1 - 390534769. We acknowledge support from the Max Planck-New York City Center for Non-Equilibrium Quantum Phenomena. Simulations were performed with computing resources granted by RWTH Aachen University. Funding was received from the European Union's Horizon 2020 research and innovation program (ERC Starting Grant, grant agreement No 757725).

K.P. and M.P. contributed equally to this work.

\begin{appendix}

\renewcommand\theequation{\Alph{section}.\arabic{equation}}

\section{Gapless case, normalization and completeness relation}
\label{app:vons_condition}

Here we show that the normalization and completeness relations (\ref{eq:u_normalization}) and (\ref{eq:u_completeness}) are consistent with the corresponding ones in the low-energy region, given by (\ref{eq:chi_p_dirac_normalization}) and (\ref{eq:chi_p_dirac_completeness}). To achieve this we first discuss the gapless case and show that (\ref{eq:chi_large_q}) reproduces the correct result for the eigenstates of $H_0$. 

For the gapless case $\delta v_m= \delta t_m = 0$ the Hamiltonian $h_k$ given by (\ref{eq:h_k_def}) has the form
\begin{align}
    \label{eq:h_k_gapless}
    h_k =  - t \sum_{j=1}^Z \left(|j+1\rangle\langle j| e^{-ika} + |j\rangle\langle j+1| e^{ika} \right) \,.
\end{align}
One obtains straightforwardly for the dispersion and the eigenfunctions
\begin{align}
    \label{eq:dispersion_gapless}
    \epsilon_{k\alpha} &= - 2t \cos(ka + \frac{2\pi}{Z}n_{k\alpha}) \,,\\
    \label{eq:_gapless}
    u_{k\alpha}(ja) &= \frac{1}{\sqrt{Z}}\, e^{i\frac{2\pi}{Z}n_{k\alpha}j}\,,
\end{align}
where $n_{k\alpha}$ are $Z$ integers in some interval of size $Z$. To achieve the ordering $\epsilon_k^{(1)}<\epsilon_k^{(2)}<\dots<\epsilon_k^{(Z)}$ in the reduced zone scheme $-\frac{\pi}{Za} < k < \frac{\pi}{Za}$, we choose $n_{k1}=0, n_{k2}=-\Sign(k), n_{k3}=\Sign(k), n_{k4}=-2 \Sign(k), n_{k5}= 2\Sign(k)$, etc., which gives
\begin{align}
    n_{k\alpha} = \frac{1}{2}\Sign(k) \begin{cases} - \alpha & \text{for} \quad \alpha \quad\text{even}\\
    \alpha - 1 & \text{for} \quad \alpha \quad\text{odd}\end{cases}\,,
\end{align}
leading to
\begin{align}
    \label{eq:_gapless_explicit}
    u_{k\alpha}(ja) = \frac{1}{\sqrt{Z}}\,\begin{cases} e^{-i\Sign(k)\frac{\pi}{Z}\alpha j} & \text{for} \quad \alpha \quad\text{even}\\
    e^{i\Sign(k)\frac{\pi}{Z}(\alpha-1) j} & \text{for} \quad \alpha \quad\text{odd}\end{cases}\,.
\end{align}
According to (\ref{eq:nu_even}) and (\ref{eq:nu_odd}) it is straightforward to show that this corresponds precisely to the choice $\chi_{q\tau p}^{(\nu)} = \delta_{p,\Sign(q\tau)}$ in Dirac theory, as stated in (\ref{eq:chi_large_q}).

To check the normalization (\ref{eq:u_normalization}) we prove (\ref{eq:dirac_lattice_identity}) rigorously for any integers $r,s\ge 0$ via the relations (\ref{eq:nu_even}) and (\ref{eq:nu_odd}). Except for the lower half of the lowest band (i.e., for $|k|<\frac{\pi}{2Za}$ and $\alpha=1$) or the upper half of the highest band (i.e., for $|k|>\frac{\pi}{2Za}$ and $\alpha=Z$), the gaps with even $\nu$ describe the region $|k|<\frac{\pi}{2Za}$, whereas the gaps with odd $\nu$ correspond to $|k|>\frac{\pi}{2Za}$. Therefore, in almost all cases, the parity of $\nu$ and $\nu'$ must be the same for given $k$. In this case we get
\begin{align}
    \nonumber
    \langle \partial_k^r u_{k\alpha}|\partial_k^s u_{k\alpha'}\rangle &=\\
    \nonumber
    &\hspace{-2cm}
    =\sum_{p,p'=\pm}\partial_q^r (\chi_{q\tau p}^{(\nu)})^* \partial_q^s \chi_{q\tau' p}^{(\nu')}
    \frac{1}{Z}\sum_{j=1}^Z e^{-i\pi(p\nu-p'\nu')j/Z}\\
    \nonumber
    &\hspace{-2cm}
    =\sum_{p,p'=\pm}\partial_q^r (\chi_{q\tau p}^{(\nu)})^* \partial_q^s \chi_{q\tau' p}^{(\nu')}
    \delta_{p\nu,p'\nu'}\\
    &\hspace{-2cm}
    =\delta_{\nu\nu'}\langle\partial_q^r \chi_{q\tau}^{(\nu)}| \partial_q^s \chi_{q\tau'}^{(\nu')}\rangle\,,
\end{align}
where we used $\delta_{p\nu,p'\nu'}=\delta_{pp'}\delta_{\nu\nu'}$ in the last step (since $\nu,\nu'>0$) and the fact that $p\nu-p'\nu'$ is always an even number (since $\nu$ and $\nu'$ have the same parity). The special case $\nu=1$ and $\nu'$ even is only possible for $\tau=-$, $p=\Sign(k)=-\Sign(q)$, and $|k|<\frac{\pi}{2Za}$. It leads to 
\begin{align}
    \nonumber
    \langle \partial_k^r u_{k1}|\partial_k^s u_{k\alpha'}\rangle &=\\
    \nonumber
    &\hspace{-2.5cm}
    =\sum_{p,p'=\pm}\partial_q^r (\chi_{q,-,p}^{(1)})^* \partial_q^s \chi_{q\tau'p}^{(\nu')}
    \frac{1}{Z}\sum_{j=1}^Z e^{-i\pi(p+\Sign(q)-p'\nu')j/Z}\,.
\end{align}
This expression must be zero since, by using $p+\Sign(q)=0$ and $\nu'$ even, we conclude that $p'\nu'$ must be zero which is not possible. Finally, the special case $\nu=Z-1$ and $\nu'$ having a different parity than $Z-1$ can be treated in a similar way and leads also to a contradiction.

To prove the completeness relationship (\ref{eq:u_completeness}) for given $k$ we note that, for each given gap $\nu$, one has to sum over both band indices $\tau=\pm$ of Dirac theory, except for $\nu=1$ or $\nu=Z-1$ when $|k|<\frac{\pi}{2Za}$ or $|k|>\frac{\pi}{2Za}$, respectively. In the latter case only the Dirac bands $\tau=-$ (for $\nu=1$ and $|k|<\frac{\pi}{2Za}$) or $\tau=+$ (for $\nu=Z-1$ and $|k|>\frac{\pi}{2Za}$) have to be taken, respectively. This means that only the eigenstates far away from the gap $\nu=1$ or $\nu=Z-1$ are involved for the lowest or highest lattice band, where we can take approximately the free solution (\ref{eq:_gapless_explicit}). The same happens for the contribution from the pairs $\tau=\pm$ since one finds from (\ref{eq:nu_even}) and (\ref{eq:nu_odd}) that the expression $\sum_\tau u_{k\alpha}(ja)u_{k\alpha}(j'a)^*$ involves always the combination
\begin{align}
    \sum_\tau \chi_{q\tau p}^{(\nu)}\left(\chi_{q\tau p'}^{(\nu)}\right)^* = \delta_{pp'}\,,
\end{align}
where we used the completeness relationship (\ref{eq:chi_p_dirac_completeness}) of the Dirac states. Since the same result arises for the free Dirac states $\chi_{q\tau p}^{(\nu)}=\delta_{p,\Sign(q\tau)}$, we find that the completeness relationship in the presence of a gap is not changed compared to the one for the free Bloch states (\ref{eq:_gapless_explicit}).

\section{Cancellation of Zak-Berry phases}
\label{app:zb_cancellation}

Here we show on quite general grounds that the sum of the Zak-Berry phases in the AF gauge of the two bands of the Dirac model is quantized in odd units of $\pi$. We first need that one can write (with $\chi_{q\tau}\equiv |\chi_{q\tau}\rangle$ defining column vectors in $p=\pm$)
\begin{align}
    \nonumber    
    \sum_\tau \gamma^{(\nu)}_\tau &= \sum_\tau\int dq \, A_{q\tau}^{(\nu)} = \sum_\tau \int dq \, \left(\chi_{q\tau}^{(\nu)}\right)^\dagger
    i\partial_q \chi_{q\tau}^{(\nu)}\\
    \nonumber
    &\hspace{-1cm}
    = \int dq\, \text{Tr}\left(M_q^{(\nu)}\right)^\dagger i\partial_q M_q^{(\nu)} = i \int \text{Tr}\left(M_q^{(\nu)}\right)^\dagger d M_q^{(\nu)}\\
    \label{eq:zak_berry_sum_dirac_winding}
    &= i \int d \ln(\text{det} M_q^{(\nu)}) \,.
\end{align}
where we defined the unitary matrix
\begin{align}
    \label{eq:M_def}
    M_q^{(\nu)} = \left(\begin{array}{cc} \chi_{q +}^{(\nu)} & \chi_{q  -}^{(\nu)} \end{array} \right) \,.
\end{align}
This means that (up to multiples of $2\pi$) the sum of the Zak-Berry phases of the two Dirac bands can be written as minus the phase difference of the determinant of $M_q^{(\nu)}$ between the asymptotic values at $q=\infty$ and $q=-\infty$
\begin{align}
    \label{eq:zak_berry_sum_dirac_phase_diff}
    \sum_\tau \gamma^{(\nu)}_\tau = - \sum_{r=1,2}\left(\varphi_{q=\infty,r}^{(\nu)} - \varphi_{q=-\infty,r}^{(\nu)} \right) 
    \quad \text{mod}\,(2\pi)\,,
\end{align}
where $e^{i\varphi_{qr}^{(\nu)}}$, with $r=1,2$, denote the two eigenvalues of the unitary matrix $M_q^{(\nu)}$. In the AF gauge the asymptotic values of $\chi_{q\tau p}^{(\nu)}$ are given by $\delta_{p,\Sign(q\tau)}$ according to (\ref{eq:chi_large_q}), leading to
\begin{align}
    \label{eq:M_asymptotics_+_infty}
    \lim_{q\rightarrow\infty} M_q^{(\nu)} &= \left(\begin{array}{cc} 1 & 0 \\ 0 & 1 \end{array}\right) \,,\\
    \label{eq:M_asymptotics_-_infty}
    \lim_{q\rightarrow -\infty} M_q^{(\nu)} &= \left(\begin{array}{cc} 0 & 1 \\ 1 & 0 \end{array}\right) \,.
\end{align}
This gives 
\begin{align}
    \label{eq:sum_phases_U_asymptocis}
    \sum_r \varphi_{q=\infty,r}^{(\nu)} = 0 \quad,\quad \sum_r \varphi_{q=-\infty,r}^{(\nu)} = \pi \,,
\end{align}
and we get from (\ref{eq:zak_berry_sum_dirac_phase_diff}) the final result
\begin{align}
    \label{eq:zak_berry_sum_dirac_phase_diff_result}
    \sum_\tau \gamma^{(\nu)}_\tau = \pi \quad \text{mod}\,(2\pi)\,,
\end{align}
consistent with (\ref{eq:zak_berry_sum_dirac}).

\section{Splitting of Wannier function}
\label{app:wannier_splitting}

Here we show how to split the lattice Wannier function for band $\alpha=2,\dots,Z-1$ into the contributions corresponding to the upper and lower halves of the band and how to extend the momentum integration to infinity, see Eqs.~(\ref{eq:wannier_splitting}), (\ref{eq:wannier_ud}), 
(\ref{eq:delta_w_AF}) and (\ref{eq:delta_w_ML}). Since the eigenfunctions of the lattice and the Dirac model are related via (\ref{eq:dirac_lattice}), (\ref{eq:psi_dirac}), (\ref{eq:psi_p_dirac}), (\ref{eq:nu_even}) and (\ref{eq:nu_odd}), we can perform this analysis
directly within the Dirac model by calculating the contributions from the region $|q|>\Lambda_\alpha$ of the Dirac Wannier functions occurring on the right hand side of (\ref{eq:wannier_AF_dirac_lattice_u}), (\ref{eq:wannier_AF_dirac_lattice_d}), (\ref{eq:wannier_ML_dirac_lattice_u}) and (\ref{eq:wannier_ML_dirac_lattice_d}). Denoting the contribution of the various Wannier functions from this region by $\delta w_{\text{u}/\text{d},\alpha}(ma)$ we get
\begin{align}
    \label{eq:delta_w_AF_u}
    \delta w_{\text{u},\alpha}(ma) &= a\sqrt{Z} \sum_p \,\delta w_{-,p}^{(\alpha)}(ma)e^{ipk_F^{(\alpha)}ma} \,,\\
    \label{eq:delta_w_AF_d}
    \delta w_{\text{d},\alpha}(ma) &= a\sqrt{Z} \sum_p \, \delta w_{+,p}^{(\alpha-1)}(ma)e^{ipk_F^{(\alpha-1)}ma}\,,\\
        \nonumber
    \delta\tilde{w}_u^{(\alpha)}(ma) &= \\ 
    \label{eq:delta_w_ML_u}
    &\hspace{-1cm}
    = a\sqrt{Z} \sum_p e^{-i\phi_{0,-}^{(\alpha)}} \delta\tilde{w}_{-,p}^{(\alpha)}(\tilde{m}_\alpha a)e^{ipk_F^{(\alpha)}ma} \,,\\
    \nonumber
    \delta\tilde{w}_d^{(\alpha)}(ma) &= \\
    \label{eq:delta_w_ML_d}
    &\hspace{-1cm}
    = a\sqrt{Z} \sum_p e^{-i\phi_{0,+}^{(\alpha-1)}} \delta\tilde{w}_{+,p}^{(\alpha-1)}(\tilde{m}_\alpha a)e^{ipk_F^{(\alpha-1)}ma}\,,
\end{align}
where
\begin{align}
    \label{eq:delta_w_p}
    \delta w^{(\nu)}_{\tau p}(x) &= \int_{|q|>\Lambda_\alpha}\frac{dq}{2\pi}\chi_{q\tau p}^{(\nu)}e^{iqx}e^{-\eta |q|}\,,\\
    \label{eq:delta_tilde_w_p}
    \delta \tilde{w}^{(\nu)}_{\tau p}(x) &= \int_{|q|>\Lambda_\alpha}\frac{dq}{2\pi}\tilde{\chi}_{q \tau p}^{(\nu)}e^{iqx}
    e^{-\eta |q|}\,,
\end{align}
with $\tilde{\chi}_{q \tau p}^{(\nu)}=e^{i\phi_{q\tau}^{(\nu)}}\chi_{q\tau p}^{(\nu)}$. We have introduced a convenient regularization of the integrals via the convergence factor $e^{-\eta|q|}$. For $|q|>\Lambda_\alpha$ we can take the free form $\chi_{q\tau p}^{(\nu)}\approx \delta_{\Sign(q),p\tau}$ cf. (\ref{eq:chi_large_q}), and get with (\ref{eq:gauge_max_loc_dirac}), (\ref{eq:zak_berry_phase_dirac}) and (\ref{eq:zb_minus_k_AF_dirac}) the following value for the phase $\phi_{q\tau}^{(\nu)}$ in the asymptotic high-momentum region
\begin{align}
    \nonumber
    \phi_{q\tau}^{(\nu)} - \phi_{0,\tau}^{(\nu)} &= \frac{1}{2}\int_{-q}^q dq'\,A_{q'\tau}^{(\nu)}\\
    \label{eq:phi_large_q}
    &\hspace{-1cm}
    \approx \frac{1}{2} \Sign(q)\int dq' \,A_{q'\tau}^{(\nu)} = \frac{1}{2}\Sign(q) \gamma^{(\nu)}_\tau\,.
\end{align}
Using these relations we obtain for (\ref{eq:delta_w_p}) and (\ref{eq:delta_tilde_w_p})
\begin{align}
    \label{eq:delta_w_p_explicit}
    \delta w^{(\nu)}_{\tau p}(x) &= \frac{i}{2\pi}\frac{e^{ip\tau\Lambda_\alpha x}}{\tau p x + i\eta}\,,\\
    \label{eq:delta_tilde_w_p_explicit}
    e^{-i\phi_{0,\tau}^{(\nu)}}\delta \tilde{w}^{(\nu)}_{\tau p}(x) &= \frac{i}{2\pi}
    \frac{e^{ip\tau(\Lambda_\alpha x + \frac{1}{2}\gamma^{(\nu)}_\tau)}}{\tau p x + i\eta}\,.
\end{align}
Leaving out the contribution proportional to the delta function $\delta(x)$ (this is the regularization we use in Section~\ref{sec:wannier_universality} to define the Dirac Wannier function), we obtain after inserting these expressions into Eqs.~(\ref{eq:delta_w_AF_u}-\ref{eq:delta_w_ML_d}) 
\begin{align}
    \label{eq:delta_w_AF_u_explicit}
    \delta w_{\text{u},\alpha}(ma) &= \frac{\sqrt{Z}}{\pi m}\sin\big\{(k_F^{(\alpha)}-\Lambda_\alpha)ma\big\} \,,\\
    \label{eq:delta_w_AF_d_explicit}
    \delta w_{\text{d},\alpha}(ma) &= -\frac{\sqrt{Z}}{\pi m}\sin\big\{(k_F^{(\alpha-1)}+\Lambda_\alpha)ma\big\}\,,\\
    \nonumber
    \delta\tilde{w}_{\text{u}}^{(\alpha)}(ma) &= \\ 
    \label{eq:delta_w_ML_u_explicit}
    &\hspace{-1cm}
    = \frac{\sqrt{Z}}{\pi \tilde{m}_\alpha}
    \sin\big\{k_F^{(\alpha)}ma-\Lambda_\alpha\tilde{m}_\alpha a -\frac{1}{2}\gamma_{-}^{(\alpha)}\big\} \,,\\
    \nonumber
    \delta\tilde{w}_{\text{d}}^{(\alpha)}(ma) &= \\
    \label{eq:delta_w_ML_d_explicit}
    &\hspace{-1cm}
    = -\frac{\sqrt{Z}}{\pi \tilde{m}_\alpha}
    \sin\big\{k_F^{(\alpha-1)}ma+\Lambda_\alpha\tilde{m}_\alpha a +\frac{1}{2}\gamma_{+}^{(\alpha-1)}\big\}\,.
\end{align}
For $\alpha=2,\dots,Z-1$ we can insert $\Lambda_\alpha=\frac{\pi}{2Za}$ and 
$\gamma_{\alpha}=\gamma_{+}^{(\alpha-1)}+\gamma_{-}^{(\alpha)}$, and get with $k_F^{(\alpha)}=\alpha\frac{\pi}{Za}$ and 
$\tilde{m}_\alpha = m-\frac{Z}{2\pi}\gamma_{\alpha}$ the final result
\begin{align}
    \delta w_{\text{u},\alpha}(ma) = - \delta w_{\text{d},\alpha}(ma) = \delta w_{\alpha}(ma)\,,\\
    \delta\tilde{w}_u^{(\alpha)}(ma) = - \delta\tilde{w}_d^{(\alpha)}(ma) = \delta\tilde{w}_{\alpha}(ma)\,,
\end{align}
where $\delta w_{\alpha}(ma)$ and $\delta\tilde{w}_{\alpha}(ma)$ are given by (\ref{eq:delta_w_AF}) and (\ref{eq:delta_w_ML}), respectively. 

In contrast, for $\alpha=1,Z$, we can use $\Lambda_{1/Z}=\frac{\pi}{Za}$, $\gamma_{1}=\gamma_-^{(1)}$, and $\gamma_{Z}=\gamma_{+}^{(Z-1)}$ with the result
\begin{align}
    \delta w_{\text{u},1}(ma) = \delta w_{\text{d},Z}(ma) = 0 \,,\\
    \delta\tilde{w}_{\text{u},1}(ma) = \delta\tilde{w}_{\text{d},Z}(ma) = 0 \,,
\end{align}
in accordance with (\ref{eq:wannier_1Z_ud}).

\section{Real momentum integral representations of Dirac Wannier functions and scaling functions}
\label{app:wannier_scaling_real_q}

In this Appendix we present the representation of the Dirac Wannier functions $w^{(\nu)}_{p\tau}(x)$ and $\tilde{w}^{(\nu)}_{p\tau}(x)$ together with the scaling functions $F_{A/B,\tau}(y;\gamma)$ and $\tilde{F}_{A/B,\tau}(y;\gamma)$ via convergent momentum integrals on the real axis. Starting from (\ref{eq:w_dirac}-\ref{eq:tilde_chi_bar_q}), we first subtract the large $|\bar{q}|\gg 1$ behavior of the integrands to get convergent integrals and to explicitly exhibit the part which remains for $\bar{\eta}\rightarrow 0$. Using (\ref{eq:chi_large_q}) and (\ref{eq:phi_large_q}) together with (\ref{eq:delta_w_p_explicit}) and (\ref{eq:delta_tilde_w_p_explicit}) (with $\Lambda_\alpha=0$) we find
\begin{align}
    \nonumber
    w^{(\nu)}_{\tau p}(x)&=\frac{i}{2\pi}\frac{1}{p\tau x+i\eta} +  \\
    \label{eq:w_real_q_conv}
    & + \frac{1}{4\pi\xi^{(\nu)}}\int d\bar{q} \,e^{i\bar{q}\bar{x}}\big[\chi_{\tau p}^{(\nu)}(\bar{q}) - \delta_{\Sign{\bar{q}},p\tau}\big]\,, \\
    \nonumber
    \tilde{w}^{(\nu)}_{\tau p}(x)&=\frac{i}{2\pi}
    \frac{e^{i\phi^{(\nu)}_{0,\tau}}e^{i\frac{1}{2}p\tau\gamma_\tau^{(\nu)}}}{p\tau x+i\eta} +  \\
    \label{eq:tilde_w_real_q_conv}
    &\hspace{-1cm} 
    + \frac{1}{4\pi\xi^{(\nu)}}\int d\bar{q} \,e^{i\bar{q}\bar{x}}\big[\tilde{\chi}_{\tau p}^{(\nu)}(\bar{q}) - 
    e^{i\phi^{(\nu)}_{0,\tau}}e^{i\frac{1}{2}p\tau\gamma_\tau^{(\nu)}}\delta_{\Sign{\bar{q}},p\tau}\big]\,. 
\end{align}
The first term on the right hand side of the Dirac Wannier functions contains an unphysical contribution $\sim\delta(x)$ which arises from large momenta and has to be subtracted in the regularization procedure. In the scaling functions this part does anyhow not appear since one multiplies with $x$. In contrast, the principal part $\sim\frac{1}{x}$ of the first term is an important contribution to the scaling functions. It should be contrasted to the contributions (\ref{eq:delta_w_p_explicit}) and (\ref{eq:delta_tilde_w_p_explicit}) which arise from extending the momentum integration to infinity and cancel in the lattice Wannier function of a certain band when adding the contributions from the upper and lower part of the band to the Wannier function, see Appendix~\ref{app:wannier_splitting}.

Using (\ref{eq:phi_0_explicit}) and (\ref{eq:zak_berry_phase_vs_gap_phase}) we note the identities
\begin{align}
    \label{eq:exp_identity_1}
    e^{i\phi^{(\nu)}_{0,\tau}}e^{i\frac{1}{2}p\tau\gamma_\tau^{(\nu)}} &= 
    \begin{cases} e^{i\gamma^{(\nu)}} & \text{for}\,p=+ \\ \tau & \text{for}\,p=-\end{cases}\,,\\
    \label{eq:exp_identity_2}
    e^{i\frac{1}{2}p\tau\gamma_\tau^{(\nu)}} &=  e^{i\frac{1}{2}p\gamma^{(\nu)}} 
    \begin{cases} 1 & \text{for}\,\tau=+ \\ -i p s_{\gamma^{(\nu)}} & \text{for}\,\tau=-\end{cases}\,.
\end{align}
Multiplying (\ref{eq:w_real_q_conv}) and (\ref{eq:tilde_w_real_q_conv}) with $x$, we obtain with (\ref{eq:exp_identity_1}) in the limit $\eta\rightarrow 0$ the useful representations
\begin{align}
    \nonumber
    x\,w^{(\nu)}_{\tau p}(x)&=i\frac{p\tau}{2\pi} +  \\
    \label{eq:x_w_real_q_conv}
    &\hspace{-1cm} + \frac{y}{2}\int \frac{d\bar{q}}{2\pi} \,e^{i\bar{q}\frac{y}{2}}\big[\chi_{\tau p}^{(\nu)}(\bar{q}) - \delta_{\Sign{\bar{q}},p\tau}\big]\,, \\
    \nonumber
    e^{-i\phi^{(\nu)}_{0,\tau}}\,x\,\tilde{w}^{(\nu)}_{\tau p}(x)&= e^{i\frac{1}{2}p\tau\gamma_\tau^{(\nu)}} 
    \Big\{ i\frac{p\tau}{2\pi} + \\
    \label{eq:x_tilde_w_real_q_conv}
    &\hspace{-2cm} 
    +  \frac{y}{2}\int \frac{d\bar{q}}{2\pi} \,e^{i\bar{q}\frac{y}{2}}\big[\frac{\sqrt{\bar{\epsilon}_{\bar{q}} + p\tau\bar{q}}}{\sqrt{2\bar{\epsilon}_{\bar{q}}}} - \delta_{\Sign{\bar{q}},p\tau}\big]\Big\}\,, 
\end{align}
where we inserted the form (\ref{eq:tilde_chi_bar_q}) for $\tilde{\chi}_{\tau p}^{(\nu)}(\bar{q})$, and used the definition $y=x/\xi^{(\nu)}=2\bar{x}$. Inserting (\ref{eq:chi_bar_q}) and using (\ref{eq:exp_identity_2}) one can calculate the scaling functions from (\ref{eq:w_A_def}-\ref{eq:tilde_F_AB_scaling}) and finds after some straightforward manipulations the compact forms
\begin{widetext}
\begin{align}
    \label{eq:F_A_comapct}
    F_{A,\tau}(y;\gamma) &= \frac{\tau}{\pi} + 
    \tau y \int_0^\infty \frac{d\bar{q}}{2\pi}\,\left[
    \frac{\bar{q}\sin\frac{\bar{q}y}{2} + \sin\gamma \cos\frac{\bar{q}y}{2}}{\sqrt{\bar{\epsilon}_{\bar{q}}(\bar{\epsilon}_{\bar{q}}+\tau\cos\gamma)}} 
    - \sin\frac{\bar{q}y}{2}\right]\,,\\
    \label{eq:F_B_comapct}
    F_{B,\tau}(y;\gamma) &=  y \int_0^\infty \frac{d\bar{q}}{2\pi}\,\cos\frac{\bar{q}y}{2}
    \left[\sqrt{1+\frac{\tau\cos\gamma}{\bar{\epsilon}_{\bar{q}}}}-1\right]\,,\\   
    \label{eq:tilde_F_A_-_comapct}
    \tilde{F}_{A,-}(y;\gamma) &=  - \frac{s_\gamma}{\pi} \left\{\sin\frac{\gamma}{2} + \frac{y}{2\sqrt{2}}\int_0^\infty d\bar{q}\,\left[\left(\sqrt{1+\frac{\bar{q}}{\bar{\epsilon}_{\bar{q}}}}-\sqrt{2}\right)\cos\frac{\bar{q}y-\gamma}{2}
    + \sqrt{1-\frac{\bar{q}}{\bar{\epsilon}_{\bar{q}}}}\,\cos\frac{\bar{q}y+\gamma}{2}
    \right]\right\}\,,\\   
    \label{eq:tilde_F_B_-_comapct}
    \tilde{F}_{B,-}(y;\gamma) &=  - \frac{s_\gamma}{\pi} \left\{\cos\frac{\gamma}{2} + \frac{y}{2\sqrt{2}}\int_0^\infty d\bar{q}\,\left[\left(\sqrt{1+\frac{\bar{q}}{\bar{\epsilon}_{\bar{q}}}}-\sqrt{2}\right)\sin\frac{\bar{q}y-\gamma}{2}
    - \sqrt{1-\frac{\bar{q}}{\bar{\epsilon}_{\bar{q}}}}\,\sin\frac{\bar{q}y+\gamma}{2}
    \right]\right\}\,,\\
    \label{eq:tilde_F_A_+_comapct}
    \tilde{F}_{A,+}(y;\gamma) &=  \frac{1}{\pi}\cos\frac{\gamma}{2} + \frac{y}{2\pi\sqrt{2}}\int_0^\infty d\bar{q}\,\left[\left(\sqrt{1+\frac{\bar{q}}{\bar{\epsilon}_{\bar{q}}}}-\sqrt{2}\right)\sin\frac{\bar{q}y+\gamma}{2}
    - \sqrt{1-\frac{\bar{q}}{\bar{\epsilon}_{\bar{q}}}}\,\sin\frac{\bar{q}y-\gamma}{2}
    \right]\,,\\   
    \label{eq:tilde_F_B_+_comapct}
    \tilde{F}_{B,+}(y;\gamma) &=  -\frac{1}{\pi}\sin\frac{\gamma}{2} + \frac{y}{2\pi\sqrt{2}}\int_0^\infty d\bar{q}\,\left[\left(\sqrt{1+\frac{\bar{q}}{\bar{\epsilon}_{\bar{q}}}}-\sqrt{2}\right)\cos\frac{\bar{q}y+\gamma}{2}
    + \sqrt{1-\frac{\bar{q}}{\bar{\epsilon}_{\bar{q}}}}\,\cos\frac{\bar{q}y-\gamma}{2}
    \right]\,,
\end{align}
\end{widetext}
with $\bar{\epsilon}_{\bar{q}}=\sqrt{1+\bar{q}^2}$ and $s_\gamma=\Sign\gamma$. These formulas can be straightforwardly evaluated numerically and have been used to produce the results shown in Fig.~\ref{fig:scaling_functions}.

\section{Integral representation on the imaginary momentum axis and asymptotics of Wannier functions in Dirac theory}
\label{app:wannier_integral}

Here we show how to obtain the integral representations (\ref{eq:w_conv}) and (\ref{eq:tilde_w_conv}) for the Wannier functions $w^{(\nu)}_{\tau p}(x)$ and $\tilde{w}^{(\nu)}_{\tau p}(x)$ in the universal regime $|x|\gg a$ of Dirac theory, together with their asymptotics at $|x|\ll\xi^{(\nu)}$ and $|x|\gg\xi^{(\nu)}$, as given by (\ref{eq:w_small_x}), (\ref{eq:tilde_w_small_x}),  (\ref{eq:w_large_x_1}-\ref{eq:w_large_x_3}), and (\ref{eq:tilde_w_large_x}). 

To obtain (\ref{eq:w_conv}) and (\ref{eq:tilde_w_conv}), we use (\ref{eq:w_dirac}) and (\ref{eq:tilde_w_dirac}), and close the integration contour in the upper or lower half of the complex plane for $s_x=\Sign(x)=\pm$, respectively. Choosing the branch cuts of the various square roots and of $e^{-\bar{\eta}|\bar{q}|}$ on the imaginary axis, we can close the integration contour around the branch cut, and obtain with $\bar{q}=is_x\kappa$ and $0<\kappa<\infty$ the form
\begin{align}
    \nonumber
    w^{(\nu)}_{\tau p}(x) &= \frac{is_x}{4\pi\xi^{(\nu)}}\int_0^\infty d\kappa \,e^{-|\bar{x}|\kappa} \\
    \label{eq:w_conv_1}
    &\hspace{-1cm}
    \left\{e^{-is_x\bar{\eta}\kappa} \chi_{\tau p}^{(\nu)}(is_x\kappa+0^+) - 
    e^{is_x\bar{\eta}\kappa} \chi_{\tau p}^{(\nu)}(is_x\kappa-0^+) \right\}\,, 
\end{align}
and an analog equation for $\tilde{w}^{(\nu)}_{\tau p}(x)$ with $\chi_{\tau p}^{(\nu)}\rightarrow\tilde{\chi}_{\tau p}^{(\nu)}$. Due to the exponentially decaying factor $e^{-|\bar{x}|\kappa}$, these integrals are convergent in the limit $\bar{\eta}\rightarrow 0$, and we obtain the results (\ref{eq:w_conv}) and (\ref{eq:tilde_w_conv}) in this limit. 

To evaluate the integrands $\delta \chi_{\tau p}^{(\nu)}(is_x\kappa)$ and $\delta \tilde{\chi}_{\tau p}^{(\nu)}(is_x\kappa)$ of (\ref{eq:w_conv}) and (\ref{eq:tilde_w_conv}), we note the following values of the various square roots for $\bar{q}=is_x\kappa\pm 0^+$
\begin{align}
    \label{eq:eps}
    \bar{\epsilon}_{\bar{q}} &= |1-\kappa^2|^{\frac{1}{2}}
    \begin{cases} 1\pm is_x 0^+ & \text{for} \quad \kappa<1 \\ \pm is_x & \text{for} \quad \kappa>1 \end{cases}\,,\\
    \label{eq:root_eps}
    \sqrt{\bar{\epsilon}_{\bar{q}}} &= |1-\kappa^2|^{\frac{1}{4}}
    \begin{cases} 1 & \text{for} \quad \kappa<1 \\ e^{\pm is_x\frac{\pi}{4}} & \text{for} \quad \kappa>1 \end{cases}\,,\\
    \label{eq:root_eps_2}
    \sqrt{\bar{\epsilon}_{\bar{q}}+p\tau\bar{q}} &= 
    \begin{cases} \sqrt{|1-\kappa^2|^{\frac{1}{2}} + ip\tau s_x\kappa} & \text{for} \quad \kappa<1 \\ 
     \sqrt{\pm i s_x |1-\kappa^2|^{\frac{1}{2}} + ip\tau s_x\kappa} & \text{for} \quad \kappa>1 \end{cases}\,.
\end{align}
Furthermore, for $\tau\cos{\gamma^{(\nu)}}>0$, we get
\begin{align}
    \nonumber
    \sqrt{\bar{\epsilon}_{\bar{q}}+\tau\cos{\gamma^{(\nu)}}} &=\\
    \label{eq:root_eps_3}
    &\hspace{-2cm}
    = \begin{cases} \sqrt{|1-\kappa^2|^{\frac{1}{2}} + |\cos{\gamma^{(\nu)}}|} & \text{for} \quad \kappa<1 \\ 
     \sqrt{\pm i s_x |1-\kappa^2|^{\frac{1}{2}} + |\cos{\gamma^{(\nu)}}|} & \text{for} \quad \kappa>1 \end{cases}\,,
\end{align}
and for $\tau\cos{\gamma^{(\nu)}}<0$
\begin{align}
    \nonumber
    \sqrt{\bar{\epsilon}_{\bar{q}}+\tau\cos{\gamma^{(\nu)}}} &= \\ 
    \label{eq:root_eps_4}
    &\hspace{-2.5cm}
    =\begin{cases} \sqrt{|1-\kappa^2|^{\frac{1}{2}} - |\cos{\gamma^{(\nu)}}|} & \text{for} \,\kappa<|\sin{\gamma^{(\nu)}}| \\ 
     \pm i s_x \sqrt{ -|1-\kappa^2|^{\frac{1}{2}} + |\cos{\gamma^{(\nu)}}|} & \text{for} \, |\sin{\gamma^{(\nu)}}|<\kappa<1 \\
     \sqrt{\pm i s_x |1-\kappa^2|^{\frac{1}{2}} - |\cos{\gamma^{(\nu)}}|} & \text{for} \, \kappa>1 
     \end{cases}\,.
\end{align}

For $|\bar{x}|\ll 1$, the integrals are dominated by the regime $\kappa\gg 1$, where we get
\begin{align}
    \label{eq:dI_large_kappa}
    \delta \chi_{\tau p}^{(\nu)}(is_x\kappa)&\rightarrow p\tau \,,\\
    \label{eq:tilde_dI_large_kappa}
    \delta \tilde{\chi}_{\tau p}^{(\nu)}(is_x\kappa)&\rightarrow p\tau 
    \begin{cases} e^{i\gamma^{(\nu)}} & \text{for} \quad p=+ \\ \tau & \text{for} \quad p=- \end{cases}\,.\\
\end{align}
Inserting these forms in (\ref{eq:w_conv}) and (\ref{eq:tilde_w_conv}), we find straightforwardly the following result for small $|x|\ll\xi^{(\nu)}$.
\begin{align}
    \label{eq:w_small_x}
    w^{(\nu)}_{\tau p}(x) &\rightarrow \frac{ip\tau}{2\pi x} \,,\\
    \label{eq:tilde_w_small_x}
    \tilde{w}^{(\nu)}_{\tau p}(x) &\rightarrow \frac{ip\tau}{2\pi x} 
    \begin{cases} e^{i\gamma^{(\nu)}} & \text{for} \quad p=+ \\ \tau & \text{for} \quad p=- \end{cases}\,.
\end{align}

For $|\bar{x}|\sim O(1)$, the integrals are of $O(1)$, leading to the universal scaling of the Wannier functions in the regime $|x|\sim \xi^{(\nu)}$
\begin{align}
    \label{eq:wannier_dirac_scaling}
    |x \,w^{(\nu)}_{\tau p}(x)| \sim |x\,\tilde{w}^{(\nu)}_{\tau p}(x)| \sim O(1) \,.
\end{align}
Due to the results(\ref{eq:w_small_x}) and (\ref{eq:tilde_w_small_x}), this scaling holds also for $|x|\ll\xi^{(\nu)}$. 

For $|\bar{x}|\gg 1$, the integrals are dominated by the regime around that branching point of the various square roots which has the smallest imaginary part. We start with $w^{(\nu)}_{\tau p}(x)$. For $\tau\cos{\gamma^{(\nu)}}>0$, there is only a single branching point at $\kappa=1$, and close to this branching point we get for small $0<\kappa'=\kappa-1\ll 1$
\begin{align}
    \label{eq:dI_small_kappa_1}
    \delta \chi_{\tau p}^{(\nu)}(is_x(1+\kappa')) \approx \frac{\tau(ip s_x + e^{ip\gamma^{(\nu)}})}{i s_x 2^{\frac{3}{4}}\sqrt{|\cos{\gamma^{(\nu)}}|}}
    (\kappa')^{-\frac{1}{4}}\,.
\end{align}
Inserting this form in (\ref{eq:w_conv}) and using 
\begin{align}
    \label{eq:gamma_integral}
    \int_0^\infty d\kappa' e^{-|\bar{x}|\kappa'} (\kappa')^r = \frac{\Gamma(1+r)}{|\bar{x}|^{1+r}}\,,
\end{align}
we find the following result for $|x|\gg\xi^{(\nu)}$ 
\begin{align}
    \nonumber
    &\underline{\tau=\Sign(\cos\gamma^{(\nu)})}:\\
    \label{eq:w_large_x_1}
    & w^{(\nu)}_{\tau p}(x)\rightarrow 
    \frac{\tau (ips_x + e^{ip\gamma^{(\nu)}})\Gamma\left(\frac{3}{4}\right)}{4\pi\sqrt{|\cos\gamma^{(\nu)}|}\,\xi^{(\nu)}}
    \left(\frac{\xi^{(\nu)}}{|x|}\right)^{\frac{3}{4}} e^{-\frac{|x|}{2\xi^{(\nu)}}}\,,
\end{align}

In contrast, for $\tau\cos{\gamma^{(\nu)}}<0$, the branching point with smallest imaginary part is located at $\kappa=|\sin{\gamma^{(\nu)}}|$. Excluding the case $\sin{\gamma^{(\nu)}}=0$ (which is treated in detail in Appendix~\ref{app:sin_gamma_zero}) we obtain close to this branching point for small $0<\kappa'=\kappa-|\sin{\gamma^{(\nu)}}|\ll 1$
\begin{widetext}
\begin{align}
    \nonumber
    \delta \chi_{\tau p}^{(\nu)}(is_x(|\sin{\gamma^{(\nu)}}|+\kappa')) &\approx 
    \frac{ip\tau|\sin{\gamma^{(\nu)}}|[s_x+ s_{\gamma^{(\nu)}}] + (-|\tan{\gamma^{(\nu)}}|+ip\tau s_x)\kappa'}{i s_x \sqrt{|\sin{\gamma^{(\nu)}}|}\sqrt{\kappa'}} \\
    \label{eq:dI_small_kappa_2}
    &\approx \begin{cases} 2p\tau \sqrt{|\sin{\gamma^{(\nu)}}|} (\kappa')^{-\frac{1}{2}} & \text{for} \quad s_x =  s_{\gamma^{(\nu)}} \\
    \frac{-|\tan{\gamma^{(\nu)}}| + ip\tau s_x}{i s_x \sqrt{|\sin{\gamma^{(\nu)}}|}} (\kappa')^{\frac{1}{2}} 
    & \text{for} \quad s_x = - s_{\gamma^{(\nu)}} \end{cases} \,.
\end{align}
Inserting this form in (\ref{eq:w_conv}) and using (\ref{eq:gamma_integral}), we find for $|x|\gg\xi^{(\nu)}$ the asymptotic behavior
\begin{align}
    \nonumber
    &\underline{\tau=-\Sign(\cos\gamma^{(\nu)})}:\\
    \label{eq:w_large_x_2}
    & w^{(\nu)}_{\tau p}(x)\rightarrow 
    \frac{ip\tau s_x - |\tan\gamma^{(\nu)}|}{2\sqrt{2\pi}\sqrt{|\sin\gamma^{(\nu)}|}\,\xi^{(\nu)}}
    \left(\frac{\xi^{(\nu)}}{|x|}\right)^{\frac{3}{2}} e^{-|\sin\gamma^{(\nu)}|\frac{|x|}{2\xi^{(\nu)}}}
    +\frac{1+s_x s_{\gamma^{(\nu)}}}{2}\frac{ip\tau s_x\sqrt{|\sin\gamma^{(\nu)}|}}{\sqrt{2\pi}\,\xi^{(\nu)}}
    \left(\frac{\xi^{(\nu)}}{|x|}\right)^{\frac{1}{2}} e^{-|\sin\gamma^{(\nu)}|\frac{|x|}{2\xi^{(\nu)}}}\,.
\end{align}
\end{widetext}

For the special case $\cos\gamma^{(\nu)}=0$, the two branching points fall together and start at $\kappa=1$. Close to this branching point we get for small $0<\kappa'=\kappa-1\ll 1$
\begin{align}
    \label{eq:dI_small_kappa_3}
    \delta \chi_{\tau p}^{(\nu)}(is_x(1+\kappa')) \approx \,p\tau\, \frac{1 + s_x s_{\gamma^{(\nu)}} + \kappa'}{\sqrt{2\kappa'}}\,.
\end{align}
Inserting this form in (\ref{eq:w_conv}) we find the following result for $|x|\gg\xi^{(\nu)}$ 
\begin{align}
    \nonumber
    &\underline{\cos\gamma^{(\nu)}=0}:\\
    \nonumber
    & w^{(\nu)}_{\tau p}(x)\rightarrow \frac{ip\tau s_x}{4\sqrt{\pi}\,\xi^{(\nu)}}\,e^{-\frac{|x|}{2\xi^{(\nu)}}}\\
    \label{eq:w_large_x_3}
    &\hspace{0.5cm}
    \times\left\{(1 + s_x\,s_{\gamma^{(\nu)}})\left(\frac{\xi^{(\nu)}}{|x|}\right)^{\frac{1}{2}} 
    + \left(\frac{\xi^{(\nu)}}{|x|}\right)^{\frac{3}{2}}\right\} \,.
\end{align}

Finally, for the Wannier function $\tilde{w}^{(\nu)}_{\tau p}(x)$, we obtain a single branching point at $\kappa=1$, and get close to this branching point for small $0<\kappa'=\kappa-1\ll 1$
\begin{align}
    \nonumber
    \delta \tilde{\chi}_{\tau p}^{(\nu)}(is_x(1+\kappa')) &\approx \\
    \label{eq:tilde_dI_small_kappa_1}
    &\hspace{-2cm}
    \approx\frac{e^{ip\tau s_x\frac{\pi}{4}}}{i s_x 2^{\frac{1}{4}}}(\kappa')^{-\frac{1}{4}}
    \begin{cases} e^{i\gamma^{(\nu)}} & \text{for} \quad p=+ \\ \tau & \text{for} \quad p=- \end{cases}\,.
\end{align}
Inserting this form in (\ref{eq:w_conv}) and using (\ref{eq:gamma_integral}), we find the following asymptotic behavior for $|x|\gg\xi^{(\nu)}$ in all cases 
\begin{align}
    \nonumber
    \tilde{w}^{(\nu)}_{\tau p}(x) &\rightarrow
    \frac{e^{ip\tau s_x\frac{\pi}{4}}\Gamma\left(\frac{3}{4}\right)}{2^{\frac{3}{2}}\pi\xi^{(\nu)}}
    \left(\frac{\xi^{(\nu)}}{|x|}\right)^{\frac{3}{4}} e^{-|x|/(2\xi^{(\nu)})}\\
    \label{eq:tilde_w_large_x}
    &\hspace{1cm}
    \times\begin{cases} e^{i\gamma^{(\nu)}} & \text{for} \quad p=+ \\ \tau & \text{for} \quad p=- \end{cases}\,.
\end{align}

\section{Properties of Dirac Wannier functions}
\label{app:dirac_wannier_properties}

From (\ref{eq:w_dirac}-\ref{eq:tilde_chi_bar_q}) one can straightforwardly derive the following useful properties for $w^{(\nu)}_{\tau p}(x)$ and $\tilde{w}^{(\nu)}_{\tau p}(x)$
\begin{align}
    \label{eq:w_-p_trafo}
    w_{\tau,-p}^{(\nu)}(x) &= w^{(\nu)}_{\tau p}(x)^* \,,\\
    \label{eq:w_-gamma_trafo}
    w_{\tau p}^{(\nu)}(x) &= w^{(\nu)}_{\tau p}(-x)^*|_{\gamma^{(\nu)}\rightarrow -\gamma^{(\nu)}} \,,\\
    \label{eq:sum_p_w_-x_trafo}
    \sum_p w^{(\nu)}_{\tau p}(-x) &= \sum_p w^{(\nu)}_{\tau p}(x)\,,\\
    \label{eq:sum_p_pw_-x_trafo}
    \sum_p p\, w^{(\nu)}_{\tau p}(-x) &= - \sum_p p\,w^{(\nu)}_{\tau p}(x)|_{\gamma^{(\nu)}\rightarrow -\gamma^{(\nu)}}\,,\\
    \label{eq:w_-tau_trafo}
    w_{-\tau,p}^{(\nu)}(x) &= w^{(\nu)}_{\tau p}(x)^*|_{\gamma^{(\nu)}\rightarrow -\gamma^{(\nu)} + \pi\,s_{\gamma^{(\nu)}}}\,,\\
    \label{eq:tilde_w_-p_trafo}
    \tilde{w}_{\tau,-p}^{(\nu)}(x) &= \tau e^{i\gamma^{(\nu)}}\tilde{w}^{(\nu)}_{\tau p}(x)^* \,,\\
    \label{eq:tilde_w_-gamma_trafo}
    \tilde{w}_{\tau p}^{(\nu)}(x) &= \tilde{w}^{(\nu)}_{\tau p}(-x)^*|_{\gamma^{(\nu)}\rightarrow -\gamma^{(\nu)}} \,,\\
    \label{eq:tilde_w_gamma_pi_trafo}
    &\hspace{-0.5cm}
    = -p \,\tilde{w}^{(\nu)}_{\tau p}(-x)^*|_{\gamma^{(\nu)}\rightarrow -\gamma^{(\nu)} + \pi s_{\gamma^{(\nu)}}} \,,\\
    \label{eq:tilde_w_-x_trafo}
    \tilde{w}^{(\nu)}_{\tau p}(-x) &= \tau e^{ip\gamma^{(\nu)}}\tilde{w}_{\tau,-p}^{(\nu)}(x) \,,\\
    \label{eq:tilde_w_-tau_trafo}
    \tilde{w}_{-\tau,p}^{(\nu)}(x) &= p\,\tilde{w}^{(\nu)}_{\tau p}(-x) \,.
\end{align}

Correspondingly, we find for the real functions $w_{A/B,\tau}^{(\nu)}(x)$ and $\tilde{w}_{A/B,\tau}^{(\nu)}(x)$, as defined by (\ref{eq:w_A_def}-\ref{eq:tilde_w_B_def}), the properties
\begin{align}
    \label{eq:w_A_-x_trafo}
    w_{A,\tau}^{(\nu)}(-x) &= - w_{A,\tau}^{(\nu)}(x)|_{\gamma^{(\nu)}\rightarrow -\gamma^{(\nu)}}\,,\\
    \label{eq:w_B_-x_trafo}
    w_{B,\tau}^{(\nu)}(-x) &= w_{B,\tau}^{(\nu)}(x) = w_{B,\tau}^{(\nu)}(x)|_{\gamma^{(\nu)}\rightarrow -\gamma^{(\nu)}}\,,\\
    \label{eq:w_A_-tau_trafo}
    w_{A,-\tau}^{(\nu)}(x) &= - w_{A,\tau}^{(\nu)}(x)|_{\gamma^{(\nu)}\rightarrow -\gamma^{(\nu)} + \pi\,s_{\gamma^{(\nu)}}}\,,\\
    \label{eq:w_B_-tau_trafo}
    w_{B,-\tau}^{(\nu)}(x) &= w_{B,\tau}^{(\nu)}(x)|_{\gamma^{(\nu)}\rightarrow -\gamma^{(\nu)} + \pi\,s_{\gamma^{(\nu)}}}\,,\\
    \label{eq:tilde_w_A_-gamma_trafo}
    \tilde{w}_{A,\tau}^{(\nu)}(-x) &= -\tilde{w}_{A,\tau}^{(\nu)}(x)|_{\gamma^{(\nu)}\rightarrow -\gamma^{(\nu)}}\\
    \label{eq:tilde_w_A_gamma_pi_trafo}
    &= \tau s_{\gamma^{(\nu)}} \tilde{w}_{B,\tau}^{(\nu)}(x)|_{\gamma^{(\nu)}\rightarrow -\gamma^{(\nu)} + \pi s_{\gamma^{(\nu)}}}\\
    \label{eq:tilde_w_A_-x_trafo}
    &\hspace{-1cm}
    = \tau\left[-\cos{\gamma^{(\nu)}}\,\tilde{w}_{A,\tau}^{(\nu)}(x) + \sin{\gamma^{(\nu)}}\,\tilde{w}_{B,\tau}^{(\nu)}(x) \right]\,,\\
    \label{eq:tilde_w_B_-gamma_trafo}
    \tilde{w}_{B,\tau}^{(\nu)}(-x) &= w_{B,\tau}^{(\nu)}(x)|_{\gamma^{(\nu)}\rightarrow -\gamma^{(\nu)}}\\
    \label{eq:tilde_w_B_gamma_pi_trafo}
    &= \tau s_{\gamma^{(\nu)}} \tilde{w}_{A,\tau}^{(\nu)}(x)|_{\gamma^{(\nu)}\rightarrow -\gamma^{(\nu)} + \pi s_{\gamma^{(\nu)}}}\\
    \label{eq:tilde_w_B_-x_trafo}
    &\hspace{-1cm}
    = \tau\left[\sin{\gamma^{(\nu)}}\,\tilde{w}_{A,\tau}^{(\nu)}(x) + \cos{\gamma^{(\nu)}}\,\tilde{w}_{B,\tau}^{(\nu)}(x) \right]\,,\\
    \label{eq:tilde_w_A_-tau_trafo}
    \tilde{w}_{A,-\tau}^{(\nu)}(x) &= -\tau\,s_{\gamma^{(\nu)}} \, \tilde{w}_{B,\tau}^{(\nu)}(-x)\,,\\
    \label{eq:tilde_w_B_-tau_trafo}
    \tilde{w}_{B,-\tau}^{(\nu)}(x) &= \tau\,s_{\gamma^{(\nu)}} \, \tilde{w}_{A,\tau}^{(\nu)}(-x)\,.
\end{align}

\section{Scaling functions for $\sin{\gamma}=0$}
\label{app:sin_gamma_zero}

In this Appendix we discuss the properties of the scaling functions $F_{A/B,\tau}(y;\gamma)$ and $\tilde{F}_{A/B,\tau}(y;\gamma)$ for the special case $\sin{\gamma}=0$. Based on the relations (\ref{eq:F_A_-y_trafo}-\ref{eq:tilde_F_B_-tau_trafo}) for $\gamma=O^\pm$ and $\gamma=\pi\pm O^+=-\pi\pm 0^+$, one finds that the scaling functions $F_{A/B,\tau}(y;\gamma)$ are continuous in $\gamma$ at $\gamma=0,\pi$ and have the following properties 
\begin{align}
    \label{eq:F_A_zero_-y_trafo}
    F_{A,\tau}(-y;0) &= F_{A,\tau}(y;0) \,,\\    
    \label{eq:F_A_pi_-y_trafo}
    F_{A,\tau}(-y;\pi) &= F_{A,\tau}(y;\pi) \,,\\    
    \label{eq:F_B_zero_-y_trafo}
    F_{B,\tau}(-y;0) &= - F_{B,\tau}(y;0) \,,\\    
    \label{eq:F_B_pi_-y_trafo}
    F_{B,\tau}(-y;\pi) &= F_{B,\tau}(y;\pi) \,,\\    
    \label{eq:F_A_zero_-tau_trafo}
    F_{A,-\tau}(y;0) &= - F_{A,\tau}(y;\pi) \,,\\    
    \label{eq:F_B_zero_-tau_trafo}
    F_{B,-\tau}(y;0) &= F_{B,\tau}(y;\pi) \,.    
\end{align}
For the scaling functions $\tilde{F}_{A/B,\tau}(y;\gamma)$ it turns out that they are continuous at $\gamma=0,\pi$ for $\tau\cos{\gamma}>0$ (i.e.,for $\gamma=\pi, \tau=-$ or for $\gamma=0, \tau=+$) but discontinuous for $\tau\cos{\gamma}<0$ (i.e., for $\gamma=0, \tau=-$ or for $\gamma=\pi,\tau=+$), with the properties
\begin{align}
    \label{eq:tilde_F_A_zero_-y_trafo}
    \tilde{F}_{A,+}(-y;0) &= \tilde{F}_{A,+}(y;0) \,,\\    
    \label{eq:tilde_F_A_pi_-y_trafo}
    \tilde{F}_{A,-}(-y;\pi) &= \tilde{F}_{A,-}(y;\pi) \,,\\    
    \label{eq:tilde_F_B_zero_-y_trafo}
    \tilde{F}_{B,+}(-y;0) &= - \tilde{F}_{B,+}(y;0) \,,\\    
    \label{eq:tilde_F_B_pi_-y_trafo}
    \tilde{F}_{B,-}(-y;\pi) &= -\tilde{F}_{B,-}(y;\pi) \,,\\    
    \label{eq:tilde_F_A_zero_-tau_trafo}
    \tilde{F}_{A,-}(y;0^\pm) &= \mp \tilde{F}_{B,+}(y;0) \,,\\    
    \label{eq:tilde_F_A_pi_-tau_trafo}
    \tilde{F}_{A,+}(y;\pi\pm 0^+) &= \mp \tilde{F}_{B,-}(y;\pi) \,,\\  
    \label{eq:tilde_F_B_zero_-tau_trafo}
    \tilde{F}_{B,-}(y;0^\pm) &= \mp\tilde{F}_{A,+}(y;0) \,\\    
    \label{eq:tilde_F_B_pi_-tau_trafo}
    \tilde{F}_{B,+}(y;\pi\pm 0^+) &= \mp\tilde{F}_{A,-}(y;\pi) \,.    
\end{align}

For $|y|\ll 1$ and $|y|\gg 1$, the values can be obtained straightforwardly from (\ref{eq:F_A_small_y}-\ref{eq:tilde_F_B_large_y}). A special case are the scaling functions $F_{A/B,-}(y;0)$ and $F_{A/B,+}(y;\pi)$ which are not exponentially decaying since the branch cuts in the complex plane start at the origin. For $\gamma=0$ and $\tau=-$ we obtain from (\ref{eq:chi_bar_q})
\begin{align}
    \chi_{-,p}^{(\nu)}(\bar{q}) = \frac{\bar{\epsilon}_{\bar{q}} - p\bar{q} - 1}{2\sqrt{\bar{\epsilon}_{\bar{q}}}\sqrt{\bar{\epsilon}_{\bar{q}}-1}}\,. 
\end{align}
Expanding around $\bar{q}=0$ with $\bar{q}=is_y\kappa\pm\eta$, we obtain
\begin{align}
    \sum_p p \,\chi_{-,p}^{(\nu)}(is_y\kappa\pm\eta) &\approx \mp\sqrt{2} \,,\\
    \sum_p \,\chi_{-,p}^{(\nu)}(is_y\kappa\pm\eta) &\approx \pm \frac{i s_y \kappa}{\sqrt{2}} \,,
\end{align}
leading to 
\begin{align}
    \sum_p p \delta\chi_{-,p}^{(\nu)}(is_y\kappa) &\approx - 2\sqrt{2} \,,\\
    \sum_p \delta\chi_{-,p}^{(\nu)}(is_y\kappa) &= i\sqrt{2} \,s_y\kappa\,.
\end{align}
Inserting in (\ref{eq:w_conv}) we obtain for the asymptotic behavior
\begin{align}
    w^{(\nu)}_{A,-}(x) &= -i \sum_p p \,w_{-,p}^{(\nu)}(x) \rightarrow - \frac{\sqrt{2}}{\pi\,x} \,,\\
    w^{(\nu)}_{B,-}(x) &= \sum_p \,w_{-,p}^{(\nu)}(x) \rightarrow - \frac{\sqrt{2}}{\pi\xi^{(\nu)}}\,\left(\frac{\xi^{(\nu)}}{x}\right)^2 \,,
\end{align}
which gives for the scaling functions the results (\ref{eq:F_A_asymptotic_sin_gam_zero}) and (\ref{eq:F_B_asymptotic_sin_gam_zero})
\begin{align}
    F_{A,-}(y;0) &= - F_{A,+}(y;\pi) \rightarrow - \frac{\sqrt{2}}{\pi} \,,\\
    F_{B,-}(y;0) &=  F_{B,+}(y;\pi) \rightarrow - \frac{\sqrt{2}}{\pi}\,\frac{1}{y} \,.
\end{align}

\section{Bloch states for $Z=2,3$}
\label{app:bloch_Z_23}

The states $u_{k\alpha}(ja)=\langle j|u_{k\alpha}\rangle$ follow from diagonalizing the Hamiltonian $h_k$ given by (\ref{eq:h_k_def}). For the particular cases $Z=2$ and $Z=3$ we obtain the matrices
\begin{align}
    \label{eq:h_k_Z=2}
    h_k = \left(\begin{array}{cc} \delta v_1 & -t_1 e^{ika} - t_2 e^{-ika} \\
    -t_1 e^{-ika} - t_2 e^{ika} & \delta v_2 \end{array}\right)\,,
\end{align}
with $\delta v_1 = - \delta v_2$ and $\frac{1}{2}(t_1+t_2)=t$, and
\begin{align}
    \label{eq:h_k_Z=3}
    h_k = \left(\begin{array}{ccc} \delta v_1 & -t_1 e^{ika} & - t_3 e^{-ika} \\
    -t_1 e^{-ika} & \delta v_2 & - t_2 e^{ika} \\
    -t_3 e^{ika} & -t_2 e^{-ika} & \delta v_3 \end{array}\right)\,,
\end{align}
with $\delta v_1 + \delta v_2 + \delta v_3 = 0$ and $\frac{1}{3}(t_1 + t_2 + t_3)=t$.

It is convenient to write 
\begin{align}
    \label{eq:u_x}
    |{u}_{k\alpha}\rangle = \frac{(-1)^{\alpha+1}}{\sqrt{N_{k\alpha}}} \underline{x}_{k\alpha}\quad,\quad
    N_{k\alpha} = \sum_{j=1}^Z |\underline{x}_{k\alpha}|^2 \,,
\end{align}
where $\underline{x}_{k\alpha}$ denotes a $Z$-dimensional vector. 

For $Z=2$ this gives the explicit formulas
\begin{align}
    \label{eq:Z=2_eigenstates}
    \underline{x}_{k\alpha} &=  
    \left(\begin{array}{c} t_1 e^{ika} + t_2 e^{-ika} \\ \delta v - \epsilon_{k\alpha} \end{array}\right) \,,\\
    \label{eq:Z=2_N}
    N_{k\alpha} &= 2\epsilon_{k\alpha}(\epsilon_{k\alpha}-\delta v)\,,\\
    \label{eq:Z=2_dispersion}
    \epsilon_{k,1/2} &= \mp \sqrt{\Delta^2 + 4 t_1 t_2 \cos^2(ka)} \,,\\
    \label{eq:Z=2_delta}
    \Delta &= \sqrt{\delta v^2 + 4\delta t^2}\,,
\end{align}
where $\delta v=\delta v_1 = -\delta v_2$, and $t_{1/2}=t\pm\delta t$. Using (\ref{eq:gap_equation}), (\ref{eq:v_fourier}), (\ref{eq:t_fourier}) and $v_F^{(1)} = 2ta$, we obtain for the gap parameter and the correlation length $\xi^{(1)}=v_F^{(1)}/(2\Delta^{(1)})$  
\begin{align}
    \label{eq:gap_parameter_Z_2}
    \Delta e^{i\gamma} &\equiv \Delta^{(1)} e^{i\gamma^{(1)}} = - \delta v - 2i \delta t \,,\\
    \label{eq:xi_Z_2}
    \xi &\equiv \xi^{(1)} = \frac{at}{\Delta} \,.
\end{align}

For $Z=3$ we get the following explicit formulas
\begin{align}
    \label{eq:Z=3_eigenstates}
    &\underline{x}_{k\alpha} =  
    \left(\begin{array}{c} \delta\bar{v}_2 t_3 e^{-ika} + t_1 t_2 e^{2ika} \\ 
    \delta\bar{v}_1 t_2 e^{ika} + t_1 t_3 e^{-2ika} \\
    \delta\bar{v}_1\delta\bar{v}_2 - t_1^2 \end{array}\right) \,,\\
    \nonumber
    & N_{k\alpha} = (t_1^2 -\delta\bar{v}_1\delta\bar{v}_2)\\
    \label{eq:Z=3_N}
    &\hspace{0.5cm}
    \times(t_1^2+t_2^2+t_3^2-\delta\bar{v}_1\delta\bar{v}_2-\delta\bar{v}_1\delta\bar{v}_3
    -\delta\bar{v}_2\delta\bar{v}_3)\,,\\
    \nonumber
    & 2t_1 t_2 t_3 \cos(3ka) = \\
    \label{eq:Z_3_dispersion}
    &\hspace{1cm} 
    =\delta\bar{v}_1\delta\bar{v}_2\delta\bar{v}_3
    -\delta\bar{v}_1 t_2^2 
    -\delta\bar{v}_2 t_3^2 -\delta\bar{v}_3 t_1^2\,,
\end{align}
where we defined
\begin{align}
    \label{eq:v_bar}
    \delta\bar{v}_j = \delta v_j - \epsilon_{k\alpha}\,.
\end{align}
We do not indicate the dependence of $\delta\bar{v}_j\equiv\delta\bar{v}_{jk\alpha}$ on $k$ and $\alpha$ since it is clear which band index has to be taken for the Abelian case. 

The two gap parameters $\Delta^{(\nu)} e^{i\gamma^{(\nu)}}$, with $\nu=1,2$, follow via (\ref{eq:gap_equation}) from the Fourier components $\delta\tilde{v}_j$ and $\delta\tilde{t}_j$. The latter can be determined from (\ref{eq:v_fourier}) and (\ref{eq:t_fourier}) as
\begin{widetext}
\begin{align}
    \label{eq:Z=3_v_fourier}
    \delta\tilde{v}_1 &= \delta\tilde{v}_2^* = -\frac{1}{2}(\delta v_1 + \delta v_2) 
    - \frac{i}{2\sqrt{3}}(\delta v_1 - \delta v_2)\,,\\
    \label{eq:Z=3_t_fourier}
    \delta\tilde{t}_1 &= \delta\tilde{t}_2^* = -\frac{1}{2}(\delta t_1 + \delta t_2) 
    - \frac{i}{2\sqrt{3}}(\delta t_1 - \delta t_2)\,.
\end{align}
Inserting this result in (\ref{eq:gap_equation}) we find after some algebra
\begin{align}
    \label{eq:Z=3_gap_parameter_1}
    \Delta^{(1)} e^{i\gamma^{(1)}} &= \delta t_1 - \frac{1}{2}(\delta v_1 + \delta v_2)
    -\frac{i}{2\sqrt{3}}(2\delta t_1 + 4 \delta t_2 + \delta v_1 - \delta v_2)\,,\\
    \label{eq:Z=3_gap_parameter_2}
    \Delta^{(2)} e^{i\gamma^{(2)}} &= -\delta t_1 - \frac{1}{2}(\delta v_1 + \delta v_2)
    -\frac{i}{2\sqrt{3}}(2\delta t_1 + 4 \delta t_2 - \delta v_1 + \delta v_2)\,.
\end{align}
\end{widetext}
The correlation lengths for $\nu=1,2$ follow from $\xi^{(\nu)} = \frac{v_F^{(\nu)}}{2\Delta^{(\nu)}}$. With $v_F^{(\nu)}=2ta\sin(k_F^{(\nu)}a)$ and $k_F^{(\nu)}=\frac{\nu\pi}{3a}$ we obtain
\begin{align}
    \label{eq:Z=3_xi}
    \xi^{(1/2)} = \frac{\sqrt{3}}{2}\frac{at}{\Delta^{(1/2)}} \,.
\end{align}

We note that we have included a sign factor $(-1)^{\alpha+1}$ to $|{u}_{k\alpha}\rangle$, such that, according to Ref.~[\onlinecite{pletyukhov_etal_prb_20}], our convention that $u_{k\alpha}(Za)>0$ is positive (see Eq.~(\ref{eq:u_k_Z_explicit})) is fulfilled.

The Zak-Berry connection can be calculated most elegantly from 
\begin{align}
    \label{eq:zb_connection_elegant}
    A_{k\alpha} = \frac{1}{N_{k\alpha}} \text{Re} \left(\underline{x}_{k\alpha}\right)^\dagger i\partial_k \underline{x}_{k\alpha} \,.
\end{align}
Inserting the forms for $\underline{x}_{k\alpha}$ from (\ref{eq:Z=2_eigenstates}) and (\ref{eq:Z=3_eigenstates}), one finds for $Z=2$
\begin{align}
    \label{eq:Z=2_zak_berry_connection}
    A_{k\alpha} = \frac{a(t_2^2 - t_1^2)}{N_{k\alpha}}\,,
\end{align}
and for $Z=3$
\begin{align}
    \nonumber
    A_{k\alpha} &= \frac{a}{N_{k\alpha}}
    \left\{2t_1^2(t_3^2-t_2^2) - \delta\bar{v}_1^2 t_2^2 + \delta\bar{v}_2^2 t_3^2 + \right.\\
    \label{eq:Z=3_zak_berry_connection}
    &\hspace{1cm} 
    \left. + (\delta\bar{v}_1 - \delta\bar{v}_2)t_1 t_2 t_3 \cos(3ka) \right\} \,.
\end{align}

For the non-Abelian case we also need the nondiagonal elements of the Zak-Berry connection which can be calculated from
\begin{align}
    \underline{\alpha\ne\beta}:\quad
    (A_k)_{\alpha\beta} = \frac{(-1)^{(\alpha+\beta)}}{\sqrt{N_{k\alpha} N_{k\beta}}} \langle x_{k\alpha} | i \partial_k x_{k\beta}\rangle \,.
\end{align}
For the special case $Z=3$ we obtain the expression
\begin{align}
    & \langle x_{k1} | i \partial_k x_{k2}\rangle = i (\delta\bar{v}_{11} \delta\bar{v}_{21} - t_1^2) (\epsilon_{k1} + 2 \epsilon_{k2}) 
    \frac{d \epsilon_{k2}}{d k} \nonumber \\
    & + a (\delta\bar{v}_{21} t_3 e^{i ka} + t_1 t_2 e^{-2i ka}) (\delta\bar{v}_{22} t_3 e^{-i ka} - 2 t_1 t_2 e^{2i ka }) \nonumber \\
    & + a (\delta \bar{v}_{11} t_2 e^{-i ka} + t_1 t_3 e^{2i ka}) (-\delta \bar{v}_{12} t_2 e^{i ka} +2 t_1 t_3 e^{-2i ka})\,,
\end{align}
where we used the notation $\delta\bar{v}_{j\alpha}=v_j - \epsilon_{k\alpha}$.

For the derivative of the dispersion we use the results obtained in Ref.~\onlinecite{pletyukhov_etal_prb_20}, which give for $Z=2$
\begin{align}
    \nonumber
    \frac{d\epsilon_{k\alpha}}{dk} &= 4a t_1 t_2 \frac{\delta\bar{v}_1}{N_{k\alpha}} \sin(2ka) \\
    \label{eq:der_eps_Z_2}
    &= - \frac{2a t_1 t_2 \sin(2ka)}{\epsilon_{k\alpha}}\,,
\end{align}
and for $Z=3$
\begin{align}
    \nonumber
    \frac{d\epsilon_{k\alpha}}{dk} &= 6a t_1 t_2 t_3 \frac{\delta\bar{v}_1\delta\bar{v}_2-t_1^2}{N_{k\alpha}} \sin(3ka) \\
    \label{eq:der_eps_Z_3}
    &= - \frac{6a t_1 t_2 t_3 \sin(3ka)}{t_1^2 + t_2^2 + t_3^2 - \delta\bar{v}_1\delta\bar{v}_2 - \delta\bar{v}_1\delta\bar{v}_3
    - \delta\bar{v}_2\delta\bar{v}_3}\,.
\end{align}

\section{Splitting of non-Abelian Wannier functions}
\label{app:nonabelian_splitting}

In this Appendix we show that extending the momentum integrations in the representations (\ref{eq:hat_w_I}) and (\ref{eq:hat_w_II}) of the Wannier functions in NA-ML gauge for the two subintervals gives only rise to delta function contributions in the total sum (\ref{eq:hat_w_splitting}) which can be disregarded. After inserting (\ref{eq:nonabelian_u_explicit}) we replace the propagators beyond the subintervals via (\ref{eq:prop_extension_I}), (\ref{eq:prop_extension_II}) and (\ref{eq:prop_splitting_II}) as 
\begin{align}
    \label{eq:prop_extension_I_1}
    \underline{(\text{I})}: & \, U^{(\nu)}(k,-\frac{\pi}{2Za}) \rightarrow 
    \begin{cases} U^{(\nu)}_{\text{I}} & \text{for}\,k>\frac{\pi}{2Za} \\
         1 &\text{for}\,k<-\frac{\pi}{2Za} \end{cases} \,,\\
    \label{eq:prop_extension_II_1}
    \underline{(\text{II})}: & \, U^{(\nu)}(k,-\frac{\pi}{2Za}) \rightarrow 
    \begin{cases} U^{(\nu)}_{\text{L}} &\text{for}\,k>\frac{3\pi}{2Za} \\ U^{(\nu)}_{\text{I}} &\text{for}\,k<\frac{\pi}{2Za} \end{cases} \,.
\end{align}
and the Bloch states are replaced with the help of (\ref{eq:_gapless_explicit}) and (\ref{eq:u_k_periodicity}) by the $k$-independent values
\begin{align}
    \label{eq:u_extension_I}
    \underline{(\text{I})}: & \, u_{k\alpha'}(ma) \rightarrow 
    \begin{cases} f_{\alpha'}(m) & \text{for}\,k>\frac{\pi}{2Za} \\
          f_{\alpha'}(m)^* &\text{for}\,k<-\frac{\pi}{2Za} \end{cases} \,,\\
    \label{eq:u_extension_II}
    \underline{(\text{II})}: & \, u_{k\alpha'}(ma) \rightarrow 
    \begin{cases} f_{\alpha'}(m)^*\,e^{-i\frac{2\pi}{Z}m} &\text{for}\,k>\frac{3\pi}{2Za} \\ 
    f_{\alpha'}(m) &\text{for}\,k<\frac{\pi}{2Za} \end{cases} \,,
\end{align}
where we defined
\begin{align}
    f_{\alpha'}(m) = \frac{1}{\sqrt{Z}} \begin{cases} e^{-i\frac{\pi}{Z}\alpha'm} & \text{for}\,\,\alpha'\,\,{\rm even} \\
    e^{i\frac{\pi}{Z}(\alpha'-1)m} & \text{for}\,\,\alpha'\,\,{\rm odd} \end{cases}\,.
\end{align}

Obviously, the two contributions $k>\frac{\pi}{2Za}$ from the extension of subinterval (I) and $k<\frac{\pi}{2Za}$ from the extension of subinterval (II) lead to the same integrand, such that the sum contains $\int dk\,e^{ik\hat{m}_\alpha^{(\nu)}a} = 2\pi\delta(\hat{m}_\alpha^{(\nu)}a)$. This are unphysical delta function contributions which we always disregard. Similar delta function contributions are generated from the high-momentum integrals in Dirac theory which cancel these terms.   

For the two remaining contributions we first shift $k\rightarrow k+\frac{2\pi}{Za}$ for the contribution $k>\frac{3\pi}{2Za}$ from the extension of subinterval (II) to get the same starting point $-\frac{2\pi}{Za}$ for the $k$-integration as the end point for the contribution $k<-\frac{\pi}{2Za}$ from the extension of subinterval (I). This shift gives an additional $e^{i\frac{2\pi}{Z}\hat{m}_\alpha^{(\nu)}}$ factor from the exponential $e^{ik\hat{m}_\alpha^{(\nu)}a}$ in the integrand. The integrands for the two contributions are then again the same since 
\begin{align}
    \label{eq:identity_1}
    e^{i\frac{2\pi}{Z}\hat{m}_\alpha^{(\nu)}}\,e^{-i\frac{2\pi}{Z}m} &= e^{-i\hat{\gamma}_\alpha^{(\nu)}} \,,\\
    \label{eq:identity_2}
    (U_{\rm L}^{(\nu)}\,V^{(\nu)})_{\alpha'\alpha} &= (V^{(\nu)})_{\alpha'\alpha}\,e^{i\hat{\gamma}_\alpha^{(\nu)}}\,,
\end{align}
such that only $(V^{(\nu)})_{\alpha'\alpha}$ remains for the product of (\ref{eq:identity_1}) and (\ref{eq:identity_2}). Here, we have used the definition $\hat{m}_\alpha^{(\nu)}=m-\frac{Z}{2\pi}\hat{\gamma}_\alpha^{(\nu)}$ for the derivation of (\ref{eq:identity_1}), and the definition (\ref{eq:V_def}) of the transformation $V^{(\nu)}$ for the derivation of (\ref{eq:identity_2}). As a result we get again a delta function from the two remaining contributions which we can disregard.

\section{The Wilson propagator in Dirac theory}
\label{app:wilson_dirac_propagator}

In this Appendix we calculate the Wilson propagator (\ref{eq:prop_dirac}) in Dirac theory and prove the central property (\ref{eq:chi_trafo_dirac_nonabelian}). We first transform the propagator via the Abelian ML gauge to
\begin{align}
    \label{eq:tilde_U_transform}
    U_q^{(\nu)} = \Lambda_q^{(\nu)} \, \tilde{U}_q^{(\nu)} \, (\Lambda_{-\infty}^{(\nu)})^\dagger \,,
\end{align}
where
\begin{align}
    \label{eq:tilde_U_dirac}
    \tilde{U}_q^{(\nu)} &= P e^{i\int_{-\infty}^q dq' \tilde{A}^{(\nu)}_{q'}} \,,\\
    \label{eq:tilde_A_dirac_abelian_trafo}
    \tilde{A}^{(\nu)}_q &= (\Lambda_q^{(\nu)})^\dagger \, A_q^{(\nu)} \, \Lambda_q^{(\nu)} 
    + i (\Lambda_q^{(\nu)})^\dagger\,\partial_q \Lambda_q^{(\nu)} \,,\\
    \label{eq:tilde_A_dirac_nonabelian}
    (\tilde{A}^{(\nu)}_q)_{\tau\tau'} &= i\langle \tilde{\chi}_{q\tau}^{(\nu)}|\partial_q\tilde{\chi}_{q\tau'}^{(\nu)}\rangle\,,\\
    \label{eq:lambda_matrix_dirac}
    \Lambda_q^{(\nu)} &= \left(\begin{array}{cc} e^{i\phi_{q+}^{(\nu)}} & 0 \\ 0 & e^{i\phi_{q-}^{(\nu)}} \end{array}\right) \,.
\end{align}
Using $\tilde{A}_{q\tau}^{(\nu)}=(\tilde{A}^{(\nu)}_q)_{\tau\tau}=0$ (which defines the Abelian ML gauge via (\ref{eq:max_loc_dirac})), we need in addition the nondiagonal matrix elements. They follow from the form (\ref{eq:chi_tilde}) of $|\tilde{\chi}_{q\tau}^{(\nu)}\rangle$ and (\ref{eq:tilde_A_dirac_nonabelian}) in dimensionless units as
\begin{align}
    \tilde{A}(\bar{q}) \equiv \frac{1}{2\xi^{(\nu)}}\,\tilde{A}_q^{(\nu)} = \frac{1}{2}\,\frac{1}{1+\bar{q}^2}\,\sigma_y \,,
\end{align}
where $\bar{q}=2\xi^{(\nu)}q$. With
\begin{align}
    \tilde{U}(\bar{q}) \equiv \tilde{U}^{(\nu)}_q = P e^{i\int_{-\infty}^{\bar{q}} d\bar{q}'\,\tilde{A}(\bar{q}')} \,,
\end{align}
we find the differential equation
\begin{align}
    \frac{d}{d\bar{q}} \tilde{U}(\bar{q}) = i \tilde{A}(\bar{q})\,\tilde{U}(\bar{q})\,, 
\end{align}
which has the solution
\begin{align}
\tilde{U}(\bar{q}) = e^{i\frac{1}{2}\sigma_y\arctan\bar{q}}\,e^{i\sigma_y\frac{\pi}{4}}\,,
\end{align}
such that the boundary condition $\tilde{U}(-\infty)=1$ is fulfilled. A straightforward analysis leads to the explicit result
\begin{align}
    \label{eq:tilde_prop_dirac_result}
    \tilde{U}(\bar{q}) = \frac{1}{\sqrt{2\bar{\epsilon}_{\bar{q}}}}\left(\begin{array}{cc} \sqrt{\bar{\epsilon}_{\bar{q}}-\bar{q}} & \sqrt{\bar{\epsilon}_{\bar{q}}+\bar{q}} \\ -\sqrt{\bar{\epsilon}_{\bar{q}}+\bar{q}} & \sqrt{\bar{\epsilon}_{\bar{q}}-\bar{q}}\end{array}\right)\,,
\end{align}
with $\bar{\epsilon}_{\bar{q}}=\sqrt{1+\bar{q}^2}$.

To prove (\ref{eq:chi_trafo_dirac_nonabelian}), we use (\ref{eq:tilde_U_transform}) and find
\begin{align}
    \hat{\chi}^{(\nu)}(\bar{q}) = \tilde{\chi}^{(\nu)}(\bar{q})\,\tilde{U}(\bar{q})\,(\Lambda_{-\infty}^{(\nu)})^\dagger\,,
\end{align}
where we defined the $2\times 2$-matrices $\big[\hat{\chi}^{(\nu)}(\bar{q})\big]_{p\tau}\equiv\hat{\chi}^{(\nu)}_{q\tau p}$ and $\big[\tilde{\chi}^{(\nu)}(\bar{q})\big]_{p\tau}\equiv\tilde{\chi}^{(\nu)}_{q\tau p}$. Using (\ref{eq:chi_tilde}), (\ref{eq:lambda_matrix_dirac}), (\ref{eq:phi_+_asymptotic}) and (\ref{eq:phi_+_asymptotic}), we find
\begin{align}
    \label{eq:tilde_chi_matrix}
    \tilde{\chi}^{(\nu)}(\bar{q}) &= \frac{1}{\sqrt{2\bar{\epsilon}_{\bar{q}}}}
    \left(\begin{array}{cc} \sqrt{\bar{\epsilon}_{\bar{q}}+\bar{q}}\,e^{i\gamma^{(\nu)}} & \sqrt{\bar{\epsilon}_{\bar{q}}-\bar{q}}\,e^{i\gamma^{(\nu)}} \\ \sqrt{\bar{\epsilon}_{\bar{q}}-\bar{q}} & -\sqrt{\bar{\epsilon}_{\bar{q}}+\bar{q}}\end{array}\right)\,,\\
    \label{eq:lambda_matrix_asymptotic}
    \Lambda_{-\infty}^{(\nu)} &= \left(\begin{array}{cc} 1 & 0 \\ 0 & e^{i\gamma^{(\nu)}} \end{array}\right)\,.
\end{align}
Multiplying (\ref{eq:tilde_chi_matrix}) with the product of (\ref{eq:tilde_prop_dirac_result}) and (\ref{eq:lambda_matrix_asymptotic}), we get
\begin{align}
    \hat{\chi}^{(\nu)}(\bar{q}) = \sigma_x \,,
\end{align}
which proves (\ref{eq:chi_trafo_dirac_nonabelian}).

\section{Spectrum of Wilson loop operator}
\label{app:spectrum_wilson_loop} 

The Wilson loop operator defining the unitary transformation $V^{(\nu)}$ can be calculated from (\ref{eq:wilson_loop_total}) as the product $U_{\text{L}}^{(\nu)} = U^{(\nu)}_{\text{II}} \,U^{(\nu)}_{\text{I}}$ of the propagators for the two subintervals. The propagators $U^{(\nu)}_{\text{I}}$ and $U^{(\nu)}_{\text{II}}$ follow from the limit $q\rightarrow\infty$ of (\ref{eq:block_structure_propagator_nu_even}) and (\ref{eq:block_structure_propagator_nu_odd}) as
\begin{widetext}
\begin{align}
    \label{eq:U_I_II_nu_even}
    \underline{\nu\,\,{\rm even}}: & \quad
    U^{(\nu)}_{\text{I}} =  
    \left(\begin{array}{cccccc} 
    - e^{-i\gamma^{(\nu)}} & & & & &  \\
    & \sigma_x & & & & \\ 
    & & \sigma_x & & & \\ 
    & & & \ddots & & \\ 
    & & & & \sigma_x & \\ 
    & & & & & 1 \\
    \end{array}\right) \quad,\quad
    U^{(\nu)}_{\text{II}} =  
    \left(\begin{array}{cccc} 
    \sigma_x & & &  \\
    & \sigma_x & & \\ 
    & & \ddots & \\ 
    & & & \sigma_x \\
    \end{array}\right) \,,\\
    \label{eq:U_I_II_nu_odd}
    \underline{\nu\,\,{\rm odd}}: & \quad
    U^{(\nu)}_{\text{I}} =  
    \left(\begin{array}{ccccc} 
    \sigma_x & & & &  \\
    & \sigma_x & & & \\ 
    & & \ddots & & \\ 
    & & & \sigma_x & \\ 
    & & & & 1 \\
    \end{array}\right) \quad,\quad
    U^{(\nu)}_{\text{II}} =  
    \left(\begin{array}{ccccc} 
    - e^{-i\gamma^{(\nu)}} & & & &  \\
    & \sigma_x & & & \\ 
    & & \sigma_x & & \\ 
    & & & \ddots & \\ 
    & & & & \sigma_x \\ 
    \end{array}\right)\,,
\end{align}
\end{widetext}
where we used 
\begin{align}
    \lim_{q\rightarrow\infty} U^{(\nu)}_{q-} &= - e^{-i\gamma^{(\nu)}} \,,\\
    \lim_{q\rightarrow\infty} U^{(\nu)}_q &= \sigma_x \,,
\end{align}
which follows from (\ref{eq:tilde_U_transform}), (\ref{eq:lambda_matrix_dirac}), (\ref{eq:tilde_prop_dirac_result}) and (\ref{eq:lambda_matrix_asymptotic}), together with (\ref{eq:phi_+_asymptotic}) and (\ref{eq:phi_-_asymptotic}). Thereby, only $\sigma_x$ are $2\times 2$-submatrices, whereas $-e^{-i\gamma^{(\nu)}}$ and $1$ denote $1\times 1$-numbers.

From (\ref{eq:U_I_II_nu_even}) and (\ref{eq:U_I_II_nu_odd}) one can see directly that the Wilson loop operator $U_{\text{L}}^{(\nu)} = U^{(\nu)}_{\text{II}} \,U^{(\nu)}_{\text{I}}$ is a unitary matrix which contains only one non-zero phase factor in each row and column. Therefore, all eigenvectors consist only of phase factors with a normalization factor $1/\sqrt{\nu}$. We conclude that the matrix elements $(V^{(\nu)})_{\nu\alpha}$ and $(U_{\rm I}^{(\nu)}\,V^{(\nu)})_{\nu\alpha}$ occurring in (\ref{eq:hat_w_I_result}) and (\ref{eq:hat_w_II_result}) are given by $1/\sqrt{\nu}$ times a phase factor. 

Furthermore, by using the special structure of the Wilson loop matrix $U_{\text{L}}^{(\nu)}$, one can easily find the eigenvalues $\lambda$. A straightforward analysis gives
\begin{align}
    \text{det} (U_{\text{L}}^{(\nu)} - \lambda\mathbbm{1}) = (-\lambda)^\nu + (-1)^\nu e^{-i\gamma^{(\nu)}} = 0 \,,
\end{align}
such that for both $\nu$ even or odd we get 
\begin{align}
    \lambda^\nu = - e^{-i\gamma^{(\nu)}} \,.
\end{align}
This equation has $\nu$ solutions for $\lambda=e^{i\gamma_\alpha^{(\nu)}}$, with $\gamma_\alpha^{(\nu)}$ given by (\ref{eq:zak_berry_phases_nonabelian_result}).

\end{appendix}

\bibliography{citations}

\begin{thebibliography}{32}%
\makeatletter
\providecommand \@ifxundefined [1]{%
 \@ifx{#1\undefined}
}%
\providecommand \@ifnum [1]{%
 \ifnum #1\expandafter \@firstoftwo
 \else \expandafter \@secondoftwo
 \fi
}%
\providecommand \@ifx [1]{%
 \ifx #1\expandafter \@firstoftwo
 \else \expandafter \@secondoftwo
 \fi
}%
\providecommand \natexlab [1]{#1}%
\providecommand \enquote  [1]{``#1''}%
\providecommand \bibnamefont  [1]{#1}%
\providecommand \bibfnamefont [1]{#1}%
\providecommand \citenamefont [1]{#1}%
\providecommand \href@noop [0]{\@secondoftwo}%
\providecommand \href [0]{\begingroup \@sanitize@url \@href}%
\providecommand \@href[1]{\@@startlink{#1}\@@href}%
\providecommand \@@href[1]{\endgroup#1\@@endlink}%
\providecommand \@sanitize@url [0]{\catcode `\\12\catcode `\$12\catcode
  `\&12\catcode `\#12\catcode `\^12\catcode `\_12\catcode `\%12\relax}%
\providecommand \@@startlink[1]{}%
\providecommand \@@endlink[0]{}%
\providecommand \url  [0]{\begingroup\@sanitize@url \@url }%
\providecommand \@url [1]{\endgroup\@href {#1}{\urlprefix }}%
\providecommand \urlprefix  [0]{URL }%
\providecommand \Eprint [0]{\href }%
\providecommand \doibase [0]{https://doi.org/}%
\providecommand \selectlanguage [0]{\@gobble}%
\providecommand \bibinfo  [0]{\@secondoftwo}%
\providecommand \bibfield  [0]{\@secondoftwo}%
\providecommand \translation [1]{[#1]}%
\providecommand \BibitemOpen [0]{}%
\providecommand \bibitemStop [0]{}%
\providecommand \bibitemNoStop [0]{.\EOS\space}%
\providecommand \EOS [0]{\spacefactor3000\relax}%
\providecommand \BibitemShut  [1]{\csname bibitem#1\endcsname}%
\let\auto@bib@innerbib\@empty
%</preamble>
\bibitem [{\citenamefont {Wannier}(1937)}]{wannier_pr_37}%
  \BibitemOpen
  \bibfield  {author} {\bibinfo {author} {\bibfnamefont {G.~H.}\ \bibnamefont
  {Wannier}},\ }\bibfield  {title} {\bibinfo {title} {The structure of
  electronic excitation levels in insulating crystals},\ }\href
  {https://doi.org/10.1103/PhysRev.52.191} {\bibfield  {journal} {\bibinfo
  {journal} {Phys. Rev.}\ }\textbf {\bibinfo {volume} {52}},\ \bibinfo {pages}
  {191} (\bibinfo {year} {1937})}\BibitemShut {NoStop}%
\bibitem [{\citenamefont {Kohn}(1959)}]{kohn_pr_59}%
  \BibitemOpen
  \bibfield  {author} {\bibinfo {author} {\bibfnamefont {W.}~\bibnamefont
  {Kohn}},\ }\bibfield  {title} {\bibinfo {title} {Analytic properties of bloch
  waves and wannier functions},\ }\href
  {https://doi.org/10.1103/PhysRev.115.809} {\bibfield  {journal} {\bibinfo
  {journal} {Phys. Rev.}\ }\textbf {\bibinfo {volume} {115}},\ \bibinfo {pages}
  {809} (\bibinfo {year} {1959})}\BibitemShut {NoStop}%
\bibitem [{\citenamefont {Marzari}\ \emph {et~al.}(2012)\citenamefont
  {Marzari}, \citenamefont {Mostofi}, \citenamefont {Yates}, \citenamefont
  {Souza},\ and\ \citenamefont {Vanderbilt}}]{marzari_etal_rmp_12}%
  \BibitemOpen
  \bibfield  {author} {\bibinfo {author} {\bibfnamefont {N.}~\bibnamefont
  {Marzari}}, \bibinfo {author} {\bibfnamefont {A.~A.}\ \bibnamefont
  {Mostofi}}, \bibinfo {author} {\bibfnamefont {J.~R.}\ \bibnamefont {Yates}},
  \bibinfo {author} {\bibfnamefont {I.}~\bibnamefont {Souza}},\ and\ \bibinfo
  {author} {\bibfnamefont {D.}~\bibnamefont {Vanderbilt}},\ }\bibfield  {title}
  {\bibinfo {title} {Maximally localized wannier functions: Theory and
  applications},\ }\href {https://doi.org/10.1103/RevModPhys.84.1419}
  {\bibfield  {journal} {\bibinfo  {journal} {Rev. Mod. Phys.}\ }\textbf
  {\bibinfo {volume} {84}},\ \bibinfo {pages} {1419} (\bibinfo {year}
  {2012})}\BibitemShut {NoStop}%
\bibitem [{\citenamefont {Bradlyn}\ \emph {et~al.}(2017)\citenamefont
  {Bradlyn}, \citenamefont {Elcoro}, \citenamefont {Cano}, \citenamefont
  {Vergniory}, \citenamefont {Wang}, \citenamefont {Felser}, \citenamefont
  {Aroyo},\ and\ \citenamefont {Bernevig}}]{bradlyn_etal_nature_17}%
  \BibitemOpen
  \bibfield  {author} {\bibinfo {author} {\bibfnamefont {B.}~\bibnamefont
  {Bradlyn}}, \bibinfo {author} {\bibfnamefont {L.}~\bibnamefont {Elcoro}},
  \bibinfo {author} {\bibfnamefont {J.}~\bibnamefont {Cano}}, \bibinfo {author}
  {\bibfnamefont {M.~G.}\ \bibnamefont {Vergniory}}, \bibinfo {author}
  {\bibfnamefont {Z.}~\bibnamefont {Wang}}, \bibinfo {author} {\bibfnamefont
  {C.}~\bibnamefont {Felser}}, \bibinfo {author} {\bibfnamefont {M.~I.}\
  \bibnamefont {Aroyo}},\ and\ \bibinfo {author} {\bibfnamefont {B.~A.}\
  \bibnamefont {Bernevig}},\ }\bibfield  {title} {\bibinfo {title} {Topological
  quantum chemistry},\ }\href {https://doi.org/10.1038/nature23268} {\bibfield
  {journal} {\bibinfo  {journal} {Nature}\ }\textbf {\bibinfo {volume} {547}},\
  \bibinfo {pages} {298–305} (\bibinfo {year} {2017})}\BibitemShut {NoStop}%
\bibitem [{\citenamefont {Po}\ \emph {et~al.}(2017)\citenamefont {Po},
  \citenamefont {Vishwanath},\ and\ \citenamefont
  {Watanabe}}]{po_etal_natcom_17}%
  \BibitemOpen
  \bibfield  {author} {\bibinfo {author} {\bibfnamefont {H.~C.}\ \bibnamefont
  {Po}}, \bibinfo {author} {\bibfnamefont {A.}~\bibnamefont {Vishwanath}},\
  and\ \bibinfo {author} {\bibfnamefont {H.}~\bibnamefont {Watanabe}},\
  }\bibfield  {title} {\bibinfo {title} {Symmetry-based indicators of band
  topology in the 230 space groups},\ }\href
  {https://doi.org/10.1038/s41467-017-00133-2} {\bibfield  {journal} {\bibinfo
  {journal} {Nature Communications}\ }\textbf {\bibinfo {volume} {8}},\
  \bibinfo {pages} {50} (\bibinfo {year} {2017})}\BibitemShut {NoStop}%
\bibitem [{\citenamefont {Po}\ \emph {et~al.}(2018)\citenamefont {Po},
  \citenamefont {Watanabe},\ and\ \citenamefont {Vishwanath}}]{po_etal_prl_18}%
  \BibitemOpen
  \bibfield  {author} {\bibinfo {author} {\bibfnamefont {H.~C.}\ \bibnamefont
  {Po}}, \bibinfo {author} {\bibfnamefont {H.}~\bibnamefont {Watanabe}},\ and\
  \bibinfo {author} {\bibfnamefont {A.}~\bibnamefont {Vishwanath}},\ }\bibfield
   {title} {\bibinfo {title} {Fragile topology and wannier obstructions},\
  }\href {https://doi.org/10.1103/PhysRevLett.121.126402} {\bibfield  {journal}
  {\bibinfo  {journal} {Phys. Rev. Lett.}\ }\textbf {\bibinfo {volume} {121}},\
  \bibinfo {pages} {126402} (\bibinfo {year} {2018})}\BibitemShut {NoStop}%
\bibitem [{\citenamefont {Resta}(1994)}]{resta_rmp_94}%
  \BibitemOpen
  \bibfield  {author} {\bibinfo {author} {\bibfnamefont {R.}~\bibnamefont
  {Resta}},\ }\bibfield  {title} {\bibinfo {title} {Macroscopic polarization in
  crystalline dielectrics: the geometric phase approach},\ }\href
  {https://doi.org/10.1103/RevModPhys.66.899} {\bibfield  {journal} {\bibinfo
  {journal} {Rev. Mod. Phys.}\ }\textbf {\bibinfo {volume} {66}},\ \bibinfo
  {pages} {899} (\bibinfo {year} {1994})}\BibitemShut {NoStop}%
\bibitem [{\citenamefont {Sgiarovello}\ \emph {et~al.}(2001)\citenamefont
  {Sgiarovello}, \citenamefont {Peressi},\ and\ \citenamefont
  {Resta}}]{sgiarovello_etal_prb_01}%
  \BibitemOpen
  \bibfield  {author} {\bibinfo {author} {\bibfnamefont {C.}~\bibnamefont
  {Sgiarovello}}, \bibinfo {author} {\bibfnamefont {M.}~\bibnamefont
  {Peressi}},\ and\ \bibinfo {author} {\bibfnamefont {R.}~\bibnamefont
  {Resta}},\ }\bibfield  {title} {\bibinfo {title} {Electron localization in
  the insulating state: Application to crystalline semiconductors},\ }\href
  {https://doi.org/10.1103/PhysRevB.64.115202} {\bibfield  {journal} {\bibinfo
  {journal} {Phys. Rev. B}\ }\textbf {\bibinfo {volume} {64}},\ \bibinfo
  {pages} {115202} (\bibinfo {year} {2001})}\BibitemShut {NoStop}%
\bibitem [{\citenamefont {Vanderbilt}(2018)}]{vanderbilt_book_18}%
  \BibitemOpen
  \bibfield  {author} {\bibinfo {author} {\bibfnamefont {D.}~\bibnamefont
  {Vanderbilt}},\ }\href {https://doi.org/110715765X} {\emph {\bibinfo {title}
  {{Berry phases in electronic structure theory: electric polarization, orbital
  magnetization and topological insulators}}}}\ (\bibinfo  {publisher}
  {Cambridge University Press},\ \bibinfo {address} {Cambridge},\ \bibinfo
  {year} {2018})\BibitemShut {NoStop}%
\bibitem [{\citenamefont {Marzari}\ and\ \citenamefont
  {Vanderbilt}(1997)}]{marzari_vanderbilt_prb_97}%
  \BibitemOpen
  \bibfield  {author} {\bibinfo {author} {\bibfnamefont {N.}~\bibnamefont
  {Marzari}}\ and\ \bibinfo {author} {\bibfnamefont {D.}~\bibnamefont
  {Vanderbilt}},\ }\bibfield  {title} {\bibinfo {title} {Maximally localized
  generalized wannier functions for composite energy bands},\ }\href
  {https://doi.org/10.1103/PhysRevB.56.12847} {\bibfield  {journal} {\bibinfo
  {journal} {Phys. Rev. B}\ }\textbf {\bibinfo {volume} {56}},\ \bibinfo
  {pages} {12847} (\bibinfo {year} {1997})}\BibitemShut {NoStop}%
\bibitem [{\citenamefont {Souza}\ \emph {et~al.}(2000)\citenamefont {Souza},
  \citenamefont {Wilkens},\ and\ \citenamefont {Martin}}]{souza_etal_prb_00}%
  \BibitemOpen
  \bibfield  {author} {\bibinfo {author} {\bibfnamefont {I.}~\bibnamefont
  {Souza}}, \bibinfo {author} {\bibfnamefont {T.}~\bibnamefont {Wilkens}},\
  and\ \bibinfo {author} {\bibfnamefont {R.~M.}\ \bibnamefont {Martin}},\
  }\bibfield  {title} {\bibinfo {title} {Polarization and localization in
  insulators: Generating function approach},\ }\href
  {https://doi.org/10.1103/PhysRevB.62.1666} {\bibfield  {journal} {\bibinfo
  {journal} {Phys. Rev. B}\ }\textbf {\bibinfo {volume} {62}},\ \bibinfo
  {pages} {1666} (\bibinfo {year} {2000})}\BibitemShut {NoStop}%
\bibitem [{\citenamefont {Park}\ \emph {et~al.}(2016)\citenamefont {Park},
  \citenamefont {Yang}, \citenamefont {Klinovaja}, \citenamefont {Stano},\ and\
  \citenamefont {Loss}}]{park_etal_prb_16}%
  \BibitemOpen
  \bibfield  {author} {\bibinfo {author} {\bibfnamefont {J.-H.}\ \bibnamefont
  {Park}}, \bibinfo {author} {\bibfnamefont {G.}~\bibnamefont {Yang}}, \bibinfo
  {author} {\bibfnamefont {J.}~\bibnamefont {Klinovaja}}, \bibinfo {author}
  {\bibfnamefont {P.}~\bibnamefont {Stano}},\ and\ \bibinfo {author}
  {\bibfnamefont {D.}~\bibnamefont {Loss}},\ }\bibfield  {title} {\bibinfo
  {title} {Fractional boundary charges in quantum dot arrays with density
  modulation},\ }\href {https://doi.org/10.1103/PhysRevB.94.075416} {\bibfield
  {journal} {\bibinfo  {journal} {Phys. Rev. B}\ }\textbf {\bibinfo {volume}
  {94}},\ \bibinfo {pages} {075416} (\bibinfo {year} {2016})}\BibitemShut
  {NoStop}%
\bibitem [{\citenamefont {Thakurathi}\ \emph {et~al.}(2018)\citenamefont
  {Thakurathi}, \citenamefont {Klinovaja},\ and\ \citenamefont
  {Loss}}]{thakurathi_etal_prb_18}%
  \BibitemOpen
  \bibfield  {author} {\bibinfo {author} {\bibfnamefont {M.}~\bibnamefont
  {Thakurathi}}, \bibinfo {author} {\bibfnamefont {J.}~\bibnamefont
  {Klinovaja}},\ and\ \bibinfo {author} {\bibfnamefont {D.}~\bibnamefont
  {Loss}},\ }\bibfield  {title} {\bibinfo {title} {From fractional boundary
  charges to quantized hall conductance},\ }\href
  {https://doi.org/10.1103/PhysRevB.98.245404} {\bibfield  {journal} {\bibinfo
  {journal} {Phys. Rev. B}\ }\textbf {\bibinfo {volume} {98}},\ \bibinfo
  {pages} {245404} (\bibinfo {year} {2018})}\BibitemShut {NoStop}%
\bibitem [{\citenamefont {Pletyukhov}\ \emph
  {et~al.}(2020{\natexlab{a}})\citenamefont {Pletyukhov}, \citenamefont
  {Kennes}, \citenamefont {Klinovaja}, \citenamefont {Loss},\ and\
  \citenamefont {Schoeller}}]{pletyukhov_etal_prbr_20}%
  \BibitemOpen
  \bibfield  {author} {\bibinfo {author} {\bibfnamefont {M.}~\bibnamefont
  {Pletyukhov}}, \bibinfo {author} {\bibfnamefont {D.~M.}\ \bibnamefont
  {Kennes}}, \bibinfo {author} {\bibfnamefont {J.}~\bibnamefont {Klinovaja}},
  \bibinfo {author} {\bibfnamefont {D.}~\bibnamefont {Loss}},\ and\ \bibinfo
  {author} {\bibfnamefont {H.}~\bibnamefont {Schoeller}},\ }\bibfield  {title}
  {\bibinfo {title} {Topological invariants to characterize universality of
  boundary charge in one-dimensional insulators beyond symmetry constraints},\
  }\href {https://doi.org/10.1103/PhysRevB.101.161106} {\bibfield  {journal}
  {\bibinfo  {journal} {Phys. Rev. B}\ }\textbf {\bibinfo {volume} {101}},\
  \bibinfo {pages} {161106(R)} (\bibinfo {year}
  {2020}{\natexlab{a}})}\BibitemShut {NoStop}%
\bibitem [{\citenamefont {Pletyukhov}\ \emph
  {et~al.}(2020{\natexlab{b}})\citenamefont {Pletyukhov}, \citenamefont
  {Kennes}, \citenamefont {Klinovaja}, \citenamefont {Loss},\ and\
  \citenamefont {Schoeller}}]{pletyukhov_etal_prb_20}%
  \BibitemOpen
  \bibfield  {author} {\bibinfo {author} {\bibfnamefont {M.}~\bibnamefont
  {Pletyukhov}}, \bibinfo {author} {\bibfnamefont {D.~M.}\ \bibnamefont
  {Kennes}}, \bibinfo {author} {\bibfnamefont {J.}~\bibnamefont {Klinovaja}},
  \bibinfo {author} {\bibfnamefont {D.}~\bibnamefont {Loss}},\ and\ \bibinfo
  {author} {\bibfnamefont {H.}~\bibnamefont {Schoeller}},\ }\bibfield  {title}
  {\bibinfo {title} {Surface charge theorem and topological constraints for
  edge states: Analytical study of one-dimensional nearest-neighbor
  tight-binding models},\ }\href {https://doi.org/10.1103/PhysRevB.101.165304}
  {\bibfield  {journal} {\bibinfo  {journal} {Phys. Rev. B}\ }\textbf {\bibinfo
  {volume} {101}},\ \bibinfo {pages} {165304} (\bibinfo {year}
  {2020}{\natexlab{b}})}\BibitemShut {NoStop}%
\bibitem [{\citenamefont {Lin}\ \emph {et~al.}(2020)\citenamefont {Lin},
  \citenamefont {Kennes}, \citenamefont {Pletyukhov}, \citenamefont {Weber},
  \citenamefont {Schoeller},\ and\ \citenamefont {Meden}}]{lin_etal_prb_20}%
  \BibitemOpen
  \bibfield  {author} {\bibinfo {author} {\bibfnamefont {Y.-T.}\ \bibnamefont
  {Lin}}, \bibinfo {author} {\bibfnamefont {D.~M.}\ \bibnamefont {Kennes}},
  \bibinfo {author} {\bibfnamefont {M.}~\bibnamefont {Pletyukhov}}, \bibinfo
  {author} {\bibfnamefont {C.~S.}\ \bibnamefont {Weber}}, \bibinfo {author}
  {\bibfnamefont {H.}~\bibnamefont {Schoeller}},\ and\ \bibinfo {author}
  {\bibfnamefont {V.}~\bibnamefont {Meden}},\ }\bibfield  {title} {\bibinfo
  {title} {Interacting rice-mele model: Bulk and boundaries},\ }\href@noop {}
  {\bibfield  {journal} {\bibinfo  {journal} {Phys. Rev. B}\ }\textbf {\bibinfo
  {volume} {102}},\ \bibinfo {pages} {085122} (\bibinfo {year}
  {2020})}\BibitemShut {NoStop}%
\bibitem [{\citenamefont {Pletyukhov}\ \emph
  {et~al.}(2020{\natexlab{c}})\citenamefont {Pletyukhov}, \citenamefont
  {Kennes}, \citenamefont {Piasotski}, \citenamefont {Klinovaja}, \citenamefont
  {Loss},\ and\ \citenamefont {Schoeller}}]{pletyukhov_etal_prr_20}%
  \BibitemOpen
  \bibfield  {author} {\bibinfo {author} {\bibfnamefont {M.}~\bibnamefont
  {Pletyukhov}}, \bibinfo {author} {\bibfnamefont {D.~M.}\ \bibnamefont
  {Kennes}}, \bibinfo {author} {\bibfnamefont {K.}~\bibnamefont {Piasotski}},
  \bibinfo {author} {\bibfnamefont {J.}~\bibnamefont {Klinovaja}}, \bibinfo
  {author} {\bibfnamefont {D.}~\bibnamefont {Loss}},\ and\ \bibinfo {author}
  {\bibfnamefont {H.}~\bibnamefont {Schoeller}},\ }\bibfield  {title} {\bibinfo
  {title} {Rational boundary charge in one-dimensional systems with interaction
  and disorder},\ }\href@noop {} {\bibfield  {journal} {\bibinfo  {journal}
  {Phys. Rev. Research}\ }\textbf {\bibinfo {volume} {2}},\ \bibinfo {pages}
  {033345} (\bibinfo {year} {2020}{\natexlab{c}})}\BibitemShut {NoStop}%
\bibitem [{\citenamefont {Weber}\ \emph {et~al.}(2021)\citenamefont {Weber},
  \citenamefont {Piasotski}, \citenamefont {Pletyukhov}, \citenamefont
  {Klinovaja}, \citenamefont {Loss}, \citenamefont {Schoeller},\ and\
  \citenamefont {Kennes}}]{weber_etal_prl_20}%
  \BibitemOpen
  \bibfield  {author} {\bibinfo {author} {\bibfnamefont {C.~S.}\ \bibnamefont
  {Weber}}, \bibinfo {author} {\bibfnamefont {K.}~\bibnamefont {Piasotski}},
  \bibinfo {author} {\bibfnamefont {M.}~\bibnamefont {Pletyukhov}}, \bibinfo
  {author} {\bibfnamefont {J.}~\bibnamefont {Klinovaja}}, \bibinfo {author}
  {\bibfnamefont {D.}~\bibnamefont {Loss}}, \bibinfo {author} {\bibfnamefont
  {H.}~\bibnamefont {Schoeller}},\ and\ \bibinfo {author} {\bibfnamefont
  {D.~M.}\ \bibnamefont {Kennes}},\ }\bibfield  {title} {\bibinfo {title}
  {Universality of boundary charge fluctuations},\ }\href
  {https://doi.org/10.1103/PhysRevLett.126.016803} {\bibfield  {journal}
  {\bibinfo  {journal} {Phys. Rev. Lett.}\ }\textbf {\bibinfo {volume} {126}},\
  \bibinfo {pages} {016803} (\bibinfo {year} {2021})}\BibitemShut {NoStop}%
\bibitem [{\citenamefont {Vanderbilt}\ and\ \citenamefont
  {King-Smith}(1993)}]{vanderbilt_kingsmith_prb_93}%
  \BibitemOpen
  \bibfield  {author} {\bibinfo {author} {\bibfnamefont {D.}~\bibnamefont
  {Vanderbilt}}\ and\ \bibinfo {author} {\bibfnamefont {R.~D.}\ \bibnamefont
  {King-Smith}},\ }\bibfield  {title} {\bibinfo {title} {Electric polarization
  as a bulk quantity and its relation to surface charge},\ }\href
  {https://doi.org/10.1103/PhysRevB.48.4442} {\bibfield  {journal} {\bibinfo
  {journal} {Phys. Rev. B}\ }\textbf {\bibinfo {volume} {48}},\ \bibinfo
  {pages} {4442} (\bibinfo {year} {1993})}\BibitemShut {NoStop}%
\bibitem [{\citenamefont {Harper}(1955)}]{harper_pps_55}%
  \BibitemOpen
  \bibfield  {author} {\bibinfo {author} {\bibfnamefont {P.~G.}\ \bibnamefont
  {Harper}},\ }\bibfield  {title} {\bibinfo {title} {Single band motion of
  conduction electrons in a uniform magnetic field},\ }\href@noop {} {\bibfield
   {journal} {\bibinfo  {journal} {Proceedings of the Physical Society. Section
  A}\ }\textbf {\bibinfo {volume} {68}},\ \bibinfo {pages} {874} (\bibinfo
  {year} {1955})}\BibitemShut {NoStop}%
\bibitem [{\citenamefont {Aubry}\ and\ \citenamefont
  {André}(1980)}]{aubry_andre_aips_80}%
  \BibitemOpen
  \bibfield  {author} {\bibinfo {author} {\bibfnamefont {S.}~\bibnamefont
  {Aubry}}\ and\ \bibinfo {author} {\bibfnamefont {G.}~\bibnamefont {André}},\
  }\bibfield  {title} {\bibinfo {title} {Analyticity breaking and anderson
  localization in incommensurate lattices},\ }\href@noop {} {\bibfield
  {journal} {\bibinfo  {journal} {Ann. Isr. Phys. Soc.}\ }\textbf {\bibinfo
  {volume} {3}},\ \bibinfo {pages} {133} (\bibinfo {year} {1980})}\BibitemShut
  {NoStop}%
\bibitem [{\citenamefont {Ganeshan}\ \emph {et~al.}(2013)\citenamefont
  {Ganeshan}, \citenamefont {Sun},\ and\ \citenamefont
  {Das~Sarma}}]{ganeshan_etal_prl_13}%
  \BibitemOpen
  \bibfield  {author} {\bibinfo {author} {\bibfnamefont {S.}~\bibnamefont
  {Ganeshan}}, \bibinfo {author} {\bibfnamefont {K.}~\bibnamefont {Sun}},\ and\
  \bibinfo {author} {\bibfnamefont {S.}~\bibnamefont {Das~Sarma}},\ }\bibfield
  {title} {\bibinfo {title} {Topological zero-energy modes in gapless
  commensurate aubry-andr\'e-harper models},\ }\href@noop {} {\bibfield
  {journal} {\bibinfo  {journal} {Phys. Rev. Lett.}\ }\textbf {\bibinfo
  {volume} {110}},\ \bibinfo {pages} {180403} (\bibinfo {year}
  {2013})}\BibitemShut {NoStop}%
\bibitem [{\citenamefont {Lahini}\ \emph {et~al.}(2009)\citenamefont {Lahini},
  \citenamefont {Pugatch}, \citenamefont {Pozzi}, \citenamefont {Sorel},
  \citenamefont {Morandotti}, \citenamefont {Davidson},\ and\ \citenamefont
  {Silberberg}}]{lahini_etal_prl_09}%
  \BibitemOpen
  \bibfield  {author} {\bibinfo {author} {\bibfnamefont {Y.}~\bibnamefont
  {Lahini}}, \bibinfo {author} {\bibfnamefont {R.}~\bibnamefont {Pugatch}},
  \bibinfo {author} {\bibfnamefont {F.}~\bibnamefont {Pozzi}}, \bibinfo
  {author} {\bibfnamefont {M.}~\bibnamefont {Sorel}}, \bibinfo {author}
  {\bibfnamefont {R.}~\bibnamefont {Morandotti}}, \bibinfo {author}
  {\bibfnamefont {N.}~\bibnamefont {Davidson}},\ and\ \bibinfo {author}
  {\bibfnamefont {Y.}~\bibnamefont {Silberberg}},\ }\bibfield  {title}
  {\bibinfo {title} {Observation of a localization transition in quasiperiodic
  photonic lattices},\ }\href@noop {} {\bibfield  {journal} {\bibinfo
  {journal} {Phys. Rev. Lett.}\ }\textbf {\bibinfo {volume} {103}},\ \bibinfo
  {pages} {013901} (\bibinfo {year} {2009})}\BibitemShut {NoStop}%
\bibitem [{\citenamefont {Kraus}\ \emph {et~al.}(2012)\citenamefont {Kraus},
  \citenamefont {Lahini}, \citenamefont {Ringel}, \citenamefont {Verbin},\ and\
  \citenamefont {Zilberberg}}]{kraus_etal_prl_12}%
  \BibitemOpen
  \bibfield  {author} {\bibinfo {author} {\bibfnamefont {Y.~E.}\ \bibnamefont
  {Kraus}}, \bibinfo {author} {\bibfnamefont {Y.}~\bibnamefont {Lahini}},
  \bibinfo {author} {\bibfnamefont {Z.}~\bibnamefont {Ringel}}, \bibinfo
  {author} {\bibfnamefont {M.}~\bibnamefont {Verbin}},\ and\ \bibinfo {author}
  {\bibfnamefont {O.}~\bibnamefont {Zilberberg}},\ }\bibfield  {title}
  {\bibinfo {title} {Topological states and adiabatic pumping in
  quasicrystals},\ }\href@noop {} {\bibfield  {journal} {\bibinfo  {journal}
  {Phys. Rev. Lett.}\ }\textbf {\bibinfo {volume} {109}},\ \bibinfo {pages}
  {106402} (\bibinfo {year} {2012})}\BibitemShut {NoStop}%
\bibitem [{\citenamefont {DeGottardi}\ \emph {et~al.}(2013)\citenamefont
  {DeGottardi}, \citenamefont {Sen},\ and\ \citenamefont
  {Vishveshwara}}]{degottardi_etal_prl_13}%
  \BibitemOpen
  \bibfield  {author} {\bibinfo {author} {\bibfnamefont {W.}~\bibnamefont
  {DeGottardi}}, \bibinfo {author} {\bibfnamefont {D.}~\bibnamefont {Sen}},\
  and\ \bibinfo {author} {\bibfnamefont {S.}~\bibnamefont {Vishveshwara}},\
  }\bibfield  {title} {\bibinfo {title} {Majorana fermions in superconducting
  1d systems having periodic, quasiperiodic, and disordered potentials},\
  }\href@noop {} {\bibfield  {journal} {\bibinfo  {journal} {Phys. Rev. Lett.}\
  }\textbf {\bibinfo {volume} {110}},\ \bibinfo {pages} {146404} (\bibinfo
  {year} {2013})}\BibitemShut {NoStop}%
\bibitem [{\citenamefont {Schreiber}\ \emph {et~al.}(2015)\citenamefont
  {Schreiber}, \citenamefont {Hodgman}, \citenamefont {Bordia}, \citenamefont
  {L{\"u}schen}, \citenamefont {Fischer}, \citenamefont {Vosk}, \citenamefont
  {Altman}, \citenamefont {Schneider},\ and\ \citenamefont
  {Bloch}}]{schreiber_etal_science_15}%
  \BibitemOpen
  \bibfield  {author} {\bibinfo {author} {\bibfnamefont {M.}~\bibnamefont
  {Schreiber}}, \bibinfo {author} {\bibfnamefont {S.~S.}\ \bibnamefont
  {Hodgman}}, \bibinfo {author} {\bibfnamefont {P.}~\bibnamefont {Bordia}},
  \bibinfo {author} {\bibfnamefont {H.~P.}\ \bibnamefont {L{\"u}schen}},
  \bibinfo {author} {\bibfnamefont {M.~H.}\ \bibnamefont {Fischer}}, \bibinfo
  {author} {\bibfnamefont {R.}~\bibnamefont {Vosk}}, \bibinfo {author}
  {\bibfnamefont {E.}~\bibnamefont {Altman}}, \bibinfo {author} {\bibfnamefont
  {U.}~\bibnamefont {Schneider}},\ and\ \bibinfo {author} {\bibfnamefont
  {I.}~\bibnamefont {Bloch}},\ }\bibfield  {title} {\bibinfo {title}
  {Observation of many-body localization of interacting fermions in a
  quasirandom optical lattice},\ }\href
  {https://doi.org/10.1126/science.aaa7432} {\bibfield  {journal} {\bibinfo
  {journal} {Science}\ }\textbf {\bibinfo {volume} {349}},\ \bibinfo {pages}
  {842} (\bibinfo {year} {2015})}\BibitemShut {NoStop}%
\bibitem [{\citenamefont {Gangadharaiah}\ \emph {et~al.}(2012)\citenamefont
  {Gangadharaiah}, \citenamefont {Trifunovic},\ and\ \citenamefont
  {Loss}}]{gangadharaiah_etal_prl_12}%
  \BibitemOpen
  \bibfield  {author} {\bibinfo {author} {\bibfnamefont {S.}~\bibnamefont
  {Gangadharaiah}}, \bibinfo {author} {\bibfnamefont {L.}~\bibnamefont
  {Trifunovic}},\ and\ \bibinfo {author} {\bibfnamefont {D.}~\bibnamefont
  {Loss}},\ }\bibfield  {title} {\bibinfo {title} {Localized end states in
  density modulated quantum wires and rings},\ }\href
  {https://doi.org/10.1103/PhysRevLett.108.136803} {\bibfield  {journal}
  {\bibinfo  {journal} {Phys. Rev. Lett.}\ }\textbf {\bibinfo {volume} {108}},\
  \bibinfo {pages} {136803} (\bibinfo {year} {2012})}\BibitemShut {NoStop}%
\bibitem [{\citenamefont {Rice}\ and\ \citenamefont
  {Mele}(1982)}]{rice_mele_prl_82}%
  \BibitemOpen
  \bibfield  {author} {\bibinfo {author} {\bibfnamefont {M.~J.}\ \bibnamefont
  {Rice}}\ and\ \bibinfo {author} {\bibfnamefont {E.~J.}\ \bibnamefont
  {Mele}},\ }\bibfield  {title} {\bibinfo {title} {Elementary excitations of a
  linearly conjugated diatomic polymer},\ }\href
  {https://doi.org/10.1103/PhysRevLett.49.1455} {\bibfield  {journal} {\bibinfo
   {journal} {Phys. Rev. Lett.}\ }\textbf {\bibinfo {volume} {49}},\ \bibinfo
  {pages} {1455} (\bibinfo {year} {1982})}\BibitemShut {NoStop}%
\bibitem [{\citenamefont {Resta}\ and\ \citenamefont
  {Sorella}(1999)}]{resta_sorella_prl_99}%
  \BibitemOpen
  \bibfield  {author} {\bibinfo {author} {\bibfnamefont {R.}~\bibnamefont
  {Resta}}\ and\ \bibinfo {author} {\bibfnamefont {S.}~\bibnamefont
  {Sorella}},\ }\bibfield  {title} {\bibinfo {title} {Electron localization in
  the insulating state},\ }\href {https://doi.org/10.1103/PhysRevLett.82.370}
  {\bibfield  {journal} {\bibinfo  {journal} {Phys. Rev. Lett.}\ }\textbf
  {\bibinfo {volume} {82}},\ \bibinfo {pages} {370} (\bibinfo {year}
  {1999})}\BibitemShut {NoStop}%
\bibitem [{\citenamefont {He}\ and\ \citenamefont
  {Vanderbilt}(2001)}]{he_vanderbilt_prl_01}%
  \BibitemOpen
  \bibfield  {author} {\bibinfo {author} {\bibfnamefont {L.}~\bibnamefont
  {He}}\ and\ \bibinfo {author} {\bibfnamefont {D.}~\bibnamefont
  {Vanderbilt}},\ }\bibfield  {title} {\bibinfo {title} {Exponential decay
  properties of wannier functions and related quantities},\ }\href
  {https://doi.org/10.1103/PhysRevLett.86.5341} {\bibfield  {journal} {\bibinfo
   {journal} {Phys. Rev. Lett.}\ }\textbf {\bibinfo {volume} {86}},\ \bibinfo
  {pages} {5341} (\bibinfo {year} {2001})}\BibitemShut {NoStop}%
\bibitem [{\citenamefont {Heeger}(2001)}]{heeger_rmp_01}%
  \BibitemOpen
  \bibfield  {author} {\bibinfo {author} {\bibfnamefont {A.~J.}\ \bibnamefont
  {Heeger}},\ }\bibfield  {title} {\bibinfo {title} {Nobel lecture:
  Semiconducting and metallic polymers: The fourth generation of polymeric
  materials},\ }\href {https://doi.org/10.1103/RevModPhys.73.681} {\bibfield
  {journal} {\bibinfo  {journal} {Rev. Mod. Phys.}\ }\textbf {\bibinfo {volume}
  {73}},\ \bibinfo {pages} {681} (\bibinfo {year} {2001})}\BibitemShut
  {NoStop}%
\bibitem [{\citenamefont {Su}\ \emph {et~al.}(1979)\citenamefont {Su},
  \citenamefont {Schrieffer},\ and\ \citenamefont {Heeger}}]{su_etal_prl_79}%
  \BibitemOpen
  \bibfield  {author} {\bibinfo {author} {\bibfnamefont {W.~P.}\ \bibnamefont
  {Su}}, \bibinfo {author} {\bibfnamefont {J.~R.}\ \bibnamefont {Schrieffer}},\
  and\ \bibinfo {author} {\bibfnamefont {A.~J.}\ \bibnamefont {Heeger}},\
  }\bibfield  {title} {\bibinfo {title} {Solitons in polyacetylene},\ }\href
  {https://doi.org/10.1103/PhysRevLett.42.1698} {\bibfield  {journal} {\bibinfo
   {journal} {Phys. Rev. Lett.}\ }\textbf {\bibinfo {volume} {42}},\ \bibinfo
  {pages} {1698} (\bibinfo {year} {1979})}\BibitemShut {NoStop}%
\end{thebibliography}%

\end{document}